\documentclass[runningheads]{llncs}
\usepackage{float}
\usepackage{caption}
\usepackage{url}
\usepackage{todonotes}
\usepackage{xspace}
\usepackage{algorithm}
\usepackage{algpseudocode}
\usepackage{amsfonts}
\usepackage{amsmath}
\usepackage{dsfont}
\usepackage{multirow}

\usepackage{tabularx}
\usepackage{placeins}

\newcommand{\tabfig}{.11}
\newcommand{\taboffset}{\vspace{-1cm}}

\usepackage{graphicx}
\graphicspath{{Figures/}}

\usepackage{array}

\newcolumntype{C}[1]{>{\centering\arraybackslash}p{#1}}
\newcolumntype{M}[1]{>{\centering\arraybackslash}m{#1}}

\begin{document}

\title{Balancing between the Local and Global Structures (LGS) in Graph Embedding}
\titlerunning{LGS}
%

\author{Jacob Miller \and Vahan Huroyan \and Stephen Kobourov}

\institute{University of Arizona}

\authorrunning{Miller et al.}

%
\maketitle              

\begin{abstract}

We present a method for balancing between the Local and Global Structures (LGS) in graph embedding, via a tunable parameter. Some embedding methods aim to capture global structures, while others attempt to preserve local neighborhoods. Few methods attempt to do both, and it is not always possible to  capture well both local and global information in two dimensions, which is where most graph drawing live. 
The choice of using a local or a global embedding for visualization depends not only on the task but also on the structure of the underlying data, which may not be known in advance. For a given graph, LGS aims to find a good balance between the local and global structure to preserve. We evaluate the performance of LGS with synthetic and real-world datasets and our results indicate that it is competitive with the state-of-the-art methods, using established quality metrics such as stress and neighborhood preservation. 
We introduce a novel quality metric, cluster distance preservation, to assess intermediate structure capture.
All source-code, datasets, experiments and analysis are available online. 

\end{abstract}

\begin{figure}
\vspace{-.8cm}
  \centering
  \includegraphics[width=\linewidth]{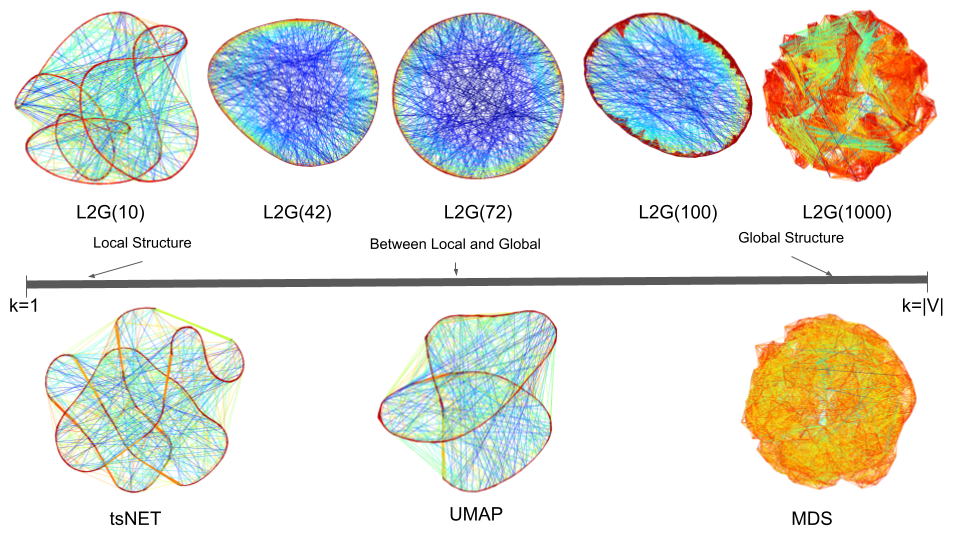}
  \caption{
    Embeddings of the connected\_watts\_1000 graph; see Sec.~\ref{sec:layouts}.
    The top row shows LGS embeddings -- from local to global -- with varying neighborhood sizes ($k$).
    The LGS(72) layout captures the correct underlying model. The bottom row shows tsNET~\cite{DBLP:journals/cgf/KruigerRMKKT17}, UMAP~\cite{mcionnes2018umap}, and MDS~\cite{DBLP:journals/tvcg/ZhengPG19} embedding of the same graph. 
  }
  \label{fig:teaser}
\end{figure}

\vspace{-.8cm}

\section{Introduction}
Graphs and networks are a powerful tool to encode relationships between objects. Graph embeddings, which map the vertices of a graph to a set of low dimensional vectors (real valued coordinates), are often used in the context of data visualization to produce node-link diagrams.
While many layout methods exist~\cite{tamassia2013handbook}, dimension reduction (DR) techniques have had success in providing desirable layouts, by capturing graph structure in reasonable computation times. 
DR methods are used to project high-dimensional data into low-dimensional space and some of these methods only rely on the relationships between the datapoints, rather than {\em datapoint coordinates} in higher dimension. These techniques are applicable for both graph embeddings and visualization.
Further, local DR algorithms attempt to preserve the local neighborhoods, while global DR algorithms attempt to retain all pairwise distances. 

Two popular techniques that are adapted in graph visualization are (metric) Multi-Dimensional Scaling (MDS)~\cite{de2005applications,kruskal1964multidimensional} and t-distributed stochastic neighbor embedding (t-SNE)~\cite{van2008visualizing}. The goals of these two  algorithms are somewhat orthogonal: 
MDS focuses on preserving all pairwise distances, 
while t-SNE aims to preserve the likelihood of points being close in the embedding if they were close in the original space.
MDS is said to preserve {\it global} 
structure, while t-SNE is said to preserve {\it local} neighborhoods~\cite{DBLP:journals/tvcg/EspadotoMKHT21}.
These ideas are directly applicable to graph visualization, where we can define 
the
distances as the graph theoretic distances, e.g., via all-pairs shortest paths (APSP) computation. 
In the graph layout literature, MDS is often referred to as stress minimization~\cite{DBLP:conf/gd/GansnerKN04,DBLP:journals/tvcg/ZhengPG19}, 
and t-SNE has been adapted to graph layout in an algorithm known as tsNET~\cite{DBLP:journals/cgf/KruigerRMKKT17} and later 
DRGraph~\cite{DBLP:journals/tvcg/ZhuCHHLZ21}.
Choosing the ``best'' graph embedding algorithm depends on
the graph structure and the task.
MDS is effective for structured/mesh-like graphs, while t-SNE works better for clustered/dense graphs.
This phenomenon 
also
applies to local and global force-directed layouts as well~\cite{DBLP:journals/cgf/KruigerRMKKT17}.

Automating the selection of the ``best'' embedding algorithm is challenging due to its dependency on graph structure.
We introduce the Local-to-Global Structures (LGS) algorithm which provides a parameter-tuneable framework that can produces embeddings that span the spectrum from local optimization to global optimization.

Smaller values of the LGS parameter prioritize local structure, while larger values emphasize global structure. LGS enables exploration of the trade-off, revealing meaningful middle ground solutions.
We introduce a new metric called {\em cluster distance} to measure how well this intermediate structure is preserved. 
Everything described in this paper is available on Github: \url{https://github.com/Mickey253/L2G}. We provide a video and additional layouts and analysis in supplemental material.


\section{Background}
\noindent
\textbf{Dimensionality Reduction (DR)} 
refers to a large family of algorithms that map a set of high-dimensional datapoints in lower-dimensionsal space.
Different DR algorithms aim to preserve various properties of the dataset, such as total variance, global distances, local distances, etc. In visualization contexts, the dataset is typically projected onto 2D or 3D Euclidean space.
DR algorithms generally accept input of two types: sample or distance.
Sample-based algorithms, such as Principal Component Analysis (PCA)~\cite{frey1978principal,jolliffe1986principal} project the high dimensional data down to the embedding space. 
For distance-based inputs, the algorithms directly work with distance metrics. In the case of graph embeddings, the graph-theoretic distance is used, often all-pairs shortest path (APSP).

Popular techniques in the local category include t-SNE~\cite{van2008visualizing}, 
UMAP~\cite{mcionnes2018umap},
LLE~\cite{roweis2000nonlinear}, 
IsoMap~\cite{tenenbaum2000global}, etc.
For global structure, methods such as PCA~\cite{frey1978principal} and MDS~\cite{de2005applications,kruskal1964multidimensional} are used. MDS has variants, but here we mean metric MDS which minimizes stress~\cite{shepard1962analysis}.
Few techniques attempt to capture both global and local structure. Chen and Buja~\cite{chen2009local} 
adapt MDS to capture local structure by selectively preserving distances between a subset of pairs using kNN.
The underlying idea is similar to 
ours
, but it does not provide a framework to cover the spectrum from local to global as our method does. 
While t-SNE's perplexity parameter aims to imitate the size of neighborhood to be preserved, in general increasing its value does not lead to a global structure preservation~\cite{wattenberg2016how}. 
Anchor-t-SNE improves the global structure preservation by anchoring a set of points to use as a skeleton for the rest of the embedding~\cite{fu2019atsne}, however, it does not provide a framework to cover the spectrum from local to global. 
UMAP~\cite{mcionnes2018umap,ghojogh2021uniform} 
also aims to preserve the local structures of a dataset. 
While UMAP claims to preserve the global structures better than t-SNE, we show that this is not universally true for graph data in Sec.~\ref{sec:experiments}.


\noindent
\textbf{Graph Embedding}
is a problem to assign 
vectors to graph vertices, capturing the graph structure. More formally, given a graph G = (V, E), find a d-dimensional vector representation of V that optimally preserves properties~\cite{DBLP:journals/tkde/CaiZC18} (e.g., pairwise distances in MDS~\cite{de2005applications,kruskal1964multidimensional}).
We restrict ourselves to 2D node-link visualization with edges represented by straight-line segments, so the problem is reduced to finding a 2D embedding for the vertices.
Aesthetic criteria are often used to evaluate the quality of a graph embedding: the number of edge crossings, average edge length, overall symmetry, etc.~\cite{DBLP:journals/vlc/Purchase02}. 
Aesthetic criteria enhance readability and task facilitation, but information \textit{faithfulness} is equally important. It ensures that the embedding accurately represents all underlying data, regardless of the task
~\cite{munzner2014visualization,DBLP:conf/apvis/NguyenEH13} and graph embeddings provide a nice benefit by directly optimizing graph structure preservation.
{\em Graph structure} is a nebulous term; referring to inherent properties of the underlying graph such as local/global distances.
{\em Global distance} preservation methods capture the graph's topological structure by closely aligning embedded distances with graph-theoretic distances. 
This approach is ideal for connectivity-based tasks and offers insights into the global scale and shape of the data.
{\em Local structure} preservation methods preserve the immediate neighborhood of each vertex, effectively capturing clusters or densely connected subgraphs. While nearby vertices in the embedding can 
be considered similar, 
distant vertices may have irrelevant distances.
This can be observed in the presence of long edges in the local embedding column in Fig.~\ref{fig:block-table}.

\paragraph{Graph embedding by dimensionality reduction:}

In a good embedding, the drawn distance should closely match the graph-theoretic distance between vertices~\cite{DBLP:journals/ipl/KamadaK89}.
This observation led to the use of  stress function, which MDS aims to optimize, to obtain a graph embedding~\cite{DBLP:conf/gd/GansnerKN04}.
Stress can be minimized by majorization~\cite{DBLP:conf/gd/GansnerKN04}, stochastic gradient descent (SGD)~\cite{DBLP:journals/tvcg/ZhengPG19}, etc. 
The MDS approach suffers from an APSP computation, which usually relies on Floyd-Warshall's  $O(|V|^3)$, or on Johnson's  $O(|V|^2\log |V| + |E||V|)$ algorithms.
The maximum entropy model (MaxEnt)~\cite{DBLP:journals/tvcg/GansnerHN13} adds a negative entropy between vertices in the graph.
The motivation for MaxEnt is to improve the asymptotic complexity. 
The MaxEnt model places neighbor nodes closer while maximizing the distance between all vertices. This is conceptually similar to the LMDS of Chen and Buja~\cite{chen2009local}.
Our approach differs from the MaxEnt model in motivation: 
Our LGS
captures local structure, global structure, or balances between the two, whereas MaxEnt is primarily concerned with speed. We cannot avoid an APSP computation, and make use of SGD to optimize our objective function in lieu of majorization.


Optimizing stress creates effective layouts, but may neglect local structures;
see Fig.~\ref{fig:block-table}. 
tsNET~\cite{DBLP:journals/cgf/KruigerRMKKT17} 
captures local structure by also adding a repulsive force between vertices to achieve cluster separation.
tsNET has been sped up by making use of negative sampling and sparse approximation to avoid the APSP computation~\cite{DBLP:journals/tvcg/ZhuCHHLZ21}. 
Nocaj et al.~\cite{DBLP:journals/jgaa/NocajOB15} achieve effects similar to tsNET by weighting edges based on ``edge embeddedness''
and perform MDS on the 
weighted graph.

\section{The Local-to-Global Structures (LGS)  Algorithm}
Local 
methods (e.g., t-SNE) preserve local neighborhoods, while global 
methods (e.g., MDS) capture all pair-wise distances.
We propose the Local-to-Global Structures (LGS) algorithm that achieves the following 3 goals:
\begin{itemize}
    \item[\textbf{G1}] 
    A single parameter controlling local-global embedding balance
    \item[\textbf{G2}] When this parameter is small, the embedding preserves local neighborhoods
    \item[\textbf{G3}] When this parameter is large, the embedding preserves the global structure
\end{itemize}
By ``local neighborhood'' of a vertex we refer to the immediate neighbors of the vertex being considered. 
If the nearest neighbors of each vertex in an embedding match well with the nearest neighbors in the actual graph, then the embedding accurately preserves the local structures.
By ``global structure'' we refer to the preservation of all pairwise graph distances (including long ones) in the embedding. 
Finally,  ``intermediate structure'' refers to capturing both local neighbors and global structure.
Fig.~\ref{fig:block-table} shows graphs exemplifying local, intermediate, and global structures and Sec.~\ref{sec:metrics} defines formal embedding measures: neighborhood error, cluster distance, and stress. 
In Sec.~\ref{sec:3.1} we explain the selection process for the balance parameter $k$ and the objective function to ensure that the solution aligns with the stated goals.
For \textbf{G1}, we modify 
MDS to preserve distances in a neighborhood 
defined by a parameter $k$).
Thus, 
preserving distances for large neighborhoods 
satisfies \textbf{G3}. 
This leaves a question for \textbf{G2}: Does applying distance preservation to a subset of pairs result in locally faithful embeddings? 




\begin{table}[t]
  \centering
  \begin{tabular}
      { c | c | c | c | c |}

      \multicolumn{5}{c}{\hspace{1cm}$\Leftarrow$\textbf{embedding}$\Rightarrow$}\\
      
      \cline{2-5} & & tsNET (local) & LGS (balanced) & MDS (global) \\

      \cline{2-5} 
      
       \multirow{3}{*}{\vspace{-3.8cm}\rotatebox[origin=c]{90}{$\Leftarrow$ \textbf{structure} $\Rightarrow$}} 
      & \raisebox{-.5\normalbaselineskip}[0pt][0pt]{\rotatebox[origin=c]{90}{\parbox{2cm}{\centering block\_2000  \\ (local)}}}
      & \parbox[c][2.2cm]{\tabfig\textwidth}{
      \includegraphics[width=\tabfig\textwidth]{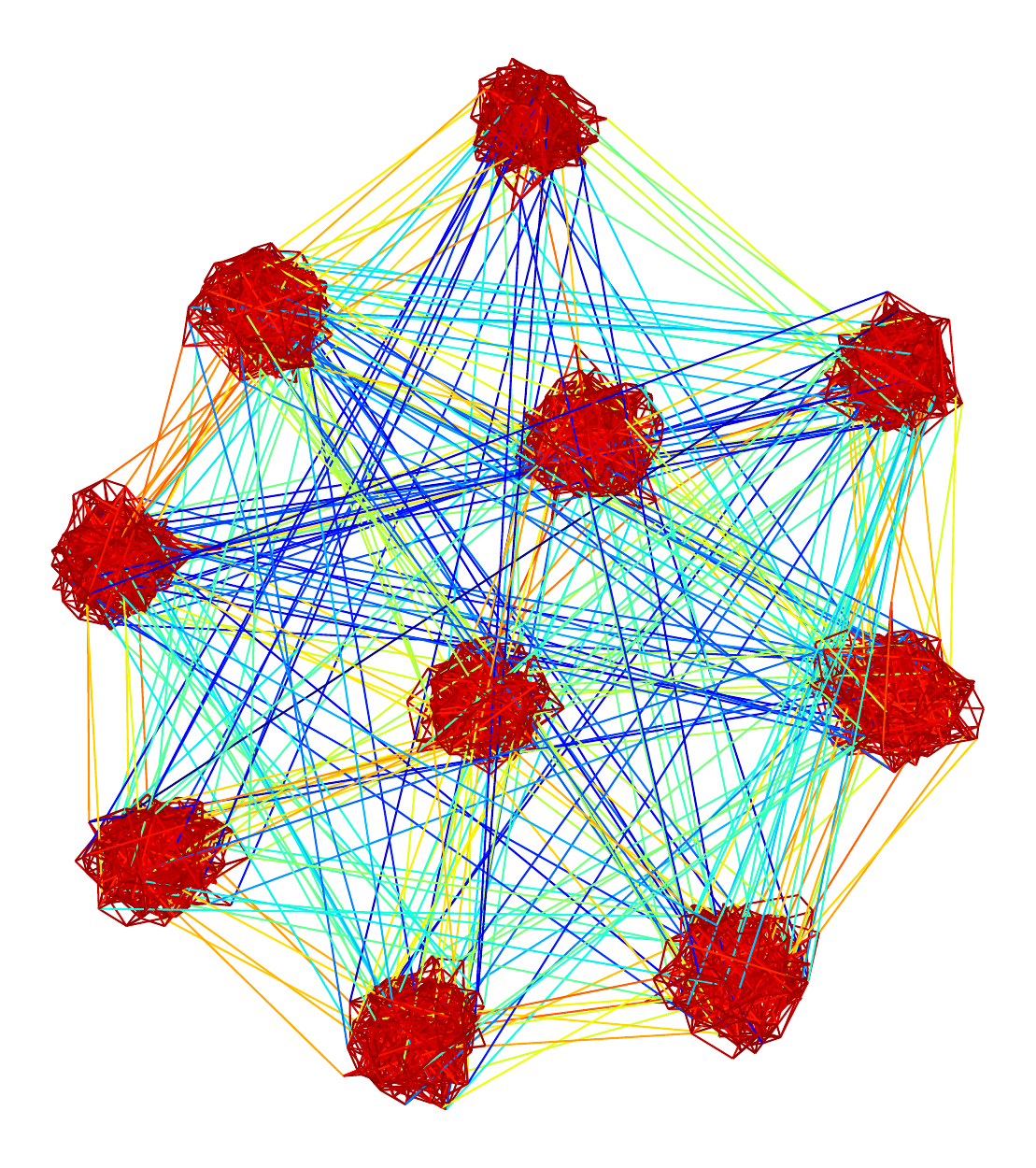}}
      & \parbox[c]{\tabfig\textwidth}{
      \includegraphics[width=\tabfig\textwidth]{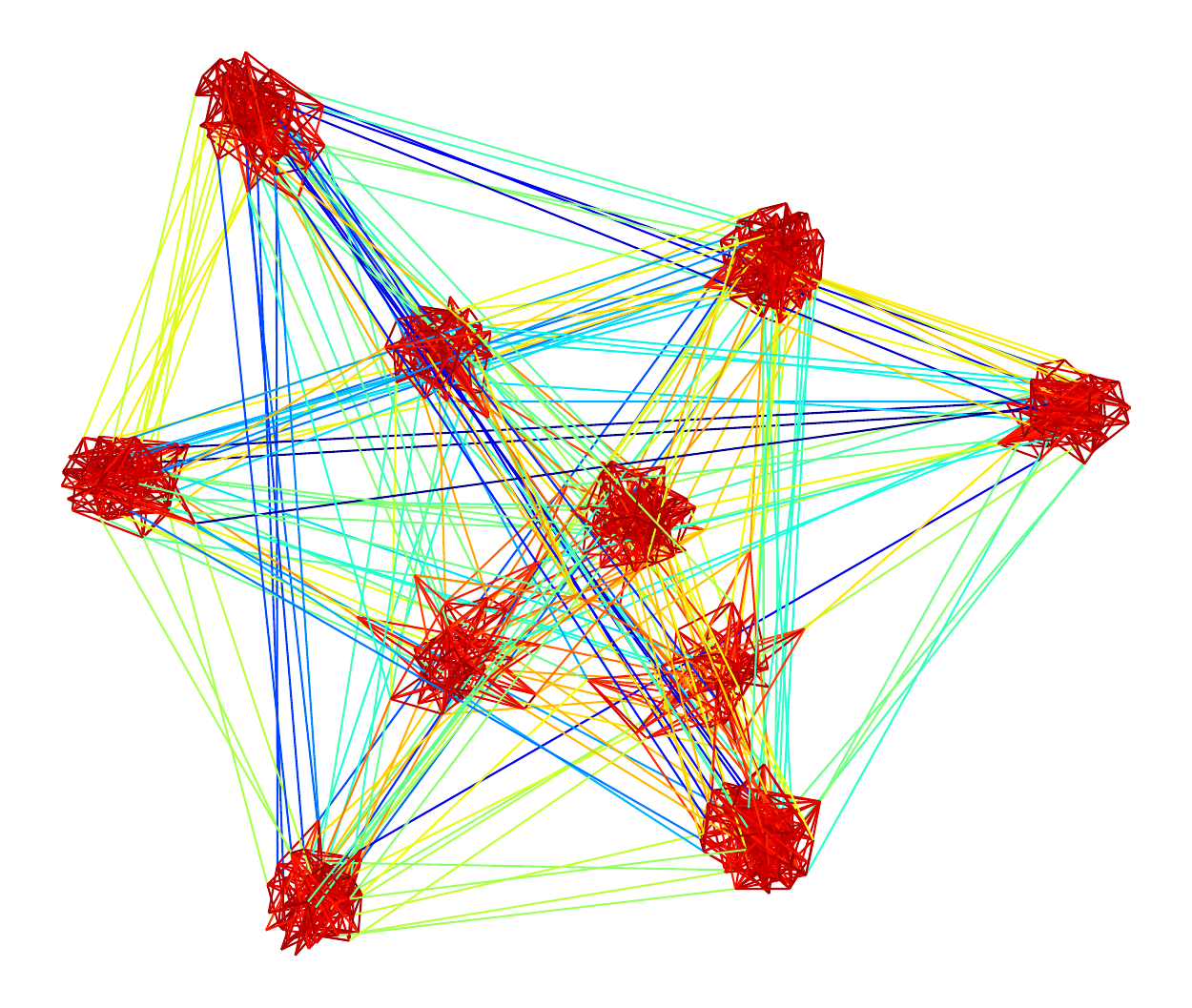}}      
       & \parbox[c]{\tabfig\textwidth}{
      \includegraphics[width=\tabfig\textwidth]{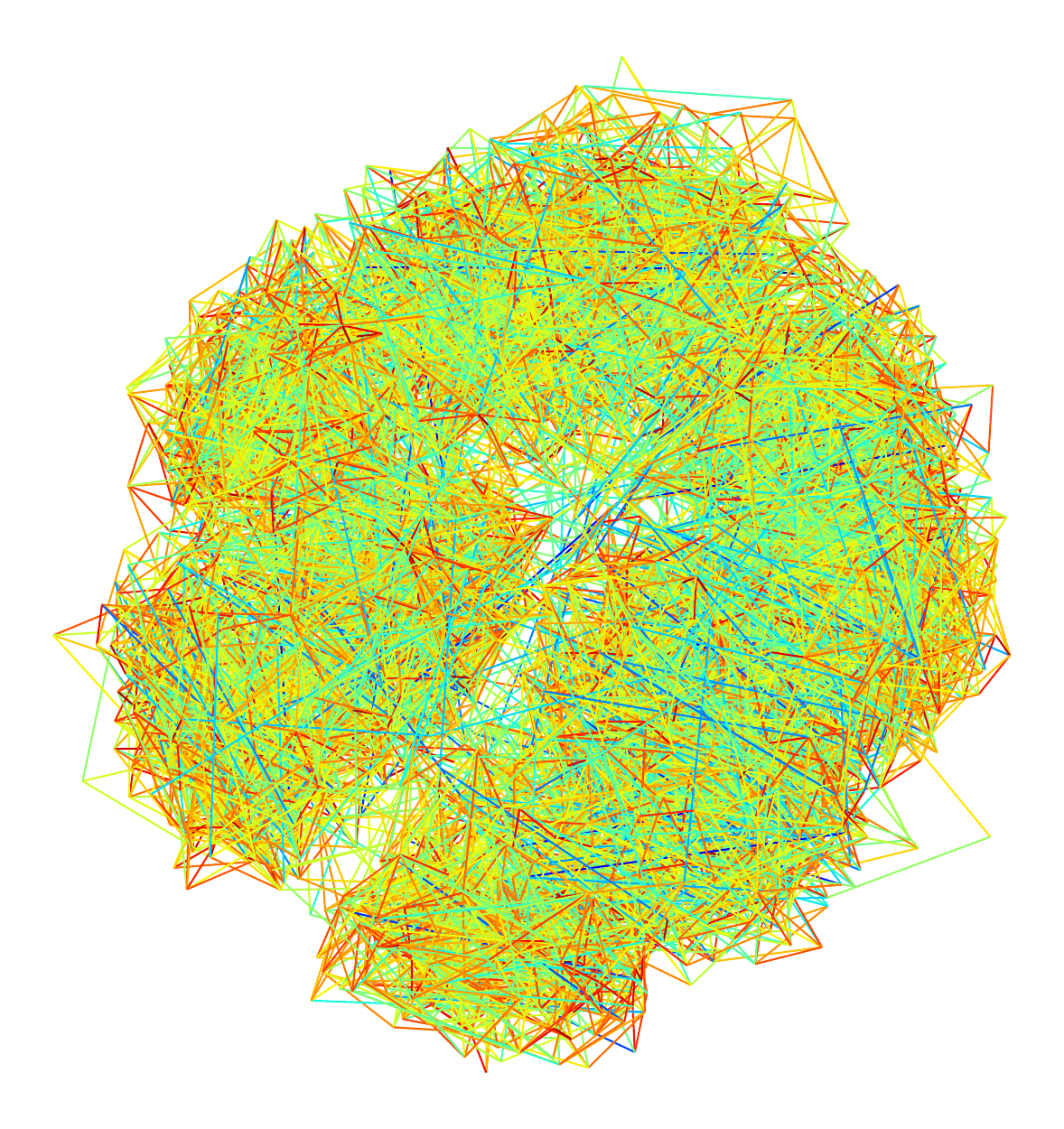}} \\
      \cline{2-5}

      & \raisebox{-.5\normalbaselineskip}[0pt][0pt]{\rotatebox[origin=c]{90}{\parbox{2cm}{\centering sierpinksi\_3d  \\ (intermediate)}}}
      & \parbox[c][2.5cm]{\tabfig\textwidth}{
      \includegraphics[width=\tabfig\textwidth]{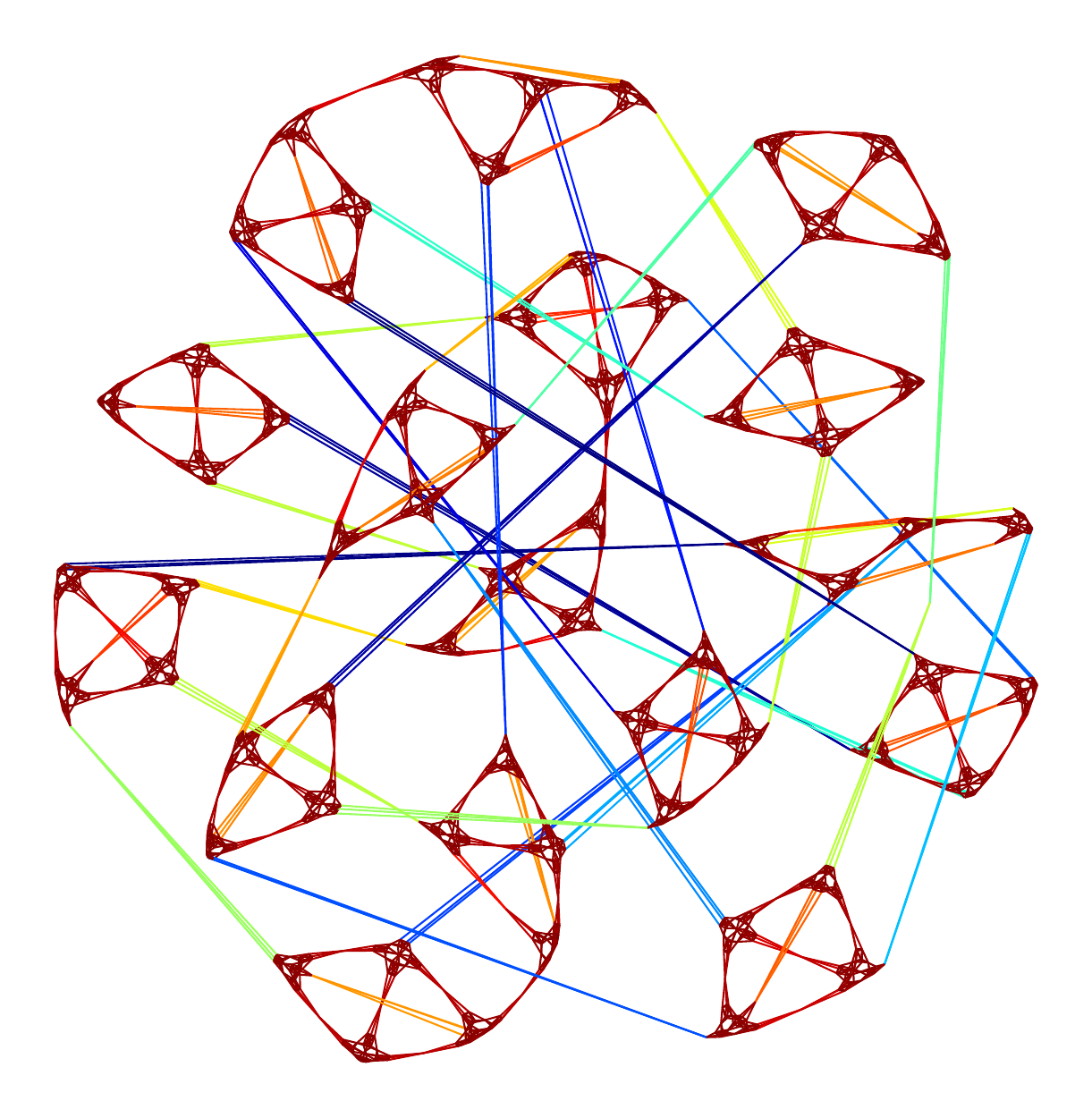}}
      & \parbox[c]{\tabfig\textwidth}{
      \includegraphics[width=\tabfig\textwidth]{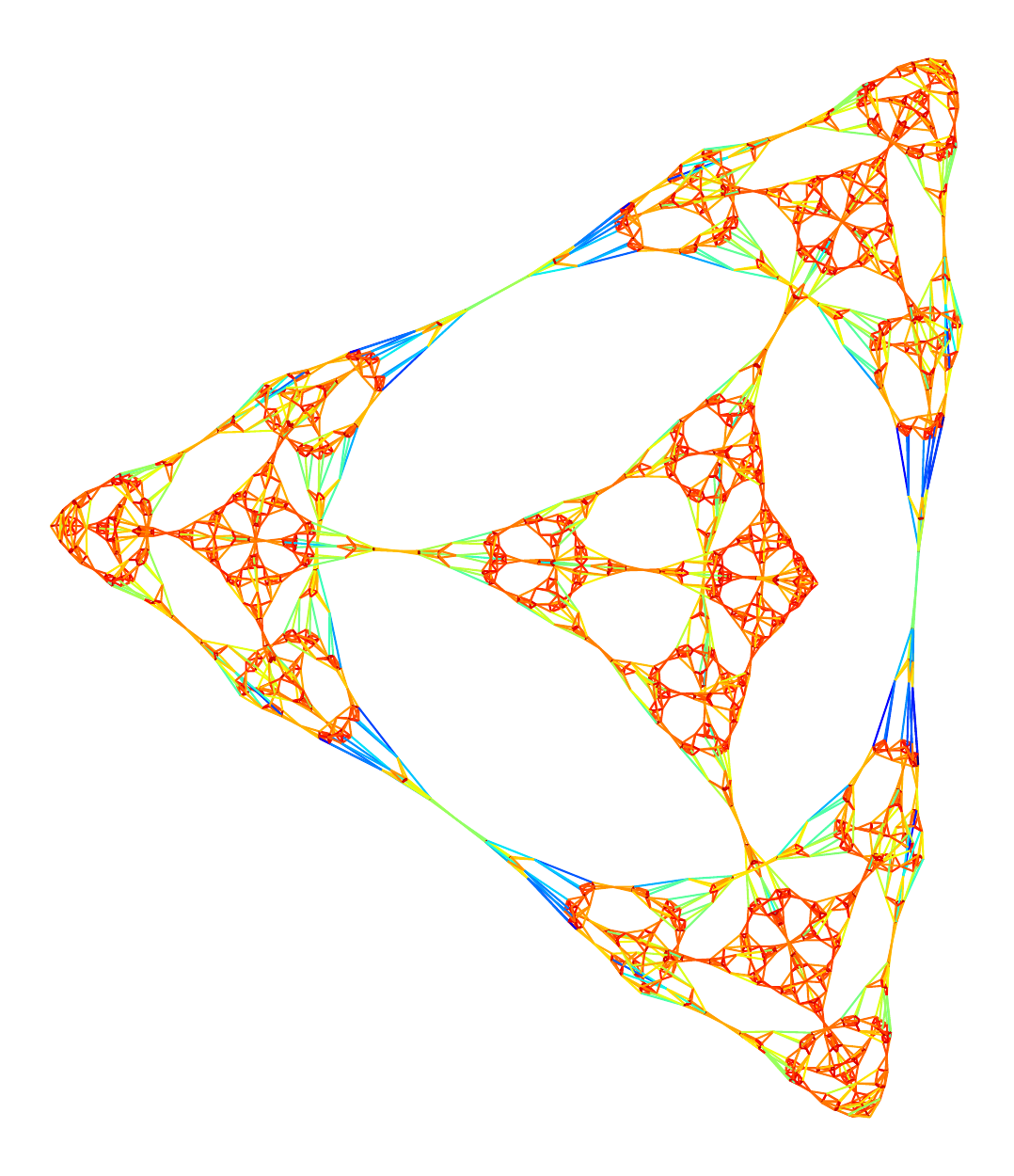}}      
      & \parbox[c]{\tabfig\textwidth}{
      \includegraphics[width=\tabfig\textwidth]{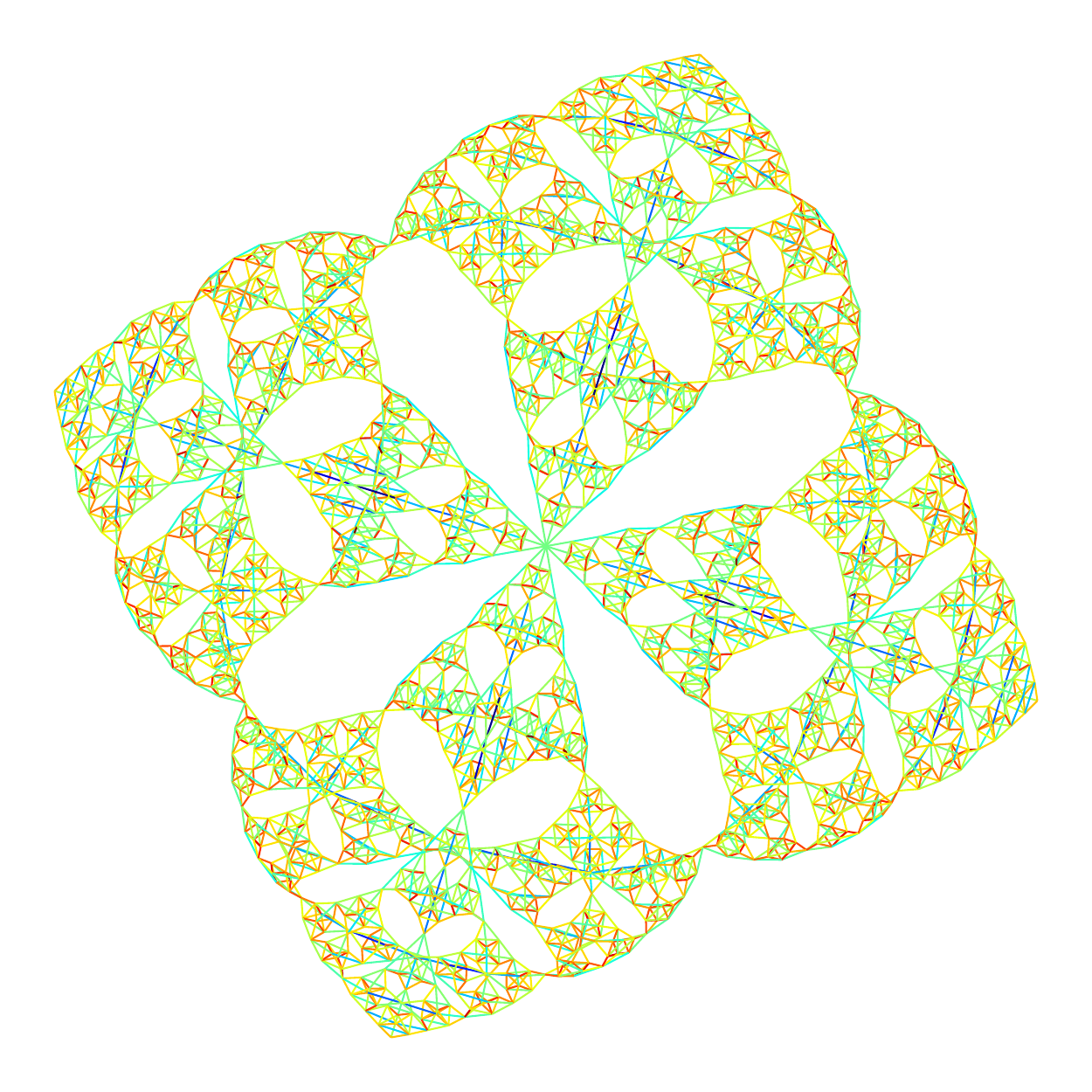}}\\
      \cline{2-5}

      & \raisebox{-.5\normalbaselineskip}[0pt][0pt]{\rotatebox[origin=c]{90}{\parbox{2cm}{\centering dwt\_1005  \\ (global)}}}
      & \parbox[c][2.2cm]{\tabfig\textwidth}{
      \includegraphics[width=\tabfig\textwidth]{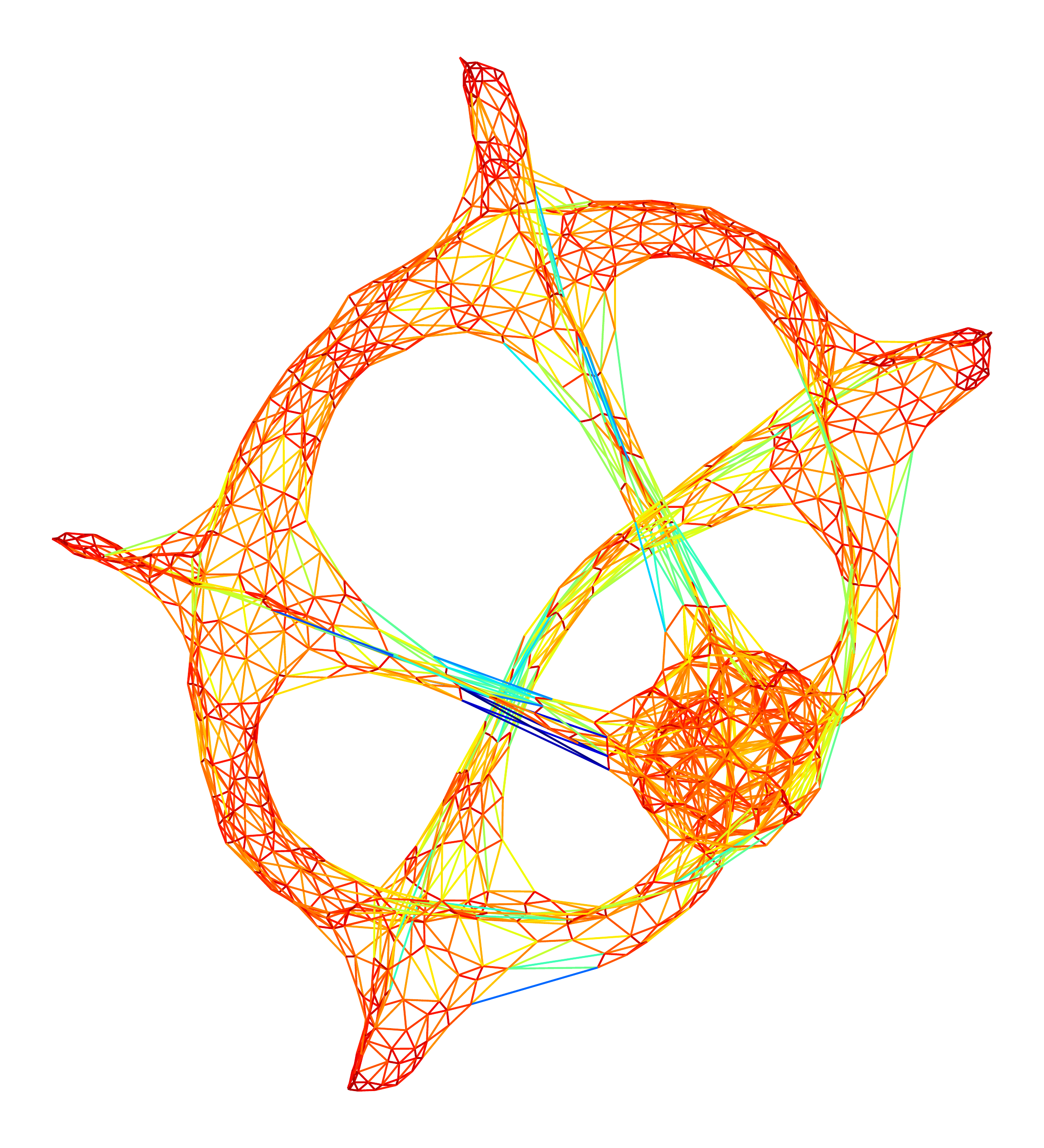}}
      & \parbox[c]{\tabfig\textwidth}{
      \includegraphics[width=\tabfig\textwidth]{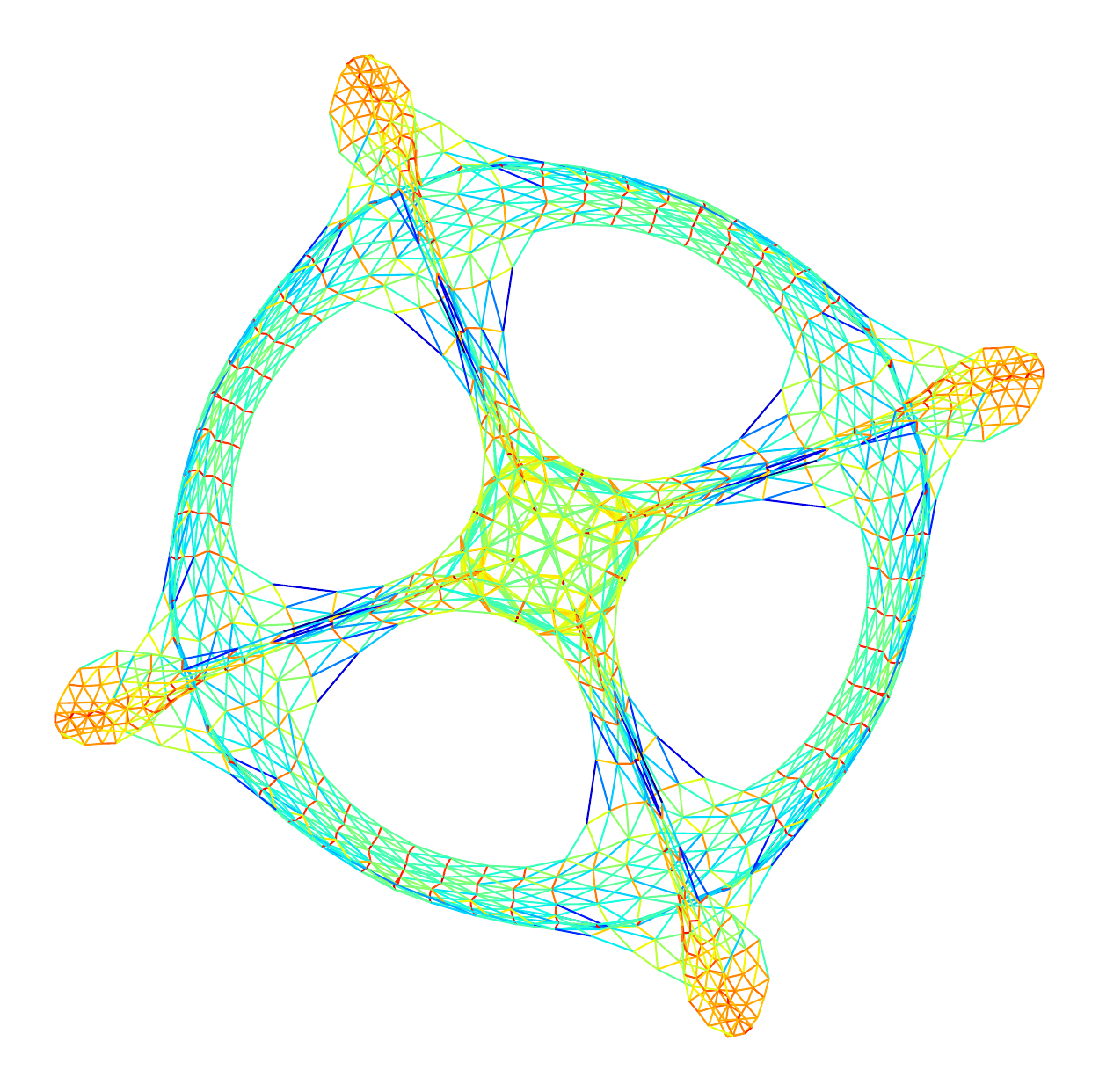}}      
      & \parbox[c]{\tabfig\textwidth}{
      \includegraphics[width=\tabfig\textwidth]{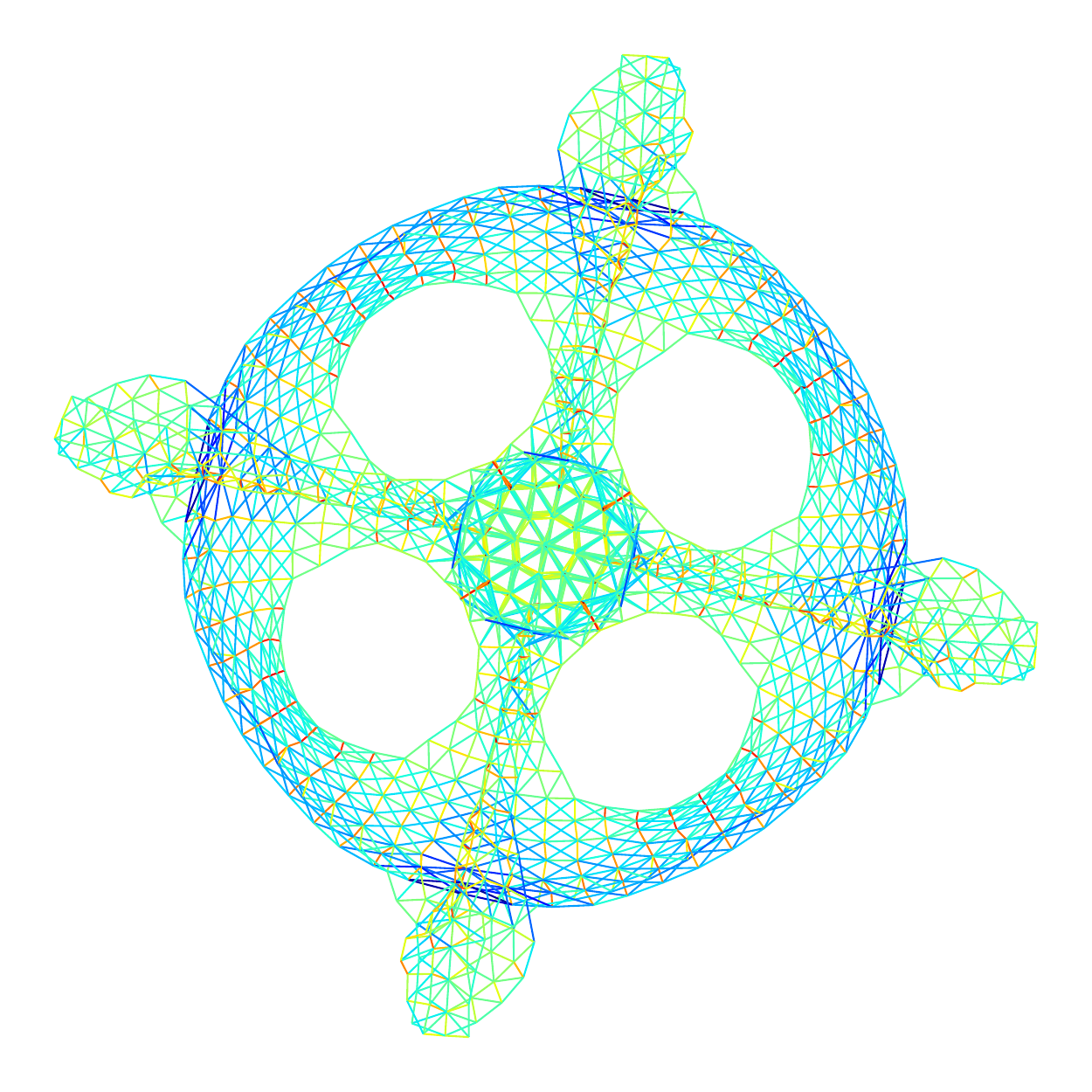}}\\
      \cline{2-5}
\end{tabular}
 \captionof{figure}{
Local embedding methods  perform well on graphs with distinct local structure (block\_2000), but they can distort the global shape of the graph (dwt\_1005). Global methods capture the overall shape (e.g., dwt\_1005), but may miss  important local structures (block\_2000). 
LGS(100) performs well for graphs with both local and global structure, such as sierpinski\_3d, allowing us to see its fractal nature.} 
\label{fig:block-table}
\end{table}

\subsection{Adapting Stress Minimization for Local Preservation}
\label{sec:3.1}

We define a parameter, $k$, that represents the size of a neighborhood surrounding each vertex. 
A straightforward approach would involve simply selecting the $k$-nearest vertices for every given vertex (as in~\cite{chen2009local}). However, the graph-theoretic distance in an undirected graph is a discrete measure, which can create complications. 
For example, consider the local structure graph (top row) in Fig.~\ref{fig:block-table}. Although the within-cluster density is high, there are many edges between different clusters. 
Unfortunately, there is no simple way to test if an edge is within cluster or out-of-cluster.
In order to produce tsNET-like embeddings, which should pay more attention to local structures, we must avoid preserving out-of-cluster edges.

\begin{figure}[t]
    \centering
    \includegraphics[width=.40\linewidth]{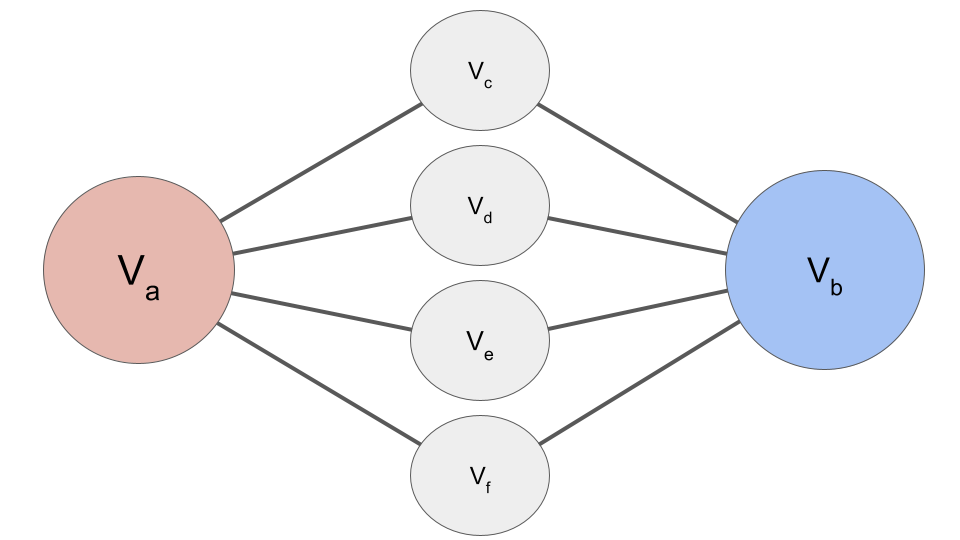}
    \caption{An example of how we may skip over immediate neighbors when selecting neighborhoods to preserve. In this case, $c=2$. There is only one unique walk of length $\leq 2$ from $v_a$ to $v_c,v_d,v_e,v_f$, but there are 4 such walks from $v_a$ to $v_b$. In this case, $v_b$ would be the first vertex added to $v_a$'s most connected neighborhood.}
    \label{fig:a_comp_example}
\end{figure}

Instead of considering distances directly, we find the top $k$ most connected vertices for each vertex based on the hypothesis that more possible walks between vertices indicate greater similarity; see Fig.~\ref{fig:a_comp_example}. 
Despite $v_A$ and $v_B$ not sharing an edge, they have the same set of neighbors. When $v_A$ and $v_B$ are both neighbors to a set of vertices, we can confidently state their similarity, confirmed by their shared proximity to $vc$, $v_d$, etc.~\cite{ertoz2002new}.
The $c$-th power of an adjacency matrix $(A_G^c)_{i,j}$ encodes the number of $c$-length walks from vertex $i$ to vertex $j$. 
To find the top $k$ ``most connected'' vertices for each vertex, follow this procedure:
Given an adjacency matrix of an undirected graph, $A_G$, raise it to the $c$-th power, take the sum of all powers 
$\mathbf{A^* = \sum_{1 \leq i \leq c} A_G^i}$,
to obtain a matrix whose $(i, j)$-th element 
shows
the number of walks from $i$ to $j$ of length less or equal than $c$.
Since each row in $\mathbf{A^*}$ corresponds to a vertex,
we find the $k$ largest values in row $i$ (by sorting). We define these top $k$ vertices to be the ``most connected'' neighborhood $N_k(v_a)$ of vertex $v_a$; see Fig.~\ref{fig:a_comp_example}. 
We further weight the power of the matrix with a decaying weight factor, $s, 0 < s < 1 $, such that $A^* = \sum_{1\leq i\leq c} s^i A^i_G$. 
We investigate a range of values for $s$, and set $s=0.1$; see the supplemental material.
We propose a procedure to reduce the number of matrix multiplications which we used in our experiments; see Appendix~\ref{sec:efficient_powers}.

\noindent
\textbf{Objective Function}
We remark, that only preserving distances of a subset of pairs will result in poor embeddings: e.g., two vertices that cannot ``see'' each other can be placed arbitrarily close with no penalty. A second term is needed in the objective function to prevent this, and we add an entropy repulsion term as in~\cite{chen2009local,DBLP:journals/tvcg/GansnerHN13}, to force pairs of vertices away from each other.
For a given pairwise distance matrix $ \left[ d_{ij} \right]_{i, j = 1}^n$
we define the following generalized stress function as an objective function:
\begin{equation}
    \label{eq:obj_func}
    \sigma(X) = \sum_{(i,j) \in N_k}(\Vert X_i - X_j \Vert - d_{ij})^2 - \alpha \sum_{(i,j)\notin N_k} \log \Vert X_i - X_j \Vert,
\end{equation}
where $X_i$ is the embedded point in $\mathbb{R}^d$, $\alpha$ is a fixed constant parameter that controls the weight of the logarithmic term, and $N_k$ corresponds to the neighborhood that we aim to preserve in the embedded space.
This objective function ensures that  distances are
preserved between the most-connected neighborhoods, while maximizing entropy. 
We use the negative logarithm of the distance between points, so that the repulsive force is relatively strong at small distances, but quickly decays (so that distant points are not forced to be too distant from each other).
While similar to LMDS~\cite{chen2009local} and MaxEnt~\cite{DBLP:journals/tvcg/GansnerHN13}, the proposed objective function in Eq.~\ref{eq:obj_func} differs in (1) how the set $N_k$ is selected (LMDS uses a kNN search and 
MaxEnt preserves distances between two vertices if and only if they share an edge) 
and (2) LMDS and MaxEnt cannot be easily parameterized to balance local and global structure preservation.
We minimize the objective function by SGD which works well for stress minimization~\cite{DBLP:conf/gd/BorsigBP20,DBLP:journals/tvcg/ZhengPG19}. 
The parameter space of the algorithm is discussed in the supplemental material, Appendix~\ref{sec:parameters}.

\begin{table*}[ht!] 

  \centering
  \begin{tabular}
      {| l | c c c c c c c|} \hline & tsNET & LGS k=16 & LGS k=32 & UMAP & LGS k=64 & LGS k=100 & MDS\\
      \hline 
     \multirow{2}{*}{\rotatebox[origin=c]{90}{\centering lesmis}}  &
      \parbox[c]{\tabfig\textwidth}{} 
      & \parbox[c]{\tabfig\textwidth}{
      \includegraphics[width=\tabfig\textwidth,height=\tabfig\textwidth]{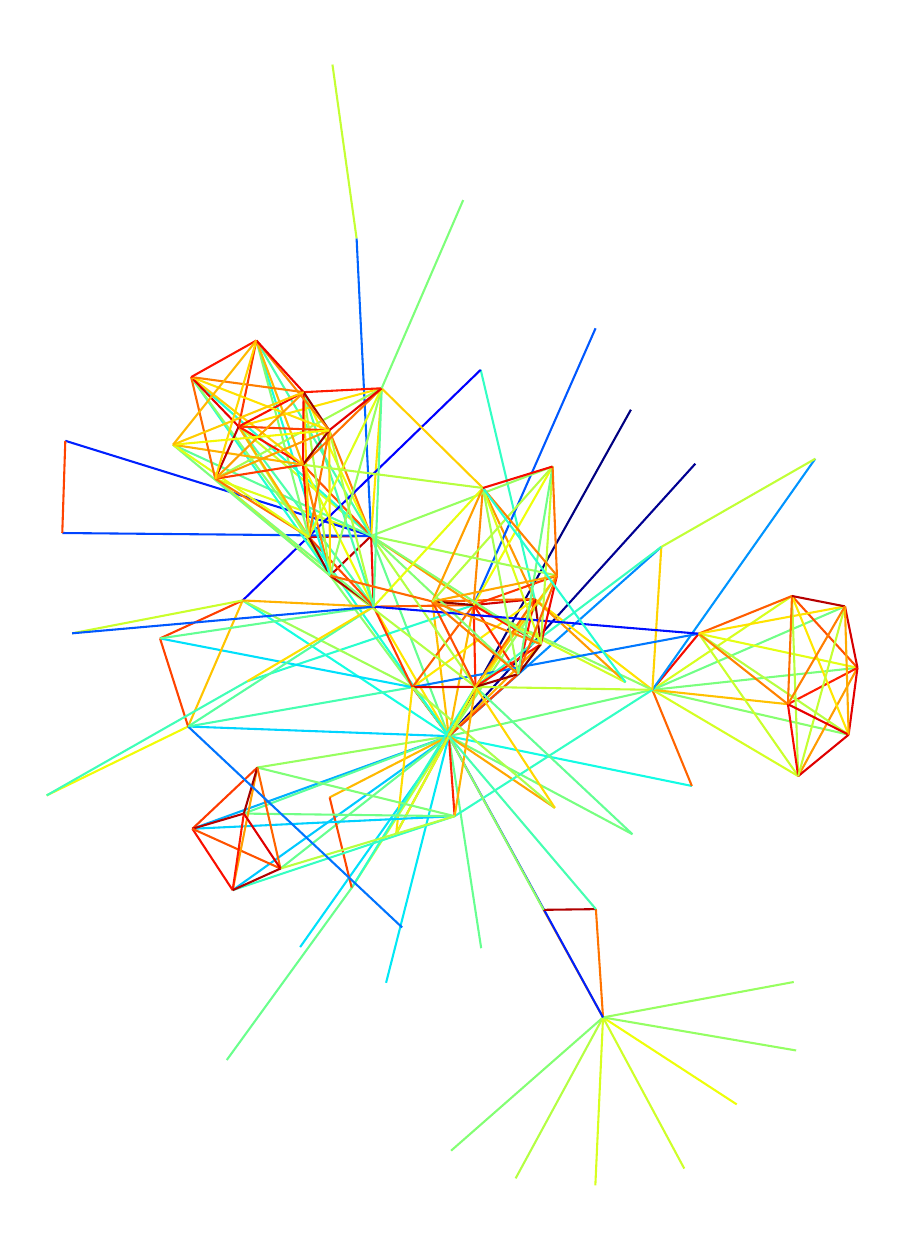}}
      & \parbox[c]{\tabfig\textwidth}{
      \includegraphics[width=\tabfig\textwidth,height=\tabfig\textwidth]{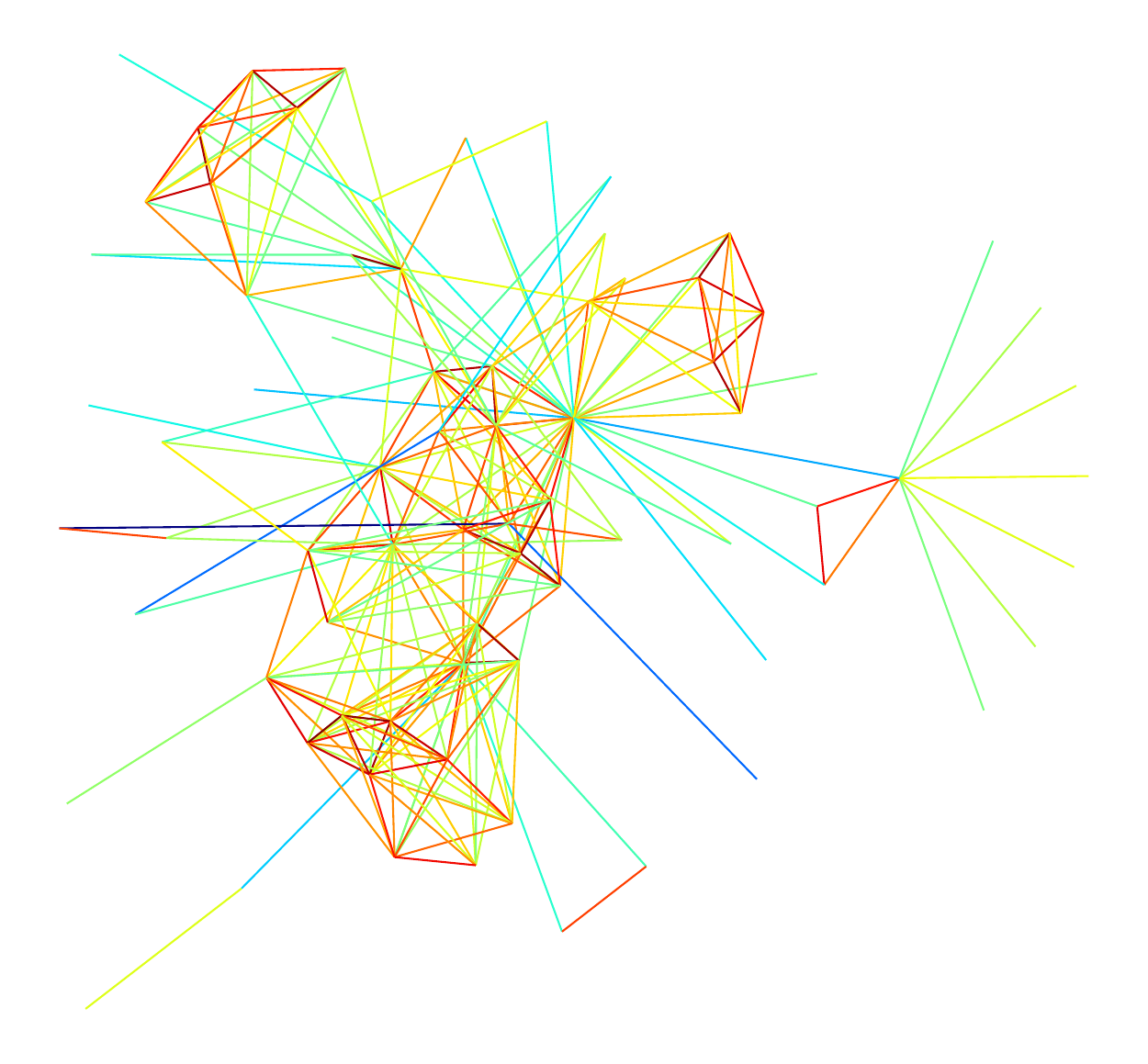}}
      & \parbox[c]{\tabfig\textwidth}{}
      & \parbox[c]{\tabfig\textwidth}{
      \includegraphics[width=\tabfig\textwidth,height=\tabfig\textwidth]{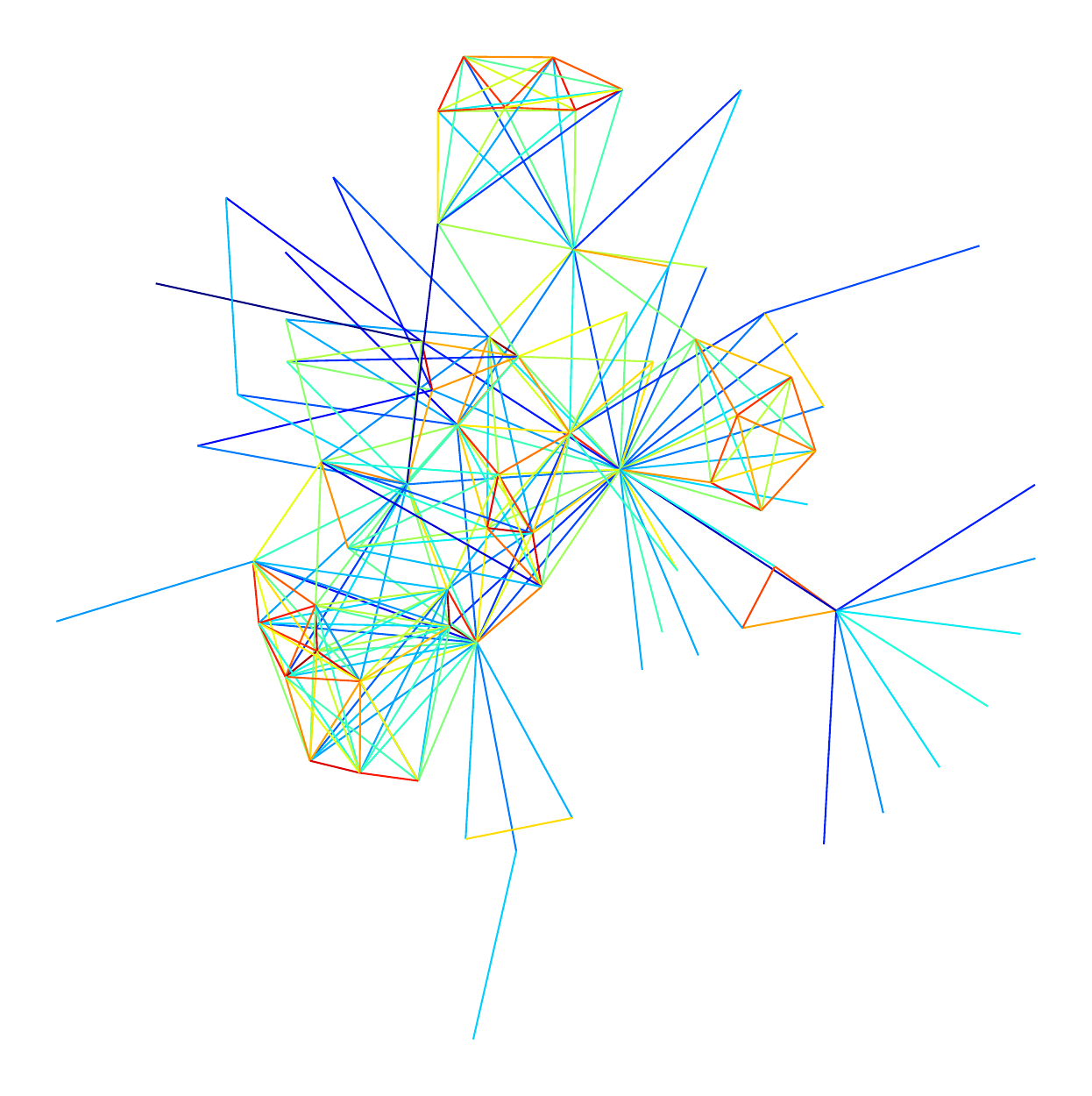}} 
      & \parbox[c]{\tabfig\textwidth}{
      \includegraphics[width=\tabfig\textwidth,height=\tabfig\textwidth]{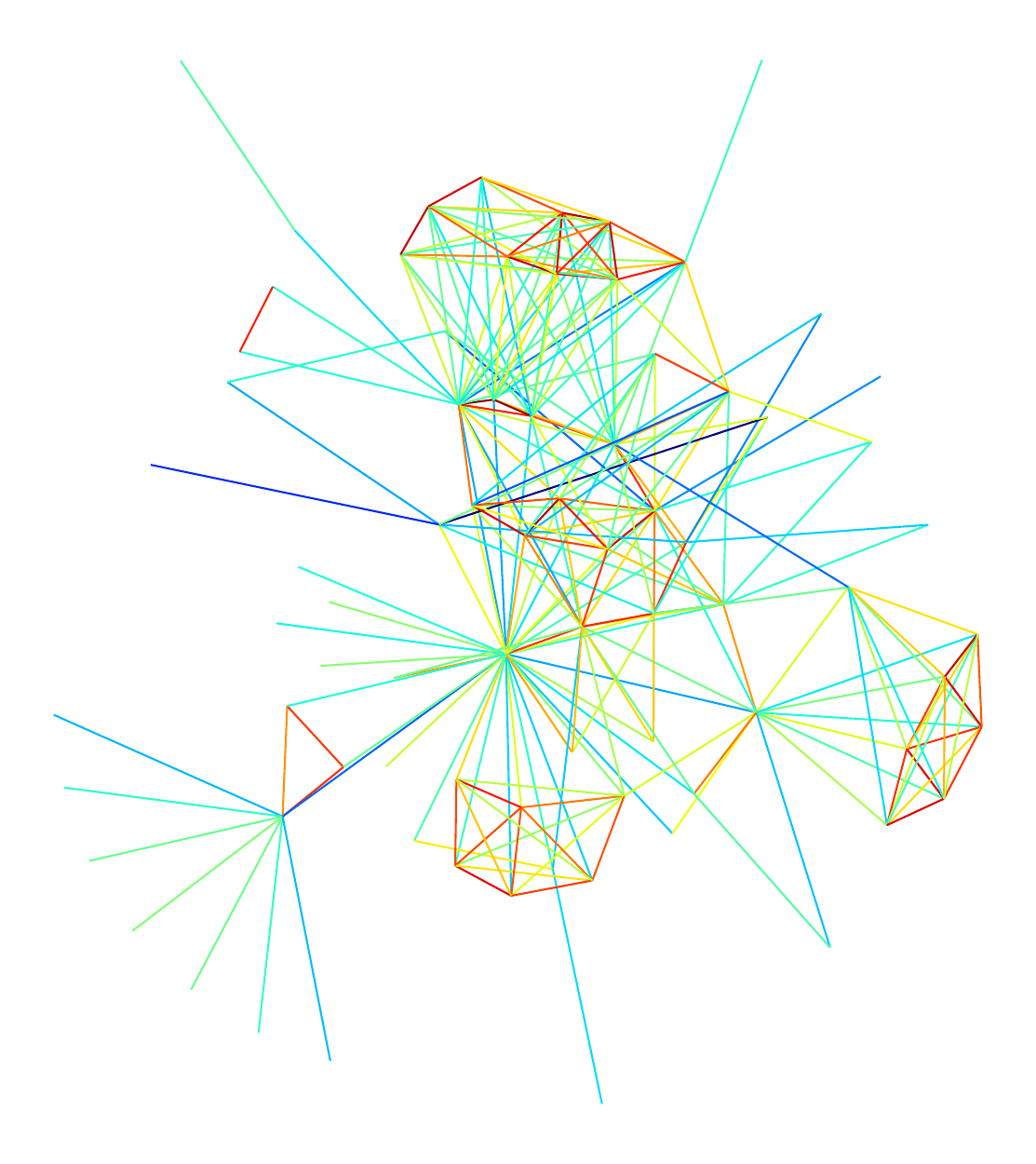}} 
      & \parbox[c]{\tabfig\textwidth}{}  \\
    
      & \parbox[c]{\tabfig\textwidth}{\taboffset
      \includegraphics[width=\tabfig\textwidth,height=\tabfig\textwidth]{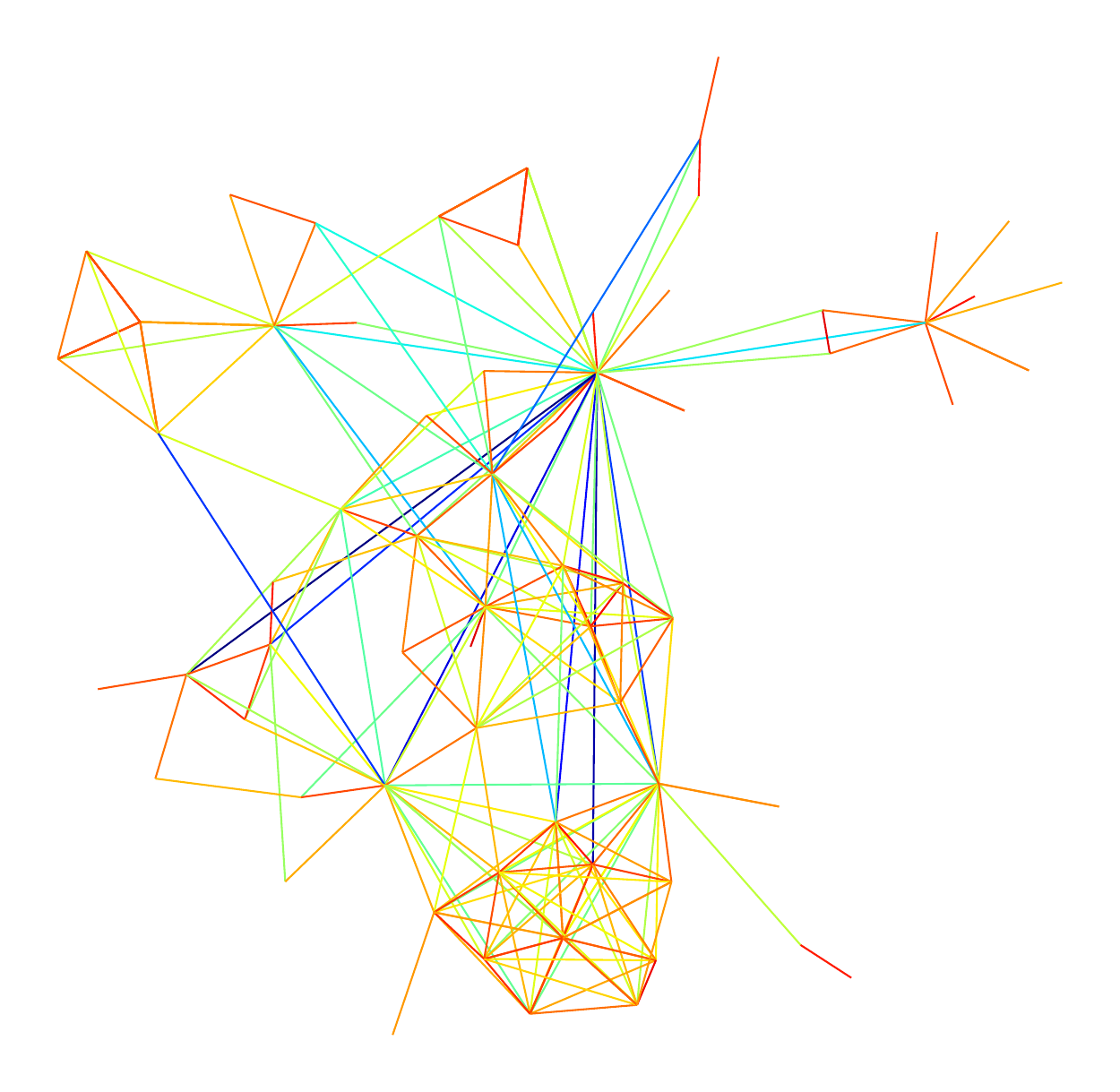}} 
      & \parbox[c]{\tabfig\textwidth}{}
      & \parbox[c]{\tabfig\textwidth}{}
      & \parbox[c]{\tabfig\textwidth}{\taboffset
      \includegraphics[width=\tabfig\textwidth,height=\tabfig\textwidth]{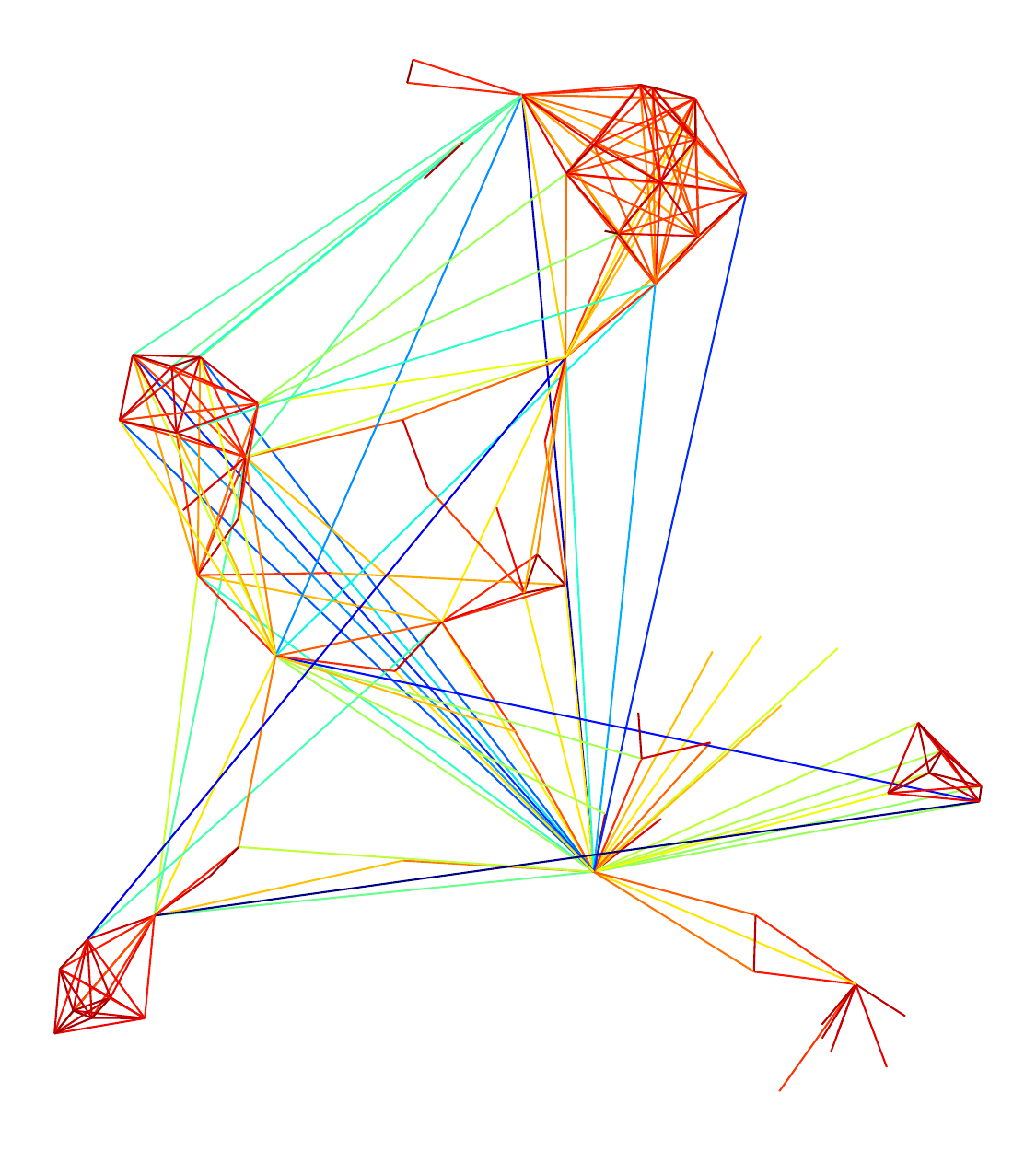}}
      & \parbox[c]{\tabfig\textwidth}{} 
      & \parbox[c]{\tabfig\textwidth}{} 
      & \parbox[c]{\tabfig\textwidth}{\taboffset
      \includegraphics[width=\tabfig\textwidth,height=\tabfig\textwidth]{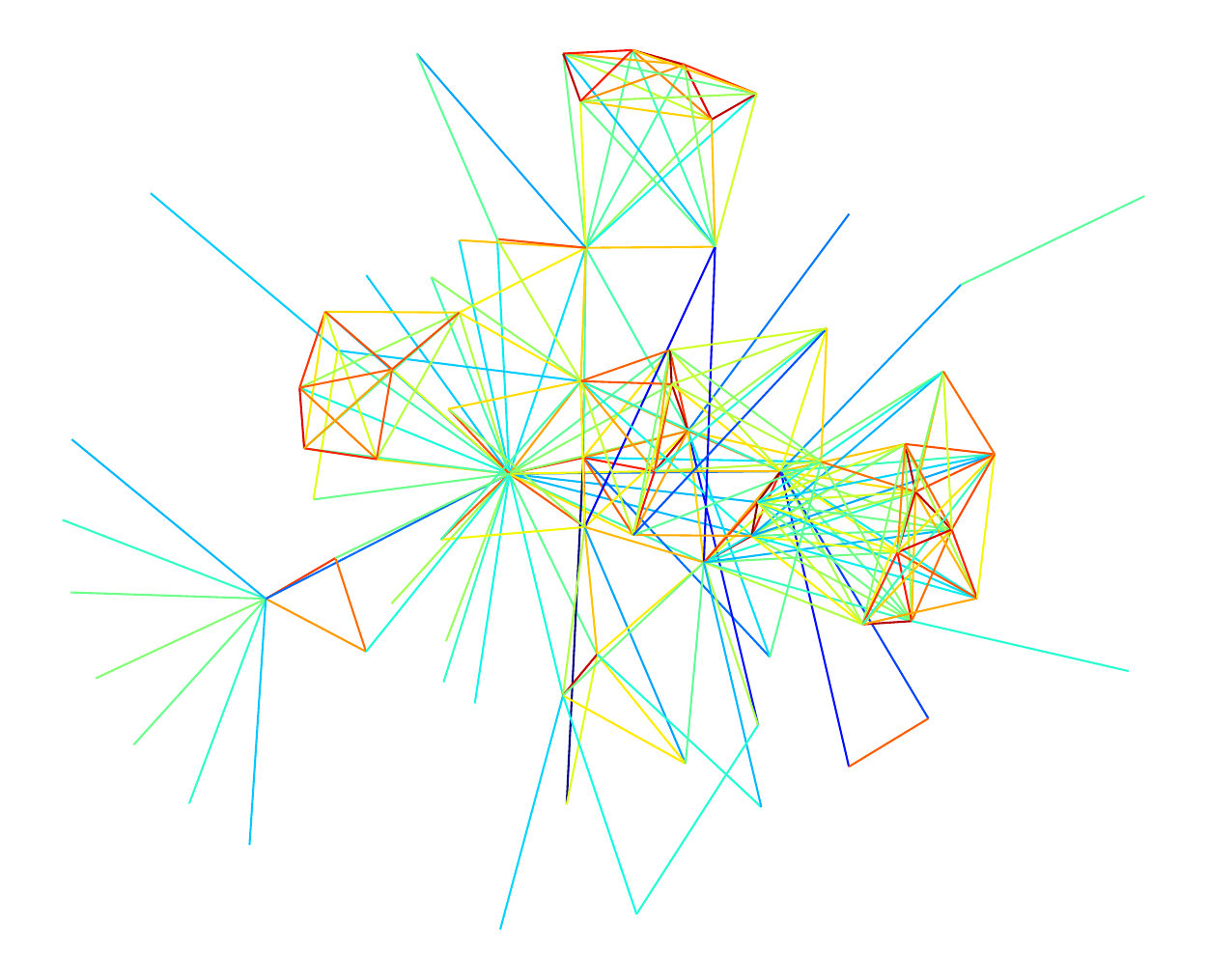}}  \\   
      \hline
      
     \multirow{2}{*}{\rotatebox[origin=c]{90}{\parbox{1.0cm}{\centering block\_500}}} &
      \parbox[c]{\tabfig\textwidth}{} 
      & \parbox[c]{\tabfig\textwidth}{
      \includegraphics[width=\tabfig\textwidth,height=\tabfig\textwidth]{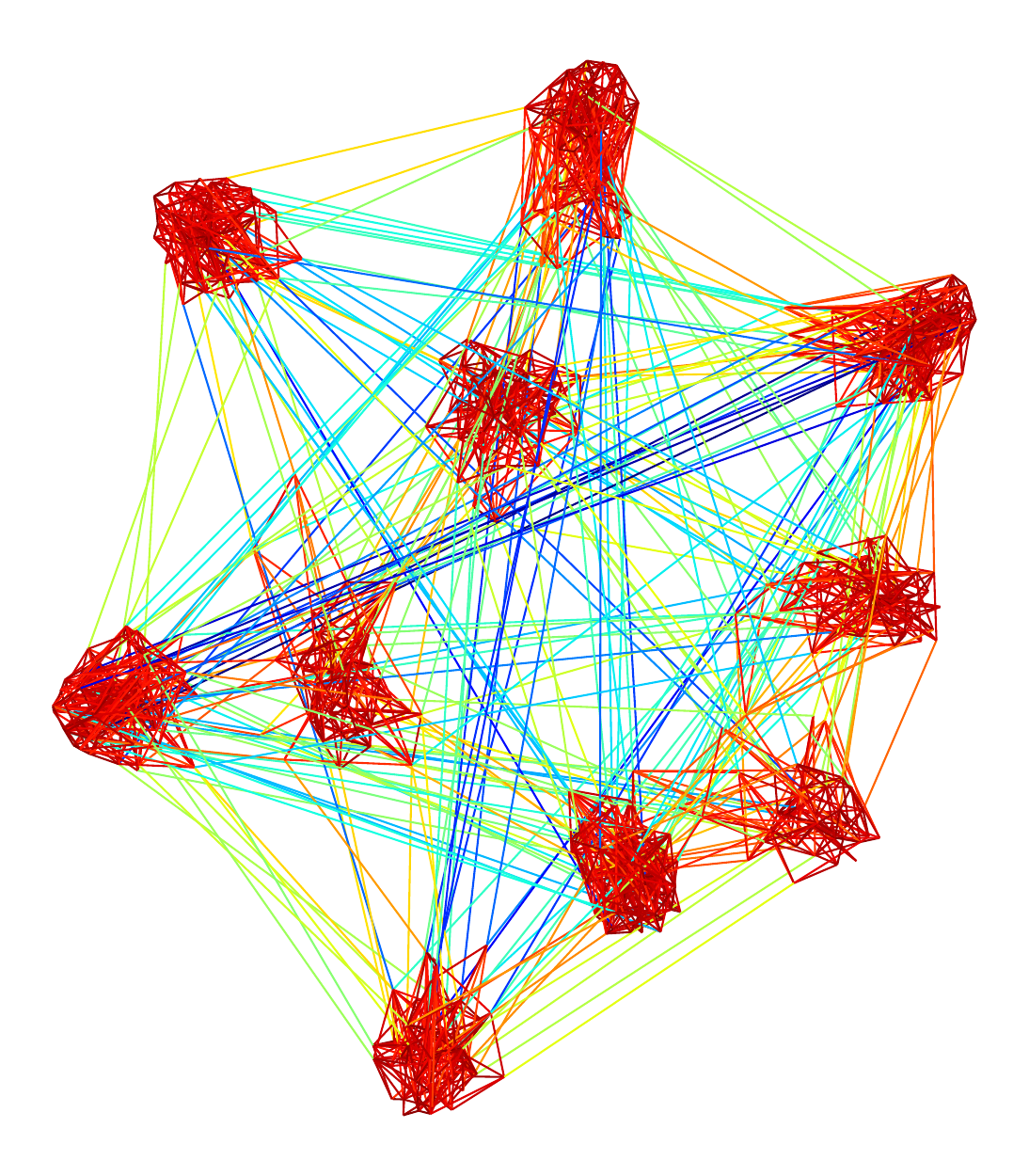}}
      & \parbox[c]{\tabfig\textwidth}{
      \includegraphics[width=\tabfig\textwidth,height=\tabfig\textwidth]{figures/2023-pdfs/L2G/block_model_500_32.pdf}}
      & \parbox[c]{\tabfig\textwidth}{}
      & \parbox[c]{\tabfig\textwidth}{
      \includegraphics[width=\tabfig\textwidth,height=\tabfig\textwidth]{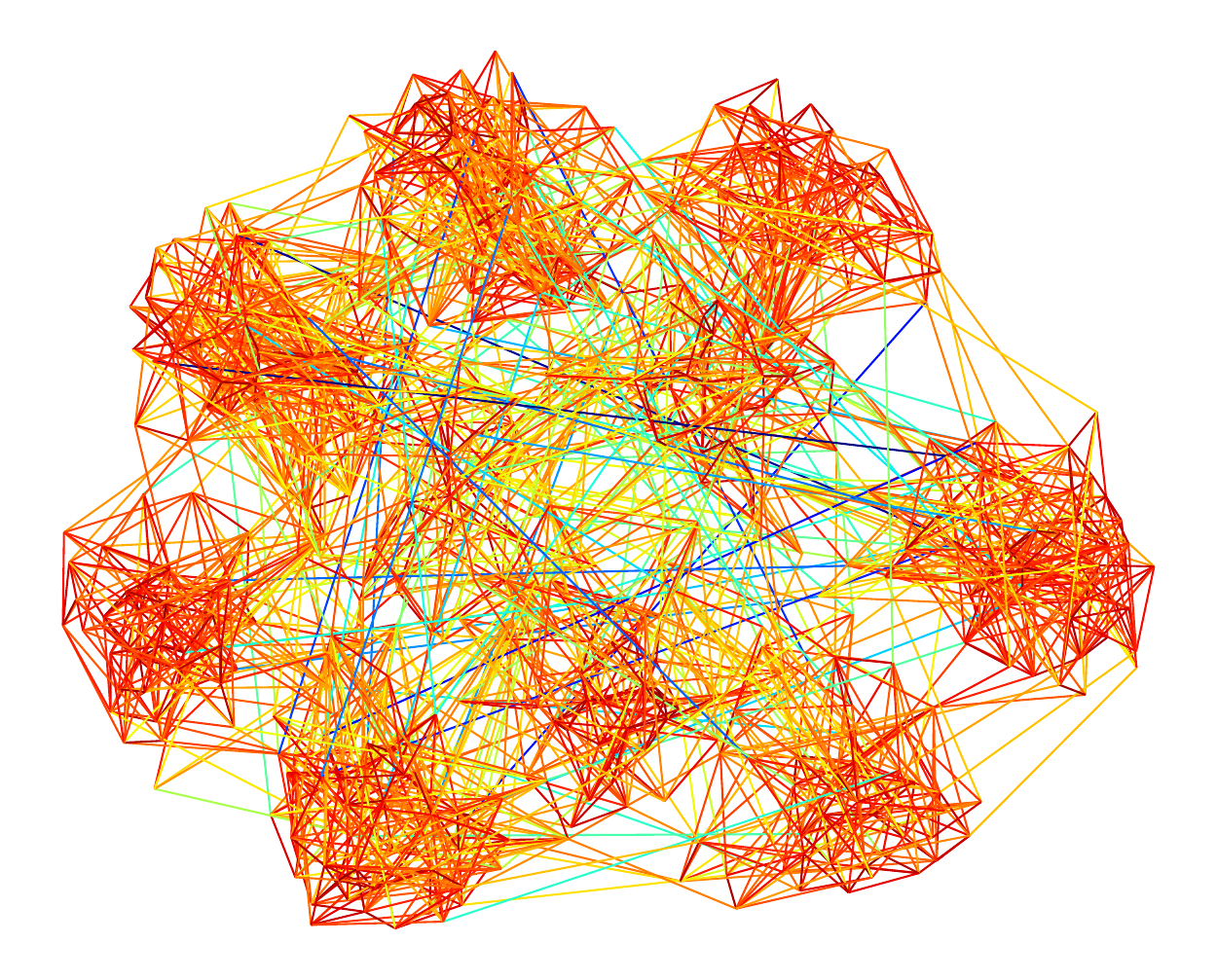}} 
      & \parbox[c]{\tabfig\textwidth}{
      \includegraphics[width=\tabfig\textwidth,height=\tabfig\textwidth]{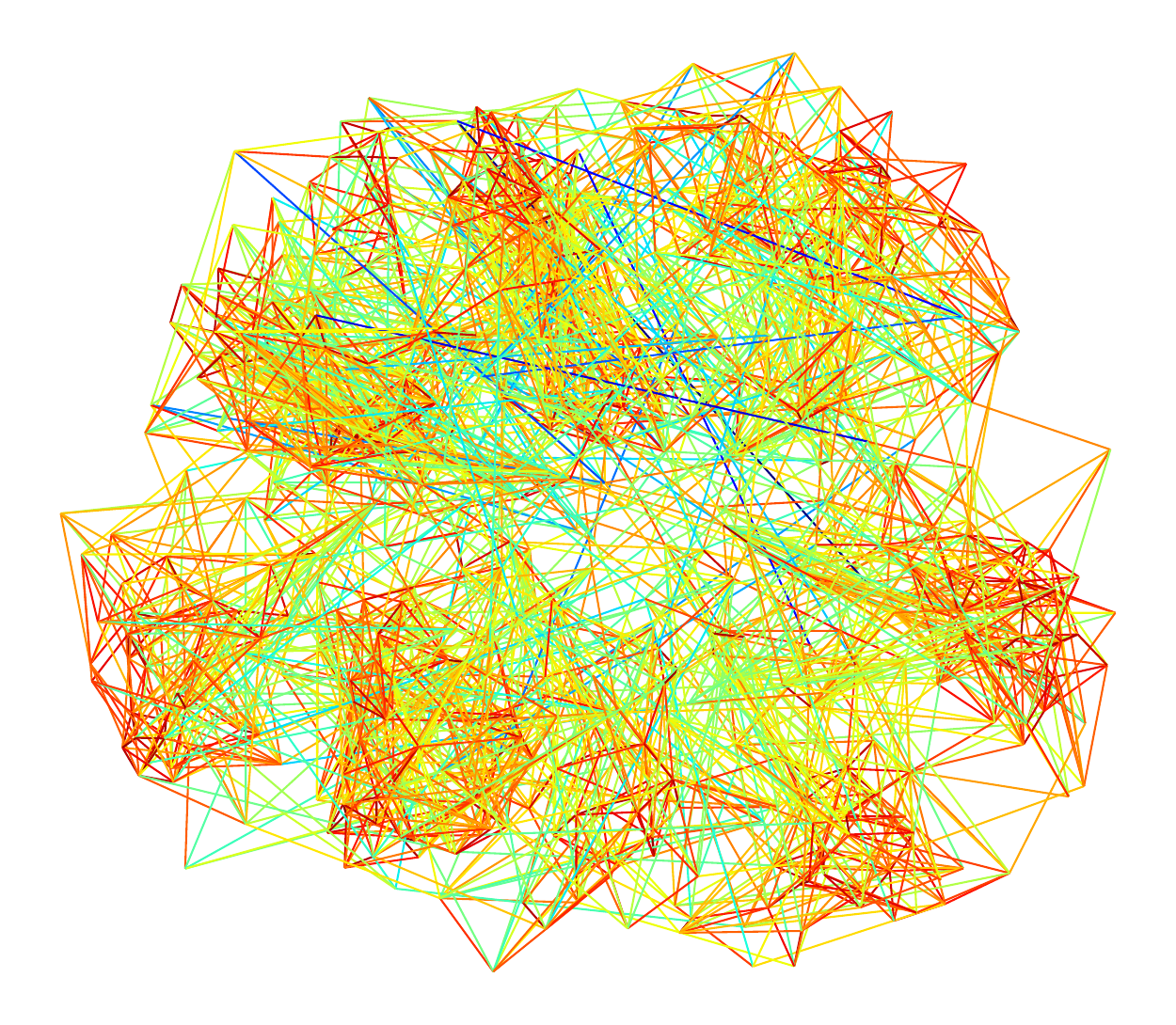}} 
      & \parbox[c]{\tabfig\textwidth}{}  \\
      
      & \parbox[c]{\tabfig\textwidth}{\taboffset\includegraphics[width=\tabfig\textwidth,height=\tabfig\textwidth]{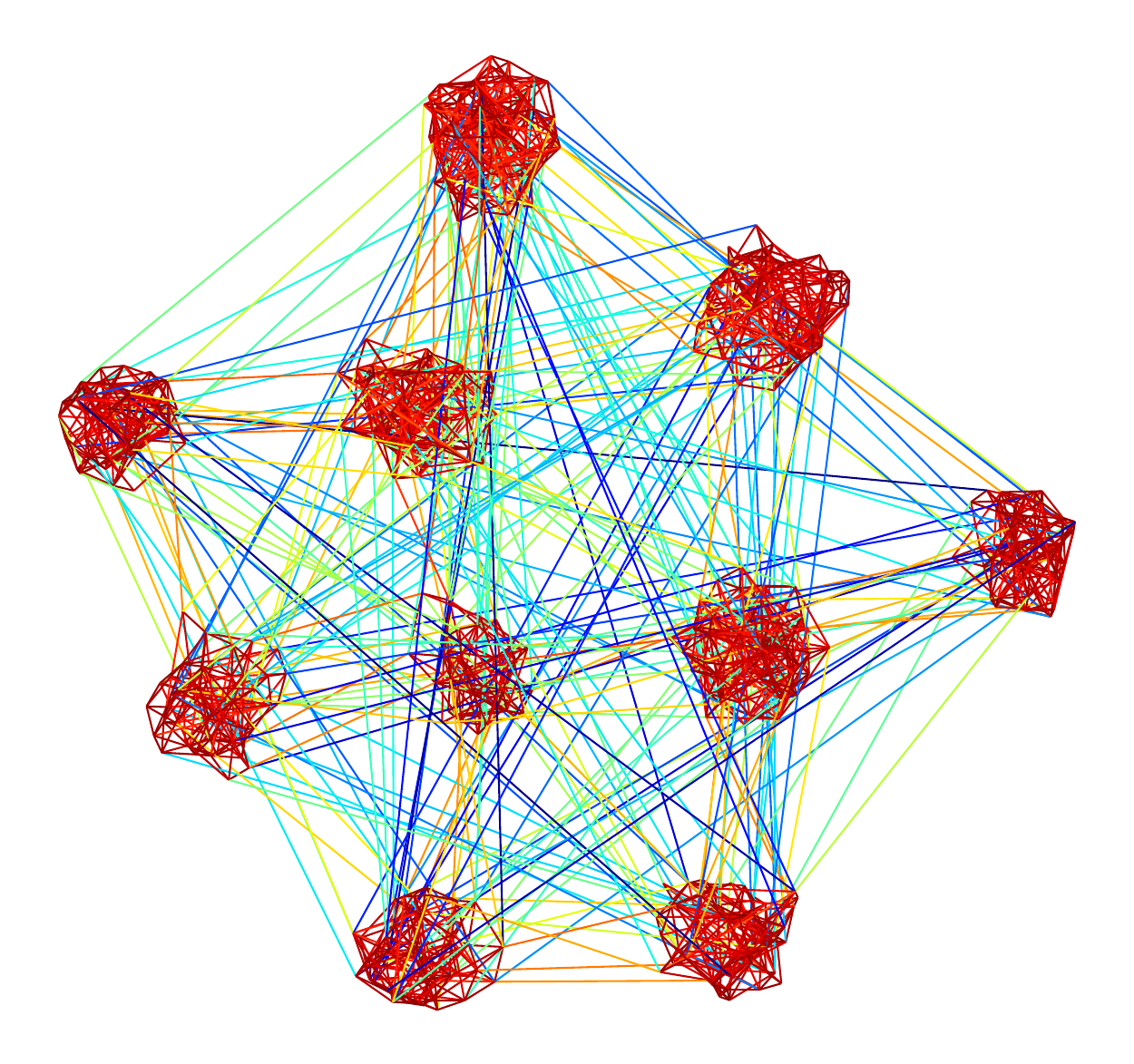}} 
      & \parbox[c]{\tabfig\textwidth}{}
      & \parbox[c]{\tabfig\textwidth}{}
      & \parbox[c]{\tabfig\textwidth}{\taboffset
      \includegraphics[width=\tabfig\textwidth,height=\tabfig\textwidth]{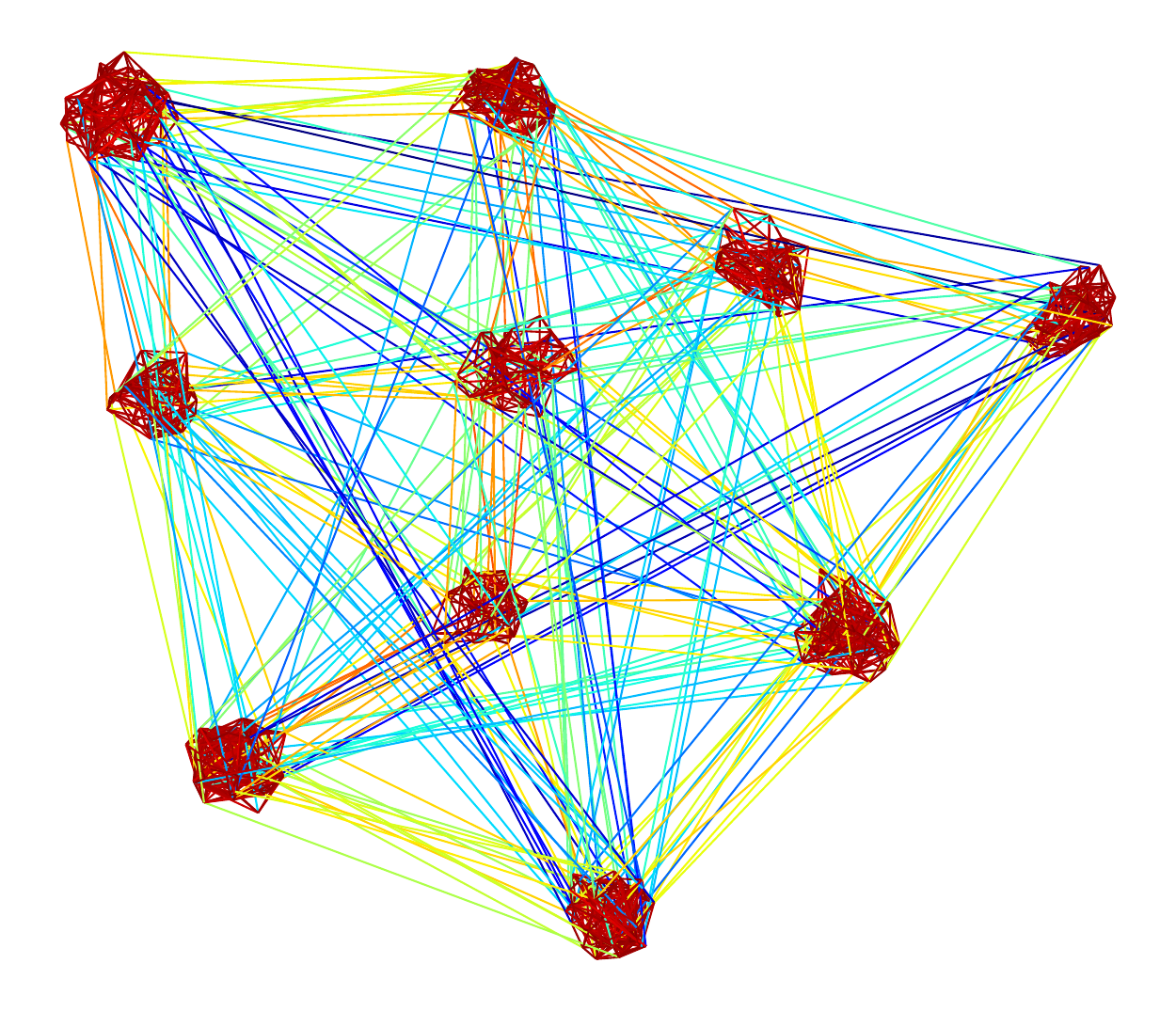}}
      & \parbox[c]{\tabfig\textwidth}{} 
      & \parbox[c]{\tabfig\textwidth}{} 
      & \parbox[c]{\tabfig\textwidth}{\taboffset
      \includegraphics[width=\tabfig\textwidth,height=\tabfig\textwidth]{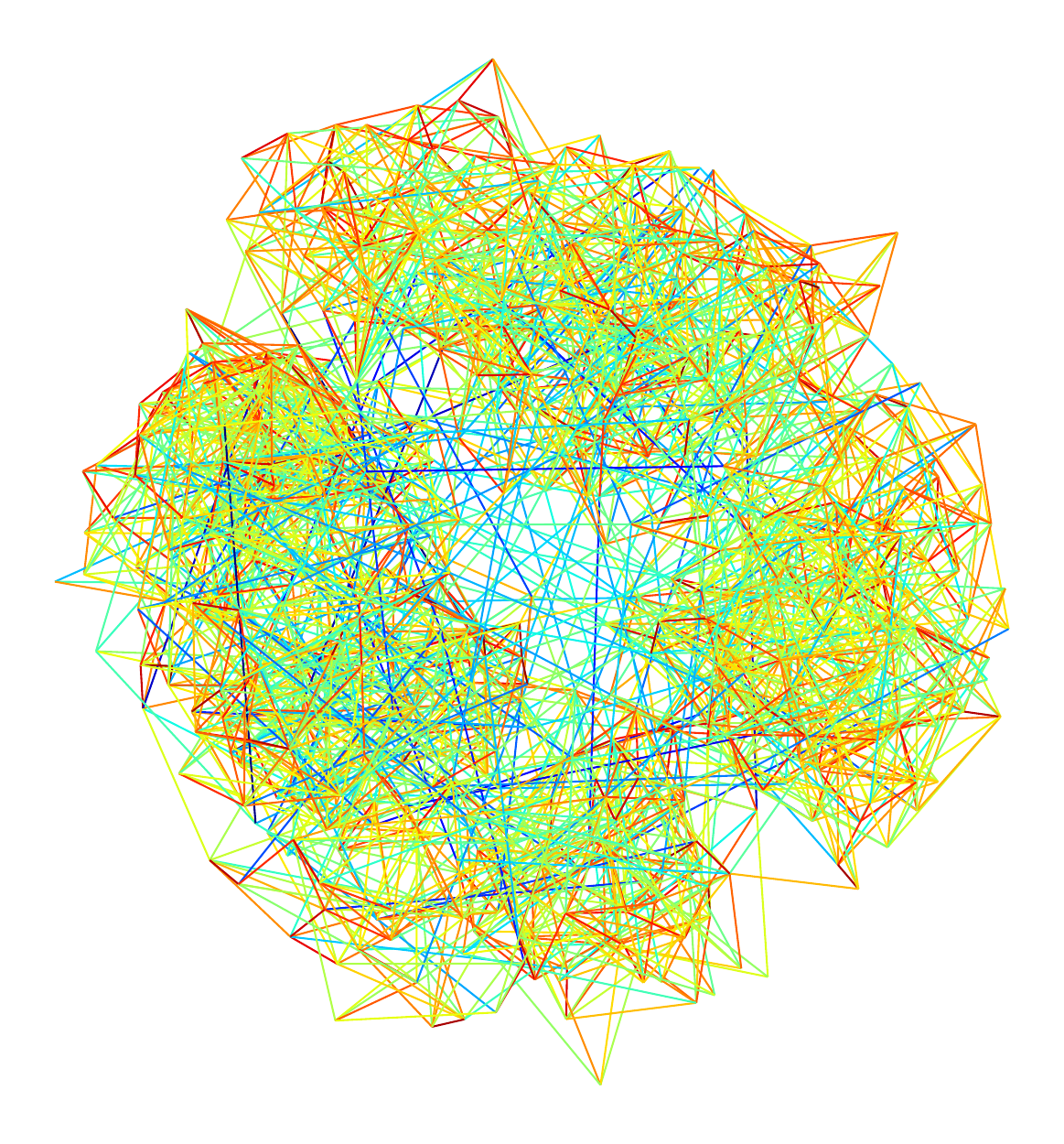}}  \\   
      \hline
      
     \multirow{2}{*}{\rotatebox[origin=c]{90}{football}} &
      \parbox[c]{\tabfig\textwidth}{} 
      & \parbox[c]{\tabfig\textwidth}{
      \includegraphics[width=\tabfig\textwidth,height=\tabfig\textwidth]{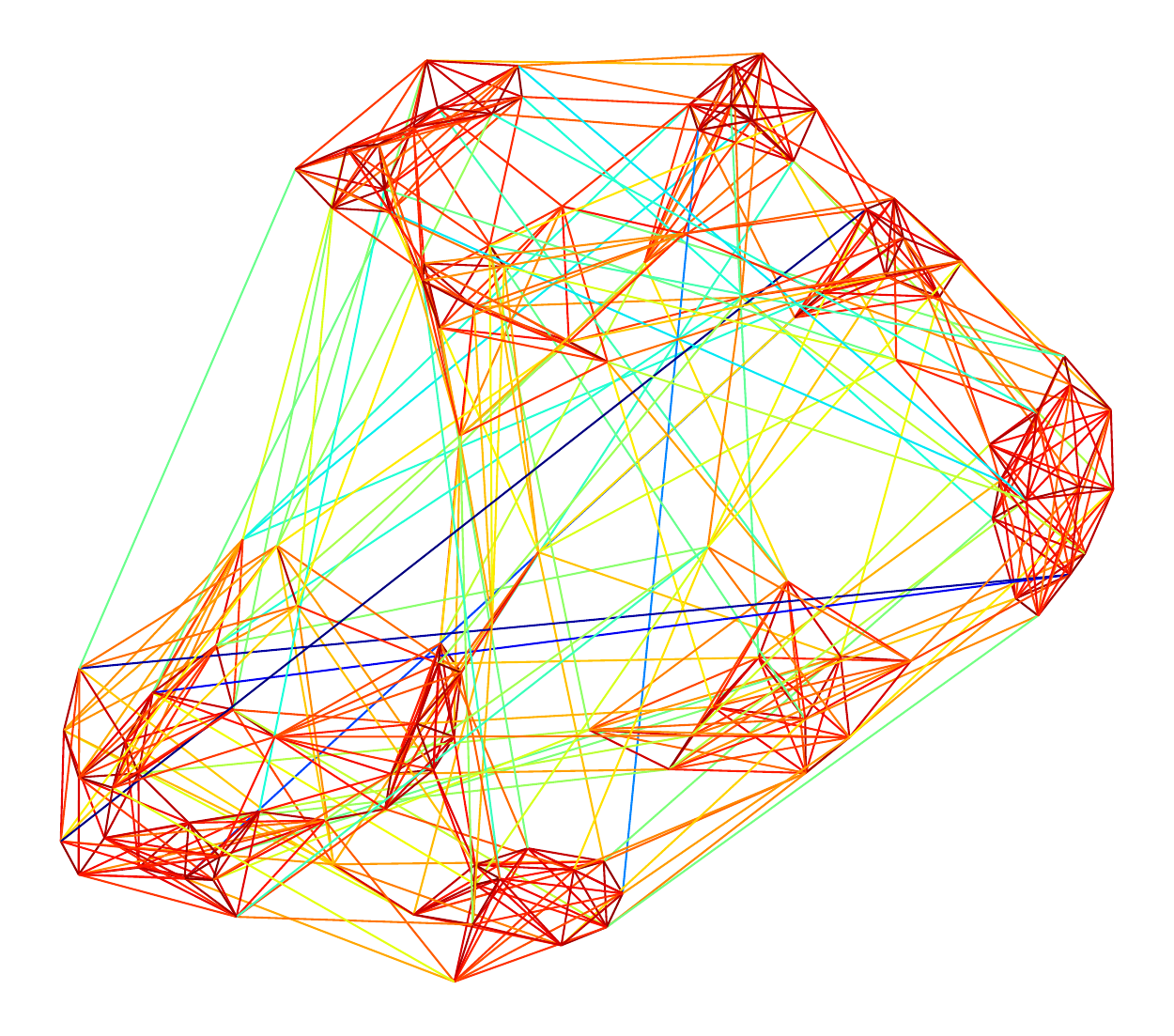}}
      & \parbox[c]{\tabfig\textwidth}{
      \includegraphics[width=\tabfig\textwidth,height=\tabfig\textwidth]{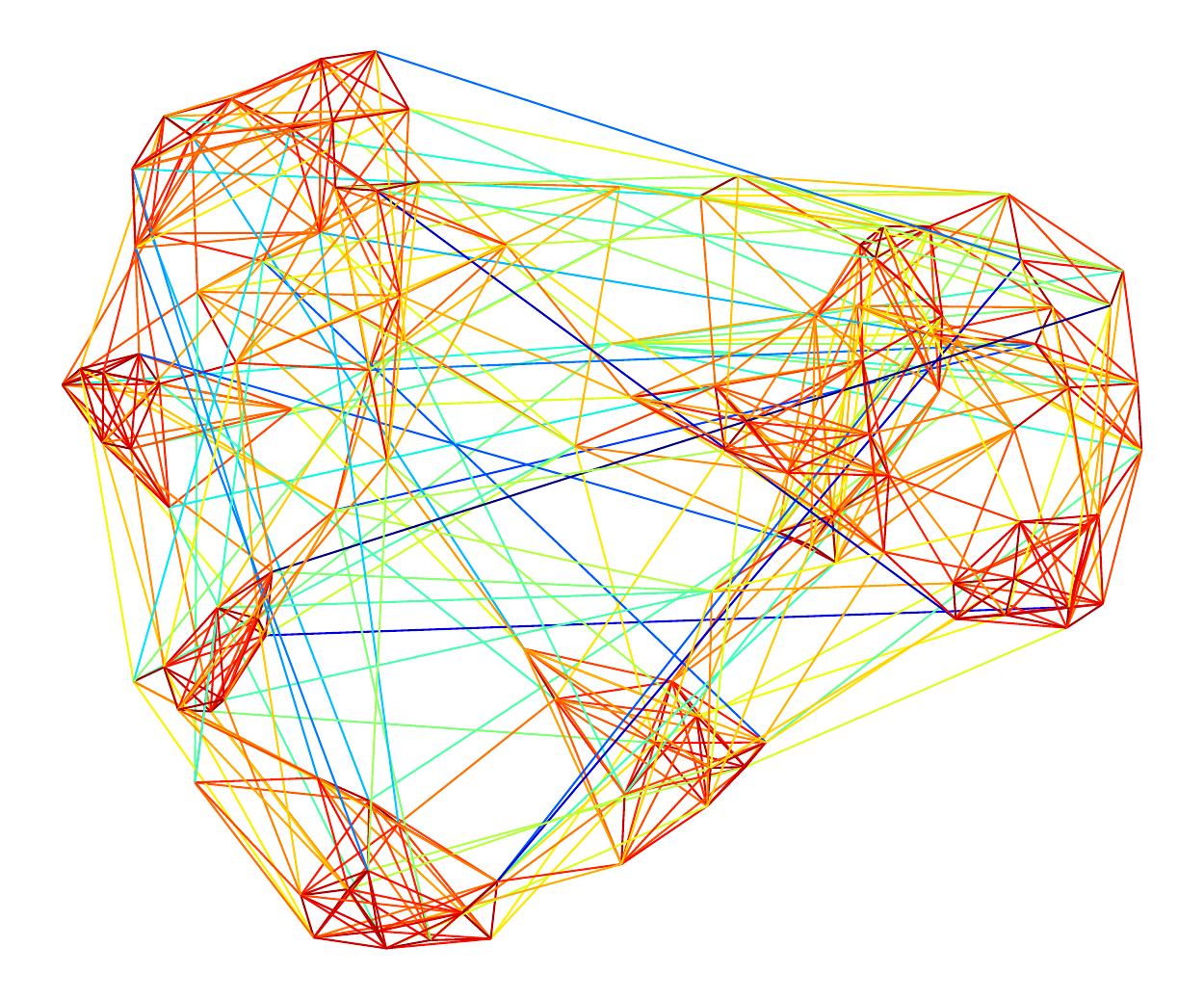}}
      & \parbox[c]{\tabfig\textwidth}{}
      & \parbox[c]{\tabfig\textwidth}{
      \includegraphics[width=\tabfig\textwidth,height=\tabfig\textwidth]{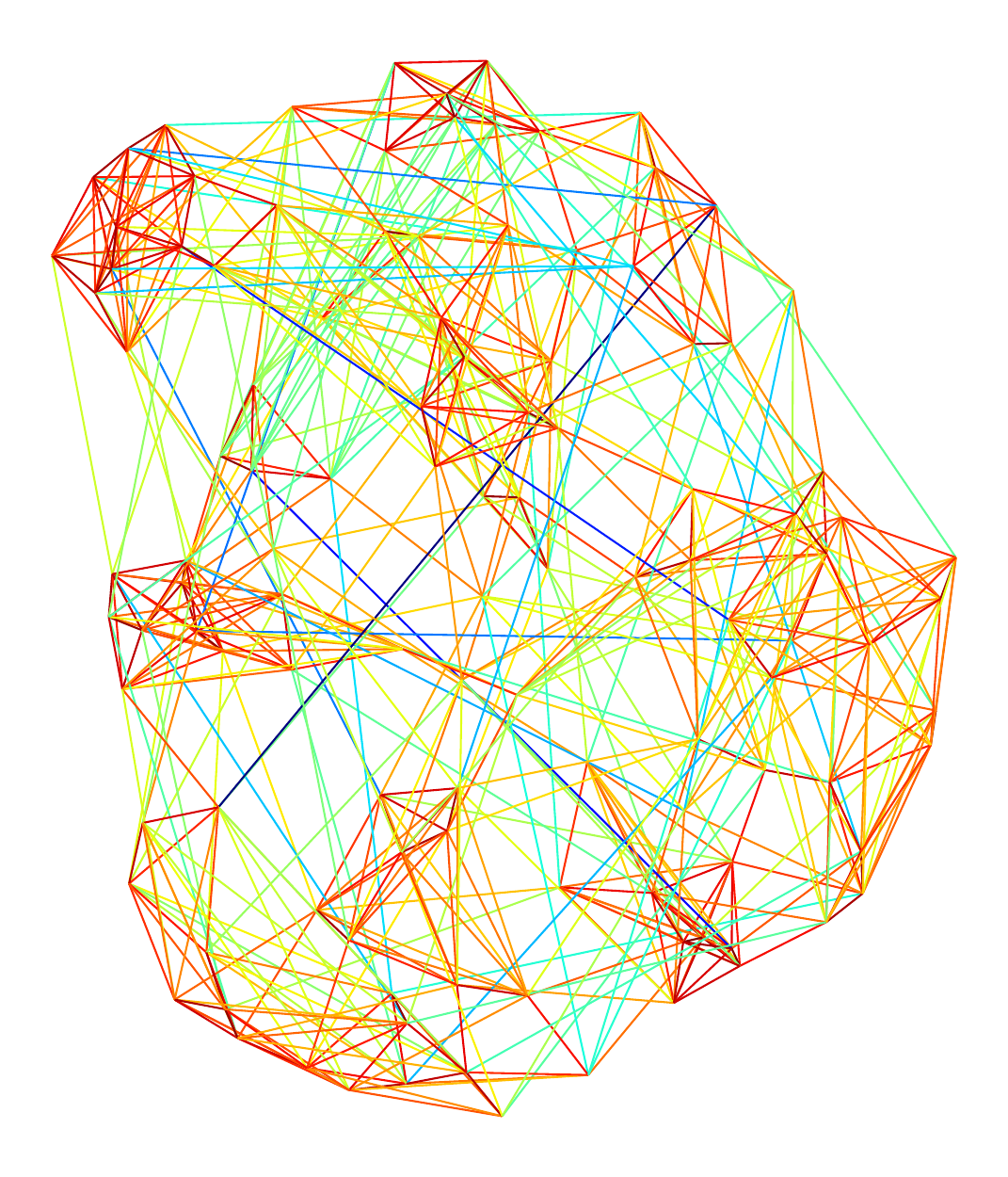}} 
      & \parbox[c]{\tabfig\textwidth}{
      \includegraphics[width=\tabfig\textwidth,height=\tabfig\textwidth]{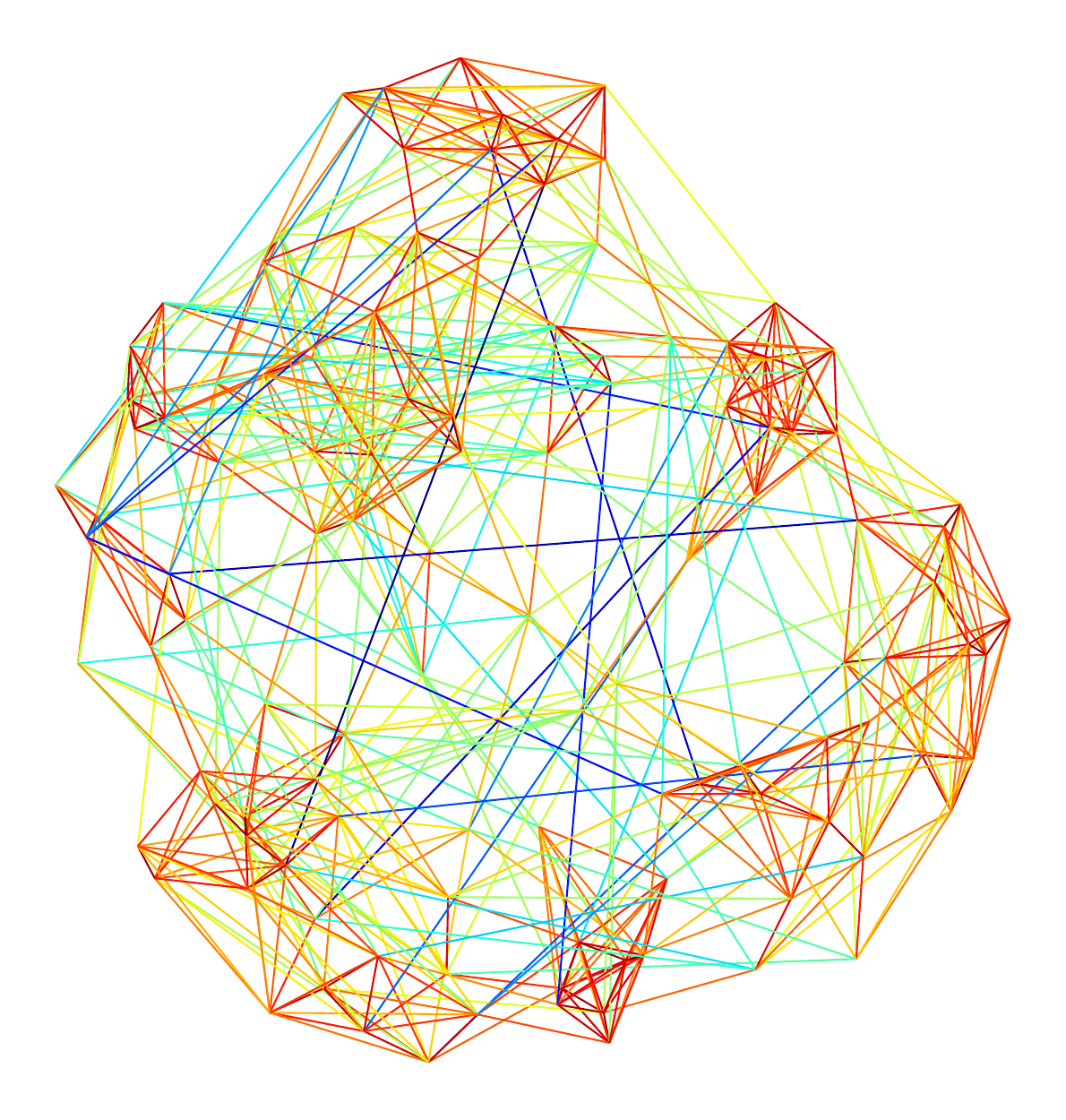}} 
      & \parbox[c]{\tabfig\textwidth}{}  \\
      & \parbox[c]{\tabfig\textwidth}{\taboffset
      \includegraphics[width=\tabfig\textwidth,height=\tabfig\textwidth]{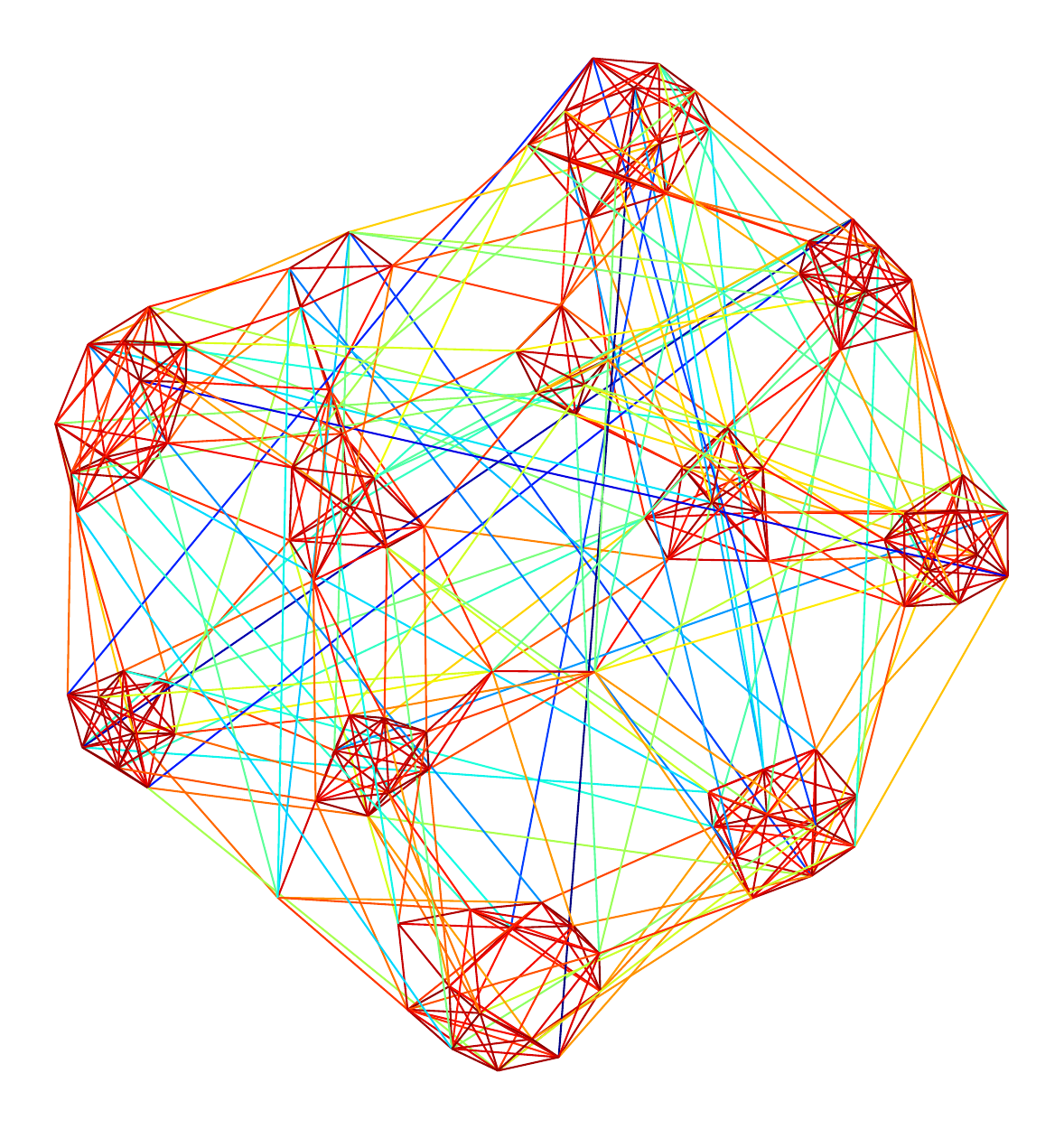}} 
      & \parbox[c]{\tabfig\textwidth}{}
      & \parbox[c]{\tabfig\textwidth}{}
      & \parbox[c]{\tabfig\textwidth}{\taboffset
      \includegraphics[width=\tabfig\textwidth,height=\tabfig\textwidth]{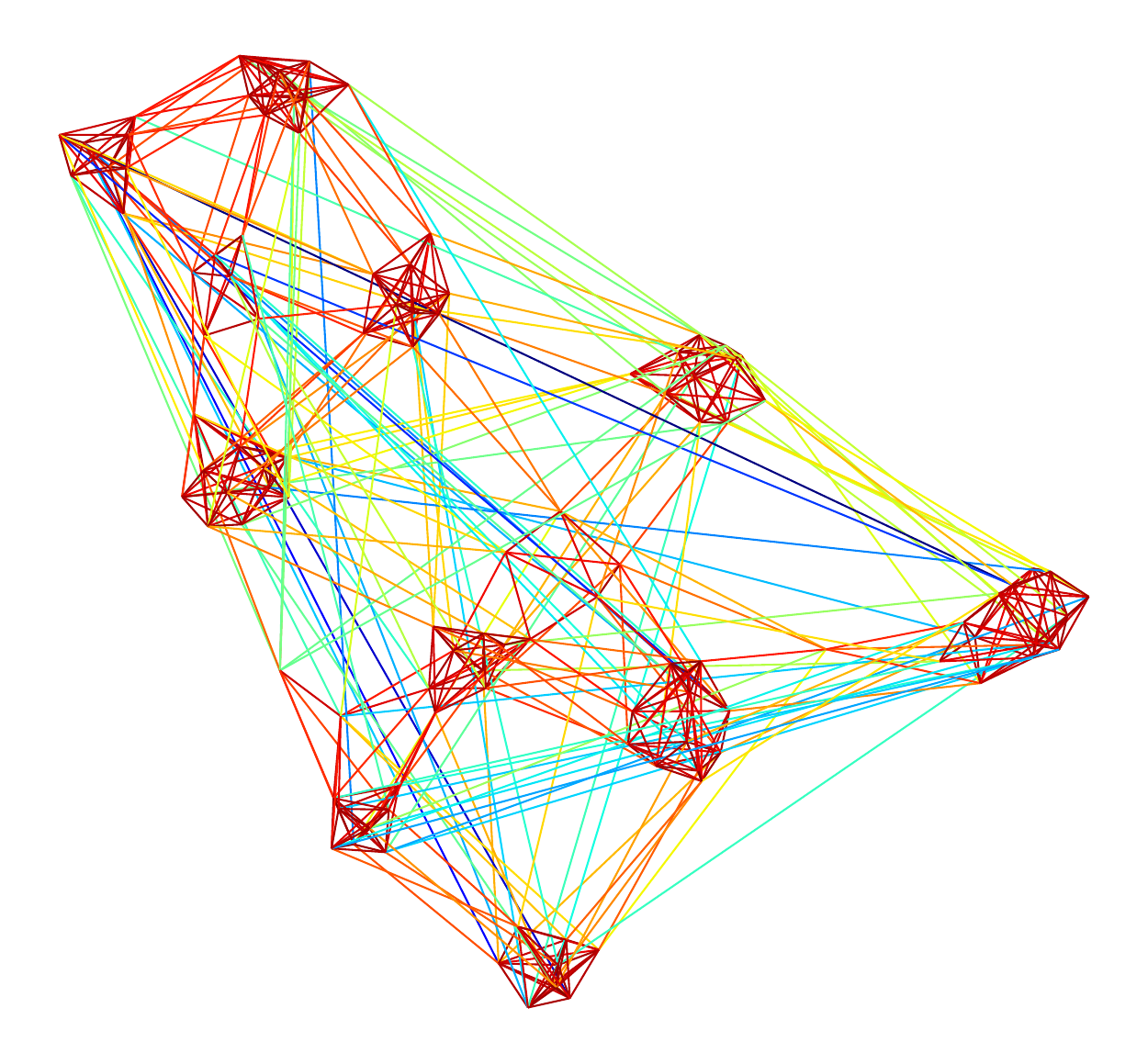}}
      & \parbox[c]{\tabfig\textwidth}{} 
      & \parbox[c]{\tabfig\textwidth}{} 
      & \parbox[c]{\tabfig\textwidth}{\taboffset
      \includegraphics[width=\tabfig\textwidth,height=\tabfig\textwidth]{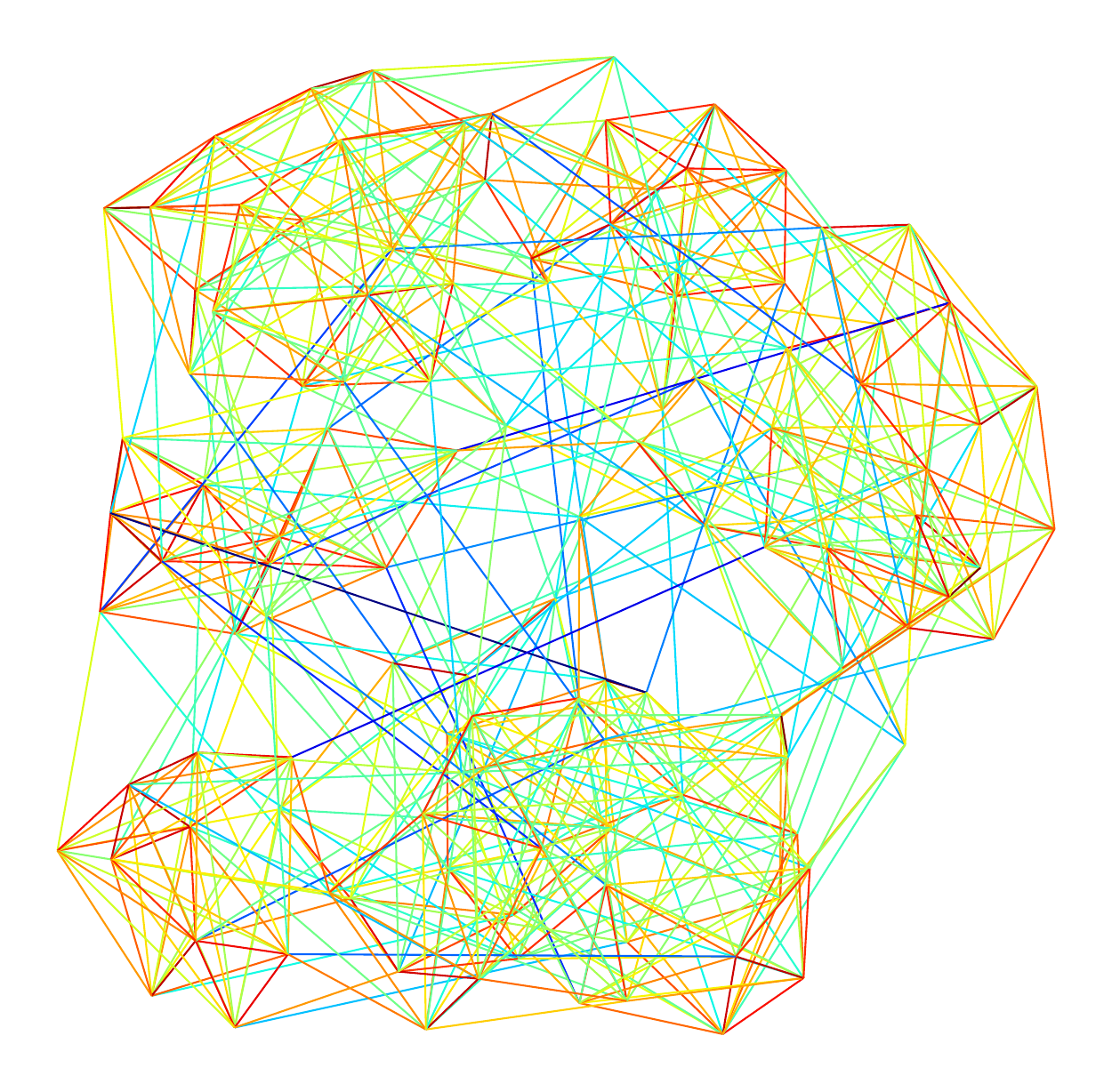}}  \\   
      \hline

      \multirow{2}{*}{\rotatebox[origin=c]{90}{dwt\_419}} &
      \parbox[c]{\tabfig\textwidth}{} 
      & \parbox[c]{\tabfig\textwidth}{
      \includegraphics[width=\tabfig\textwidth,height=\tabfig\textwidth]{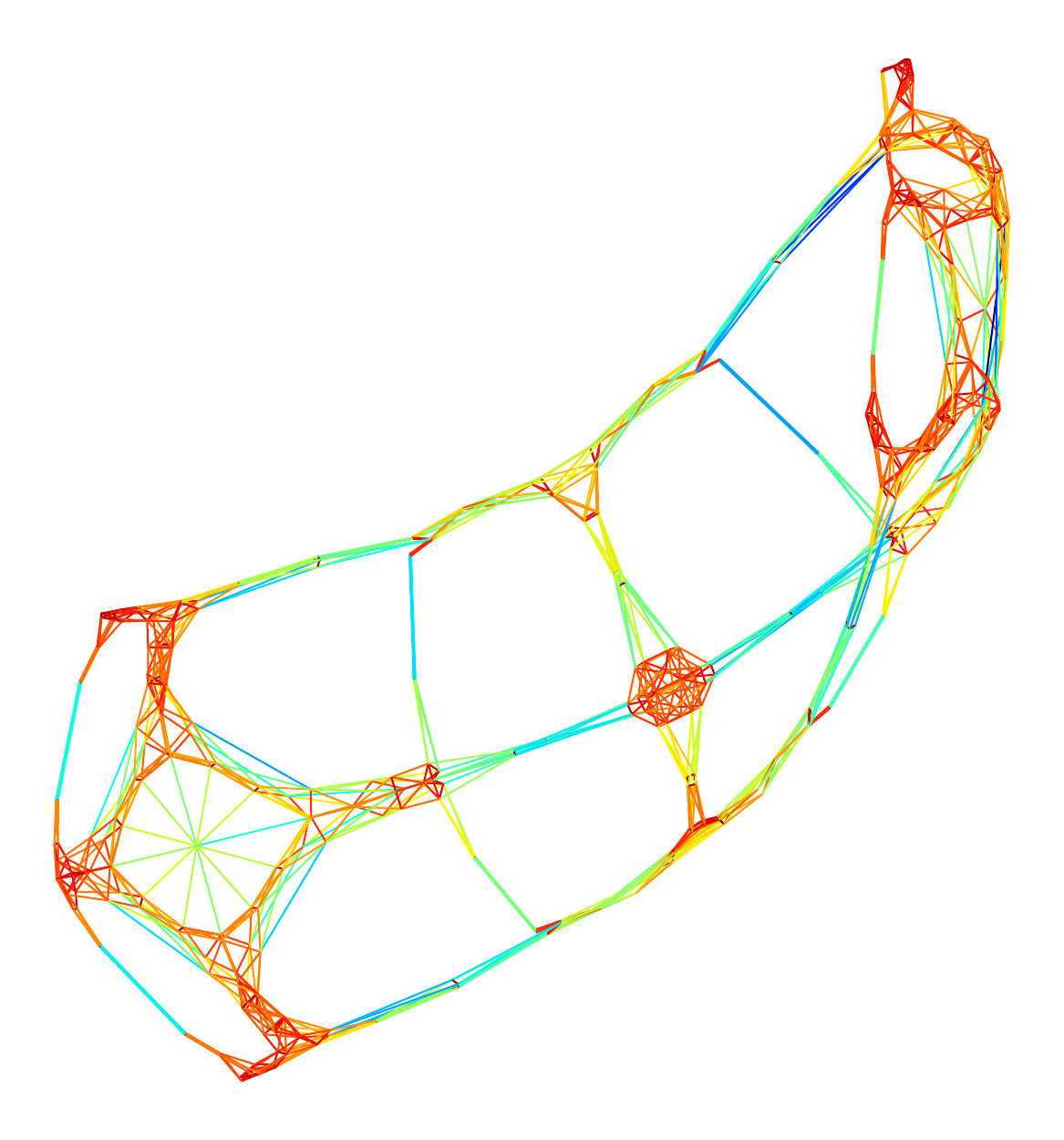}}
      & \parbox[c]{\tabfig\textwidth}{
      \includegraphics[width=\tabfig\textwidth,height=\tabfig\textwidth]{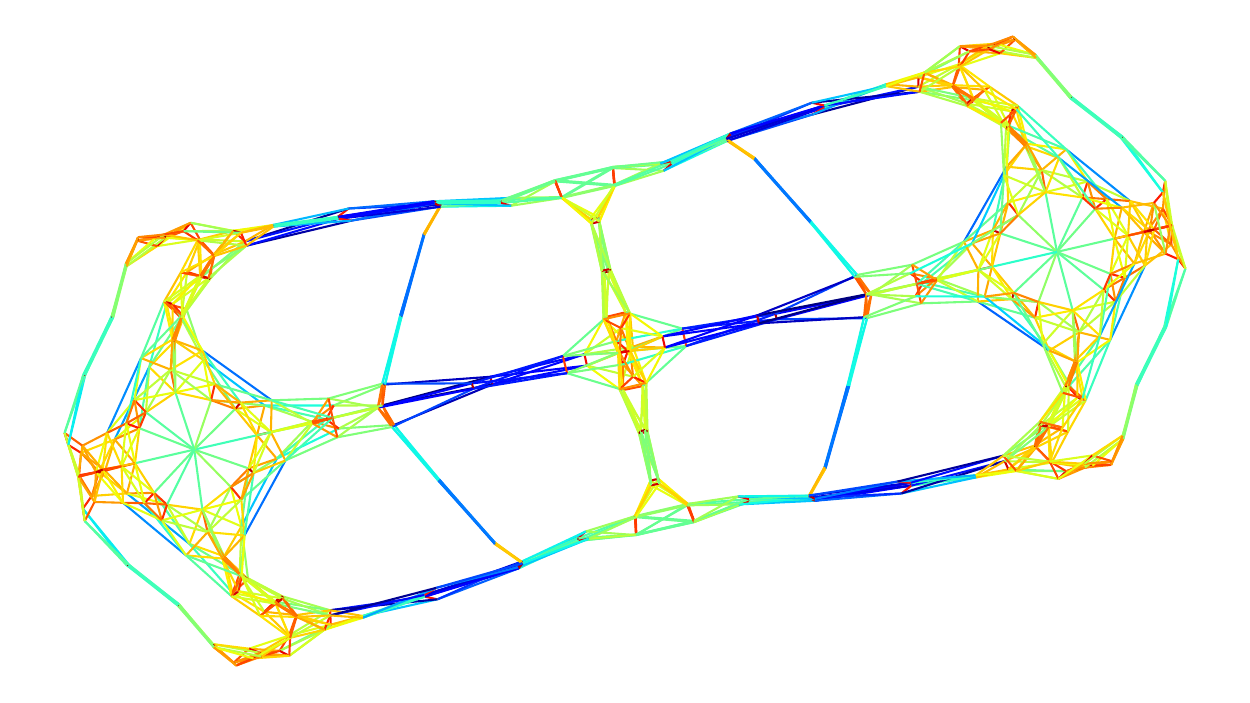}}
      & \parbox[c]{\tabfig\textwidth}{}
      & \parbox[c]{\tabfig\textwidth}{
      \includegraphics[width=\tabfig\textwidth,height=\tabfig\textwidth]{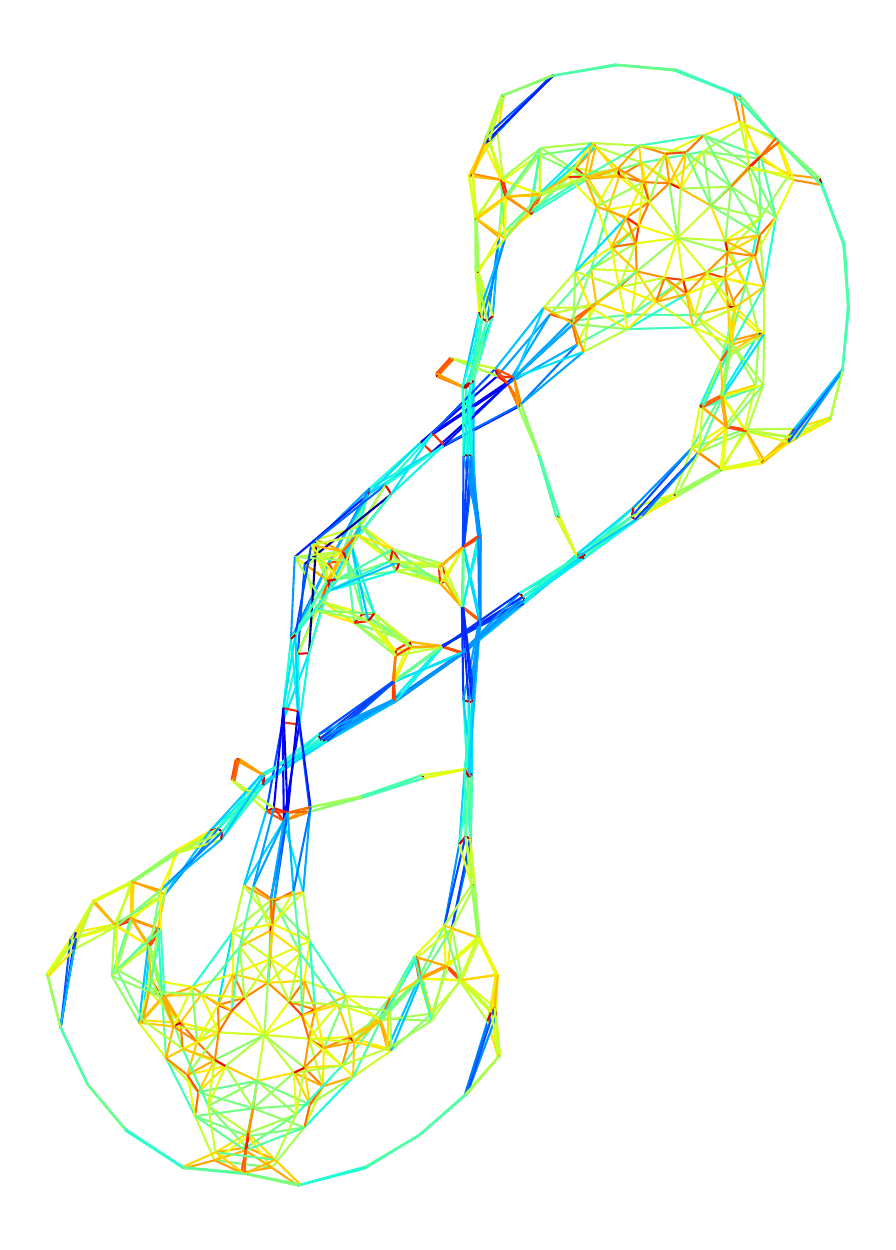}} 
      & \parbox[c]{\tabfig\textwidth}{
      \includegraphics[width=\tabfig\textwidth,height=\tabfig\textwidth]{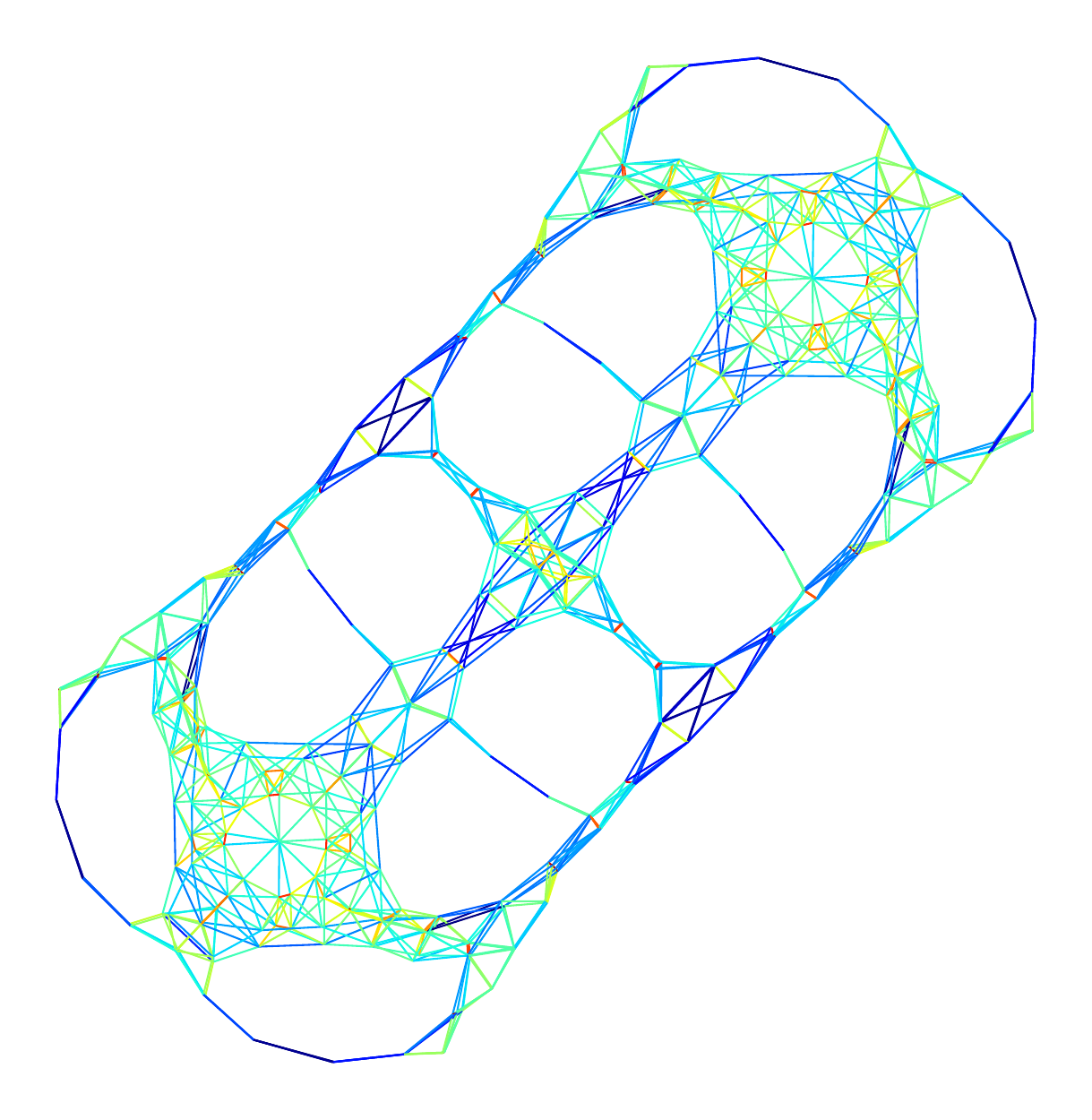}} 
      & \parbox[c]{\tabfig\textwidth}{}  \\
      
      & \parbox[c]{\tabfig\textwidth}{\taboffset
      \includegraphics[width=\tabfig\textwidth,height=\tabfig\textwidth]{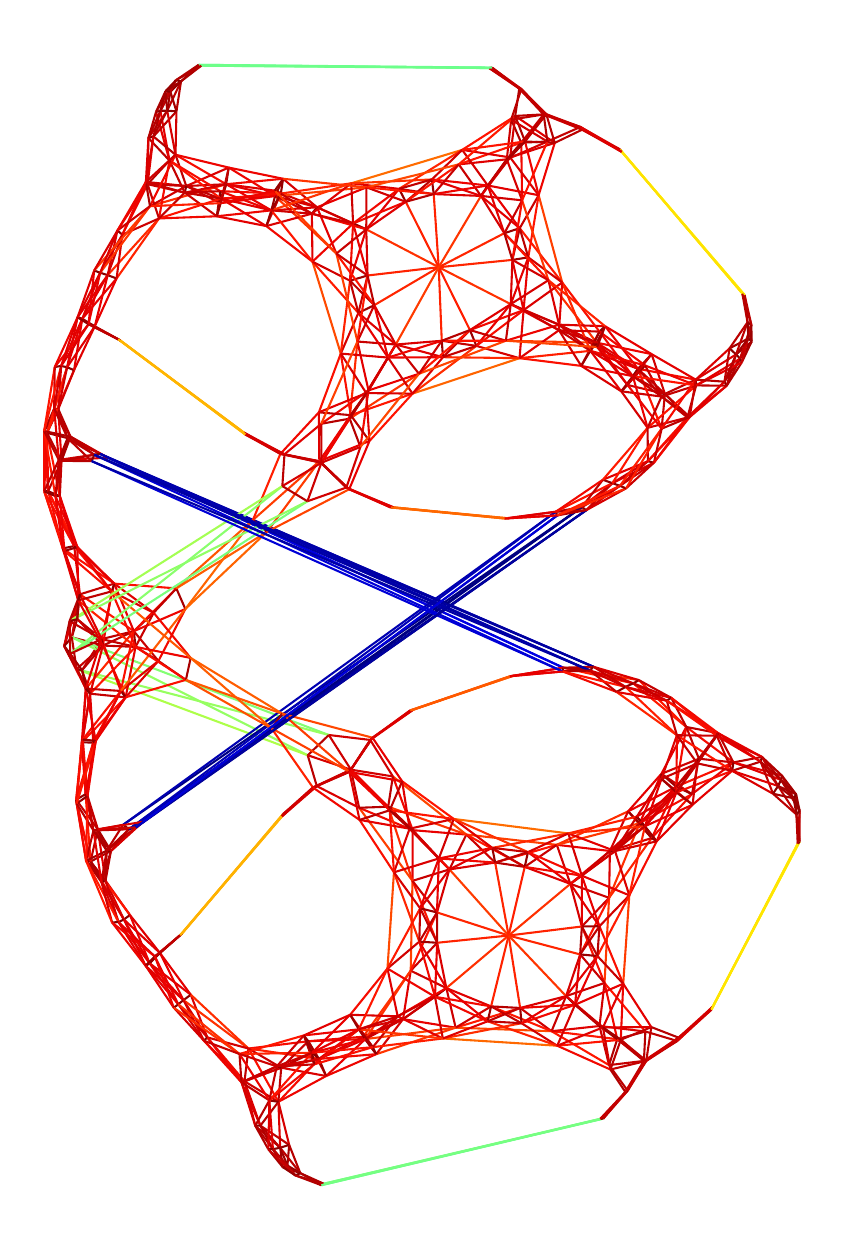}} 
      & \parbox[c]{\tabfig\textwidth}{}
      & \parbox[c]{\tabfig\textwidth}{}
      & \parbox[c]{\tabfig\textwidth}{\taboffset
      \includegraphics[width=\tabfig\textwidth,height=\tabfig\textwidth]{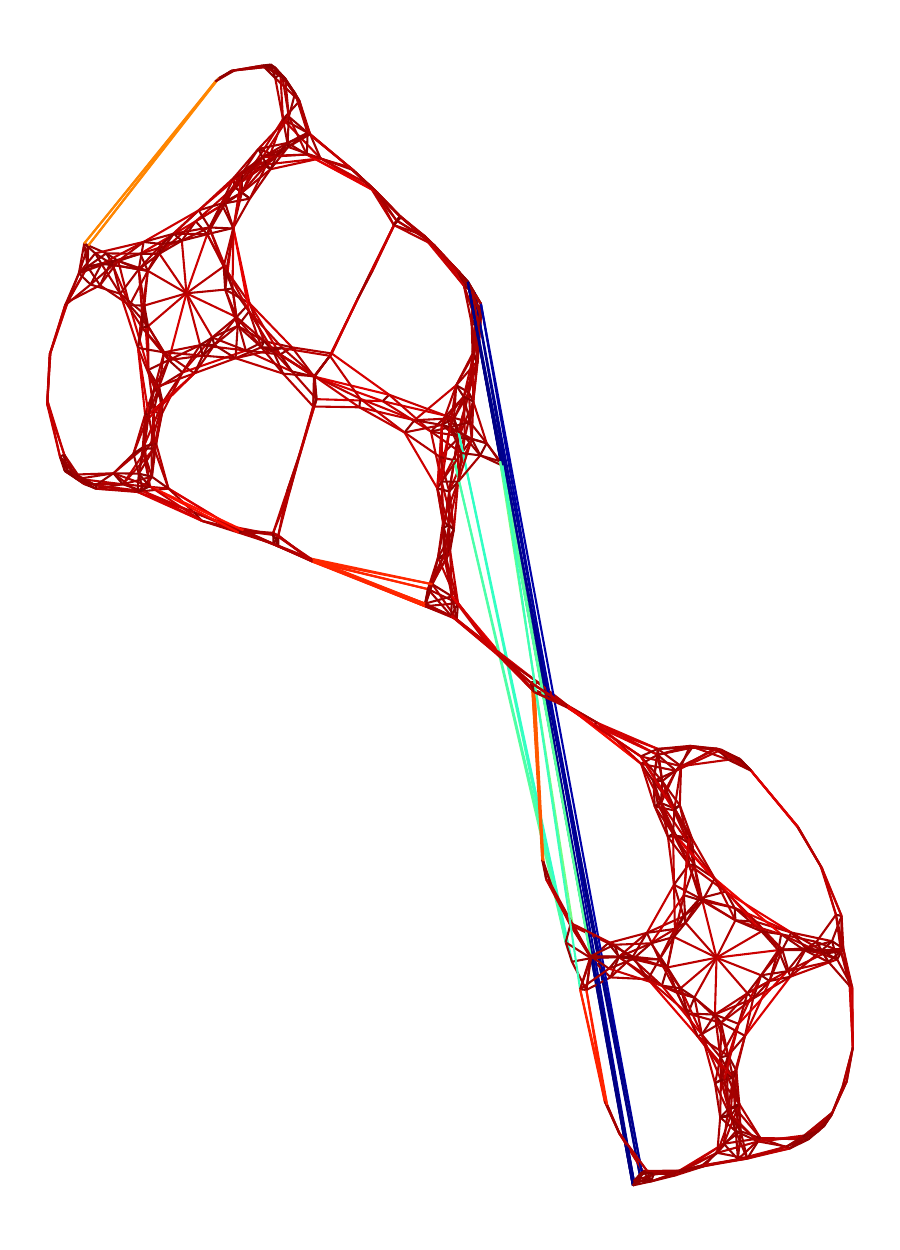}}
      & \parbox[c]{\tabfig\textwidth}{} 
      & \parbox[c]{\tabfig\textwidth}{} 
      & \parbox[c]{\tabfig\textwidth}{\taboffset
      \includegraphics[width=\tabfig\textwidth,height=\tabfig\textwidth]{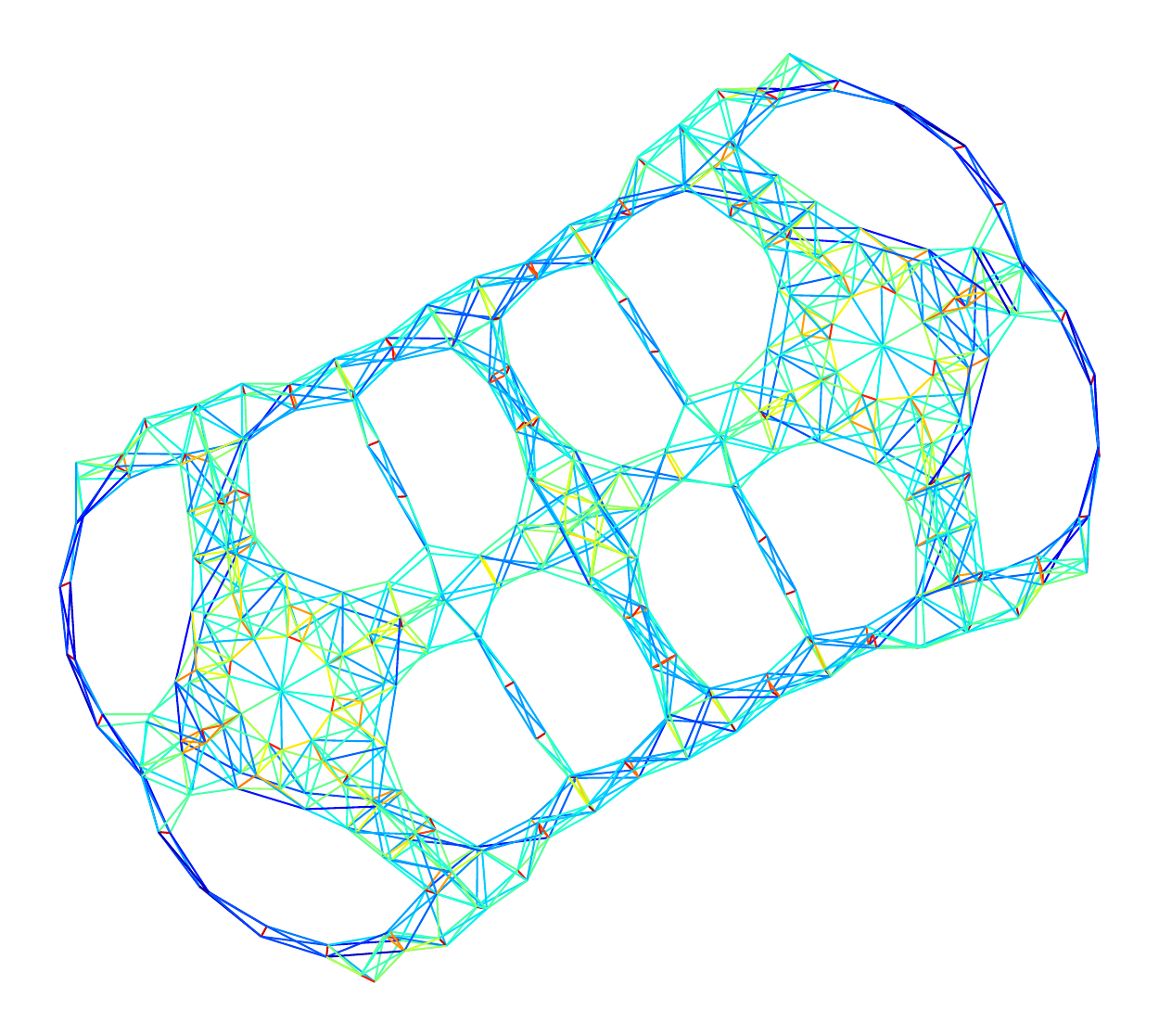}}  \\   
      \hline           

      \multirow{2}{*}{\rotatebox[origin=c]{90}{qh882}} &
      \parbox[c]{\tabfig\textwidth}{} 
      & \parbox[c]{\tabfig\textwidth}{
      \includegraphics[width=\tabfig\textwidth,height=\tabfig\textwidth]{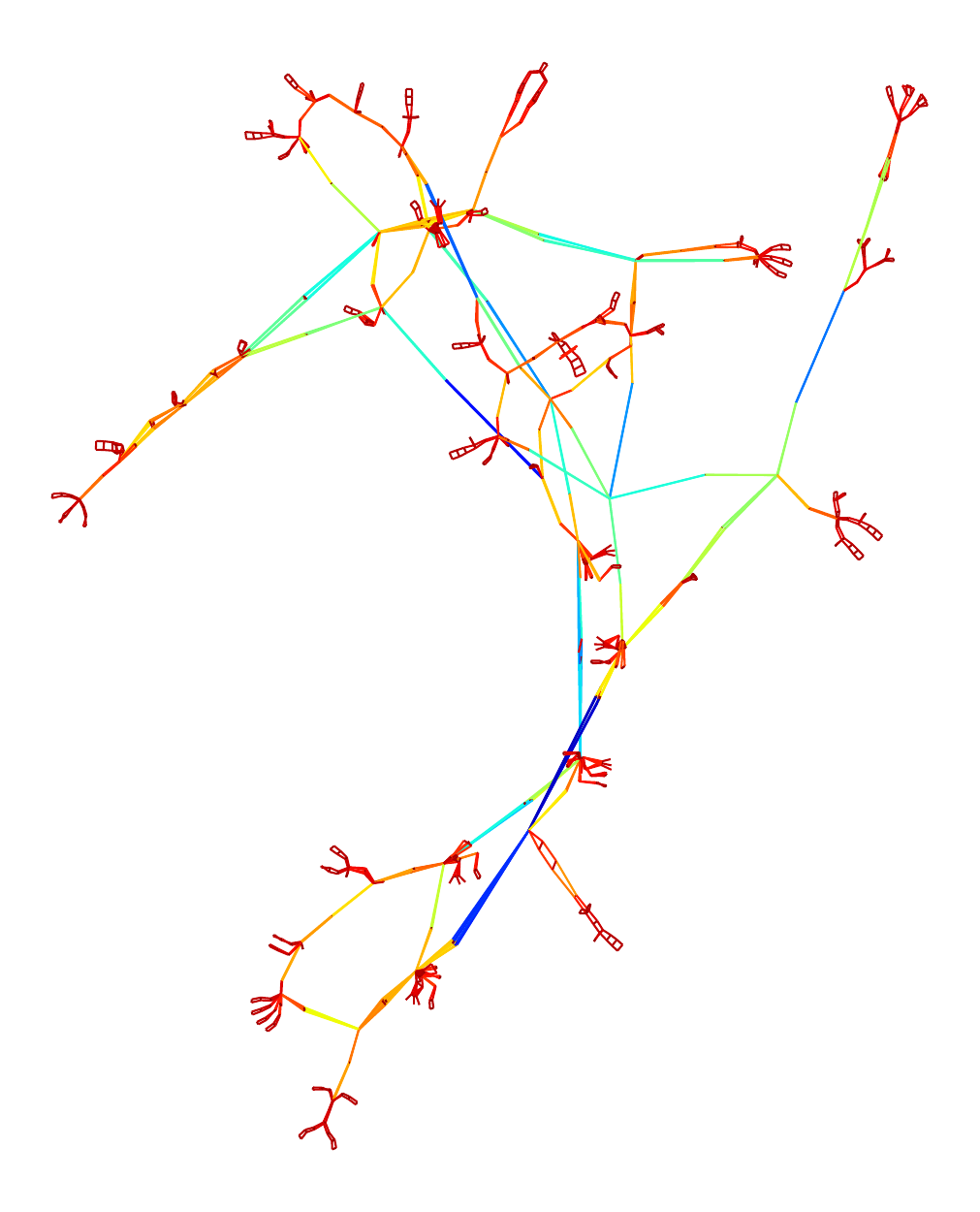}}
      & \parbox[c]{\tabfig\textwidth}{
      \includegraphics[width=\tabfig\textwidth,height=\tabfig\textwidth]{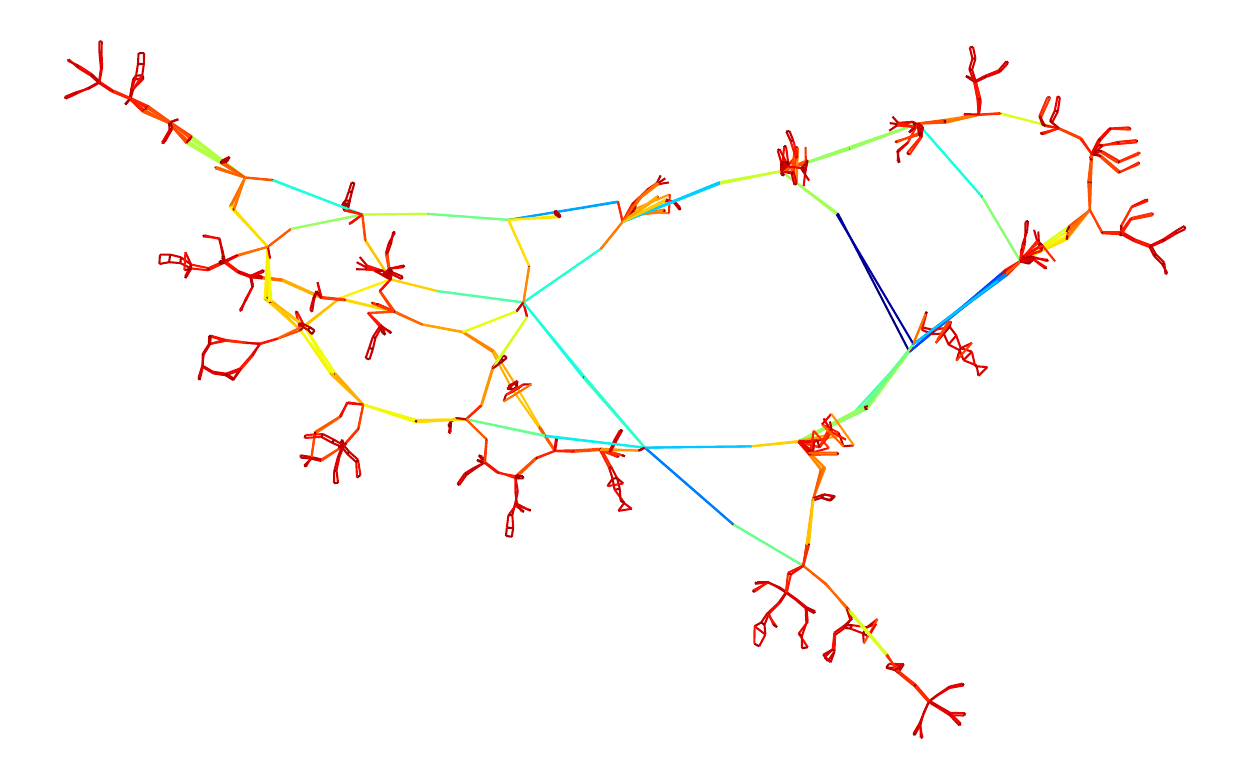}}
      & \parbox[c]{\tabfig\textwidth}{}
      & \parbox[c]{\tabfig\textwidth}{
      \includegraphics[width=\tabfig\textwidth,height=\tabfig\textwidth]{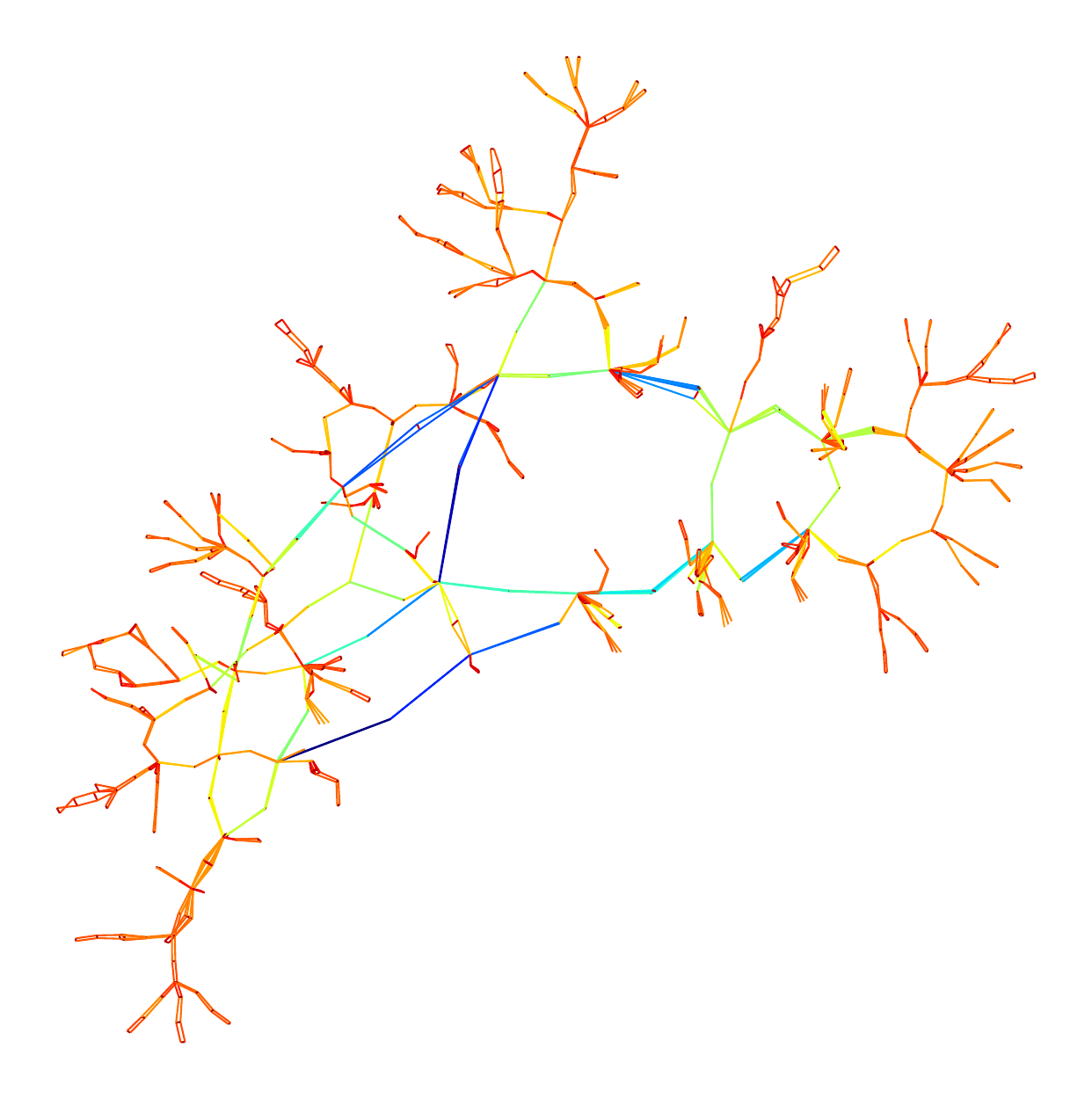}} 
      & \parbox[c]{\tabfig\textwidth}{
      \includegraphics[width=\tabfig\textwidth,height=\tabfig\textwidth]{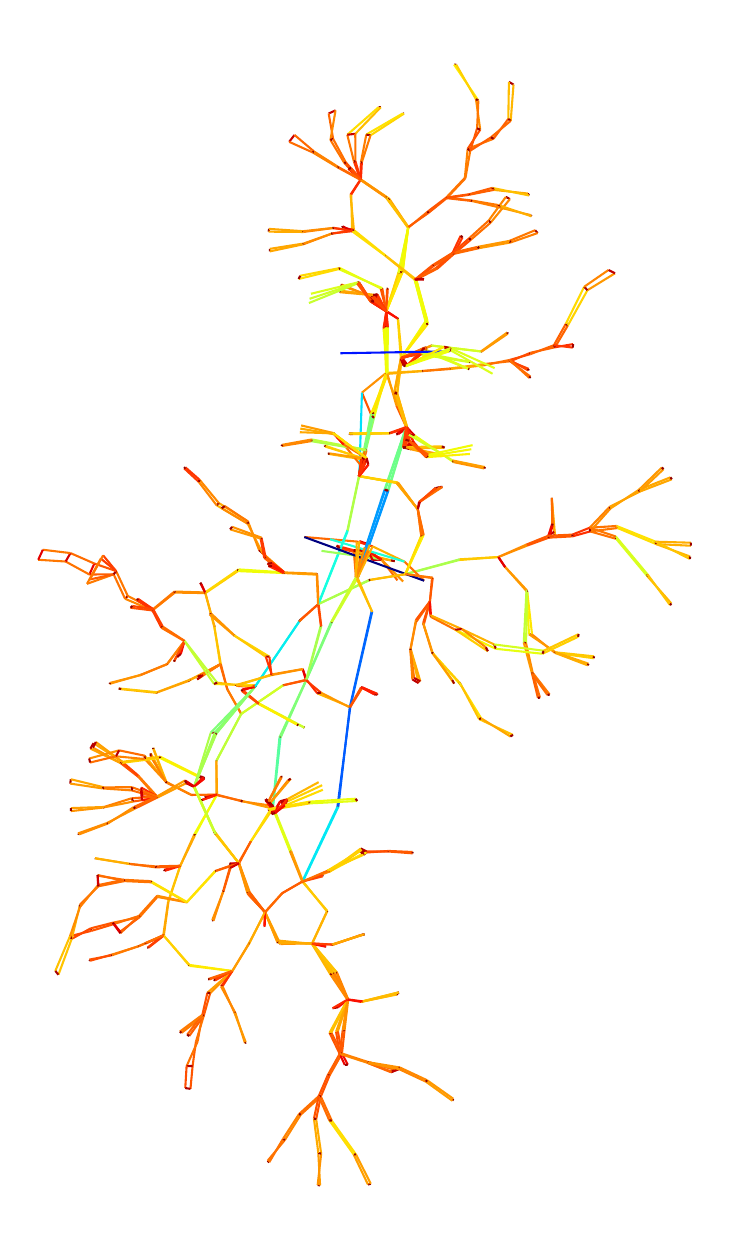}} 
      & \parbox[c]{\tabfig\textwidth}{}  \\
      
      & \parbox[c]{\tabfig\textwidth}{\taboffset \includegraphics[width=\tabfig\textwidth,height=\tabfig\textwidth]{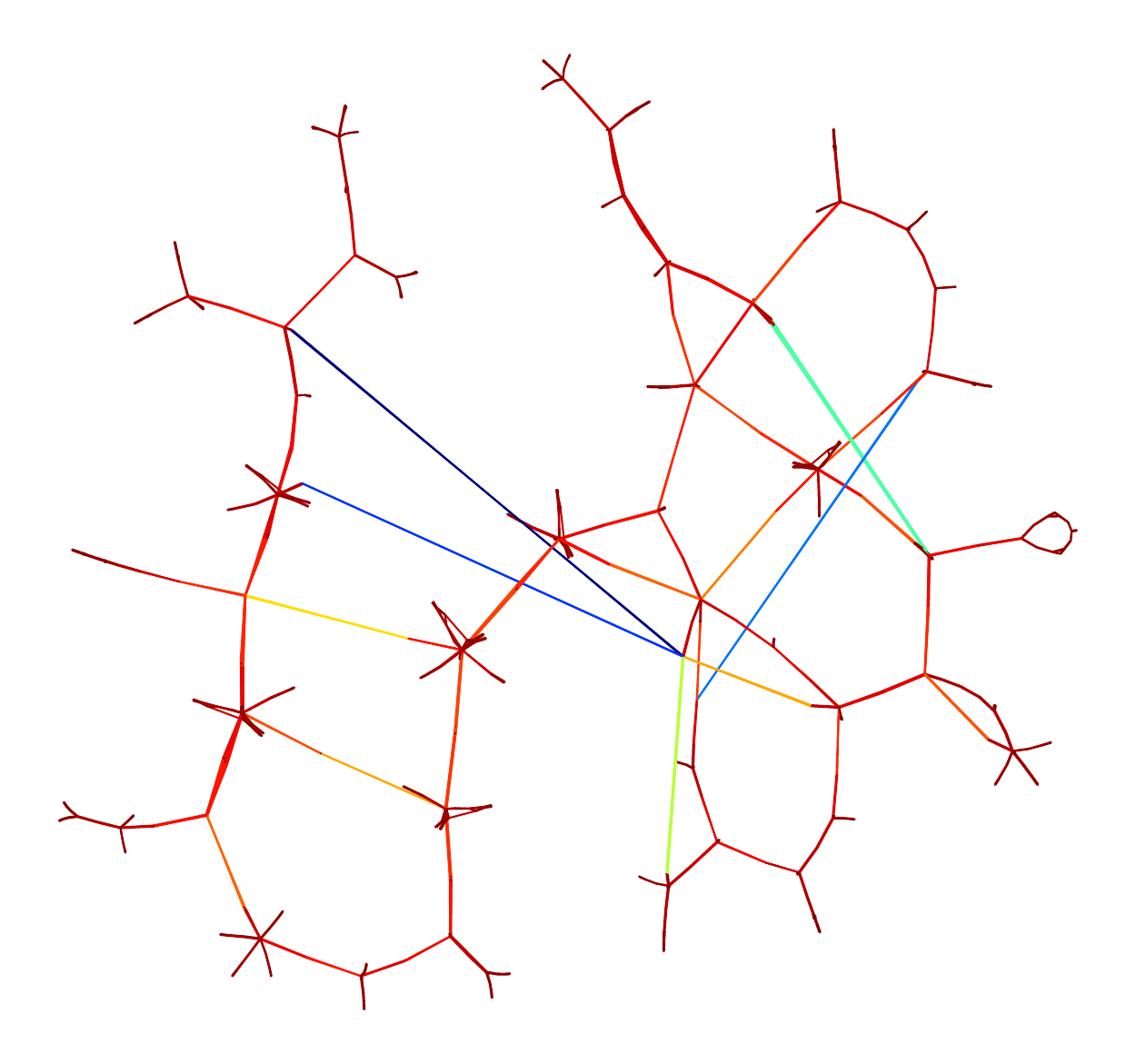}} 
      & \parbox[c]{\tabfig\textwidth}{}
      & \parbox[c]{\tabfig\textwidth}{}
      & \parbox[c]{\tabfig\textwidth}{\taboffset
      \includegraphics[width=\tabfig\textwidth,height=\tabfig\textwidth]{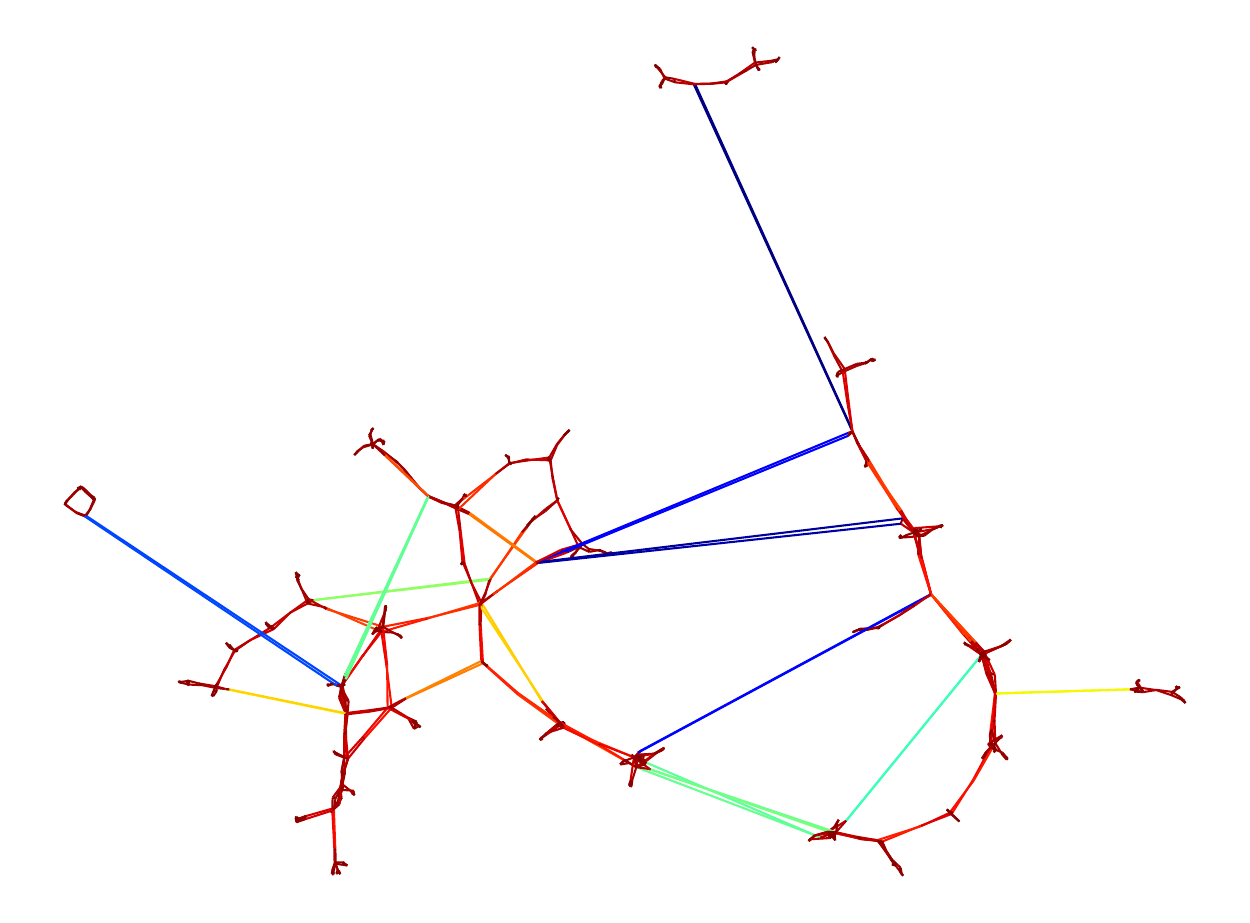}}
      & \parbox[c]{\tabfig\textwidth}{} 
      & \parbox[c]{\tabfig\textwidth}{} 
      & \parbox[c]{\tabfig\textwidth}{\taboffset
      \includegraphics[width=\tabfig\textwidth,height=\tabfig\textwidth]{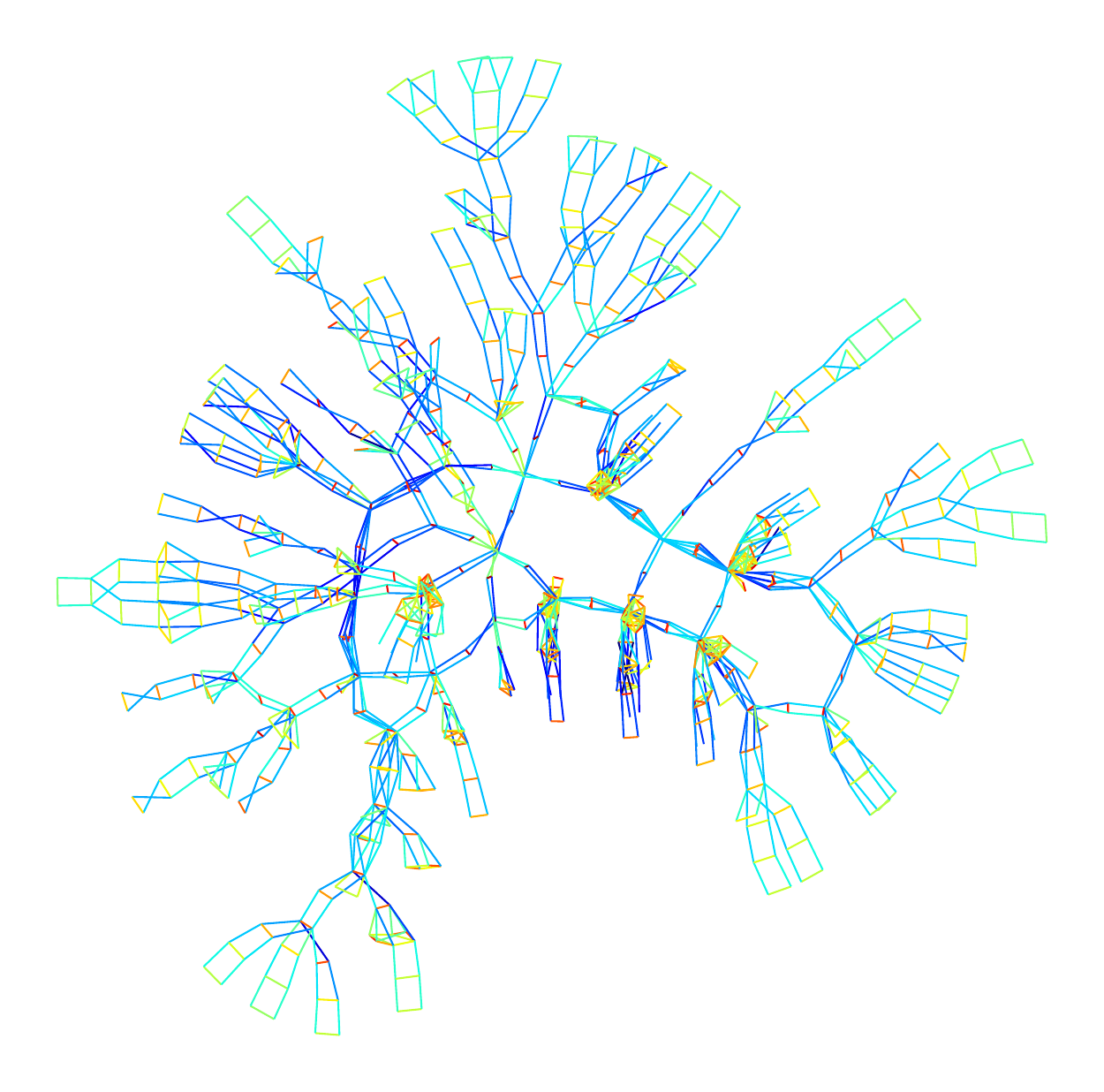}}  \\   
      \hline

     \multirow{2}{*}{\rotatebox[origin=c]{90}{\parbox{1.0cm} {\centering netscience}}}  &
      \parbox[c]{\tabfig\textwidth}{} 
      & \parbox[c]{\tabfig\textwidth}{
      \includegraphics[width=\tabfig\textwidth,height=\tabfig\textwidth]{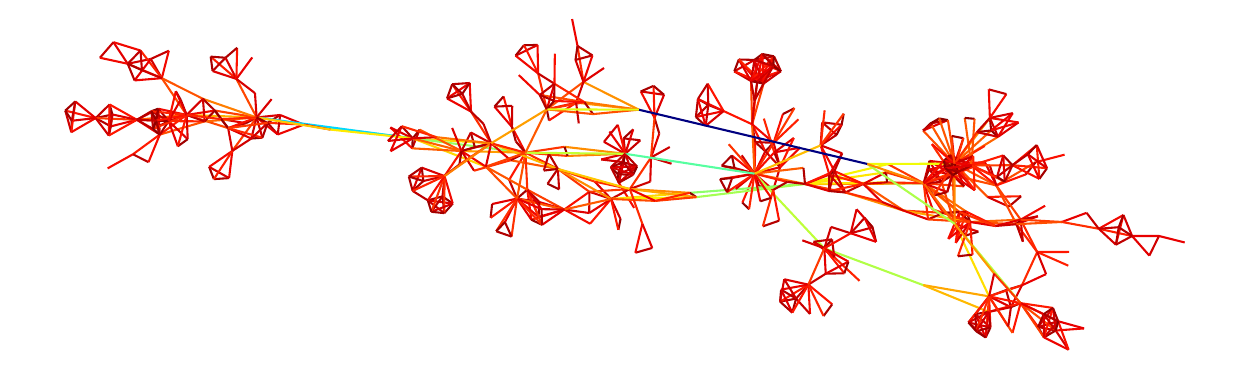}}
      & \parbox[c]{\tabfig\textwidth}{
      \includegraphics[width=\tabfig\textwidth,height=\tabfig\textwidth]{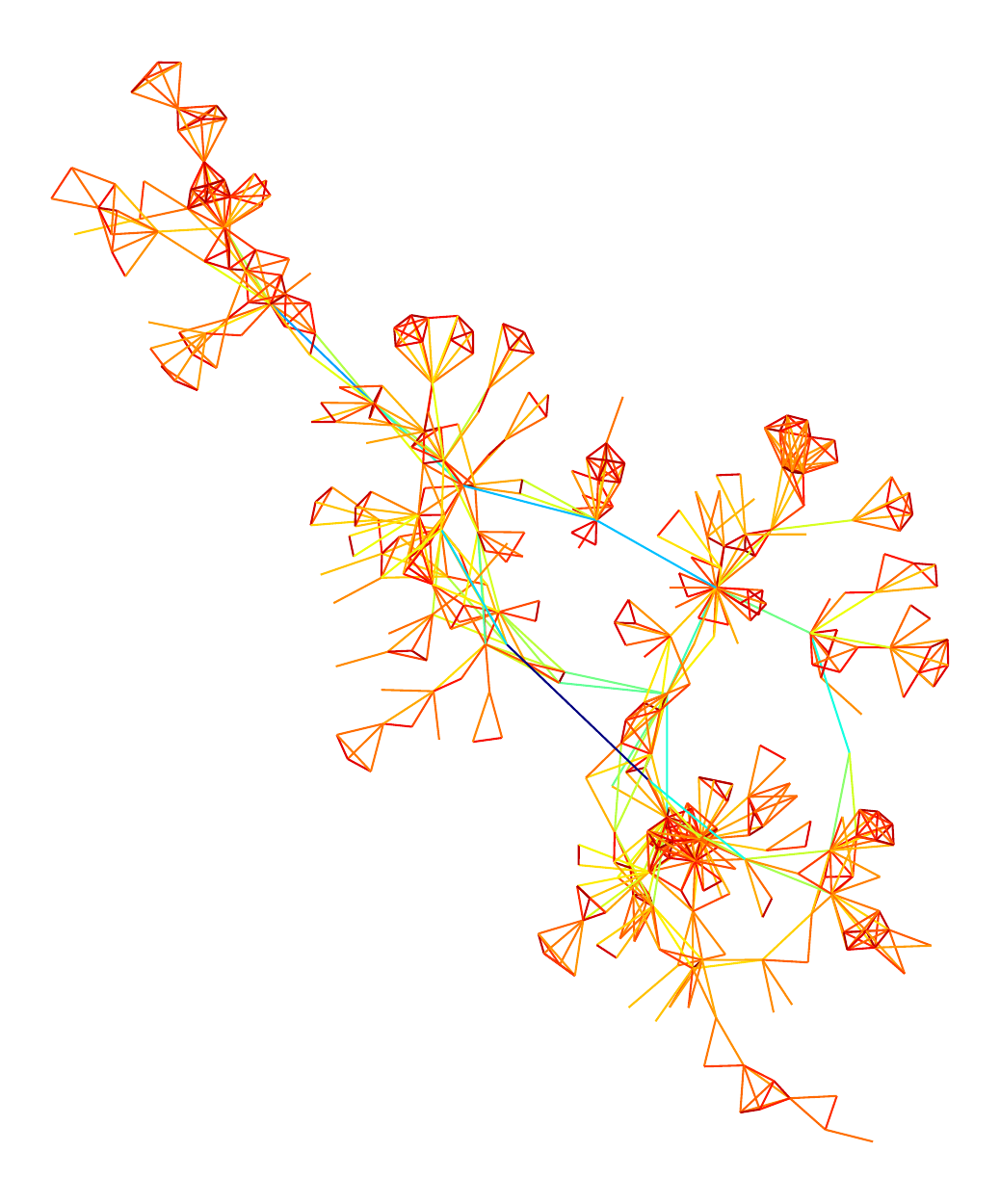}}
      & \parbox[c]{\tabfig\textwidth}{}
      & \parbox[c]{\tabfig\textwidth}{
      \includegraphics[width=\tabfig\textwidth,height=\tabfig\textwidth]{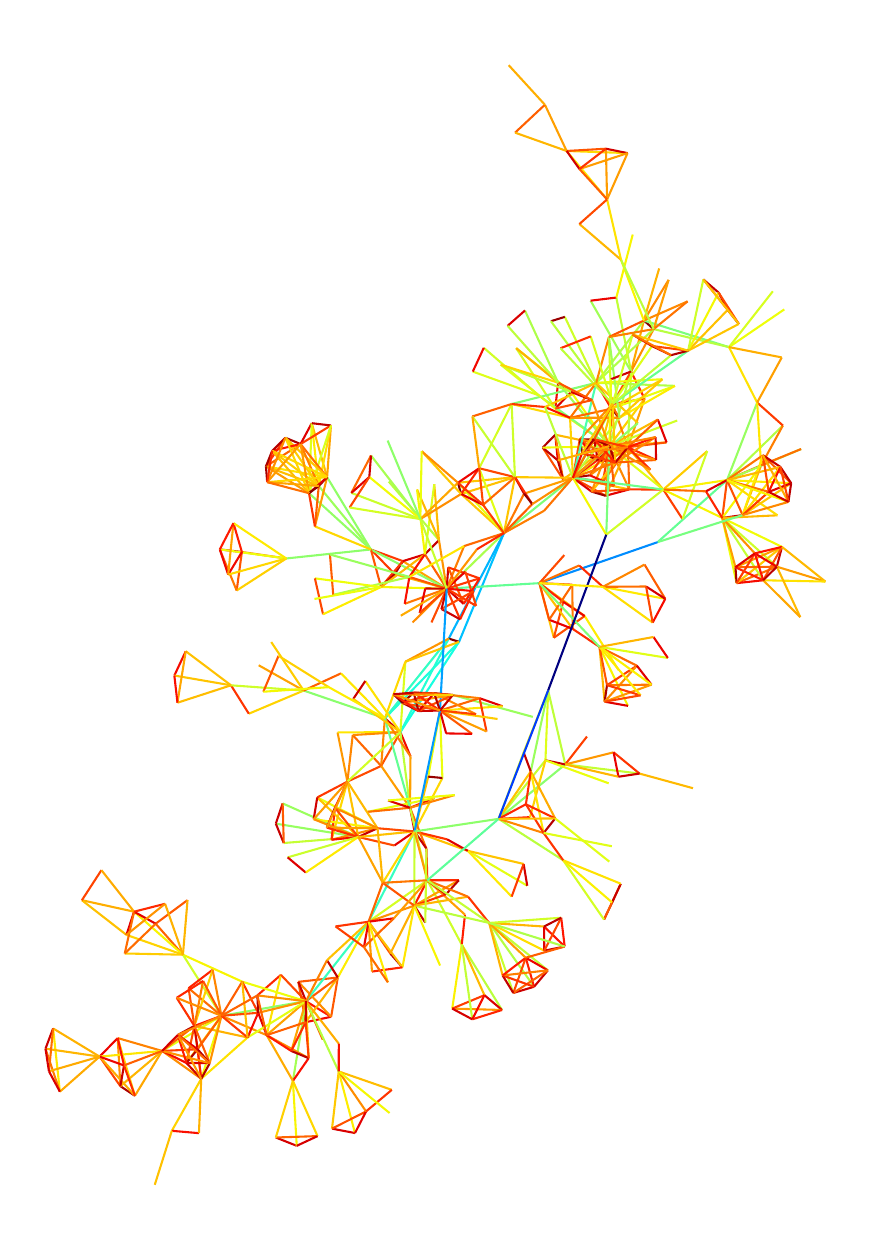}} 
      & \parbox[c]{\tabfig\textwidth}{
      \includegraphics[width=\tabfig\textwidth,height=\tabfig\textwidth]{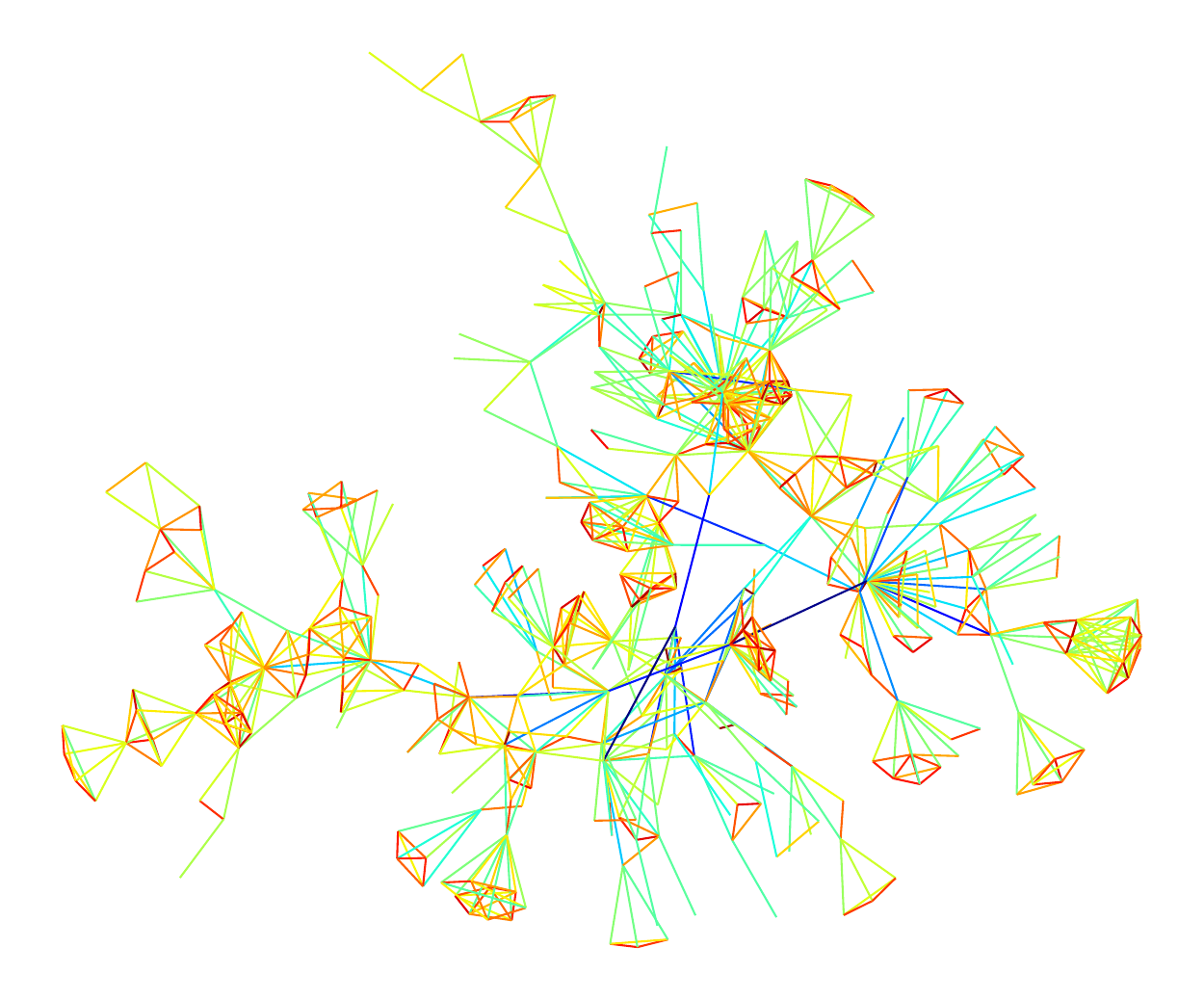}} 
      & \parbox[c]{\tabfig\textwidth}{}  \\
      
      & \parbox[c]{\tabfig\textwidth}{\taboffset\includegraphics[width=\tabfig\textwidth,height=\tabfig\textwidth]{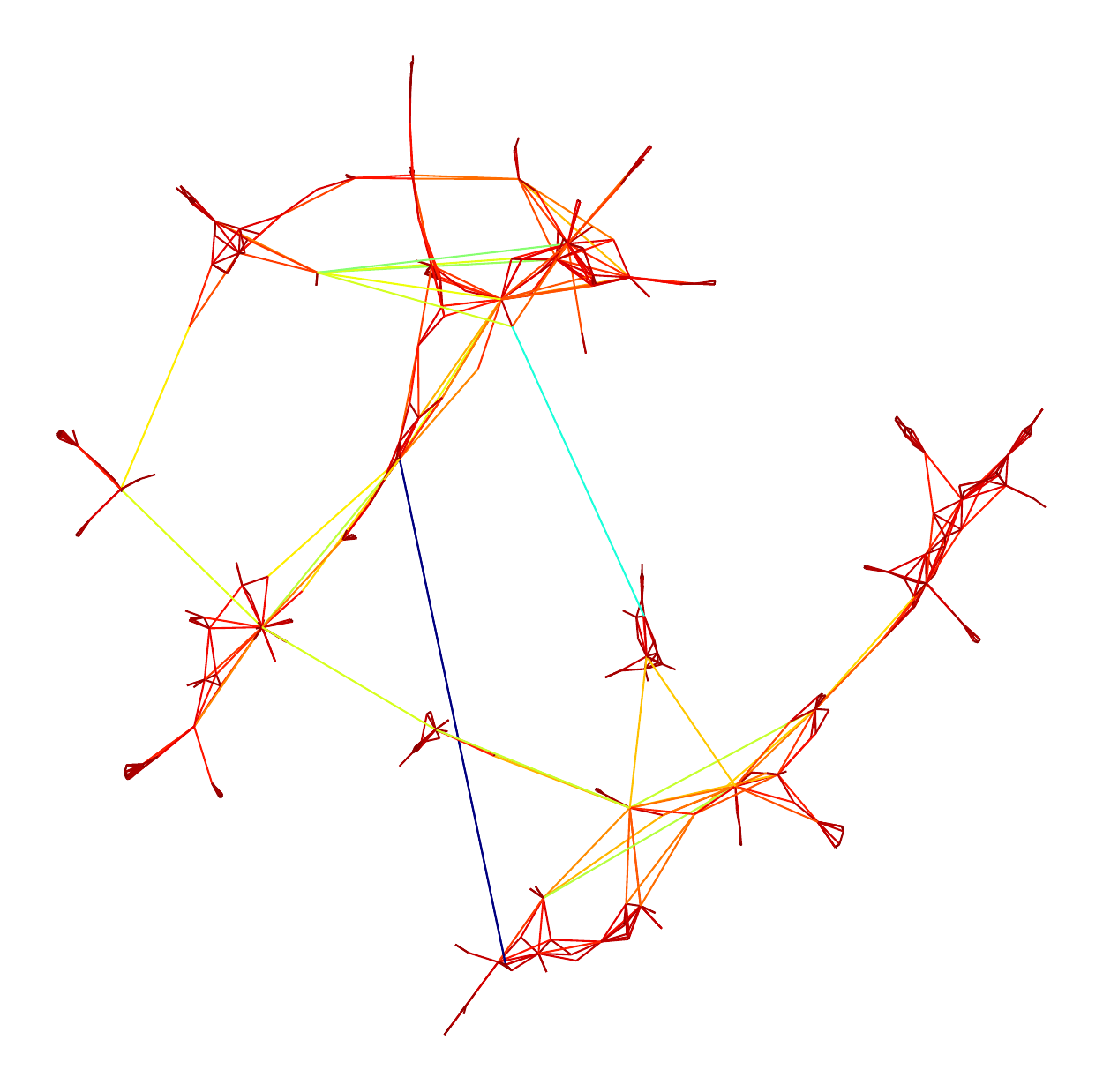}} 
      & \parbox[c]{\tabfig\textwidth}{}
      & \parbox[c]{\tabfig\textwidth}{}
      & \parbox[c]{\tabfig\textwidth}{\taboffset
      \includegraphics[width=\tabfig\textwidth,height=\tabfig\textwidth]{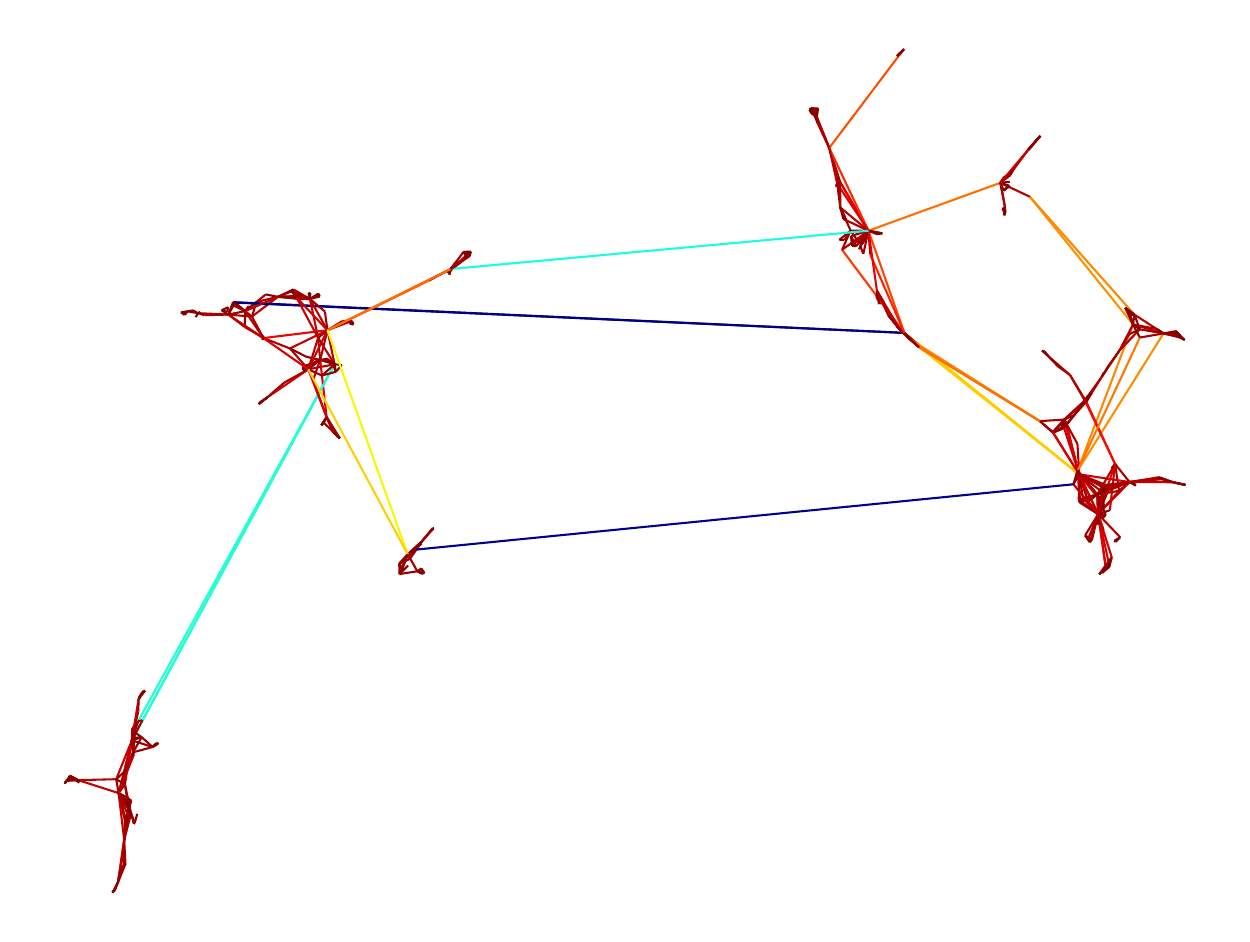}}
      & \parbox[c]{\tabfig\textwidth}{} 
      & \parbox[c]{\tabfig\textwidth}{} 
      & \parbox[c]{\tabfig\textwidth}{\taboffset
      \includegraphics[width=\tabfig\textwidth,height=\tabfig\textwidth]{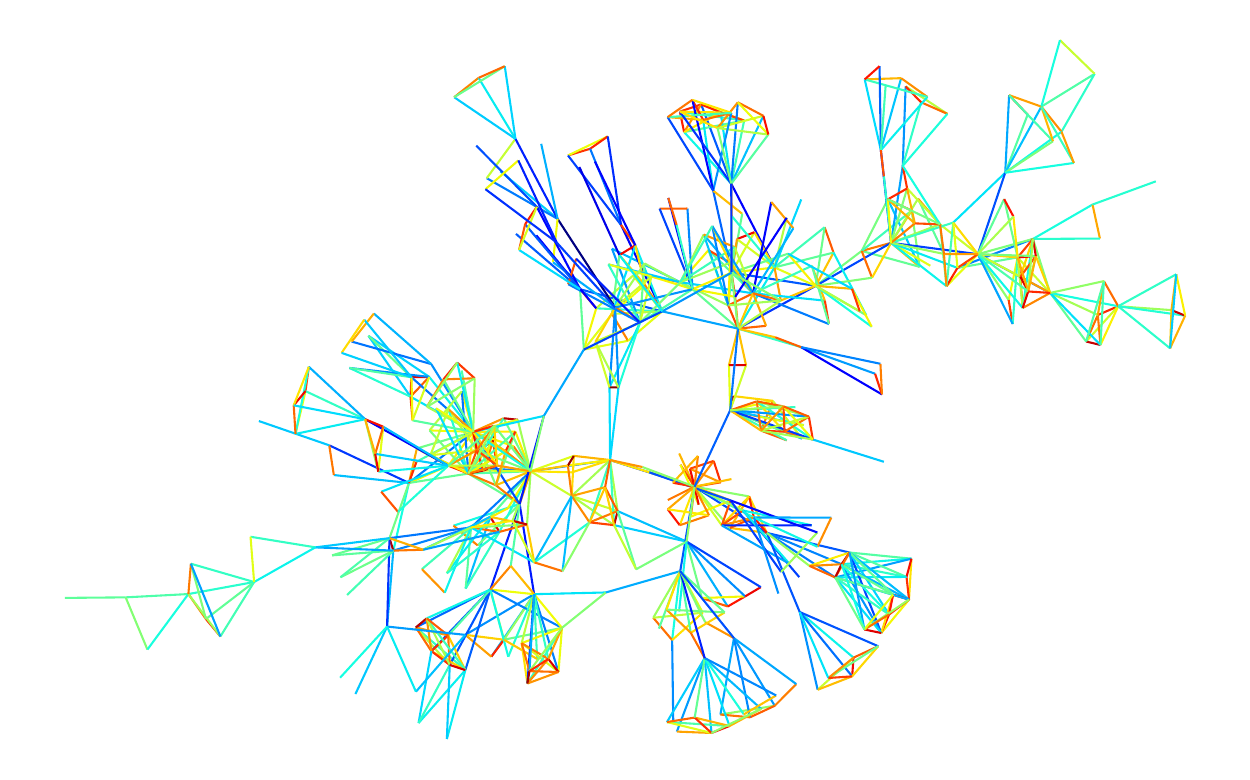}}  \\   
      \hline

     \multirow{2}{*}{\rotatebox[origin=c]{90}{EVA}}  &
      \parbox[c]{\tabfig\textwidth}{} 
      & \parbox[c]{\tabfig\textwidth}{
      \includegraphics[width=\tabfig\textwidth,height=\tabfig\textwidth]{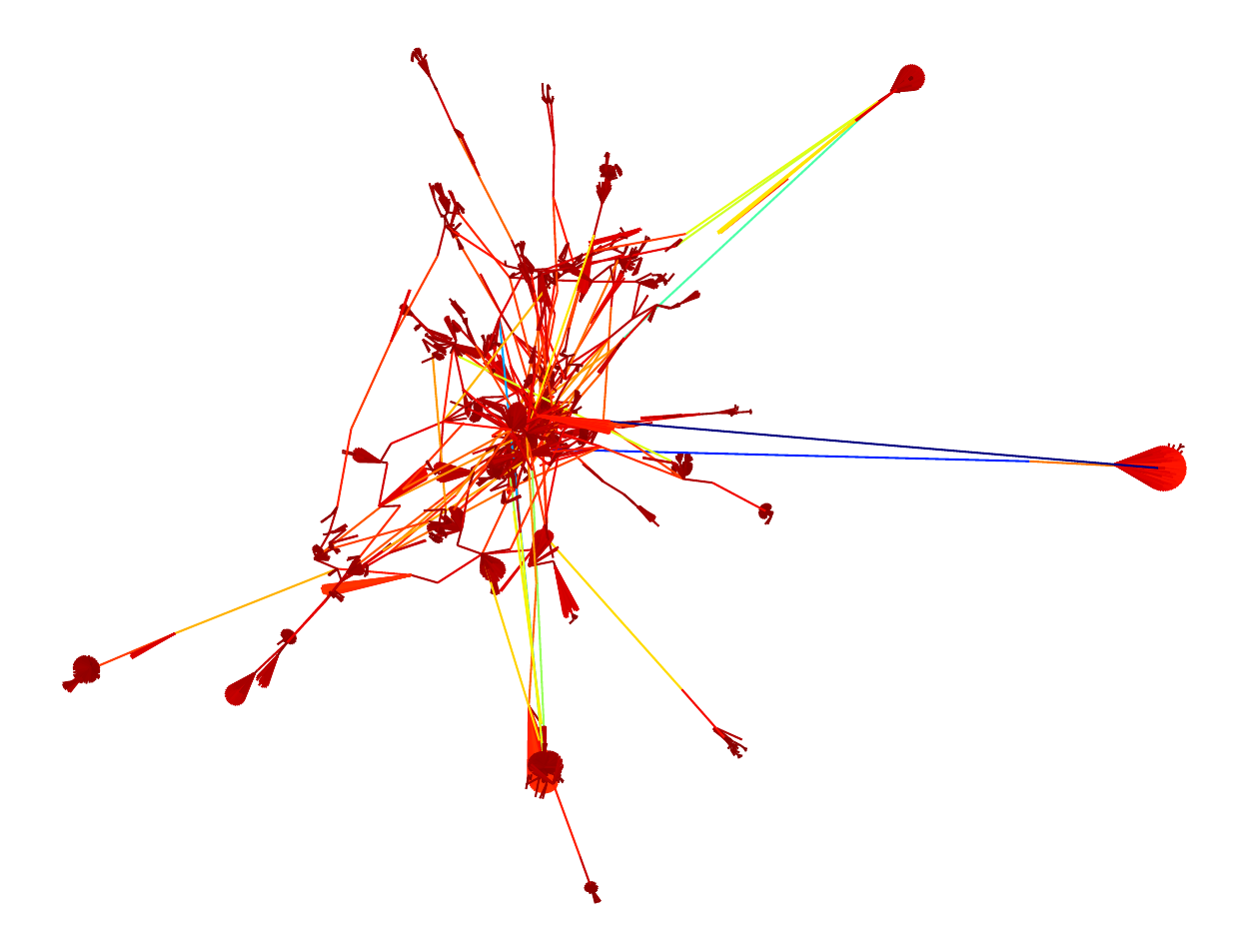}}
      & \parbox[c]{\tabfig\textwidth}{
      \includegraphics[width=\tabfig\textwidth,height=\tabfig\textwidth]{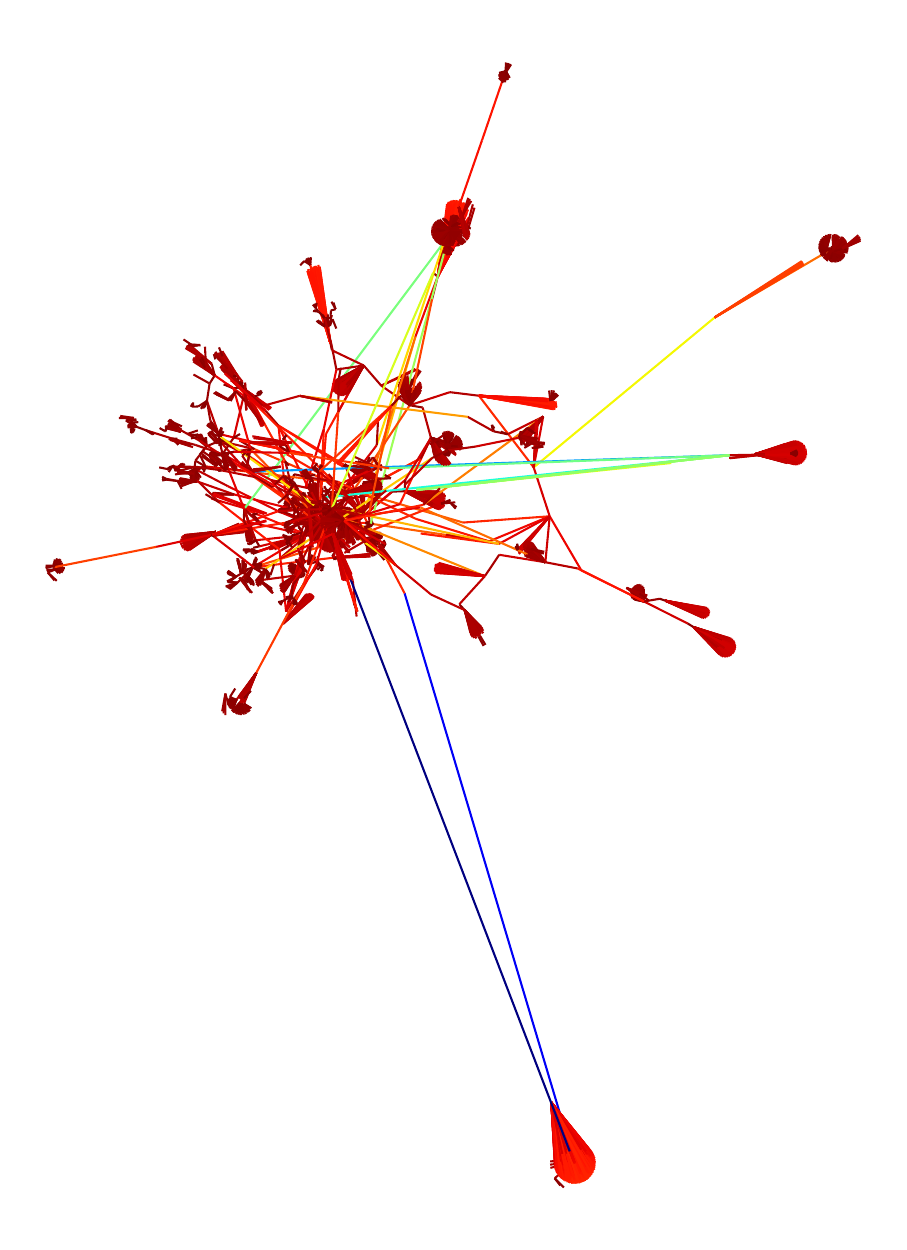}}
      & \parbox[c]{\tabfig\textwidth}{}
      & \parbox[c]{\tabfig\textwidth}{
      \includegraphics[width=\tabfig\textwidth,height=\tabfig\textwidth]{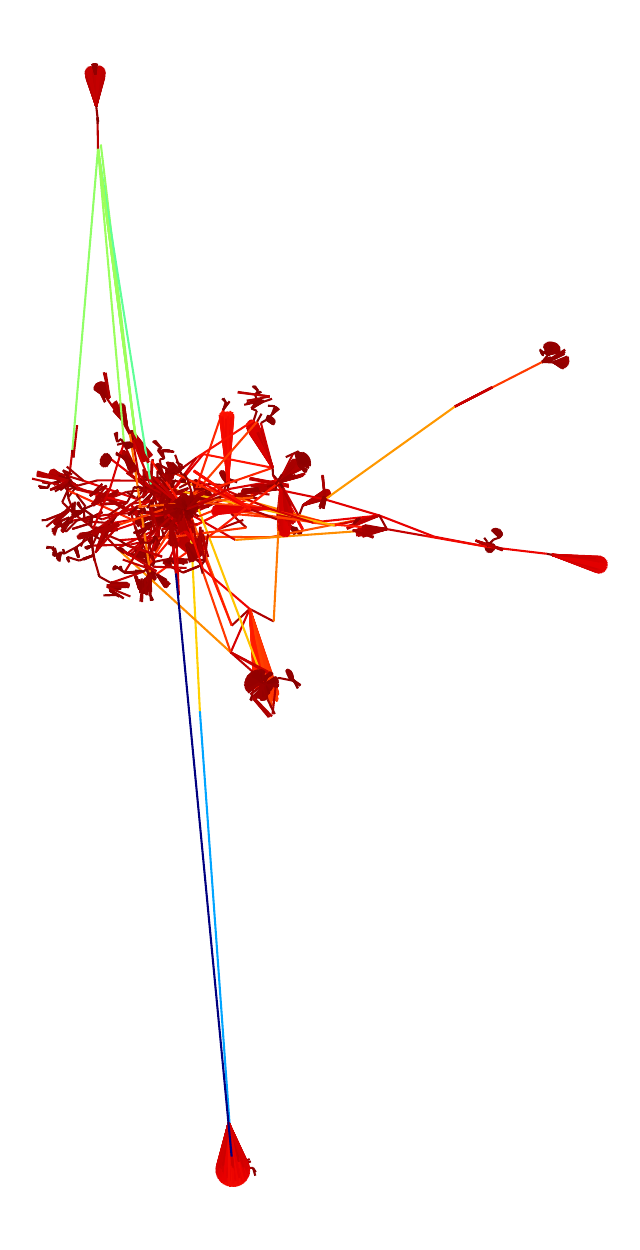}} 
      & \parbox[c]{\tabfig\textwidth}{
      \includegraphics[width=\tabfig\textwidth,height=\tabfig\textwidth]{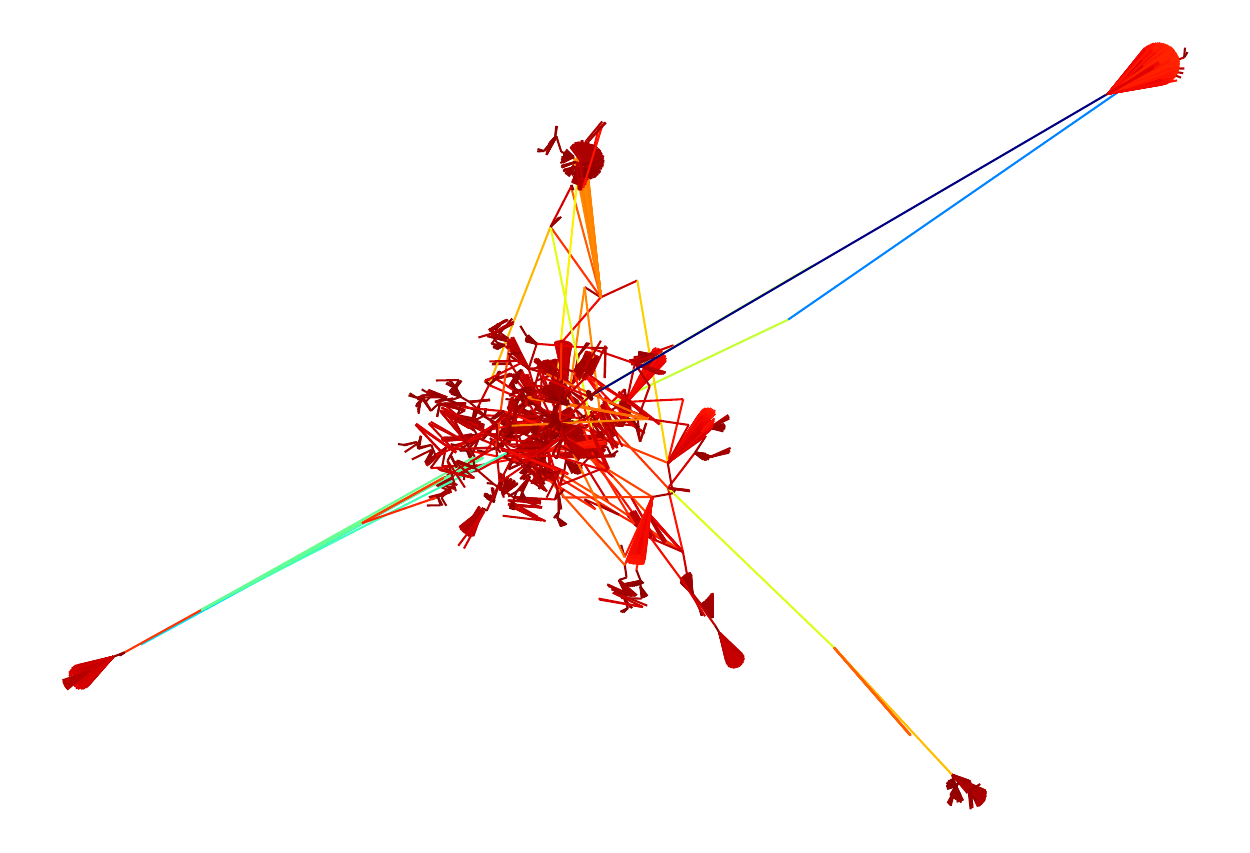}} 
      & \parbox[c]{\tabfig\textwidth}{}  \\
      
       &
      \parbox[c]{\tabfig\textwidth}{
      \taboffset\includegraphics[width=\tabfig\textwidth,height=\tabfig\textwidth]{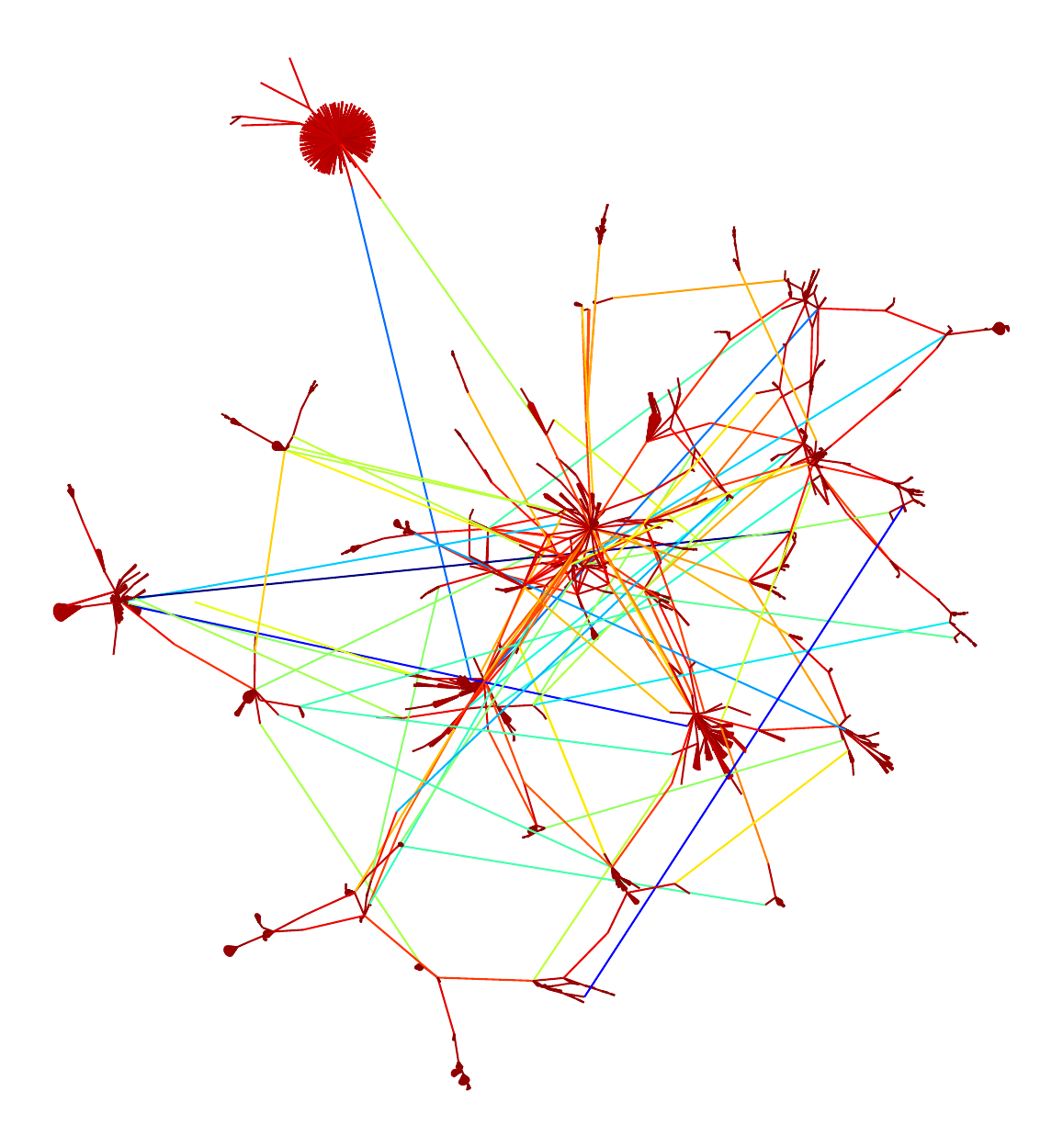}} 
      & \parbox[c]{\tabfig\textwidth}{}
      & \parbox[c]{\tabfig\textwidth}{}
      & \parbox[c]{\tabfig\textwidth}{\taboffset
      \includegraphics[width=\tabfig\textwidth,height=\tabfig\textwidth]{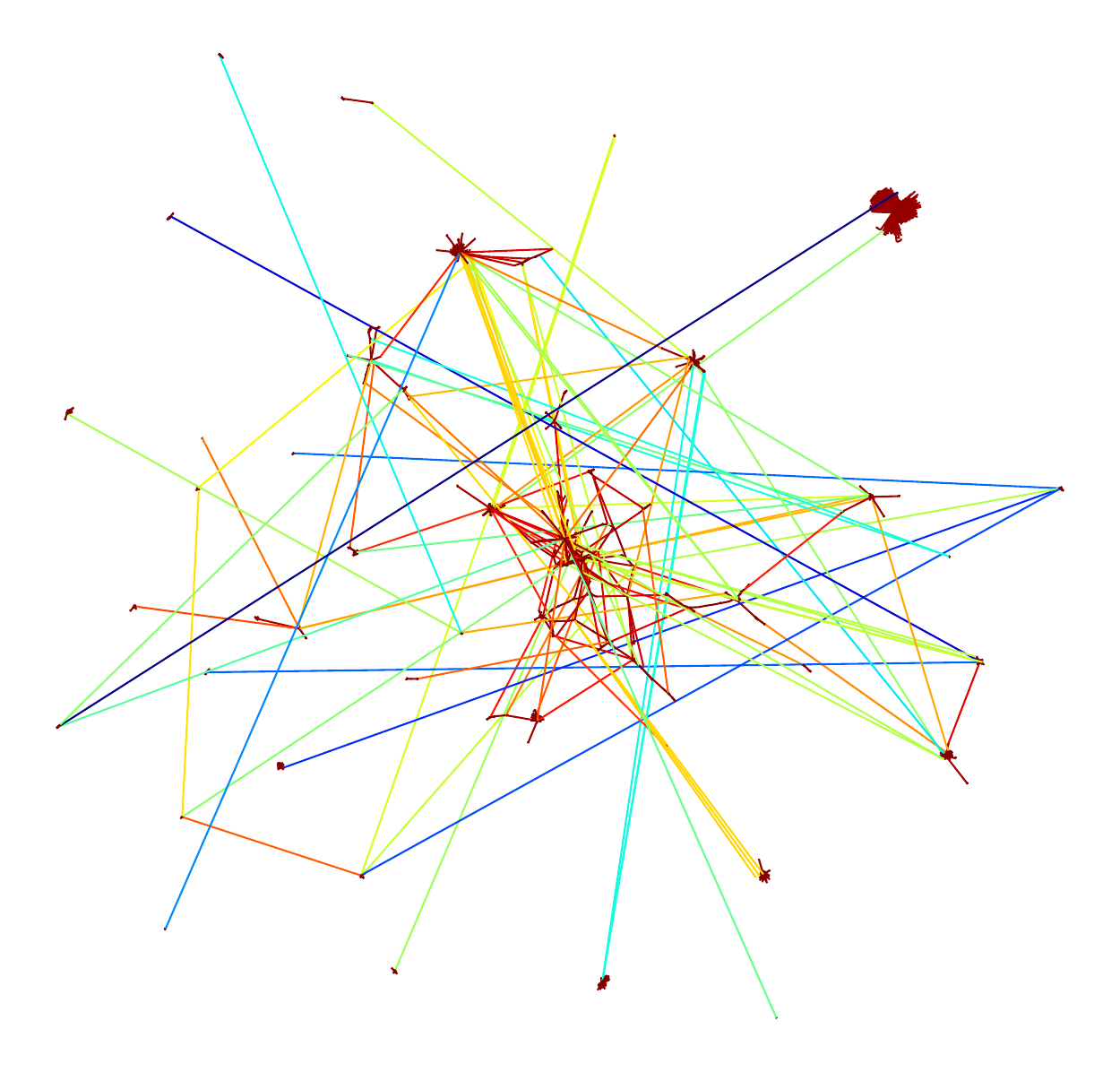}}
      & \parbox[c]{\tabfig\textwidth}{} 
      & \parbox[c]{\tabfig\textwidth}{} 
      & \parbox[c]{\tabfig\textwidth}{\taboffset
      \includegraphics[width=\tabfig\textwidth,height=\tabfig\textwidth]{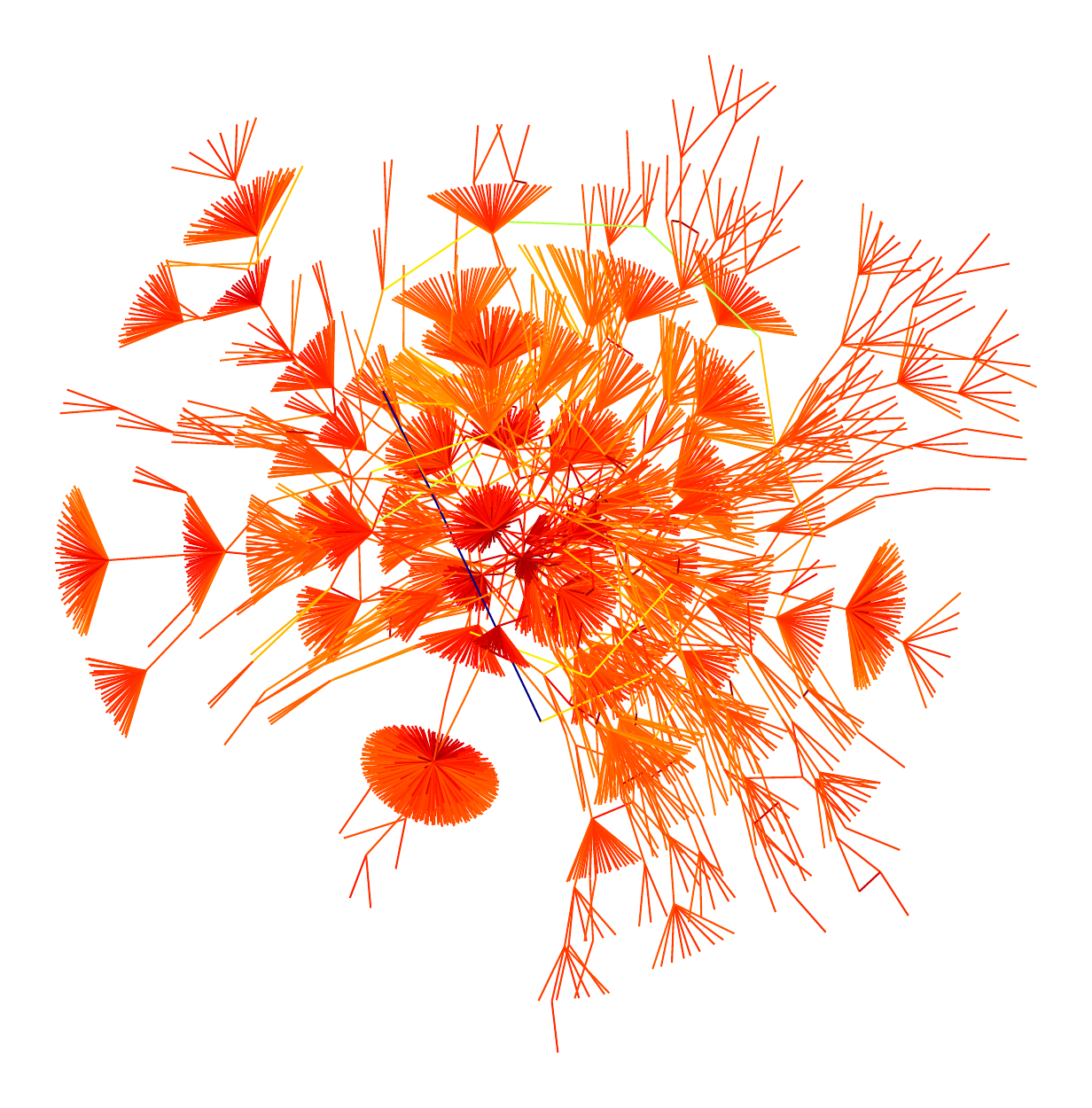}}  \\   
      \hline
  \end{tabular}
  \captionof{figure}{Example embeddings. The first and last columns show the two extremes tsNET (local) and MDS (global); the middle column shows UMAP. The remaining columns show a gradual increase of LGS's $k$ parameter, moving from local to global distance preservation (left to right). Note LGS outputs are vertically higher in each row.} 
  \label{tab:drawings1}
\end{table*}


\begin{figure}
    \begin{center}
        \includegraphics[width=0.22\linewidth]{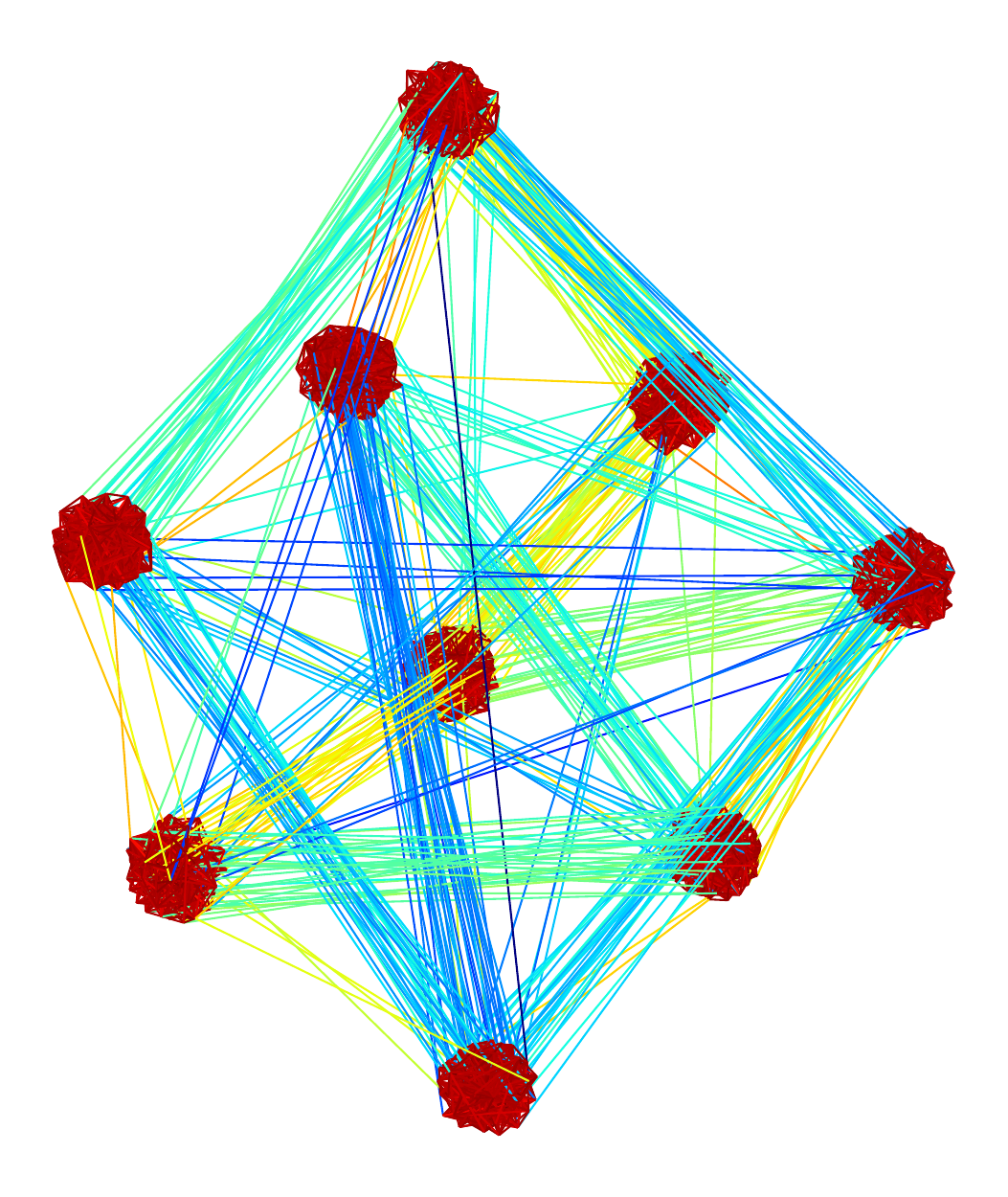}
        \includegraphics[width=0.22\linewidth]{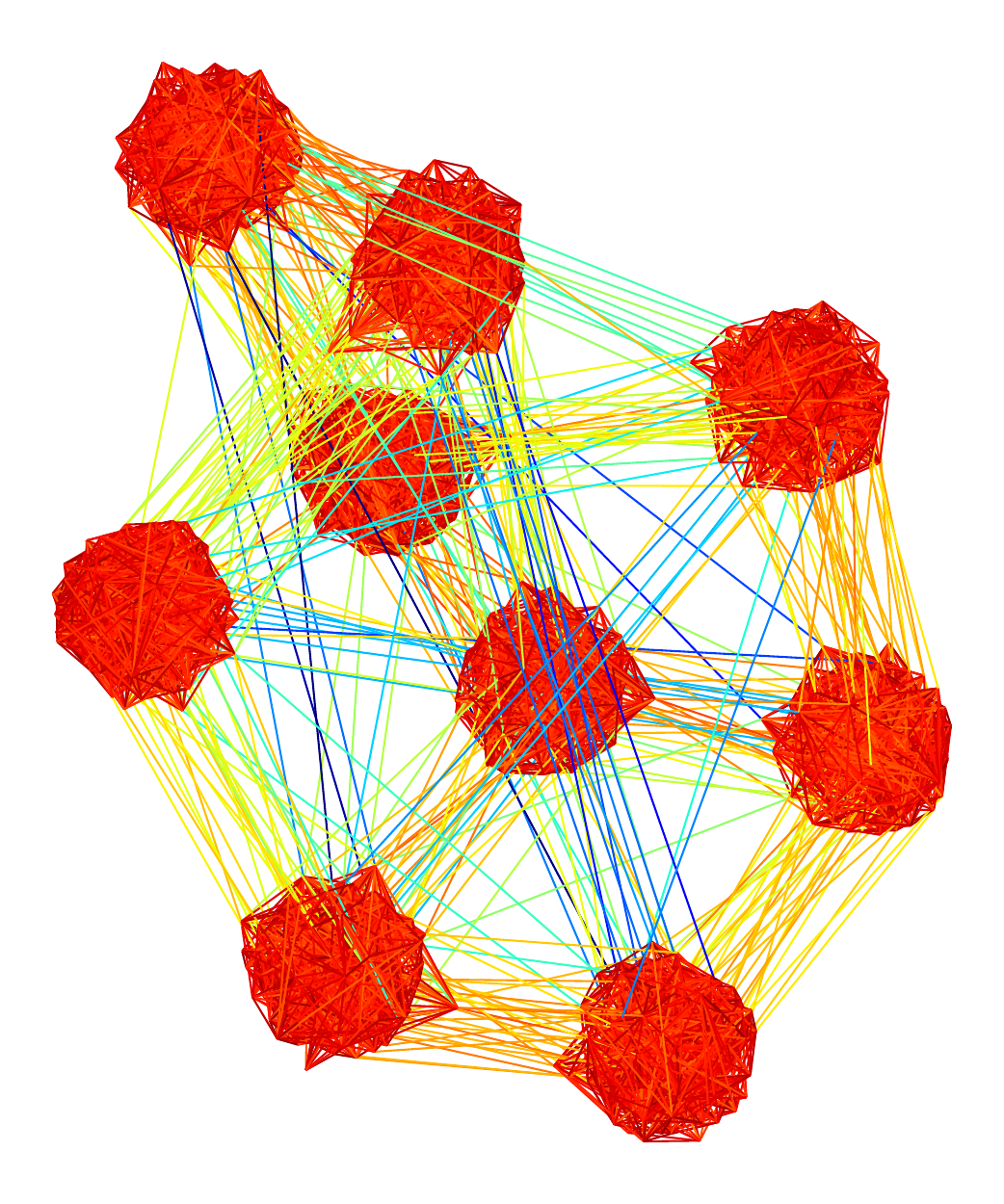} 
        \includegraphics[width=0.22\linewidth]{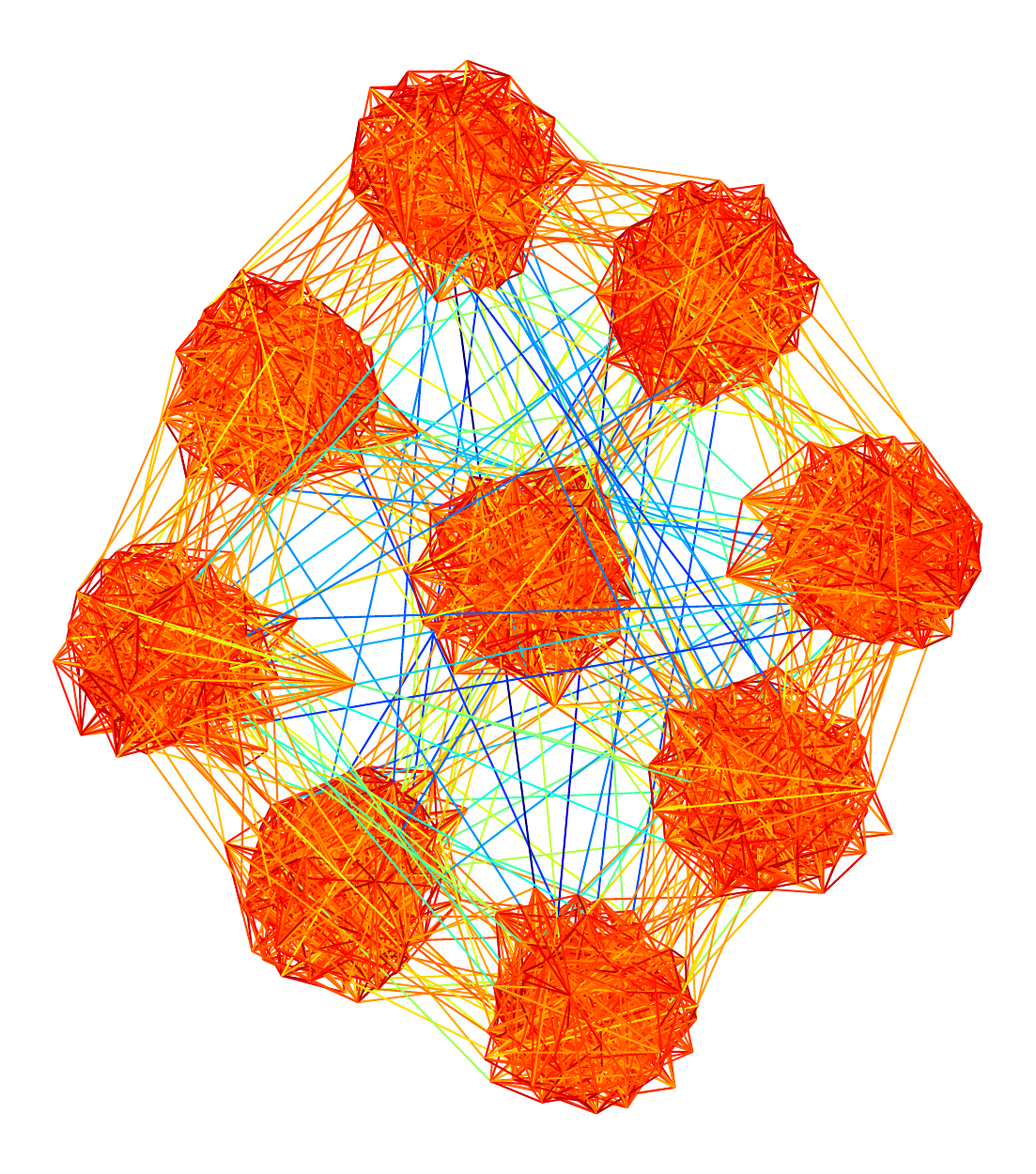} 
        \includegraphics[width=0.22\linewidth]{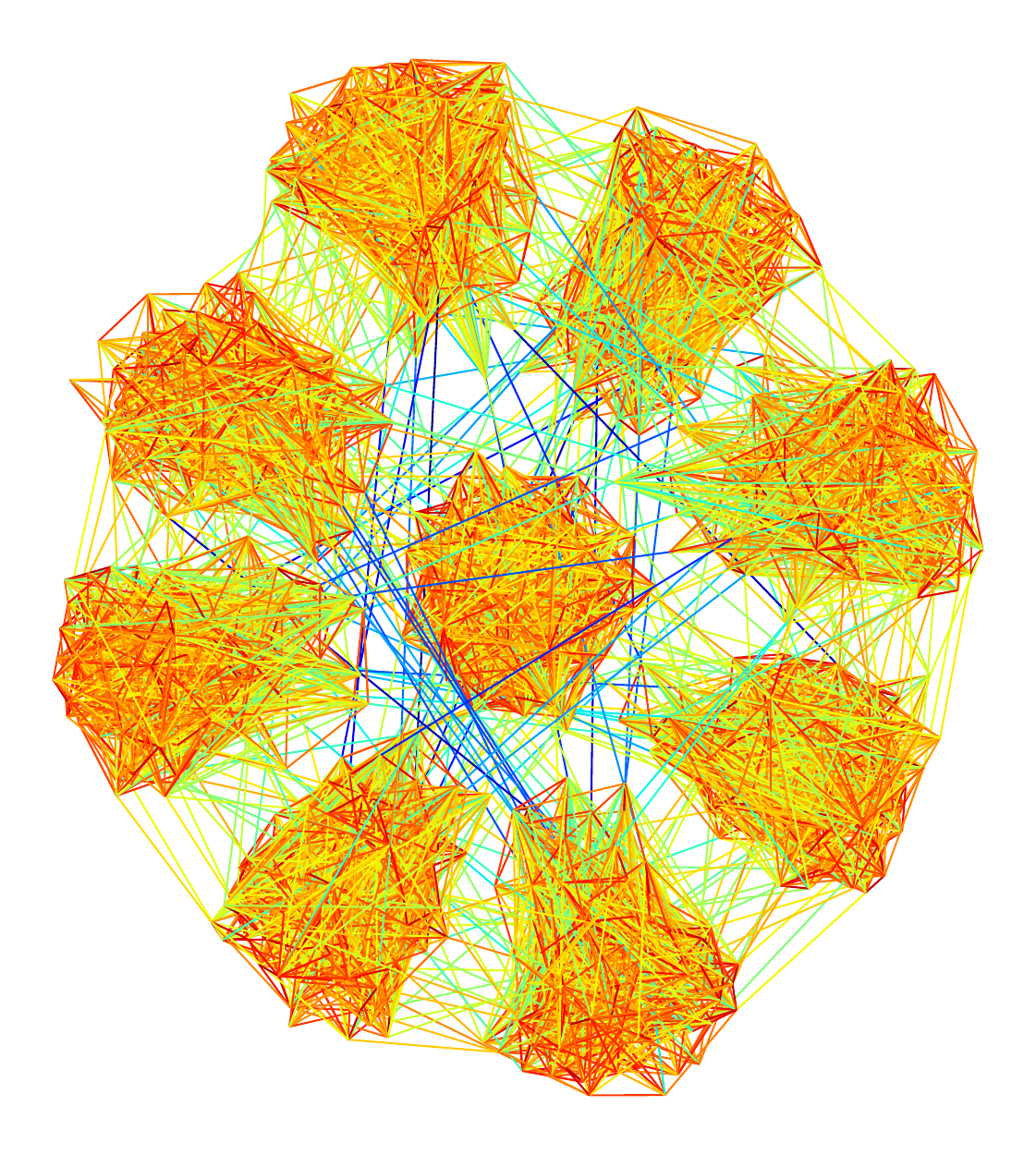}        
    
        \parbox[c]{0.22\linewidth}{\centering LGS(32)}
        \parbox[c]{0.22\linewidth}{\centering LGS(64)}
        \parbox[c]{0.22\linewidth}{\centering LGS(100)}
        \parbox[c]{0.22\linewidth}{\centering LGS(200)}
    
        \includegraphics[width=0.32\linewidth]{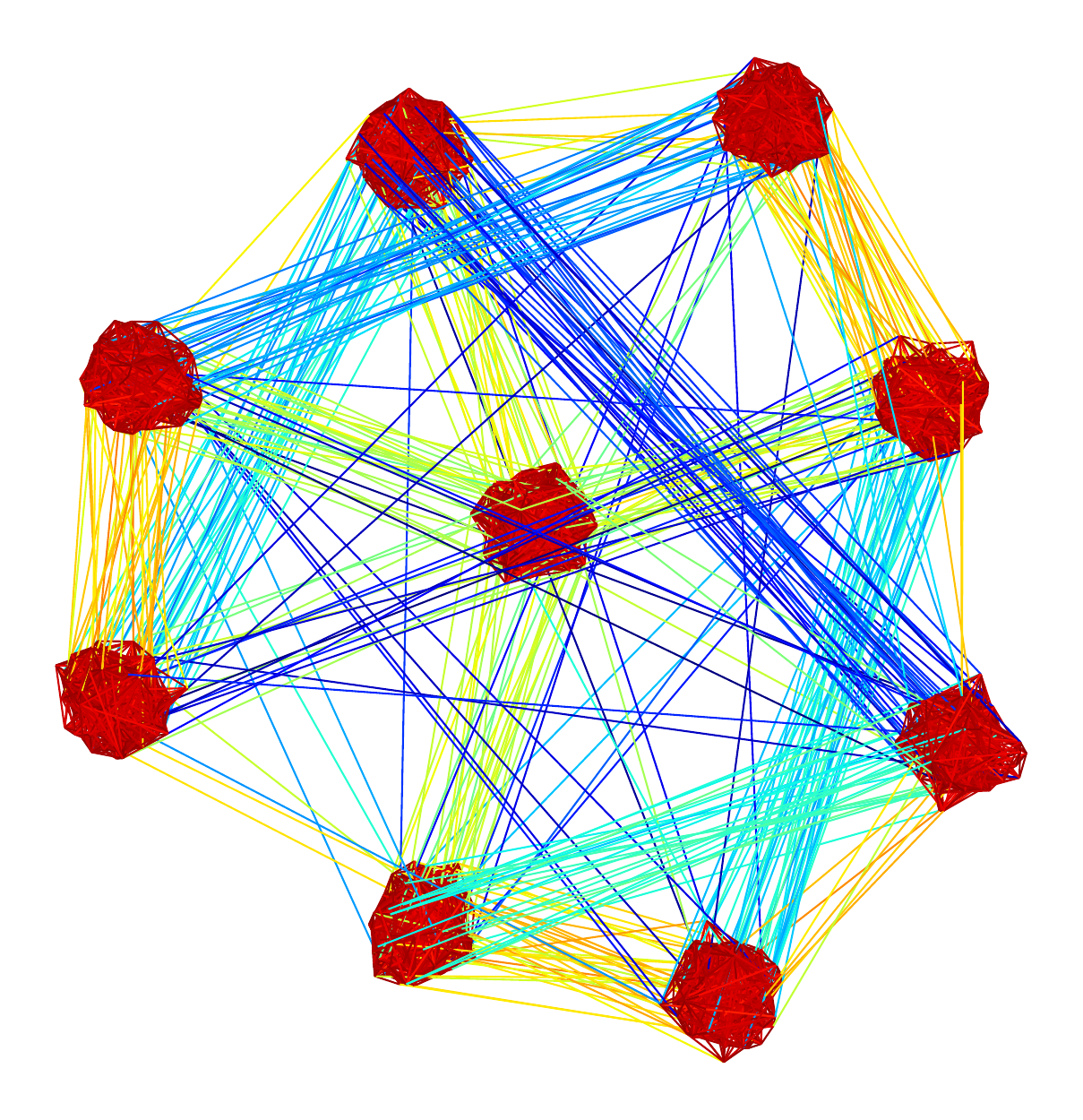}
        \includegraphics[width=0.32\linewidth]{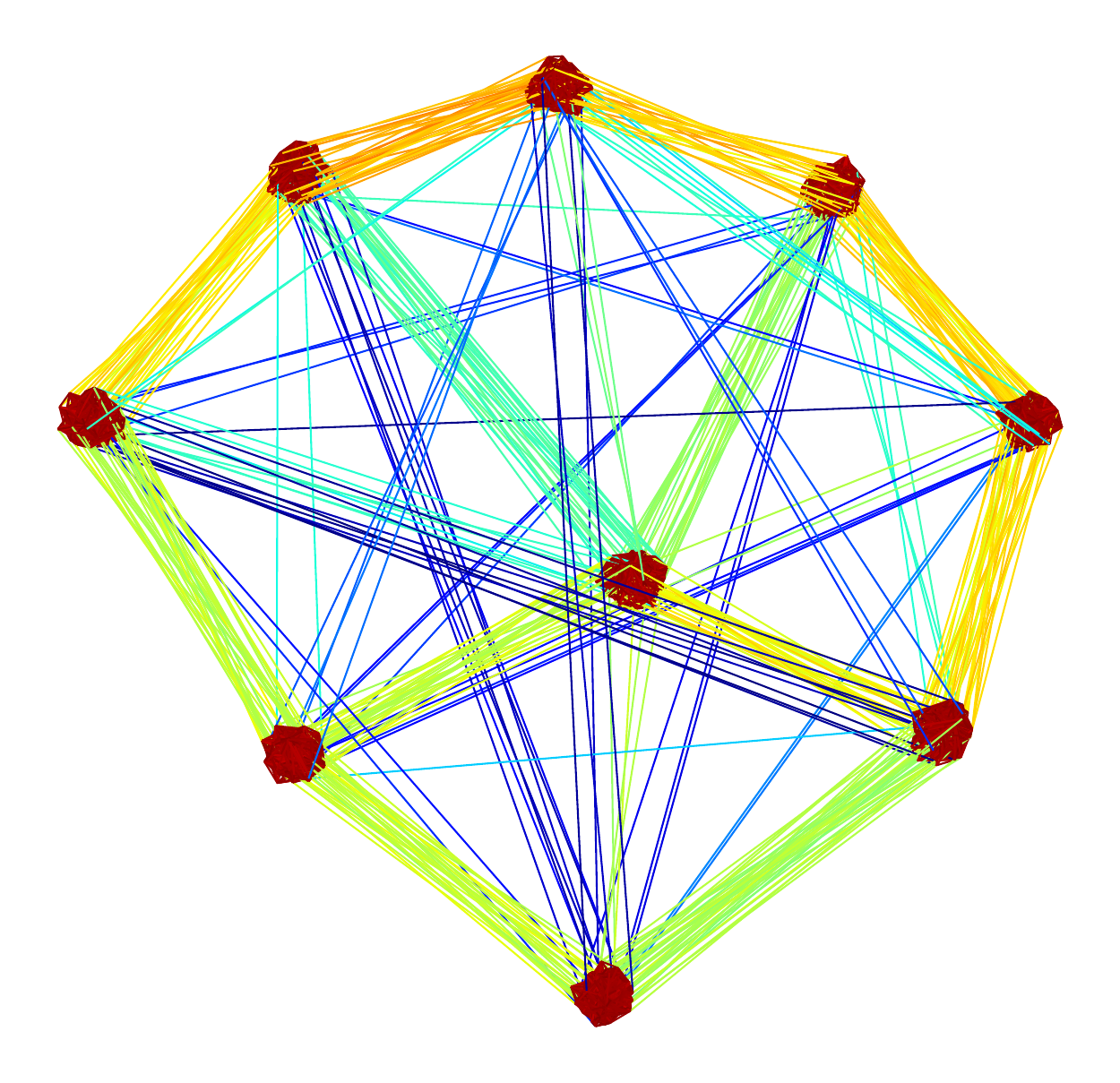} 
        \includegraphics[width=0.32\linewidth]{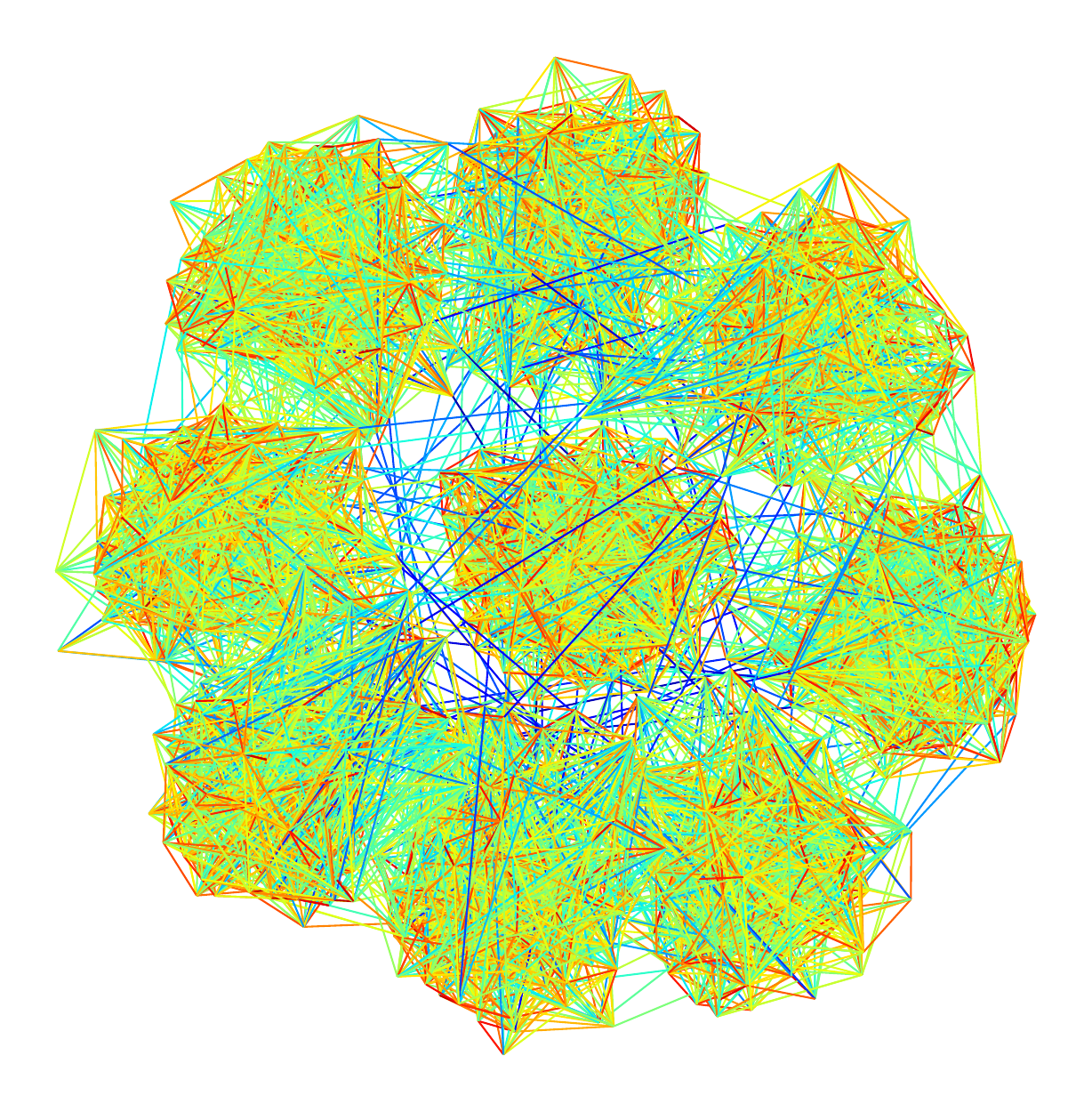} 

        \parbox[c]{0.26\linewidth}{\centering tsNET}
        \parbox[c]{0.26\linewidth}{\centering UMAP}
        \parbox[c]{0.26\linewidth}{\centering MDS}
        
    \end{center}
    
    \caption{
    The grid\_cluster graph is generated so that each cluster has many out-of-cluster edges to its neighbors in a $3\times3$ lattice, providing a recognizable intermediate structure. tsNET and UMAP do not place clusters on a grid, MDS mixes the clusters;  LGS(100) captures the $3\times3$ grid and shows distinct clusters.}
    \label{fig:grid-cluster}
\end{figure}

\subsection{Evaluation Metrics}
\label{sec:metrics}
We discuss the evaluation metrics for embedding algorithms: local neighborhood error (NE) score, intermediate structure (CD), and global distances (Stress).
\subsubsection{NE Metric:}
Neighborhood hits (NH) measures how well an embedding preserves local structures~\cite{chen2009local,DBLP:journals/tvcg/EspadotoMKHT21}. NH is the average Jaccard similarity of the neighbors in the high-dimensional and low-dimensional embedding.
Let $Y$ be an $n\times d$ dimensional dataset, $X$ be its $n \times 2$-dimensional embedding, and a radius $r$ defines the size of the neighborhood one intends to measure. NH is defined as:
\begin{equation}
    \label{eq:neighbor-hit}
    NH(Y,X,r) = \frac{1}{n} \sum_{i=1}^{n} \frac{| N_Y (p_i,r) \cap N_X(p_i,r) |}{| N_Y (p_i,r) \cup N_X(p_i,r) |}
\end{equation}
where $N_Y(p_i,r)$ denotes the $r$ nearest points to point $p_i$ in $Y$ and  $N_X(p_i,r)$ the $r$ nearest points to point $p_i$ in $X$.
For graph embeddings, this notion is called neighborhood preservation (NP)~\cite{DBLP:journals/cgf/KruigerRMKKT17,DBLP:journals/tvcg/ZhuCHHLZ21,DBLP:journals/tvcg/GansnerHN13}, with the main difference being that the radius $r$ now refers to graph-theoretic distance: all vertices with shortest path distance $\leq r$ from vertex $v_i$. Specifically, NP measures the average Jaccard similarity of a vertex's graph-theoretic neighborhood of radius $r$ and an equally sized neighborhood of that vertex's closest embedded neighbors. Since NH and NP measure accuracy, it is desirable to maximize these values. 
To facilitate comparison with the other two metrics (where lower scores mean better embeddings), we use Jaccard dissimilarity instead and refer to it as  Neighborhood Error (NE). 

\subsubsection{Cluster Distance Metric:}
We introduce a new metric to measure how well intermediate structures are captured in an embedding. Since the distances between clusters in t-SNE cannot be interpreted as actual distances
~\cite{wattenberg2016how}, while clusters in MDS embeddings are often poorly separated, we measure how faithful the relative distances between cluster centers are represented in the embedding.
When cluster labels are given as part of the input (e.g., labels, classes), we can use them to define distances between the clusters. When cluster information is not given, we 
use
$k$-means clustering in the high-dimensional data case, and modularity clustering in the graph case. The distances between clusters 
in the high-dimensional case is given by the Euclidean distance between the cluster centers. 
For graphs, we measure the distance between clusters by first taking the normalized count of edges between them, then subtracting the normalized count to convert similarity into dissimilarity. 
This produces a cluster-distance matrix, $\delta$. Let $C_1, \dots ,C_n$ be the set of vertices belonging to cluster $1, \dots, n$, then
\begin{equation*}
\label{eq:cluster-connectivity}
\delta_{i,j} = 1 - \frac{1}{|E|}\sum_{u\in C_i, v\in C_j}\mathds{1}(u,v\in E)
\end{equation*}
where $\mathds{1}$ is the indicator function (1 if $(u,v)$ is an edge and 0 otherwise).
Once $\delta$ is computed, 
we compute the geometric center of each embedded cluster 
and compute the cluster-level stress between the graph-level-cluster and realized-cluster distances. 
This measure is small when similar clusters 
are placed 
closer
and dissimilar clusters are placed far apart. 
The cluster distance (CD) is: 
\begin{equation}
    \label{eq:cluster-distance}
    CD(\delta, \chi) = \sum_{i,j}\left( \frac{\delta_{i,j} - ||\chi_i - \chi_j ||}{\delta_{i,j}}\right)^2
\end{equation}
where $\delta_{i,j}$ is the dissimilarity measure between cluster $i$ and cluster $j$ and $\chi_i$ is the geometric center of cluster $i$ in the embedding.
Although there are several existing metrics to measure cluster accuracy, such as silhouette distance and between/within-cluster sum of squares, they are not well suited to measure the quality of intermediate embeddings. Ideally, we would need a measure that checks how well the clusters are preserved and also verifies that the relative placement of the clusters is meaningful. The CD metric verifies meaningful cluster placements by measuring all pairwise distances between cluster centers.
We remark that the CD metric works best when the clusters have convex shapes (or shapes similar to spheres). For arbitrary non-convex shapes, such as half-moons or donuts, the CD metric might not provide meaningful insights.

\subsubsection{Stress Metric:}

Stress has been used in many graph embedding evaluations. 
\begin{equation}
    \label{eq:norm-stress}
    \text{stress}(d,X) = \sum_{i,j} \left( \frac{d_{i,j} - ||X_i - X_j ||}{d_{i,j}}\right)^2
\end{equation}
where $d$ 
is the given distance matrix
and $X$ is the embedding. 
Embeddings are scaled to ensure fair comparisons in computing stress~\cite{DBLP:journals/tvcg/GansnerHN13,DBLP:journals/cgf/KruigerRMKKT17,DBLP:journals/tvcg/ZhuCHHLZ21}.


\section{LGS Embedding of Graphs}
\label{sec:layouts}

\begin{figure}[h!]
    \centering
    \includegraphics[width=0.49\linewidth]{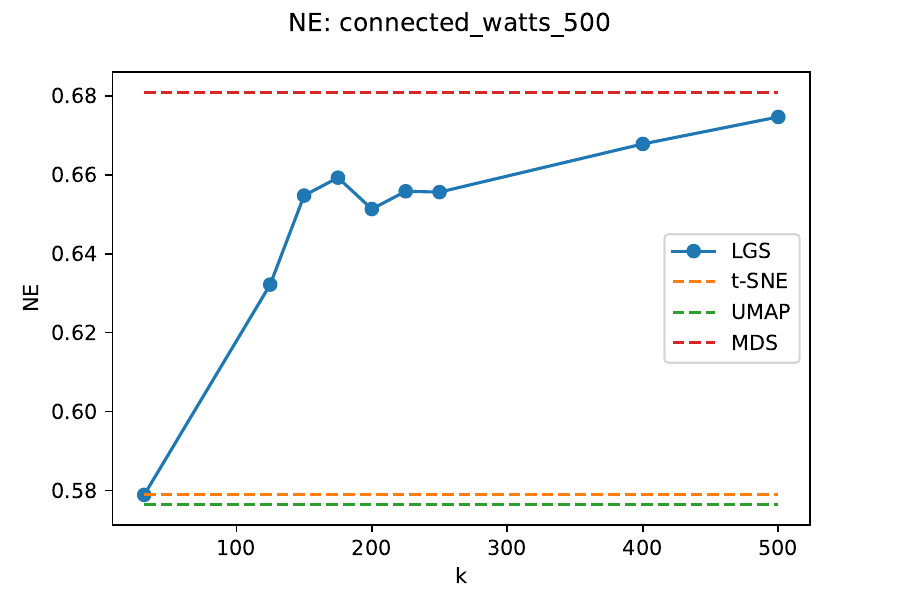}
    \includegraphics[width=0.49\linewidth]{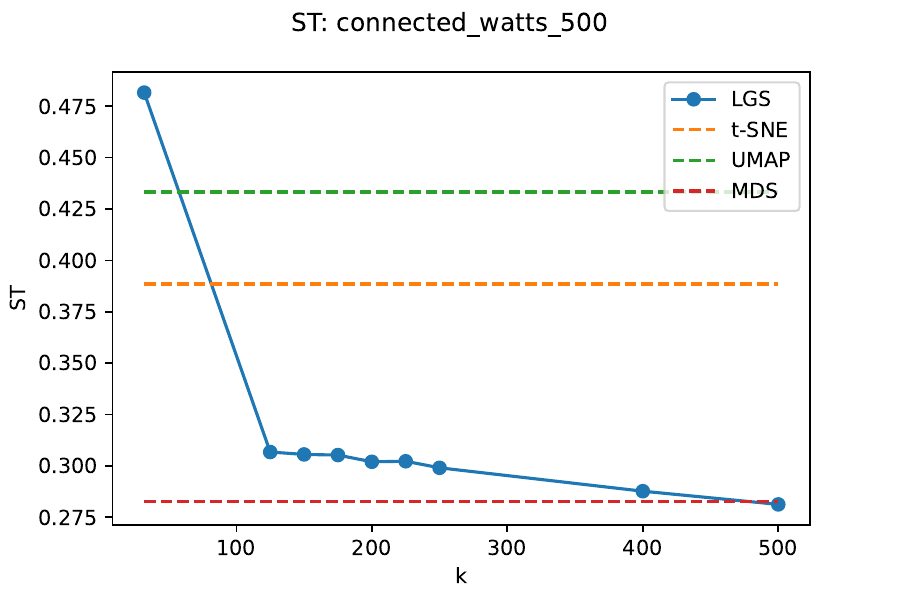}

    \includegraphics[width=0.49\linewidth]{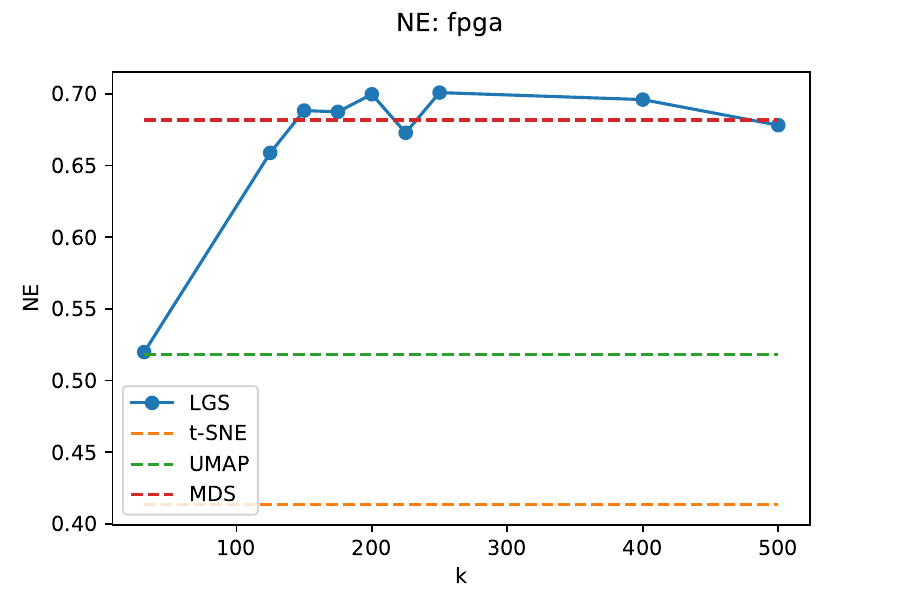}
    \includegraphics[width=0.49\linewidth]{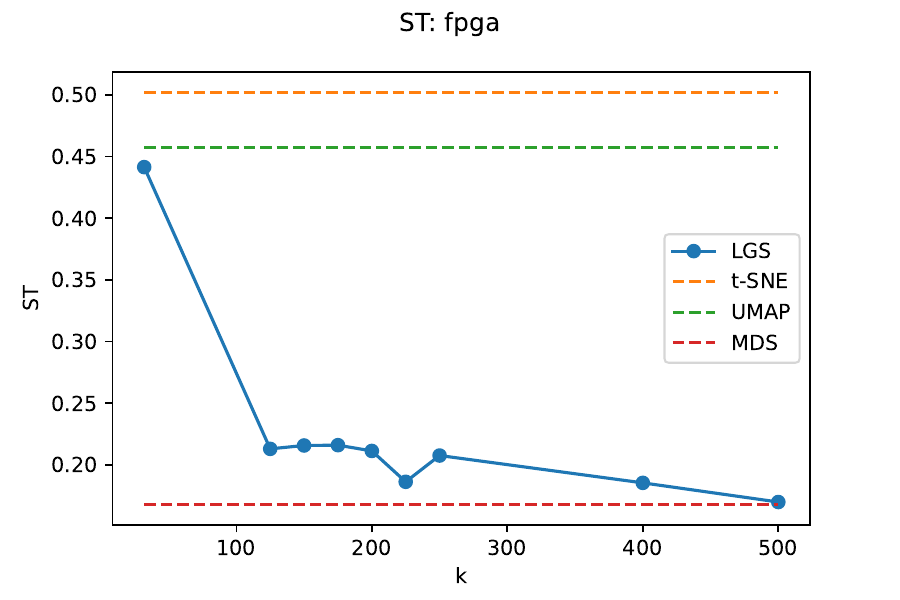}

    \caption{
    Behavior of NE and stress: as $k$ increases  NE gets worse and stress gets better (LGS transitions from preserving local to global structure); tsNET, UMAP, and MDS values are shown as dotted lines for comparison.
    Note that in general, we expect to see an upward trend in NE, a downward trend for stress, and a parabola shape for CD.}
    \label{fig:crossing-metrics}
\end{figure}

\begin{figure}
    \centering
    \includegraphics[width=0.32\textwidth]{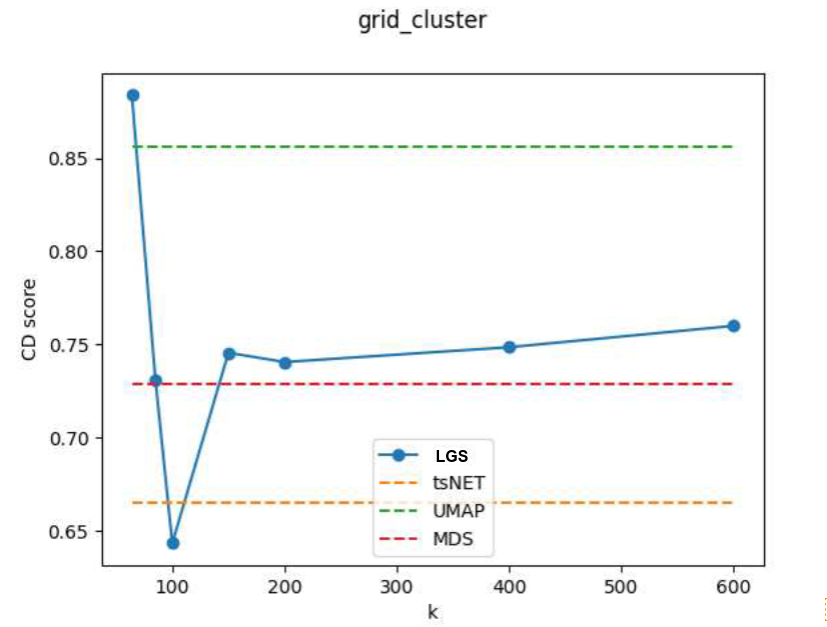}
    \includegraphics[width=0.32\textwidth]{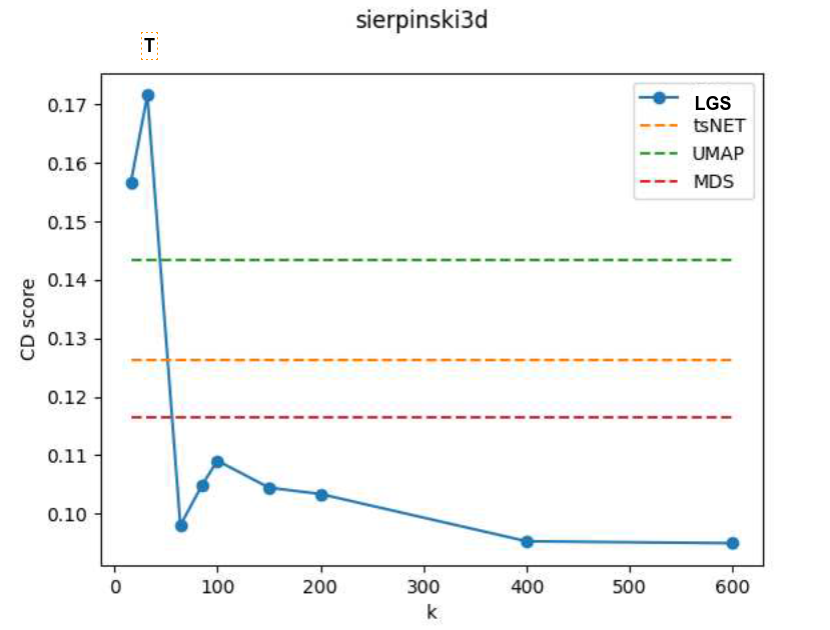}
    \includegraphics[width=0.32\textwidth]{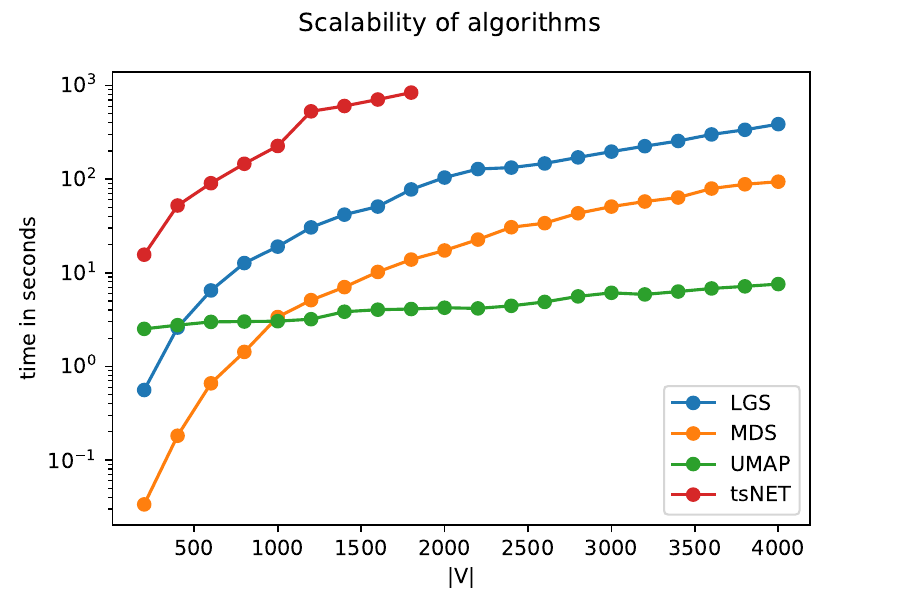}  

    \parbox[c]{0.32\textwidth}{\centering (a)}
    \parbox[c]{0.32\textwidth}{\centering (b)} 
    \parbox[c]{0.32\textwidth}{\centering (c)}    
    
    \caption{
    (a-b) CD metric on the grid\_cluster and sierpinkski3d graphs. Note that in these examples there are values of $k$ which outperform competing algorithms. (d) Running time of each tested algorithm.
    }
    \label{fig:cd-metrics}
\end{figure}

\begin{figure}[h]
    \centering
    \includegraphics[angle=30,width=0.22\linewidth]{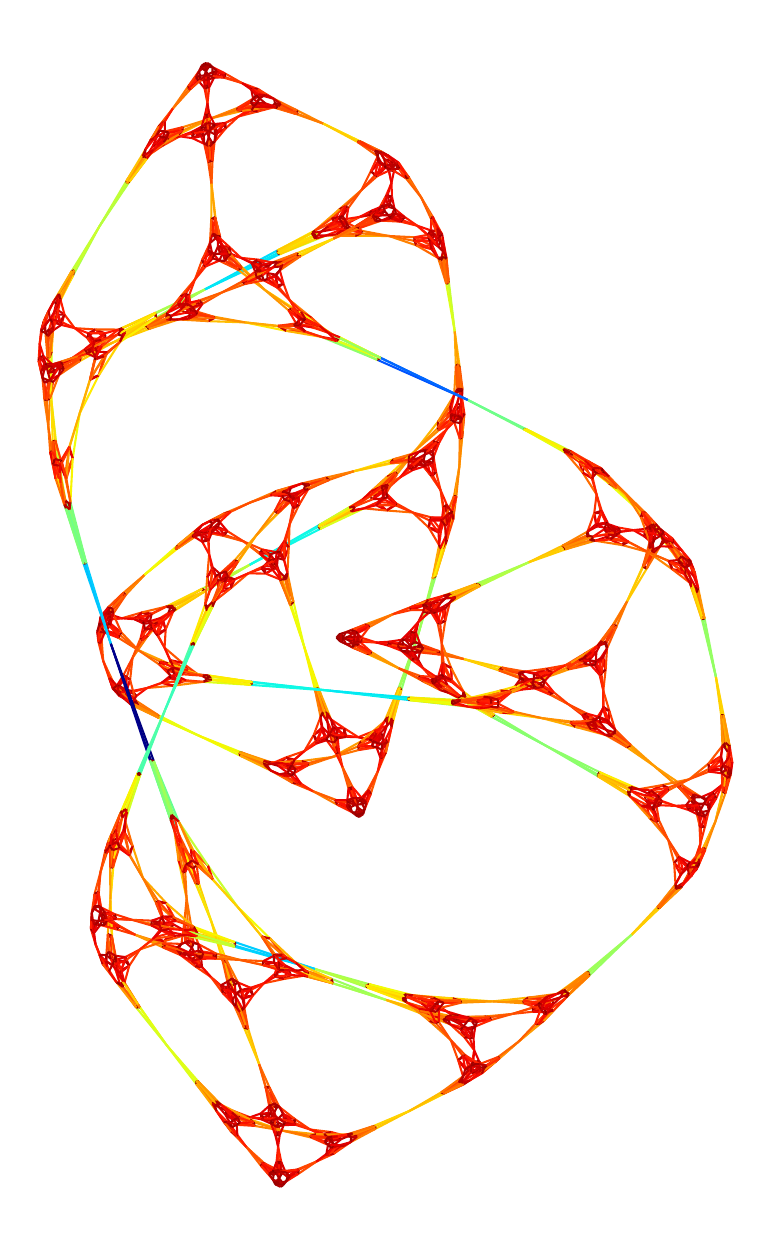}
    \includegraphics[width=0.22\linewidth]{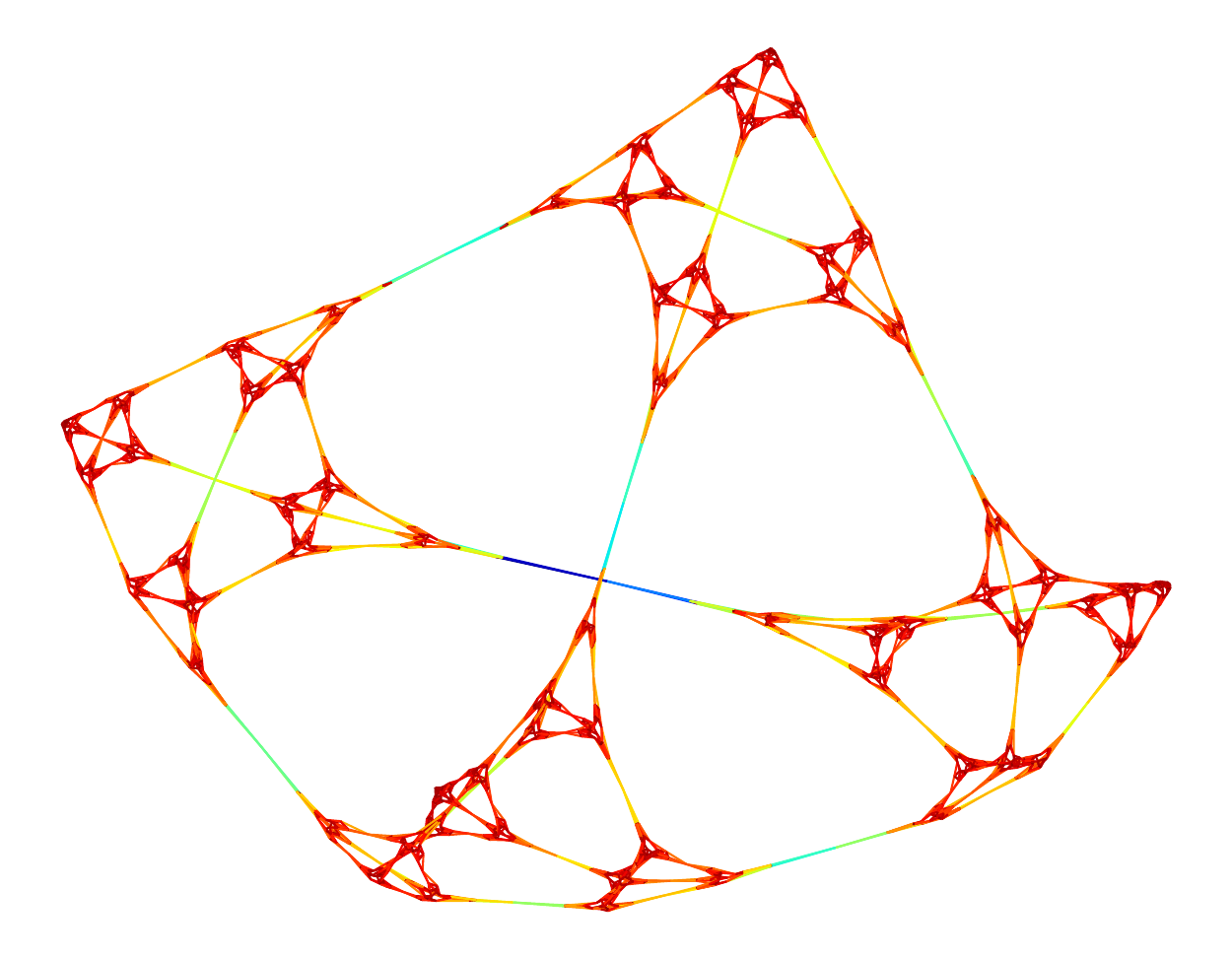}
    \includegraphics[width=0.22\linewidth]{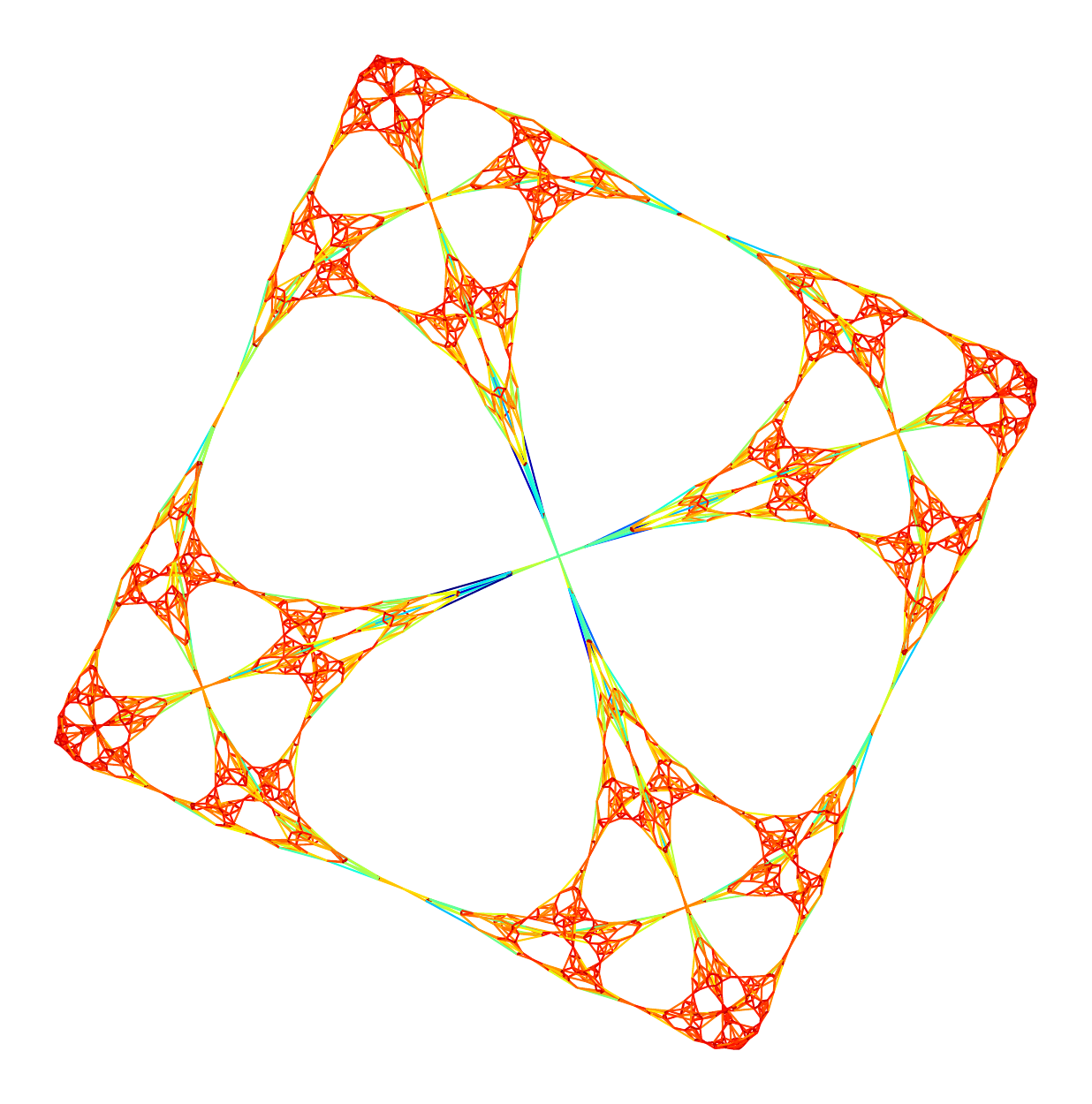}
    \includegraphics[width=0.22\linewidth]{figures/2023-pdfs/L2G/sierpinski3d_100.pdf}

    \parbox[c]{0.22\linewidth}{\centering LGS(16)}
    \parbox[c]{0.22\linewidth}{\centering LGS(32)}
    \parbox[c]{0.22\linewidth}{\centering LGS(64)}
    \parbox[c]{0.22\linewidth}{\centering LGS(100)}
    
    \includegraphics[width=0.28\linewidth]{figures/2023-pdfs/tsnet/sierpinski3d.pdf}
    \includegraphics[width=0.28\linewidth]{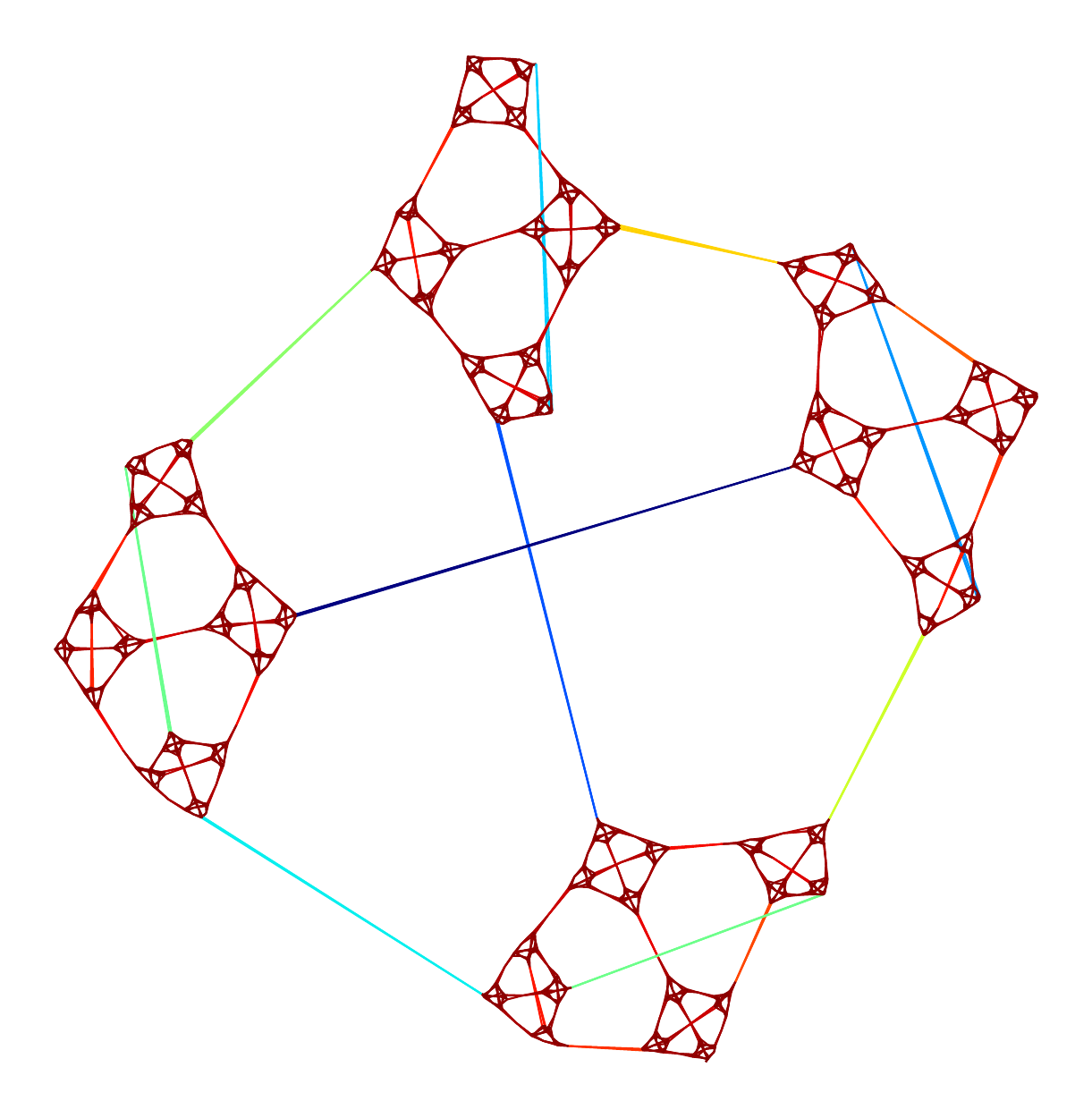}
    \includegraphics[width=0.28\linewidth]{figures/2023-pdfs/mds/sierpinski3d.pdf}

    \parbox[c]{0.28\linewidth}{\centering tsNET}
    \parbox[c]{0.28\linewidth}{\centering UMAP}
    \parbox[c]{0.28\linewidth}{\centering MDS}

    \caption{Sierpinksi3d graph is a fractal with regular local and global structure. LGS manages to capture the recursive 
    nature of the underlying structure. tsNET, UMAP miss the global placement of pyramids and MDS stretches them.}
    \label{fig:sierpinski}
\end{figure}

We start with a visual analysis and discussion of layouts produced by LGS. Following the convention, several embeddings of the same graph are displayed side-by-side with increasing the value of $k$ from left to right, going from local to global. 
Underneath the LGS embeddings, we place t-SNE, UMAP, and MDS embeddings of the same graph.
For all graph embeddings provided in this paper, we use the jet color scheme to encode edge length. An edge length of 1 (ideal for unweighted graphs) is drawn in green, while red indicates that edge has been compressed (length $<1$) and blue indicates the edge is stretched (length $>1$). This makes clusters easy to spot as bundles of red edges, and global structure preservation apparent when most edges are green.
Similar to tsNET, low values of $k$ capture local neighborhoods well, by allowing some longer edges. As a result, clusters tend to be well separated. Note that tsNET allows even longer edges in an embedding, occasionally breaking the topology; see Fig.~\ref{fig:teaser}.
Higher $k$ values make LGS similar to MDS, with more uniform edge lengths. This reveals global structures (e.g., mesh, grid, lattice) but may overlook clusters.




\noindent \textbf{Grid\_cluster}
is a synthetic example with 900 vertices (9 clusters of size 100 each) and 10108 edges, 
created by stochastic block model (SBM)
to illustrate the notion of cluster distance preservation. 
Within cluster edges are created with probability 0.8. We distinguish between two types of out-of-cluster edges. Clusters are first placed on a lattice. Out-of-cluster edges are created with probability 0.01 if they are adjacent in the lattice (no diagonals) and 0.001 otherwise.  
The layouts of this graph are in Fig.~\ref{fig:grid-cluster}. Note the visual similarities between LGS(32) and tsNET; both seperate each cluster into dense sub-regions and place them seemingly randomly in the plane. Also note the similarities between LGS(200) and MDS, where both methods tend to miss the clusters. UMAP also fails to capture the intermediate structure built into this network: while there is a single cluster placed in the middle of the other eight, the surrounding shape is  not a square. 
LGS(100) accurately places each cluster in the appropriate position, making it the only one that clearly shows the 3x3 underlying lattice. 
Although less dense and separable the clusters are more faithful in terms of placement.

\noindent \textbf{Connected\_watts\_1000}
is a Watts-Strogatz random graph on a 1000 vertices and 11000 edges. It first assigns the vertices evenly spaced around a cycle with the nearest (7) vertices connected by an edge. Then, with low probability, some random `chords' of the cycle are added by rewiring some of the local edges to other random vertices. 
This type of graph models the small-world phenomenon 
seen in real-world examples, such as social networks. 
The embeddings of connected\_watts\_1000, obtained by  LGS, tsNET, UMAP, and MDS are  in Fig.~\ref{fig:teaser}.
We observe that tsNET and UMAP embeddings accurately capture the existence of a one-dimensional structure, but twist and break the circle to varying degrees. Meanwhile, MDS overcrowds the space, forming a classic `hairball' where there is no discernible structure. 
For intermediate values of $k$ in LGS, the circular structure in the data and the numerous chord connections become clearly visible.


\noindent \textbf{Sierpinksi\_3d }
 models the Sierpinski pyramid with 2050 vertices and 6144 edges -- a finite fractal object with recursively smaller recurring patterns (the pyramid itself is built out of smaller pyramids). 
These fractal properties are ideal for showcasing the LGS algorithm at work, as small local structures build upon each other to create a global shape; see Fig.~\ref{fig:sierpinski}.
We observe that tsNET captures the smallest structures well but places them arbitrarily in the embedding space. UMAP does better at placing the local structures in context but still creates long edges and twists not present in the data. While MDS visually captures the fractal motifs, it `squishes' local structures. LGS can be used to balance these extremes.

\begin{figure*}[ht!]
    \centering
    \includegraphics[width=0.7\linewidth]{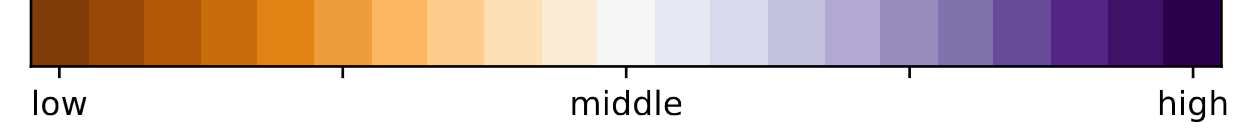}
    \includegraphics[width=1\linewidth]{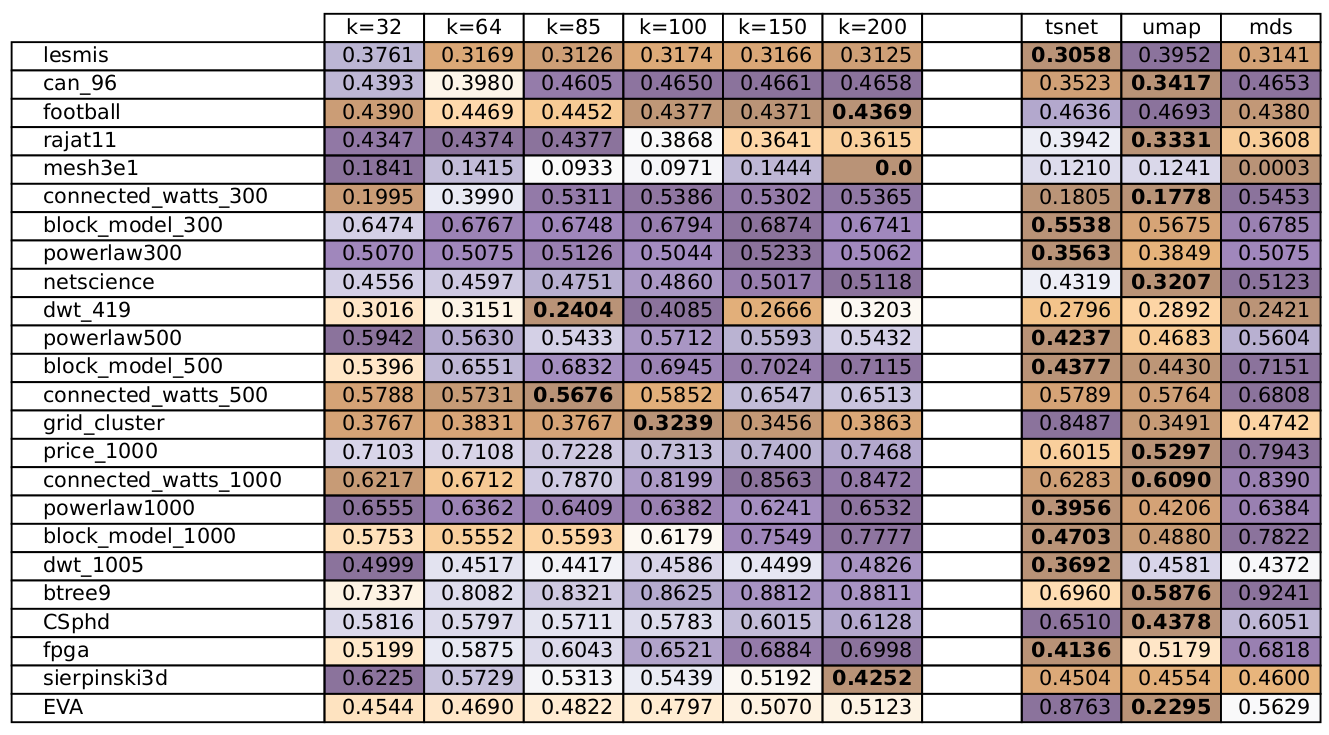}
    


    
    \captionof{table}{
    NE scores on LGS for varying values of $k$ (left) and on competing algorithms (right). The colormap is normalized by row with dark orange representing the lowest score (best) and dark purple representing the highest (worst).
    Bold text indicates the lowest score in that row.}
    \label{tab:ne-table}
\end{figure*}

We demonstrate more examples in Table~\ref{tab:drawings1}, with additional embeddings available in the supplemental material. 
Note that for lower values of $k$ the embedding obtained by LGS visually resembles the output of tsNET, while for larger values of $k$  the obtained embedding is more similar to the outputs of MDS. 
We see this reflected numerically in many graphs; see Fig.~\ref{fig:crossing-metrics} and Table~\ref{tab:ne-table}.
For intermediate values of $k$, LGS often outperforms tsNET, UMap and MDS with respect to the intermediate structure preservation, measured by  cluster distance (CD).


\section{Evaluation}
\label{sec:experiments}


We test 
LGS
on a selection of real-world and synthetic graphs from~\cite{DBLP:journals/toms/DavisH11,DBLP:journals/cgf/KruigerRMKKT17,DBLP:journals/tvcg/ZhuCHHLZ21}; A full list can be found in the appendix.

\begin{figure}[t!]
    \centering
    \includegraphics[width=0.7\linewidth]{figures/2023-pdfs/cmap.png}    
    \includegraphics[width=\textwidth]{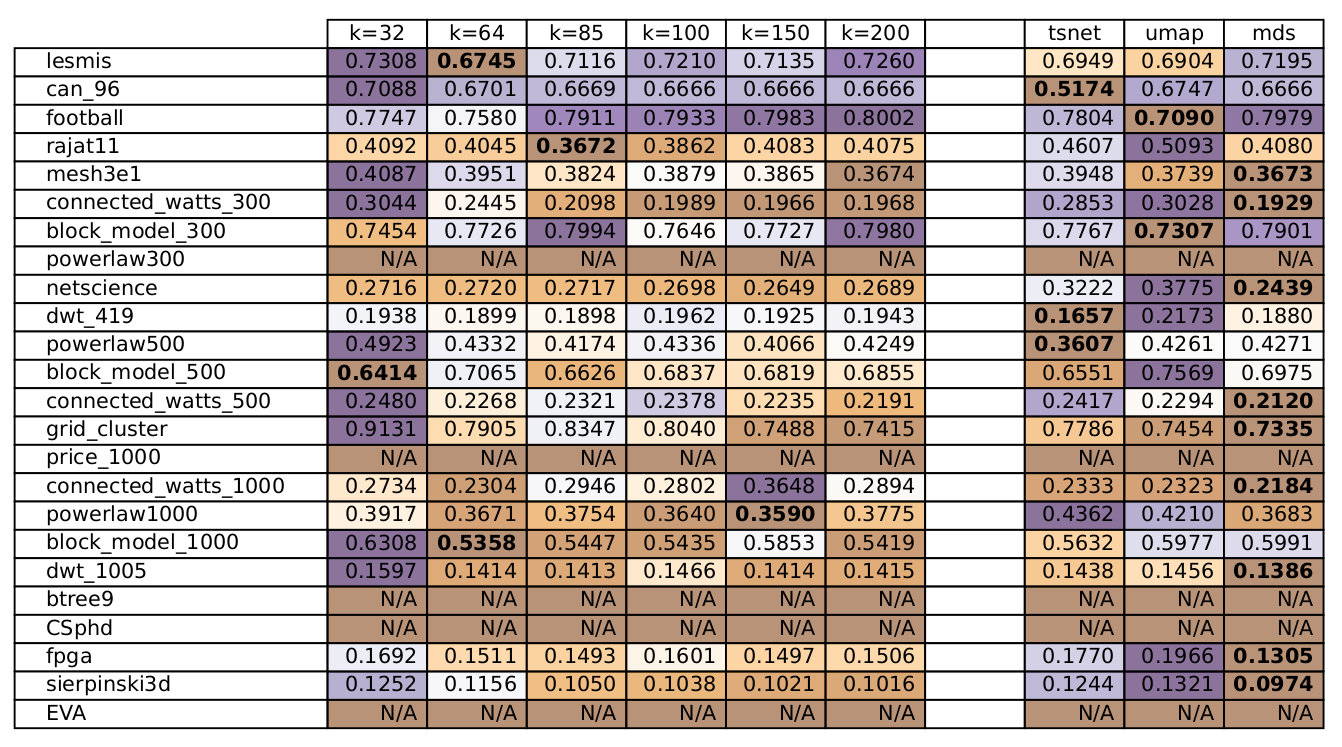}
    \captionof{table}{CD scores following the same scheme as Table~\ref{tab:ne-table}. NA indicates no clusters present in the data.}
    \label{tab:cd-table}
\end{figure}

\begin{figure}[t!]
    \centering
    \includegraphics[width=\textwidth]{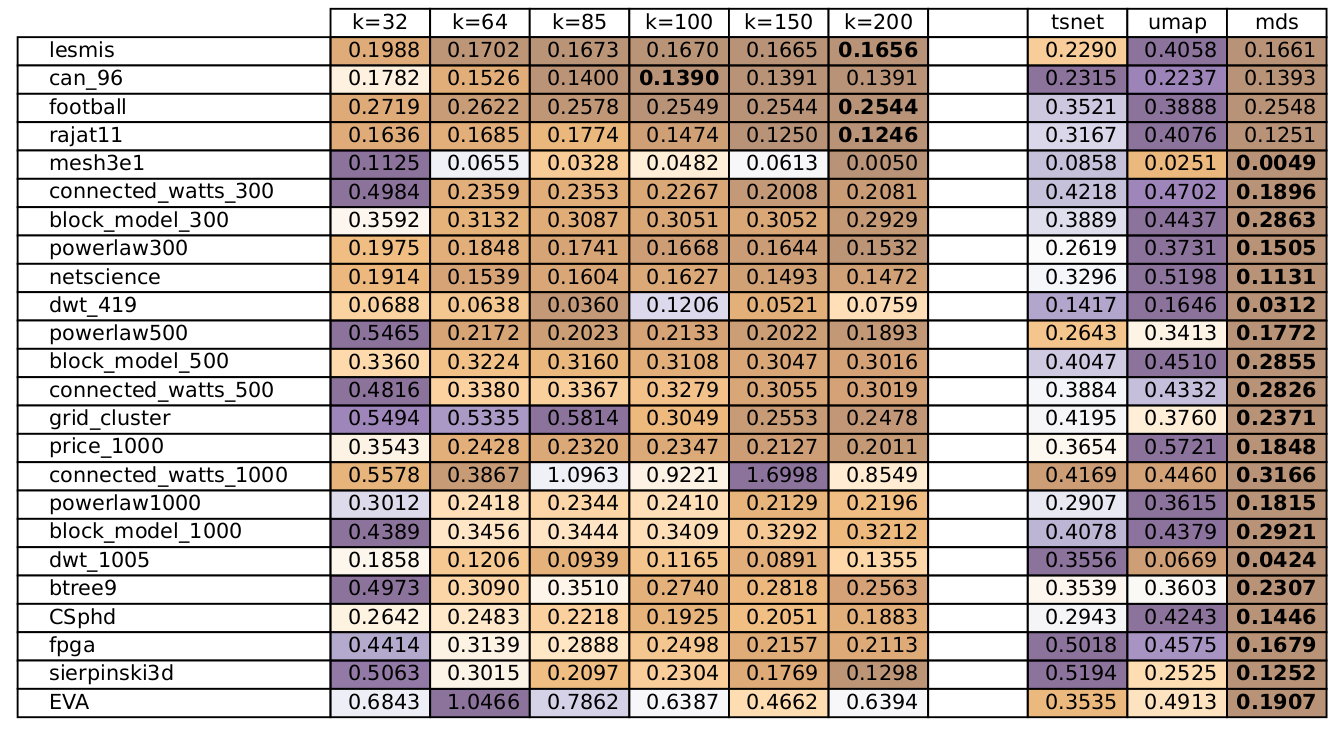}
    \captionof{table}{Stress scores following the same scheme as Table~\ref{tab:ne-table}}
    \label{tab:stress-table}
\end{figure}

We compare LGS against state-of-the-art techniques for local and global embeddings: 
tsNET from the repository linked in~\cite{DBLP:journals/cgf/KruigerRMKKT17}, UMAP from the umap python library written by the authors of~\cite{mcionnes2018umap}, and  MDS via the python bindings from~\cite{DBLP:journals/tvcg/ZhengPG19}
with default parameters.
Our implementation of LGS is available online.
The experiments were performed on an Intel® Core™ i7-3770 machine (CPU @ 3.40GHz × 8 with 32 GB of RAM) running Ubuntu 20.04.3 LTS. 

\subsubsection{NE, CD and Stress Values and Trends}

To evaluate how well LGS preserves local neighborhoods, we compare the average NE scores over several runs and present our results in Table~\ref{tab:ne-table}. 
We can see a general trend: although, tsNET performs better with respect to  NE values, LGS has consistently lower NE values than MDS. Additionally, as we increase the size of the neighborhood parameter, the NE values tend to increase by bringing the layouts closer to those of MDS. Interestingly, UMAP also tends to fall somewhere between tsNET and MDS on this metric. 
As expected, in many cases, LGS  `transitions' from tsNET to UMAP then finally to MDS as one goes from left to right, increasing $k$. 

Next, we report the average CD scores for the graphs in our benchmark in Table~\ref{tab:cd-table}.
Unlike the NE values which increase as we increase $k$ and the stress values which decrease as we increase $k$, the best CD values are obtained for intermediate values of $k$. This confirms that a balance between local and global optimization is needed to capture intermediate structures.

We compute and report the averaged stress scores in Table~\ref{tab:stress-table}.
MDS is consistently good at minimizing the stress,
but we see a salient trade-off between the stress scores of LGS's tsNET-like embeddings with low $k$ values and LGS's MDS-like embeddings with high $k$ values. When we look at small neighborhoods such as $k=16$, we tend to see high stress values, however, the values decrease as we expand the neighborhoods. UMAP does not seem to capture global structure well for these graphs, often having the highest stress values. 




\subsubsection{Effect of $k$ on Evaluation Metrics:}

To visually explain LGS behavior, we plot examples NE, CD, and stress with respect to $k$. In Fig.~\ref{fig:crossing-metrics}, we demonstrate two separate plots for each graph. It can be seen that we often fall in between the values of NE and stress that tsNET and MDS reach. These plots show what we expect to see: as $k$ increases NE increases and stress decreases. 
In Fig.~\ref{fig:cd-metrics}(a-b), we plot the CD values of our layout with tsNET, UMAP, MDS for comparison. In many layouts 
LGS indeed has the lowest CD score. Values of $k$ were chosen to be representative of the local-global tradeoff. 

\section{Discussion and Limitations}
We described LGS: an adaptable algorithmic framework for embeddings that can prioritize local neighborhoods, global structure, or a balance between the two. 
LGS provides flexible structure preservation choices with comparable embedding quality to previous single-purpose methods (local or global),
while also outperforming state-of-the-art methods in preserving intermediate structures.



There are several limitations: 
Our results
are based on a small number of graphs.
Additional systematic experimentation would further support the usefulness of LGS
Some experiments for high-dimensional datasets are in Appendix~\ref{sec:hd-data}.
LGS modifies MDS's objective function to accommodate varying neighborhood sizes, similarly, one could adapt the KL divergence cost function of t-SNE.
Note that t-SNE's perplexity parameter ostensibly controls the size of a neighborhood, but high perplexity values do  not result in global structure preservation~\cite{wattenberg2016how}. 

LGS algorithm has several hyperparameters, including $c$, $\alpha$, and $k$. 
We provide default values for $c$ and $\alpha$ based on experiments,
and leave $k$ as a true hyperparameter. Our intention is for a visualization designer to adjust $k$ as needed; to generate a spectrum of embeddings to get a sense of both local and global properties of a dataset. An interactive LGS version is not yet available.
While LGS runs in seconds for graphs with a few thousand vertices, the running time can become untenable for larger instances, due to the $O(|V|^2)$ optimization per epoch, and pre-processing with APSP. 
While LGS's runtime is comparable with those of tsNET and MDS (see Fig.~\ref{fig:cd-metrics}(c)) both 
can be sped up through the use of approximations~\cite{fu2019atsne,DBLP:journals/jgaa/OrtmannKB17,DBLP:journals/tvcg/ZhuCHHLZ21},
Speeding up LGS is a potential future work.

\bibliographystyle{splncs04}
\bibliography{main}

\begin{thebibliography}{10}
\providecommand{\url}[1]{\texttt{#1}}
\providecommand{\urlprefix}{URL }
\providecommand{\doi}[1]{https://doi.org/#1}

\bibitem{anguita2013public}
Anguita, D., Ghio, A., Oneto, L., Parra, X., Reyes{-}Ortiz, J.L.: A public
  domain dataset for human activity recognition using smartphones. In: 21st
  European Symposium on Artificial Neural Networks, {ESANN} 2013, Bruges,
  Belgium, April 24-26, 2013 (2013)

\bibitem{DBLP:conf/gd/BorsigBP20}
B{\"{o}}rsig, K., Brandes, U., P{\'{a}}sztor, B.: Stochastic gradient descent
  works really well for stress minimization. In: Auber, D., Valtr, P. (eds.)
  Graph Drawing and Network Visualization - 28th International Symposium, {GD}
  2020. Revised Selected Papers. Lecture Notes in Computer Science, vol. 12590,
  pp. 18--25. Springer (2020)

\bibitem{DBLP:series/lncs/Bottou12}
Bottou, L.: Stochastic gradient descent tricks. In: Montavon, G., Orr, G.B.,
  M{\"{u}}ller, K. (eds.) Neural Networks: Tricks of the Trade - Second
  Edition, Lecture Notes in Computer Science, vol.~7700, pp. 421--436. Springer
  (2012)

\bibitem{DBLP:journals/tkde/CaiZC18}
Cai, H., Zheng, V.W., Chang, K.C.: A comprehensive survey of graph embedding:
  Problems, techniques, and applications. {IEEE} Trans. Knowl. Data Eng.
  \textbf{30}(9),  1616--1637 (2018)

\bibitem{chen2009local}
Chen, L., Buja, A.: Local multidimensional scaling for nonlinear dimension
  reduction, graph drawing, and proximity analysis. Journal of the American
  Statistical Association  \textbf{104}(485),  209--219 (2009)

\bibitem{DBLP:journals/toms/DavisH11}
Davis, T.A., Hu, Y.: The {U}niversity of {F}lorida sparse matrix collection.
  {ACM} Trans. Math. Softw.  \textbf{38}(1),  1:1--1:25 (2011)

\bibitem{de2005applications}
De~Leeuw, J.: Applications of convex analysis to multidimensional scaling
  (2005)

\bibitem{DBLP:journals/toms/DuffGL89}
Duff, I.S., Grimes, R.G., Lewis, J.G.: Sparse matrix test problems. {ACM}
  Trans. Math. Softw.  \textbf{15}(1),  1--14 (1989)

\bibitem{ertoz2002new}
Ertoz, L., Steinbach, M., Kumar, V.: A new shared nearest neighbor clustering
  algorithm and its applications. In: Workshop on clustering high dimensional
  data and its applications at 2nd SIAM international conference on data
  mining. vol.~8 (2002)

\bibitem{DBLP:journals/tvcg/EspadotoMKHT21}
Espadoto, M., Martins, R.M., Kerren, A., Hirata, N.S.T., Telea, A.C.: Toward a
  quantitative survey of dimension reduction techniques. {IEEE} Trans. Vis.
  Comput. Graph.  \textbf{27}(3),  2153--2173 (2021)

\bibitem{frey1978principal}
Frey, D., Pimentel, R.: Principal component analysis and factor analysis (1978)

\bibitem{fu2019atsne}
Fu, C., Zhang, Y., Cai, D., Ren, X.: Atsne: Efficient and robust visualization
  on gpu through hierarchical optimization. In: Proceedings of the 25th ACM
  SIGKDD International Conference on Knowledge Discovery \& Data Mining. pp.
  176--186 (2019)

\bibitem{DBLP:journals/tvcg/GansnerHN13}
Gansner, E.R., Hu, Y., North, S.C.: A maxent-stress model for graph layout.
  {IEEE} Trans. Vis. Comput. Graph.  \textbf{19}(6),  927--940 (2013)

\bibitem{DBLP:conf/gd/GansnerKN04}
Gansner, E.R., Koren, Y., North, S.C.: Graph drawing by stress majorization.
  In: Pach, J. (ed.) Graph Drawing, 12th International Symposium, {GD} 2004.
  Revised Selected Papers. Lecture Notes in Computer Science, vol.~3383, pp.
  239--250. Springer (2004)

\bibitem{ghojogh2021uniform}
Ghojogh, B., Ghodsi, A., Karray, F., Crowley, M.: Uniform manifold
  approximation and projection {(UMAP)} and its variants: Tutorial and survey.
  CoRR  \textbf{abs/2109.02508} (2021)

\bibitem{gurbuzbalaban2021random}
G{\"u}rb{\"u}zbalaban, M., Ozdaglar, A., Parrilo, P.A.: Why random reshuffling
  beats stochastic gradient descent. Mathematical Programming  \textbf{186}(1),
   49--84 (2021)

\bibitem{hagberg2008exploring}
Hagberg, A., Swart, P., S~Chult, D.: Exploring network structure, dynamics, and
  function using networkx. Tech. rep., Los Alamos National Lab.(LANL), Los
  Alamos, NM (United States) (2008)

\bibitem{holland1983stochastic}
Holland, P.W., Laskey, K.B., Leinhardt, S.: Stochastic blockmodels: First
  steps. Social networks  \textbf{5}(2),  109--137 (1983)

\bibitem{jolliffe1986principal}
Jolliffe, I.T.: Principal Component Analysis. Springer Series in Statistics,
  Springer (1986)

\bibitem{DBLP:journals/ipl/KamadaK89}
Kamada, T., Kawai, S.: An algorithm for drawing general undirected graphs. Inf.
  Process. Lett.  \textbf{31}(1),  7--15 (1989)

\bibitem{kim2002eva}
Kim~Norlen, G.L., Gebbie, M., Chuang, J.: Eva: Extraction, visualization and
  analysis of the telecommunications and media ownership network. In:
  International Telecommunications Society 14th Biennial Conference (ITS2002),
  Seoul Korea (2002)

\bibitem{DBLP:journals/cgf/KruigerRMKKT17}
Kruiger, J.F., Rauber, P.E., Martins, R.M., Kerren, A., Kobourov, S.G., Telea,
  A.C.: Graph layouts by t-sne. Comput. Graph. Forum  \textbf{36}(3),  283--294
  (2017)

\bibitem{kruskal1964multidimensional}
Kruskal, J.B.: Multidimensional scaling by optimizing goodness of fit to a
  nonmetric hypothesis. Psychometrika  \textbf{29}(1),  1--27 (1964)

\bibitem{van2008visualizing}
Van~der Maaten, L., Hinton, G.: Visualizing data using t-sne. Journal of
  machine learning research  \textbf{9}(11) (2008)

\bibitem{mcionnes2018umap}
McInnes, L., Healy, J.: {UMAP:} uniform manifold approximation and projection
  for dimension reduction. CoRR  \textbf{abs/1802.03426} (2018)

\bibitem{miller2022spherical}
Miller, J., Huroyan, V., Kobourov, S.G.: Spherical graph drawing by
  multi-dimensional scaling. In: Graph Drawing and Network Visualization - 30th
  International Symposium, {GD} 2022, Tokyo, Japan, September 13-16, 2022,
  Revised Selected Papers. Lecture Notes in Computer Science, vol. 13764, pp.
  77--92. Springer (2022)

\bibitem{miller2022browser}
Miller, J., Kobourov, S.G., Huroyan, V.: Browser-based hyperbolic visualization
  of graphs. In: 15th {IEEE} Pacific Visualization Symposium, PacificVis 2022,
  Tsukuba, Japan, April 11-14, 2022. pp. 71--80. {IEEE} (2022)

\bibitem{munzner2014visualization}
Munzner, T.: Visualization analysis and design. CRC press (2014)

\bibitem{DBLP:conf/apvis/NguyenEH13}
Nguyen, Q.H., Eades, P., Hong, S.: On the faithfulness of graph visualizations.
  In: Carpendale, S., Chen, W., Hong, S. (eds.) {IEEE} Pacific Visualization
  Symposium, PacificVis 2013, February 27 2013-March 1, 2013, Sydney, NSW,
  Australia. pp. 209--216. {IEEE} Computer Society (2013)

\bibitem{noack2003energy}
Noack, A.: An energy model for visual graph clustering. In: International
  symposium on graph drawing. pp. 425--436. Springer (2003)

\bibitem{DBLP:journals/jgaa/NocajOB15}
Nocaj, A., Ortmann, M., Brandes, U.: Untangling the hairballs of
  multi-centered, small-world online social media networks. J. Graph Algorithms
  Appl.  \textbf{19}(2),  595--618 (2015)

\bibitem{DBLP:journals/jgaa/OrtmannKB17}
Ortmann, M., Klimenta, M., Brandes, U.: A sparse stress model. J. Graph
  Algorithms Appl.  \textbf{21}(5),  791--821 (2017)

\bibitem{peixoto_graph-tool_2014}
Peixoto, T.P.: The graph-tool python library. figshare  (2014).
  \doi{10.6084/m9.figshare.1164194}

\bibitem{DBLP:journals/vlc/Purchase02}
Purchase, H.C.: Metrics for graph drawing aesthetics. J. Vis. Lang. Comput.
  \textbf{13}(5),  501--516 (2002)

\bibitem{roweis2000nonlinear}
Roweis, S.T., Saul, L.K.: Nonlinear dimensionality reduction by locally linear
  embedding. {S}cience  \textbf{290}(5500),  2323--2326 (2000)

\bibitem{shepard1962analysis}
Shepard, R.N.: The analysis of proximities: multidimensional scaling with an
  unknown distance function. i. Psychometrika  \textbf{27}(2),  125--140 (1962)

\bibitem{tamassia2013handbook}
Tamassia, R.: Handbook of graph drawing and visualization. CRC press (2013)

\bibitem{tenenbaum2000global}
Tenenbaum, J.B., Silva, V.d., Langford, J.C.: A global geometric framework for
  nonlinear dimensionality reduction. {S}cience  \textbf{290}(5500),
  2319--2323 (2000)

\bibitem{wattenberg2016how}
Wattenberg, M., Viégas, F., Johnson, I.: How to use t-sne effectively. Distill
   (2016). \doi{10.23915/distill.00002}

\bibitem{xiao2017/online}
Xiao, H., Rasul, K., Vollgraf, R.: Fashion-mnist: a novel image dataset for
  benchmarking machine learning algorithms (2017)

\bibitem{DBLP:journals/tvcg/ZhengPG19}
Zheng, J.X., Pawar, S., Goodman, D.F.M.: Graph drawing by stochastic gradient
  descent. {IEEE} Trans. Vis. Comput. Graph.  \textbf{25}(9),  2738--2748
  (2019)

\bibitem{DBLP:journals/tvcg/ZhuCHHLZ21}
Zhu, M., Chen, W., Hu, Y., Hou, Y., Liu, L., Zhang, K.: Drgraph: An efficient
  graph layout algorithm for large-scale graphs by dimensionality reduction.
  {IEEE} Trans. Vis. Comput. Graph.  \textbf{27}(2),  1666--1676 (2021)

\end{thebibliography}

\newpage

\appendix 

\section{Optimization Parameters}
\label{sec:param}
The parameters that we use to optimize the objective function~\eqref{eq:obj_func} by using SGD include: {\bf 
Learning rate}, {\bf Stopping condition} and {\bf Randomization}.
We start by defining an `epoch' as a complete partial gradient calculation of all $\binom{n}{2}$ pairs of vertices in random order.

For the learning rate, while an exponential regime was shown to work well for SGD in the full stress model, the added entropy potentially allows for exploding gradients. However, we did not observe this to become a problem on the graphs that were tested with LGS, as the objective function stabilizes within 10-15 epochs.
We adopt a learning rate schedule similar to that of~\cite{DBLP:journals/tvcg/ZhengPG19}: 
large learning rates initially, with an exponential decay schedule, followed by a $\Theta(1/t)$ schedule, to satisfy the necessary properties for converge to a stationary point~\cite{DBLP:series/lncs/Bottou12}.

The stopping condition for LGS is triggered when either a pre-specified maximum number of epochs is reached, or when convergence occurs before reaching the maximum number.
We experimentally determine the default  maximum number of epochs to be 60.
To declare that the algorithm has converged we need a specific convergence threshold value. As we are using SGD, we want to avoid computing the full cost function (which is computationally expensive). Therefore, we record the maximum distance any vertex moves in an epoch and when it falls below a certain threshold over the entire epoch, we declare convergence. By default, we set  this convergence threshold to a small value $10^{-7}.$ 

Randomization refers to the method used to select the order of pairs updated in an epoch, which has a non-trivial effect on the resulting embedding. While sampling with replacement was originally used for SGD schemes~\cite{holland1983stochastic}, it has recently been shown that 
random reshuffling leads to higher quality in fewer epochs~\cite{gurbuzbalaban2021random}, so we implement random reshuffling for LGS.


\section{Efficient Similarity Computation Between the Nodes}
\label{sec:efficient_powers}

When the number of vertices is large, computing the powers of $\mathbf{A}_G$ is time-consuming, as matrix multiplication is costly. 
We propose the following procedure to reduce the number of matrix multiplications to compute 
$\mathbf{A}^* = \sum_{1 \leq i \leq c} \mathbf{A}_G^i$. 
For a given symmetric affinity matrix $\mathbf{A}_G$ we compute its eigendecomposition $[\mathbf{Q}, \Lambda] = eig(\mathbf{A}_G)$, where the rows of $\mathbf{Q}$ contain the eigenvectors and the diagonal elements of $\Lambda$ contain the corresponding eigenvalues of $\mathbf{A}_G$, $\lambda_1, \lambda_2, \dots, \lambda_n$. Next, we use the well-known fact from linear algebra, that for any $1 \le i$, $\mathbf{A}_G$ and $\mathbf{A}^i_G$ share the same set of eigenvectors with corresponding eigenvalues $\lambda^i_1, \lambda^i_2, \dots, \lambda^i_n$. Next, we observe that, 
\begin{equation*}
\mathbf{A}^* = \sum_{i = 1}^c \mathbf{A}_G^i = \sum_{1 \leq i \leq c} s^i \mathbf{Q} \Lambda^i \mathbf{Q}^T = s^i \mathbf{Q} \left( \sum_{i = 1}^c \Lambda^i \right) \mathbf{Q}^T
\end{equation*}
Computing $A^*$ naively requires $c$ matrix multiplications, but in this way we only use two. We remark, that $\Lambda$ is a diagonal matrix and one can compute its $i$-th power simply by considering the diagonal matrix $\Lambda^i$ with diagonal entries $\lambda^i_1, \lambda^i_2, \dots, \lambda^i_n$.

\section{Algorithmic Parameters}
\label{sec:parameters}
\begin{figure}
    \centering

    \includegraphics[width=0.40\linewidth]{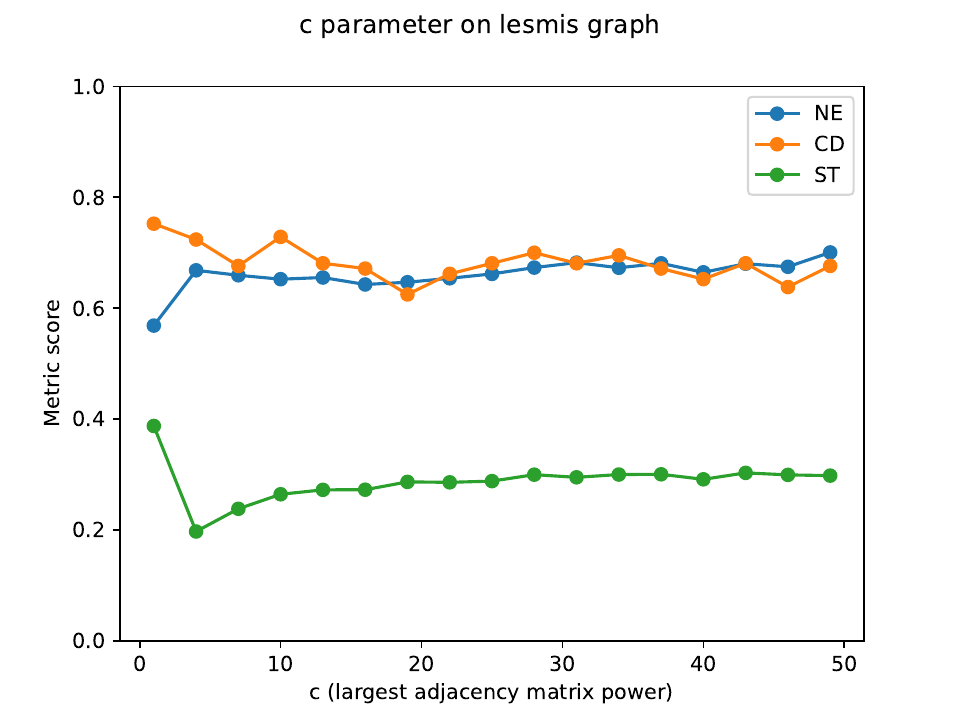}
    \includegraphics[width=0.40\linewidth]{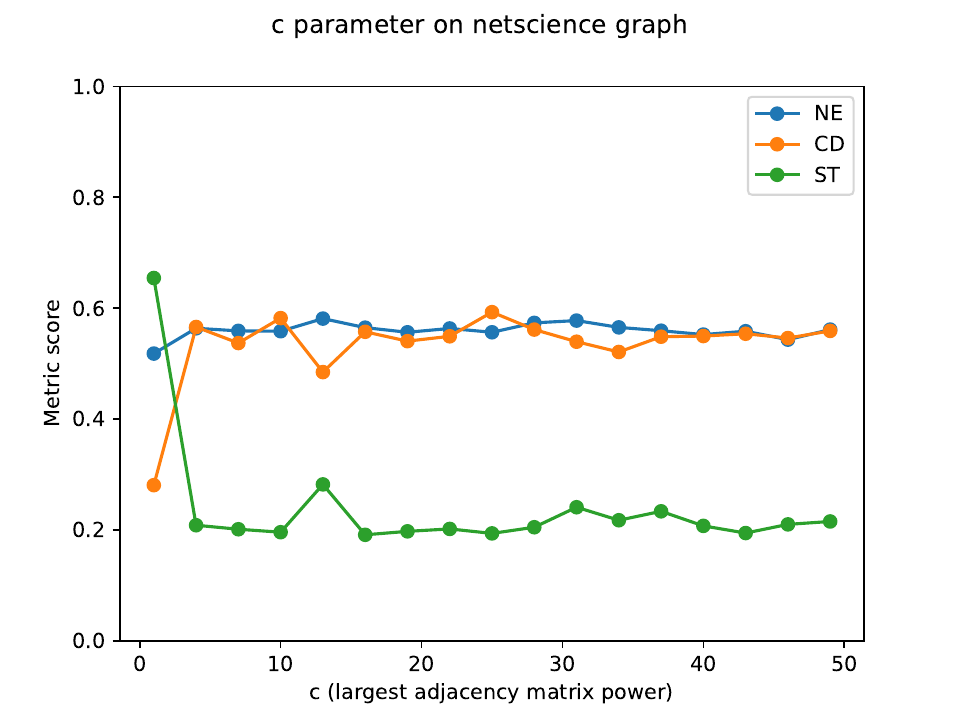}

    \includegraphics[width=0.40\linewidth]{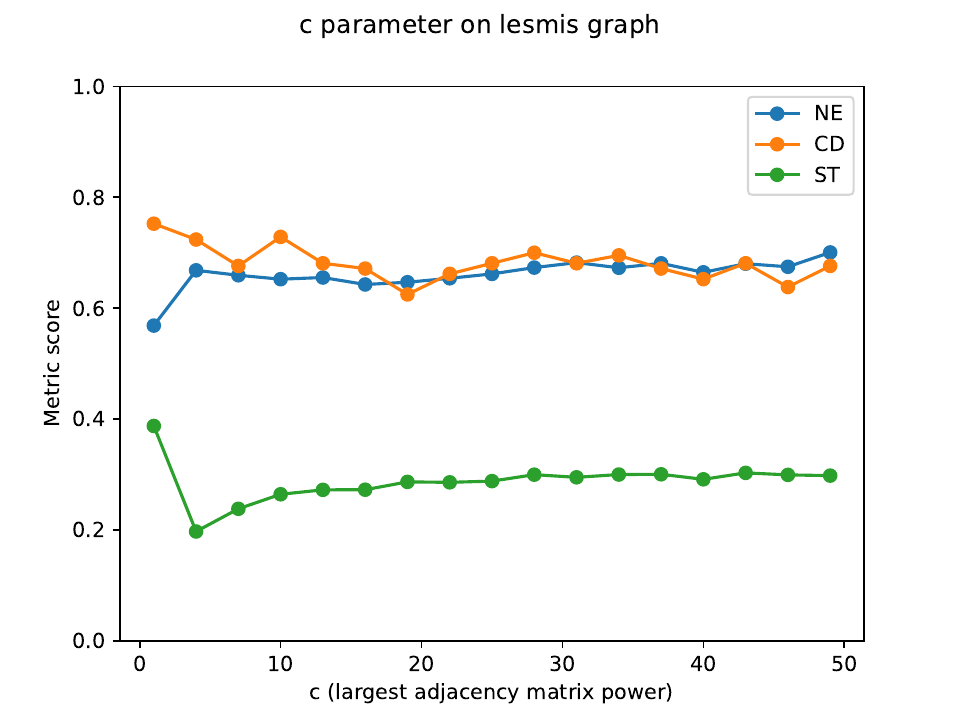}
    \includegraphics[width=0.40\linewidth]{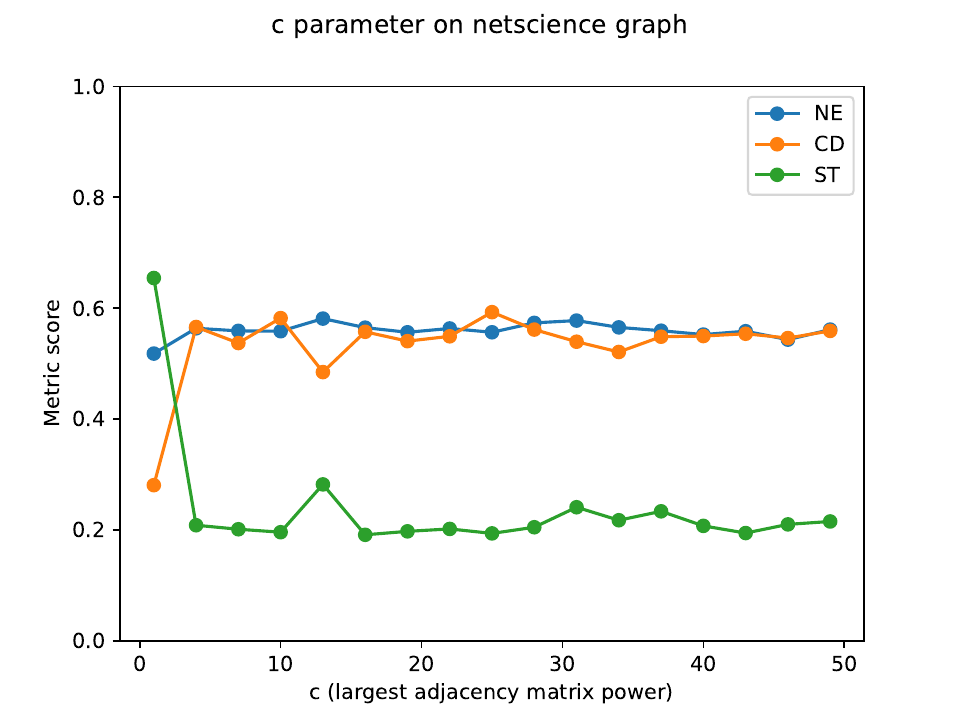}

    \includegraphics[width=0.40\linewidth]{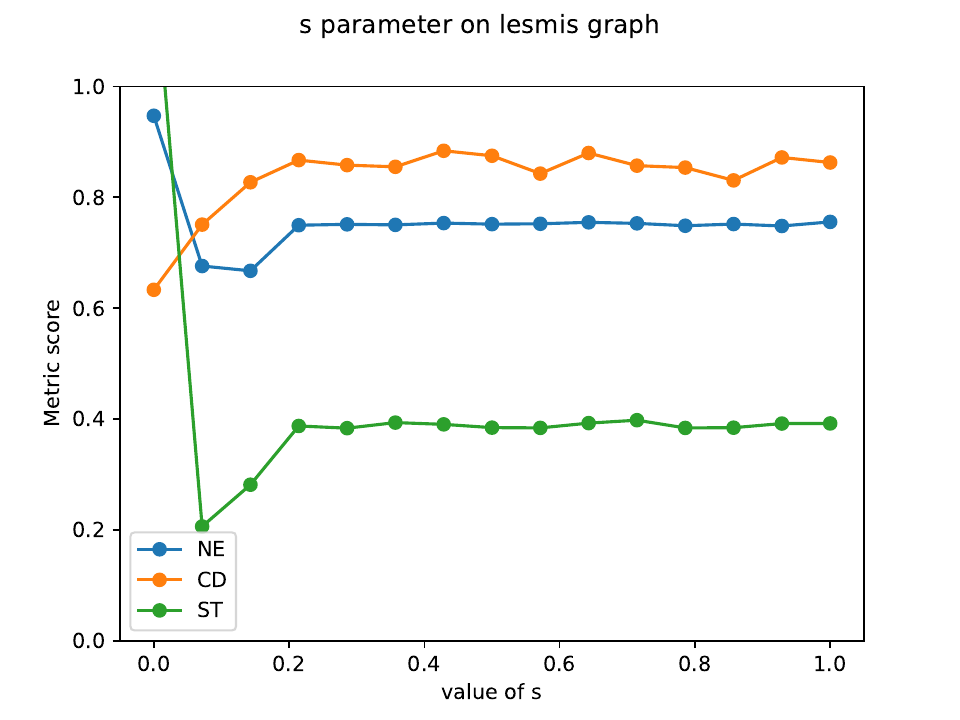}
    \includegraphics[width=0.40\linewidth]{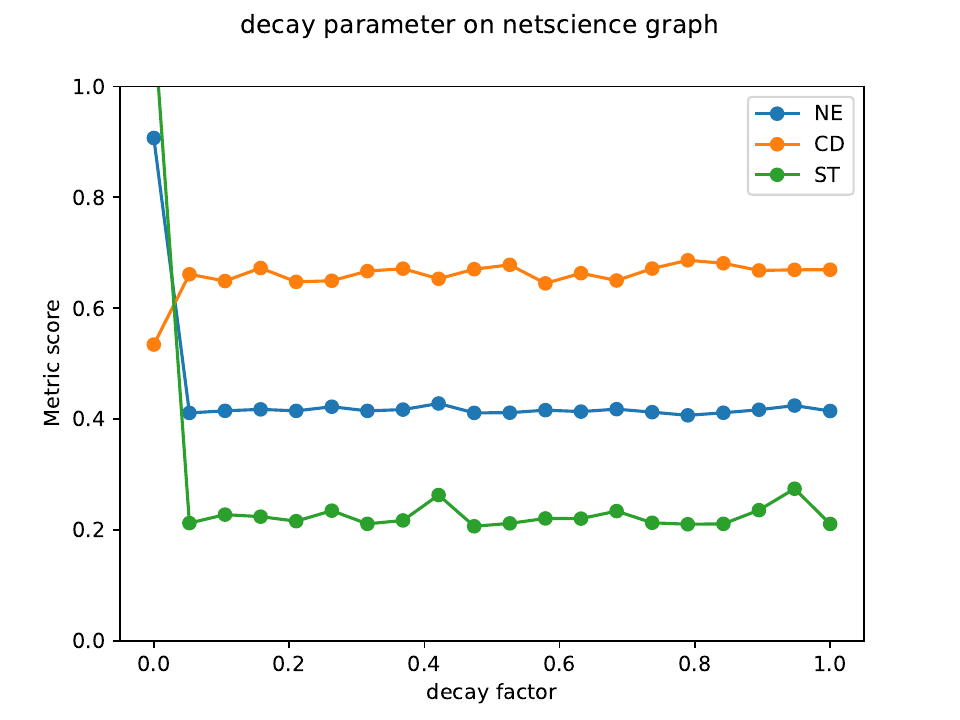}

    \caption{(Top): The effects of the $c$ parameter with $k=25$ on different graphs. The effects appear to be consistent after around 10 or 20, so we set the default to 10. 
    (Middle): The effects of the $\alpha$ parameter with $k=25$ on different graphs. As long as $\alpha > 0$, there is no large effect on the layout. We set the default to 0.2
    (Bottom): The effects of the $s$ parameter or decay factor. A value of 0.1 gave us the best aggregate of all metrics, so we use that as default.}
    \label{fig:c-alpha-exp}
\end{figure}

We investigate LGS's parameters $c$, as defined in Sec.~\ref{sec:3.1}
and $\alpha$, as defined in equation~\eqref{eq:obj_func}, that have an impact on our objective function:
{\bf c}, the power we raise the adjacency matrix to, in pre-processing and
  $\mathbf{\alpha}$, the repulsion term's strength in the modified stress equation.
The power $c$,
has a non-trivial impact on the layout. 
We experimented with the graphs in our benchmark to determine suitable values for $c$, and found that large values have diminishing returns on the quality of the layout; see Fig.~\ref{fig:c-alpha-exp}~(Top). With this in mind, we set the default value of $c$ to 10, but note that small values may result in better local structure (NE) for for non-dense graphs.  

While the MaxEnt model proposes a decaying repulsive force, we fix the 
coefficient of the 
repulsive force, similar to tsNET. As Fig.~\ref{fig:c-alpha-exp}~(Bottom) demonstrates, the value of $\alpha$ does not have a large effect on the resulting layout as long as it is positive. By default, we set $\alpha$ to $0.2$.

\section{LGS Embeddings of High-Dimensional Data}
\label{sec:hd-data}
We can extend the idea of LGS from graph data to high dimensional data. 
Similar to t-SNE~\cite{van2008visualizing} and UMAP~\cite{mcionnes2018umap}, for a given dataset we can construct a weighted graph that represents the given dataset.
The vertices of the graph correspond to datapoints and pairwise as 
\begin{equation*}
w_{ij} = e^{-\frac{d^2(x_i, x_j)}{\sigma^2}}
\end{equation*}
where $\sigma^2$ is a dataset specific constant (can be the sample variance or just set to be 1) and $d(x_i, x_j)$ is the Euclidean distance between the $i$-th and $j$-th datapoints. We further normalize  the weights so that the rows of the corresponding weight matrix add to 1.

This leaves us with a matrix where the weights for small distances approach 1 and weights for large distances approach 0.
We can interpret these as the probability of randomly moving from datapoint $i$ to $j$ in one `hop.' Similar to a graph adjacency matrix, the $c^{th}$ power of the matrix represents the probability of making that jump in $c$ hops. We can then apply the LGS method using the transformed probability matrix in  place of the adjacency matrix; see more details in Sec.~\ref{sec:hd-data}.

To illustrate LGS on non-graph data, we use a high-dimensional dataset. Specifically, we consider the Human Activity Recognition (HAR)~\cite{anguita2013public} dataset: a collection of experimental data of subjects performing 6 different tasks (walking, walking upstairs, walking downstairs, sitting, standing, laying) while wearing a smartphone sensor to collect data. We use the pre-processed data available from~\cite{DBLP:journals/tvcg/EspadotoMKHT21} which has 735 samples with 561 features (dimensions). A typical task for this type of data is to identify which physical activity is being performed based on the data.
We use LGS to produce a low-dimensional (2D) embedding where the activity classes can clearly be seen, and where we can also see the similarity between the classes.


The embeddings shown in Fig.~\ref{fig:har-table} suggest the presence of three distinct clusters within the data: one related to walking activities, another comprising of sitting and standing, and a final one representing lying down.
LGS successfully captures the similarity between the clusters corresponding to sitting/standing and laying by placing them near each other without merging them, as observed in MDS. Interestingly, t-SNE and UMAP do not show any clear organization of the three clusters. 

There are a small number of outliers in the data, shown in green. They are distinctly apparent and separated in all of the LGS embeddings, and we can also see the points individually due to local distance preservation. Of the other algorithms, only UMAP makes these outliers apparent and even then UMAP condenses these data to an overlapping point so it is not as visually clear. 

\begin{figure}
    \begin{center}
        \includegraphics[width=0.32\linewidth]{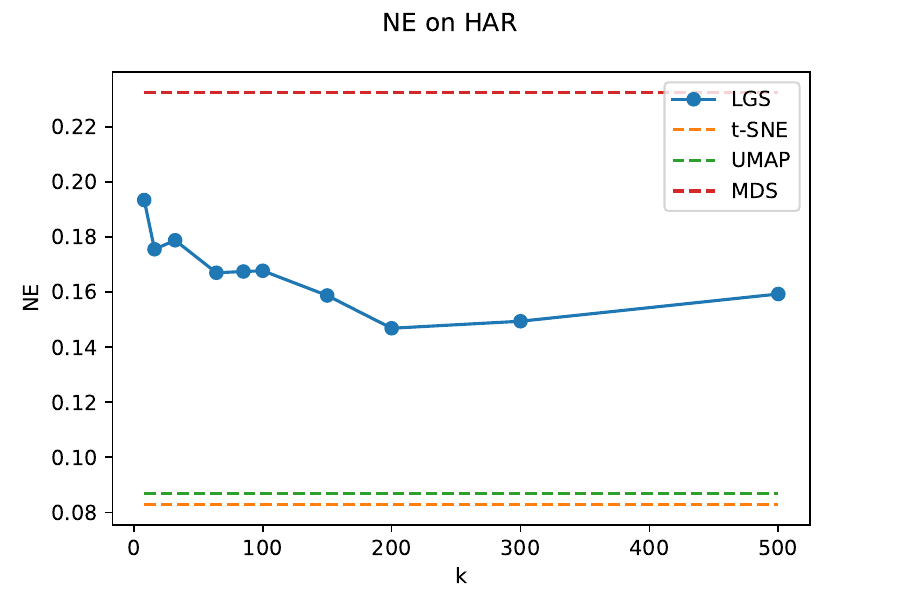}
        \includegraphics[width=0.32\linewidth]{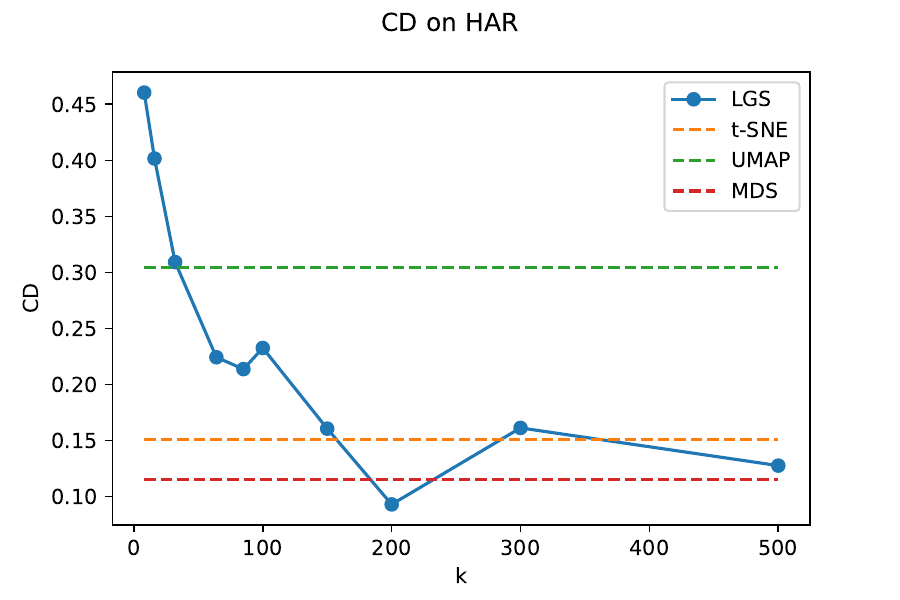}
        \includegraphics[width=0.32\linewidth]{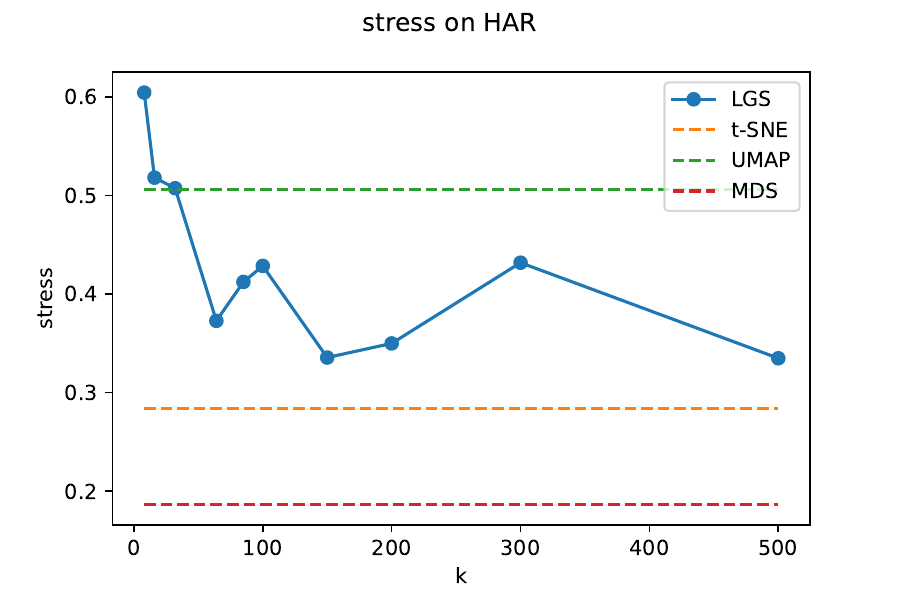}
    \end{center}
    \caption{Metric values for the HAR dataset. The x axis corresponds to $k$, the number of nearest neighbors we consider in LGS. t-SNE, UMAP, and MDS are fixed with default parameters and shown as dotted lines.}
    \label{fig:HAR-metric}
    \vspace{-0.5cm}
\end{figure}

We can verify that LGS captures this intermediate structure appropriately by observing the metrics; see Fig.~\ref{fig:HAR-metric}. LGS manages to be no worse NE than MDS and no worse than UMAP in stress, but is able to achieve consistently lower CD values on the HAR dataset. 
For additional experiments with real-world high dimensional datasets see the supplemental material.

\begin{table*}[ht]
  \centering
  \begin{tabular}
      { |c|c|c|c|}
        \hline
       \parbox[c]{0.2\textwidth}{
      \includegraphics[width=0.2\textwidth]{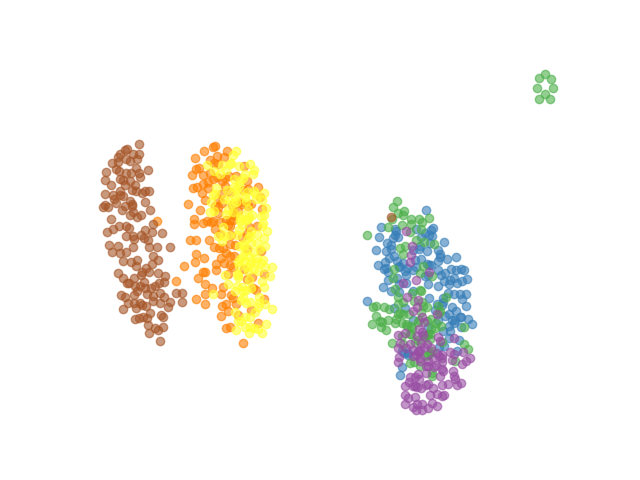}}
       & \parbox[c]{0.2\textwidth}{
      \includegraphics[width=0.2\textwidth]{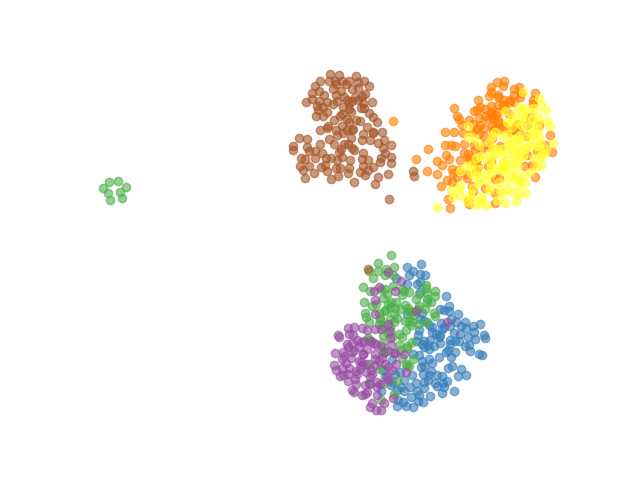}}
       & \parbox[c]{0.2\textwidth}{
      \includegraphics[width=0.2\textwidth]{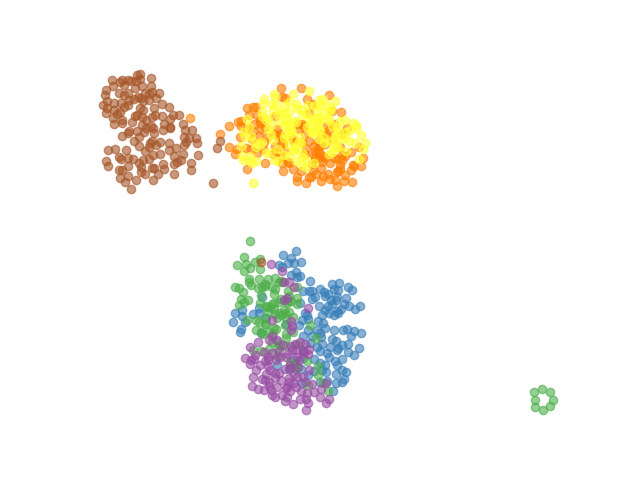}}
       & \parbox[c]{0.2\textwidth}{
      \includegraphics[width=0.2\textwidth]{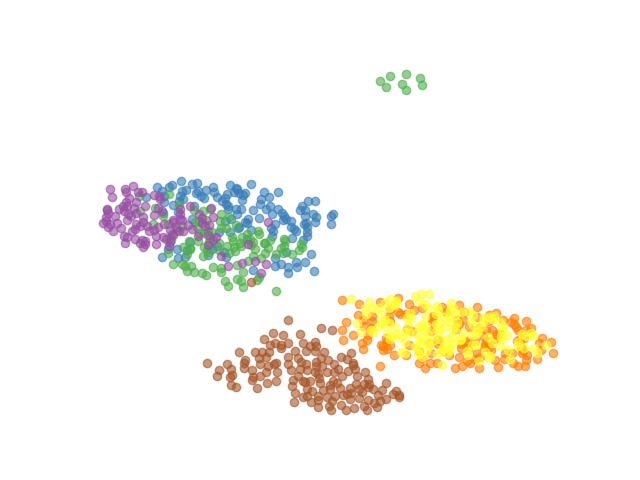}} \\

      LGS(32) & LGS(64) & LGS(85) & LGS(200)\\
      \hline

       \parbox[c]{0.2\textwidth}{
       \centering
      \includegraphics[width=\linewidth]{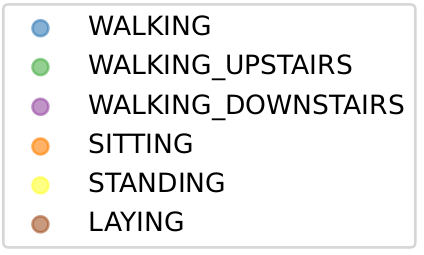}}
       & \parbox[c]{0.2\textwidth}{
      \includegraphics[width=0.2\textwidth]{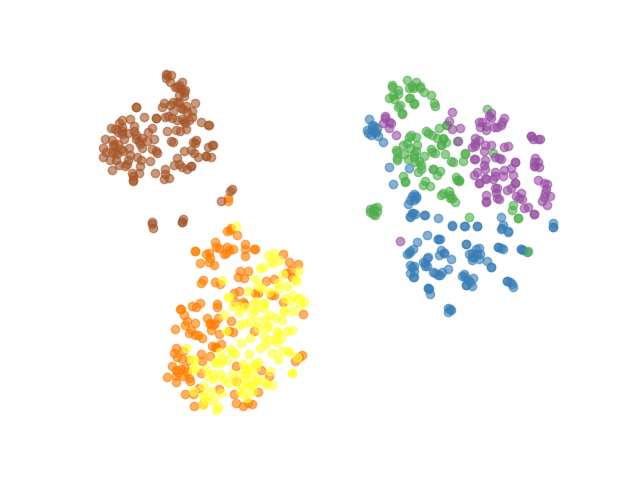}}
       & \parbox[c]{0.2\textwidth}{
      \includegraphics[width=0.2\textwidth]{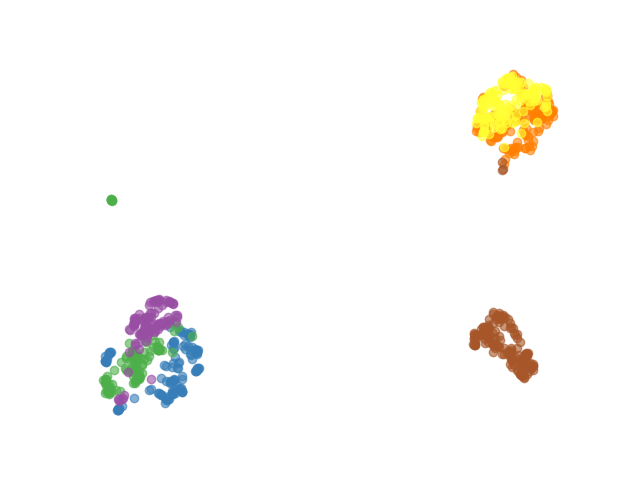}}
       & \parbox[c]{0.2\textwidth}{
      \includegraphics[width=0.2\textwidth]{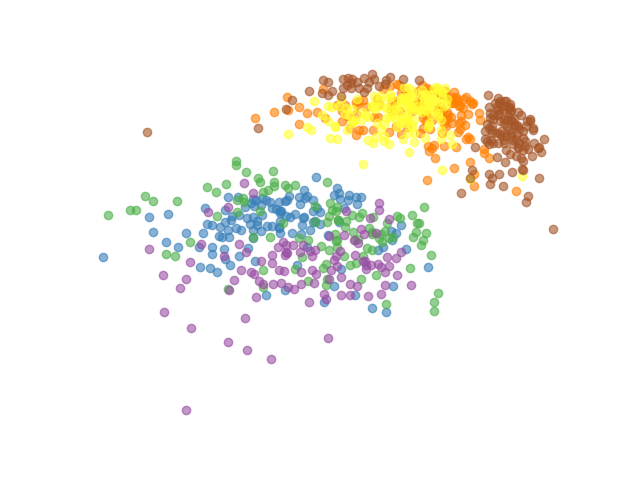}} \\
      & t-SNE & UMAP & MDS\\
      \hline

\end{tabular}
 \captionof{figure}{Embeddings of the HAR dataset by LGS (Top) and t-SNE, UMAP, MDS (Bottom). MDS mixes the sitting/standing and laying clusters, but all other methods capture the three separable clusters present in the data. Note that t-SNE and UMAP clusters are placed arbitrarily far apart, but LGS correctly places the sitting/standing and laying clusters nearby but separable. } \label{fig:har-table}
\end{table*}

\section{Benchmark data}

We list all graphs we drew, experimented, and evaluated on in table~\ref{tab:benchmark}. Source indicates the location the graph can be found or generated from. 

\begin{table}[h!]
    \centering
    \begin{tabularx}{\linewidth}{l X X l X}
        \hline \hline
        Graph & $|V|$ & $|E|$ & Description & Source \\ 
        \hline
        
        lesmis & 77 & 254 & Social & ~\cite{DBLP:journals/toms/DavisH11}\\ 
        
        can\_96 & 96 & 336 & Mesh & ~\cite{DBLP:journals/toms/DuffGL89}\\ 
        
        football & 115 & 613 & Social & ~\cite{noack2003energy}\\ 
        
        rajat11 & 135 & 377 & Social & ~\cite{DBLP:journals/toms/DavisH11}\\ 
        
        mesh3e1 & 289 & 800 & Mesh & ~\cite{DBLP:journals/toms/DavisH11}\\ 
        
        connected\_watts\_300 & 300 & 2400 & WS & ~\cite{hagberg2008exploring}\\ 
        
        powerlaw300 & 300 & 596 & PL & ~\cite{hagberg2008exploring}\\ 
        
        block\_model\_300 & 300 & 1462 & SBM & ~\cite{peixoto_graph-tool_2014}\\ 
        
        netscience & 379 & 914 & Collaboration & ~\cite{DBLP:journals/toms/DavisH11}\\ 
        
        dwt\_419 & 419 & 1572 & Mesh & ~\cite{DBLP:journals/toms/DavisH11}\\ 
        
        block\_model\_500 & 500 & 2494 & SBM & ~\cite{peixoto_graph-tool_2014}\\ 
        
        powerlaw500 & 500 & 1491 & PL & ~\cite{hagberg2008exploring}\\ 
        
        connected\_watts\_500 & 500 & 6500 & WS & ~\cite{hagberg2008exploring}\\ 
        
        grid\_cluster & 900 & 10108 & Synthetic & Sec. 4\\ 
        
        powerlaw1000 & 1000 & 1996 & PL & ~\cite{hagberg2008exploring}\\ 
        
        connected\_watts\_1000 & 1000 & 11000 & WS & ~\cite{hagberg2008exploring}\\ 
        
        block\_model\_1000 & 1000 & 5030 & SBM & ~\cite{peixoto_graph-tool_2014}\\ 
        
        price\_1000 & 1000 & 999 & Synthetic & ~\cite{peixoto_graph-tool_2014}\\ 
        
        dwt\_1005 & 1005 & 3808 & Mesh & ~\cite{DBLP:journals/toms/DavisH11}\\ 
        
        btree9 & 1023 & 1022 & Tree & ~\cite{DBLP:journals/toms/DavisH11}\\ 
        
        CSphd & 1025 & 1043 & Collaboration & ~\cite{DBLP:journals/toms/DavisH11}\\ 
        
        fpga & 1220 & 2807 & Circuit & ~\cite{DBLP:journals/toms/DavisH11}\\ 
        
        block\_2000 & 2000 & 9912 & SBM & ~\cite{DBLP:journals/cgf/KruigerRMKKT17}\\ 
        
        sierpinski3d & 2050 & 6144 & Structural & ~\cite{DBLP:journals/cgf/KruigerRMKKT17}\\ 
        
        EVA & 4475 & 4652 & Collaboration & ~\cite{kim2002eva}\\     
        
        \hline
        \hline
        
    \end{tabularx}
    \caption{Graphs used in our benchmark, listed by number of vertices. All graphs are publicly available as indicated through the source. Synthetic graph source indicates what was used to generate them, and are provided through our GitHub repository for convenience.
    }
    \vspace{-0.5cm}
    \label{tab:benchmark}
\end{table}

\section{fashion-MNIST}

\begin{figure}
    \begin{center}
        \includegraphics[width=0.24\linewidth]{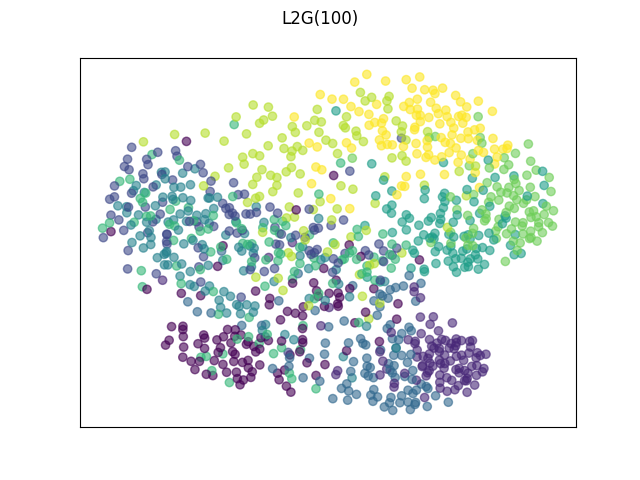}
        \includegraphics[width=0.24\linewidth]{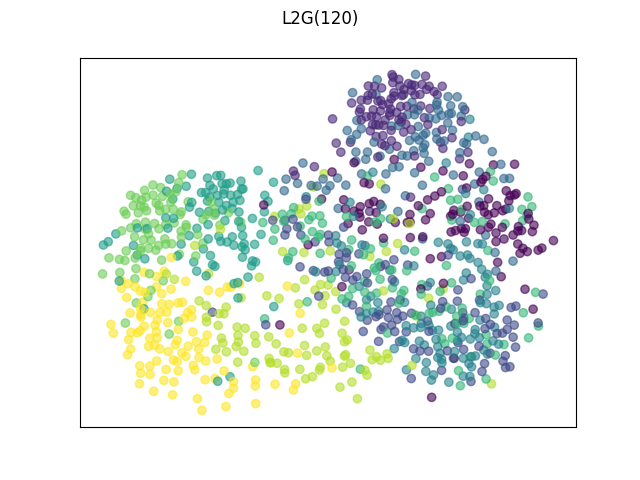} 
        \includegraphics[width=0.24\linewidth]{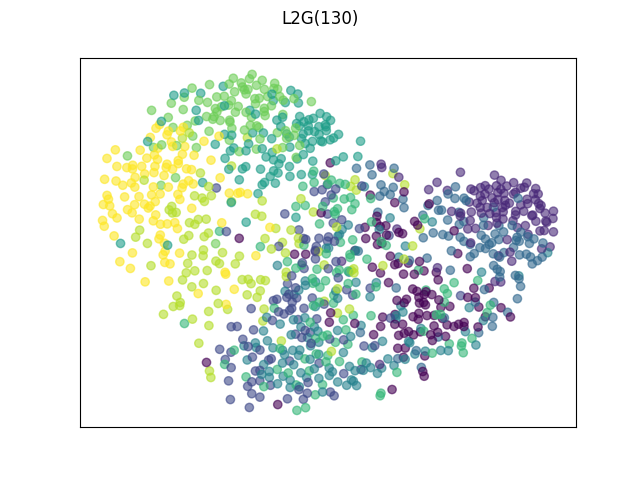} 
        \includegraphics[width=0.24\linewidth]{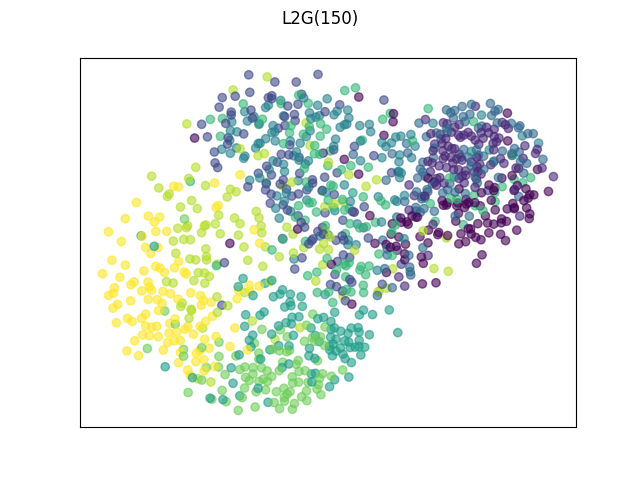}        

    \parbox[c]{0.24\linewidth}{\centering L2G(100)}
    \parbox[c]{0.24\linewidth}{\centering L2G(120)}
    \parbox[c]{0.24\linewidth}{\centering L2G(130)}
    \parbox[c]{0.24\linewidth}{\centering L2G(150)}    
    \end{center}

    \begin{center}
          \includegraphics[width=0.32\linewidth]{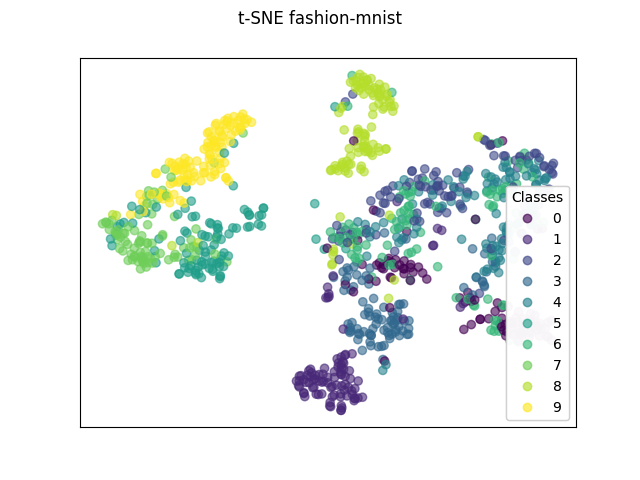}
        \includegraphics[width=0.32\linewidth]{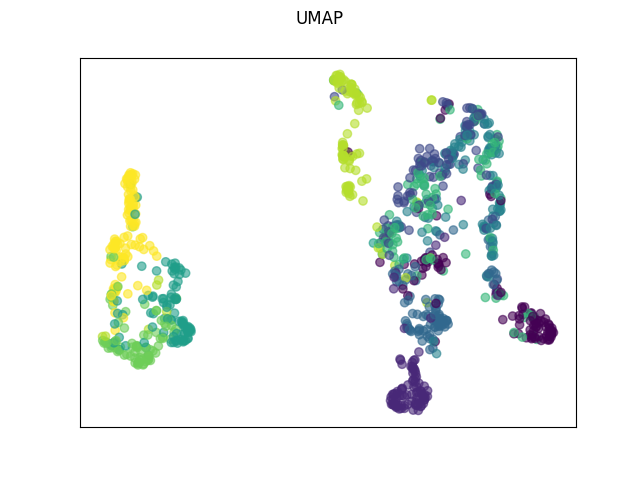} 
        \includegraphics[width=0.32\linewidth]{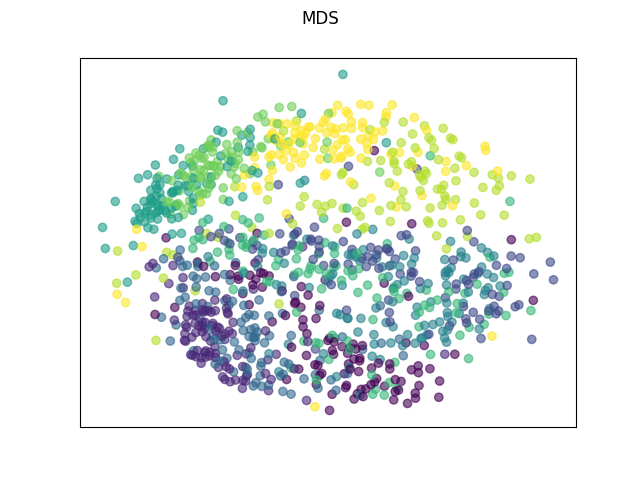}       
    \parbox[c]{0.32\linewidth}{\centering t-SNE}
    \parbox[c]{0.32\linewidth}{\centering UMAP}
    \parbox[c]{0.32\linewidth}{\centering MDS}         
    \end{center}

    \caption{fashion-MNIST displayed by each embedding technique. We see that t-SNE adn UMAP seperate clusters out into islands, while MDS mixes and obscures them. L2G finds some middle ground, with clusters being seperable, but with little white space between them, which is faithful to the underlying data.}
    \label{fig:fashion-mnist}
\end{figure}

\noindent
\textbf{fashion-MNIST}
To give an example of applying L2G to high-dimensional data, we investigate how L2G performs on the MNIST-fashion~\cite{xiao2017/online} dataset compared to t-SNE, UMAP, and MDS. MNIST-fashion is a 784 dimensional set of grey-scale images consisting of clothing items. There are 10 classes corresponding to type of clothing (i.e. trousers, dress, sneakers). 

MNIST-fashion has distinct classes that one might expect should be related at a cluster level. For instance, `t-shirts' and `shirts' would seem more similar than `t-shirts' and `sneakers'. This makes it an ideal use case for L2G where one might expect local methods to visually seperate clusters arbitrarily while global methods may not seperate clusters well at all.  

t-SNE and UMAP create `islands' of clusters, which is not entirely accurate to the underlying data. For instance the light green-yellow island (class 7) near the top of the t-SNE and UMAP plots are sneakers and have been placed with visual separation from the yellow class of ankle boots. While it would seem that these classes are similar yet distinct, t-SNE and UMAP have seperated them. L2G on the other hand, places them nearby in all shown embeddings; See Fig.~\ref{fig:fashion-mnist}.

\section{Additional embeddings}
We provide the remaining embeddings from our benchmark graphs that did not appear in the paper in Fig.~\ref{tab:drawings2} and Fig.~\ref{tab:drawings3}. They appear just as they do in the main body, with  tsNET on the far left, MDS on the far right, UMAP in the middle and the remaining being differing values of L2G. 

\begin{table*}
  [ht] \caption{Additional embeddings} \label{tab:drawings2}
  \vspace{0.5cm}
  \centering
  \begin{tabular}
      {| l | c c c c c c c|} \hline & tsNET & L2G k=16 & L2G k=32 & UMAP & L2G k=64 & L2G k=100 & MDS\\
      \hline 

     \multirow{2}{*}{\vspace{-1cm}\rotatebox[origin=c]{90}{can\_96}}  &
      \parbox[c]{\tabfig\textwidth}{} 
      & \parbox[c]{\tabfig\textwidth}{
      \includegraphics[width=\tabfig\textwidth,height=\tabfig\textwidth]{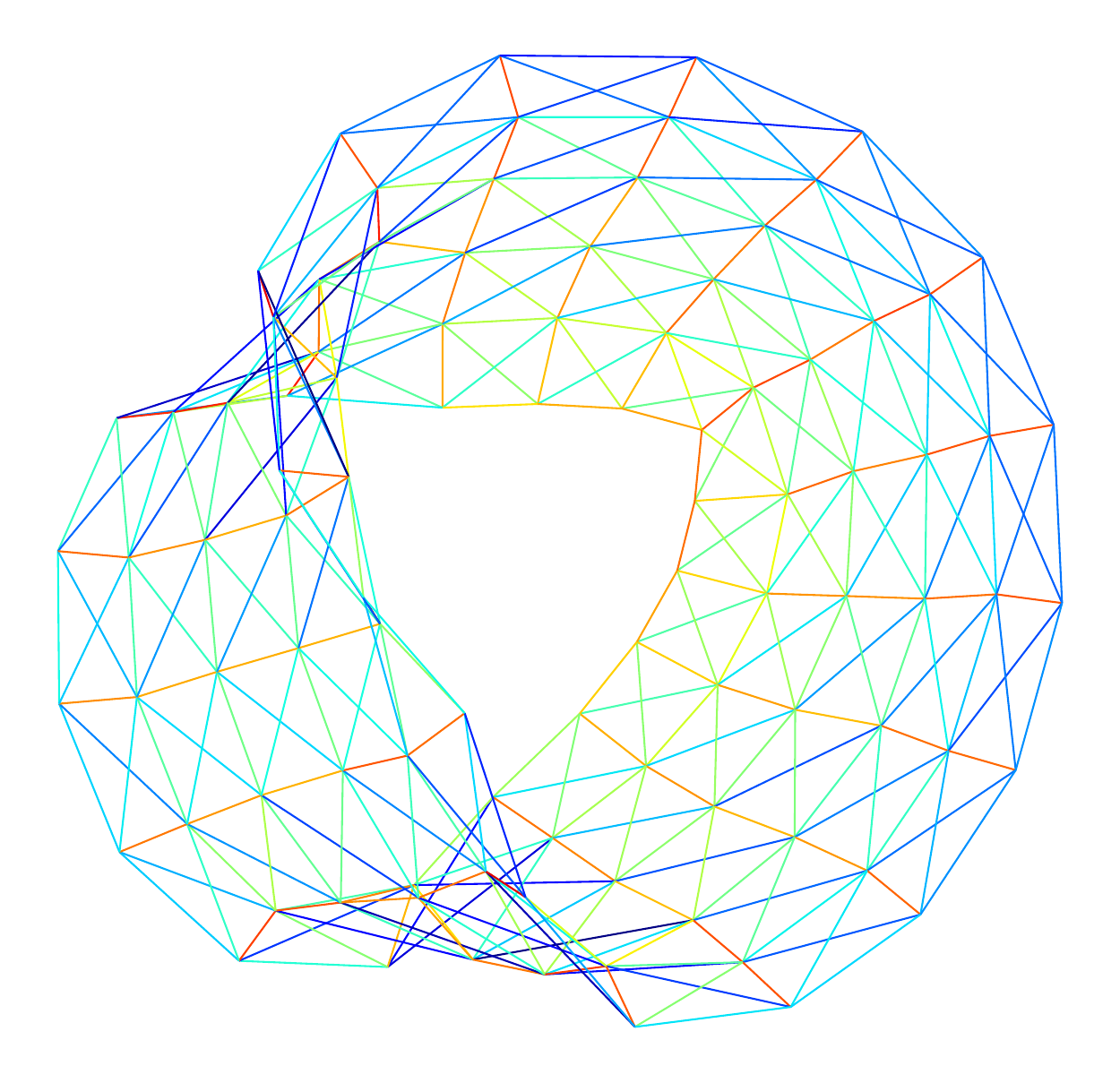}}
      & \parbox[c]{\tabfig\textwidth}{
      \includegraphics[width=\tabfig\textwidth,height=\tabfig\textwidth]{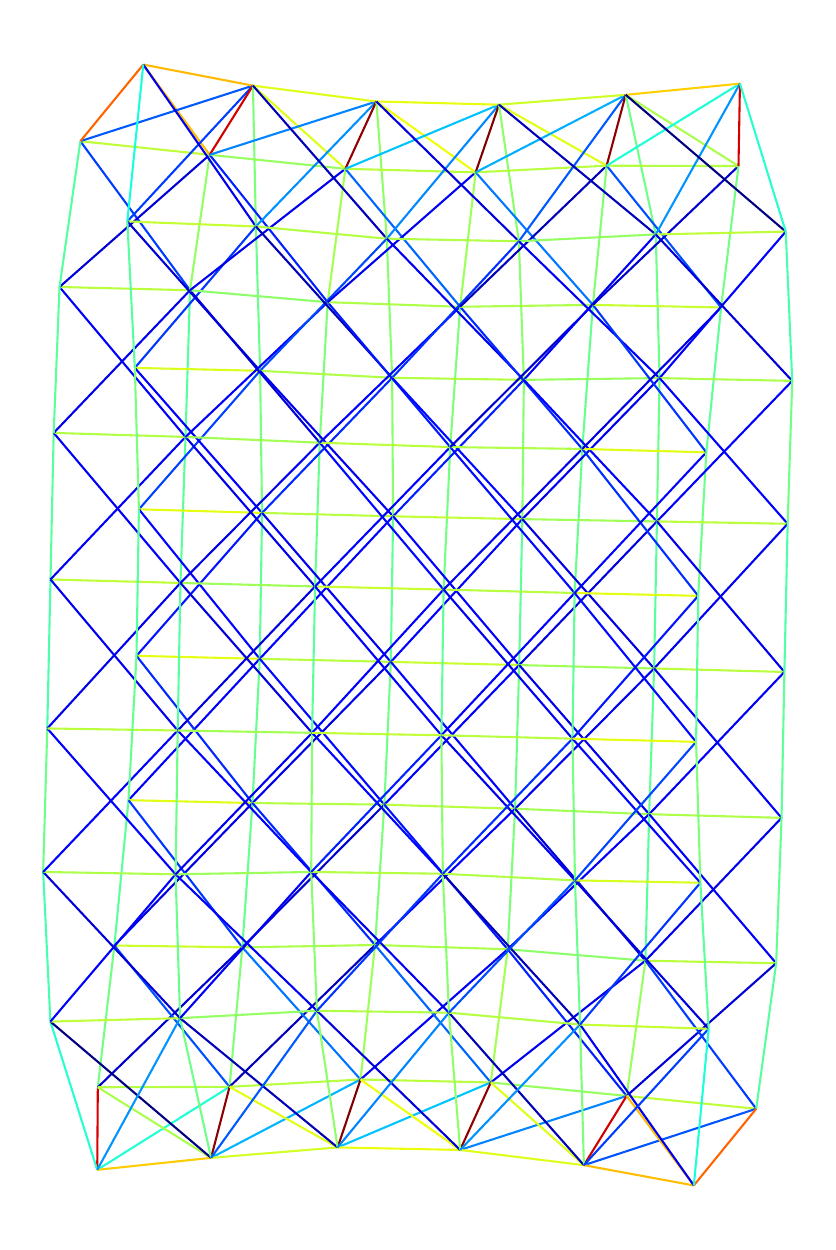}}
      & \parbox[c]{\tabfig\textwidth}{}
      & \parbox[c]{\tabfig\textwidth}{
      \includegraphics[width=\tabfig\textwidth,height=\tabfig\textwidth]{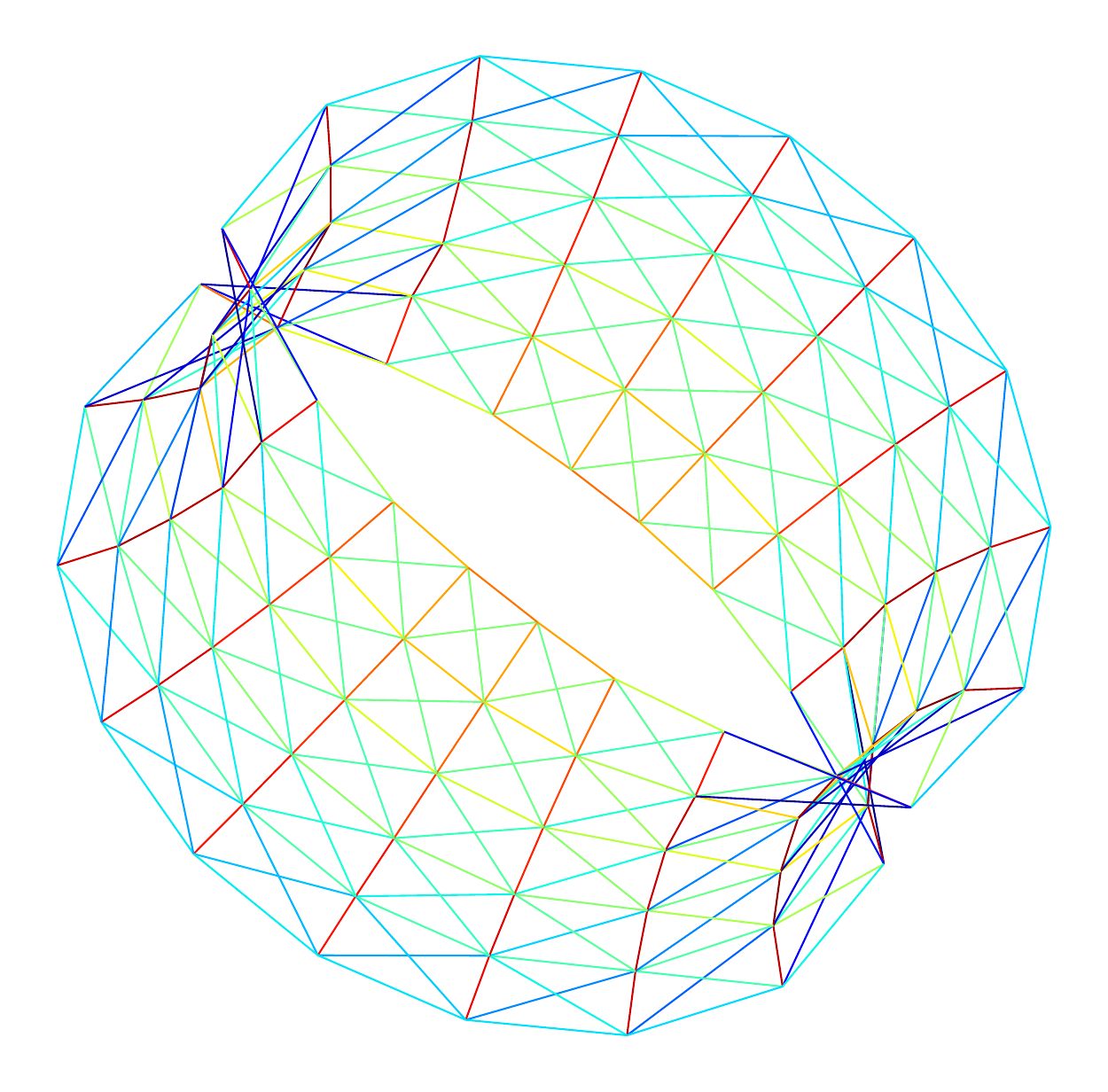}} 
      & \parbox[c]{\tabfig\textwidth}{
      \includegraphics[width=\tabfig\textwidth,height=\tabfig\textwidth]{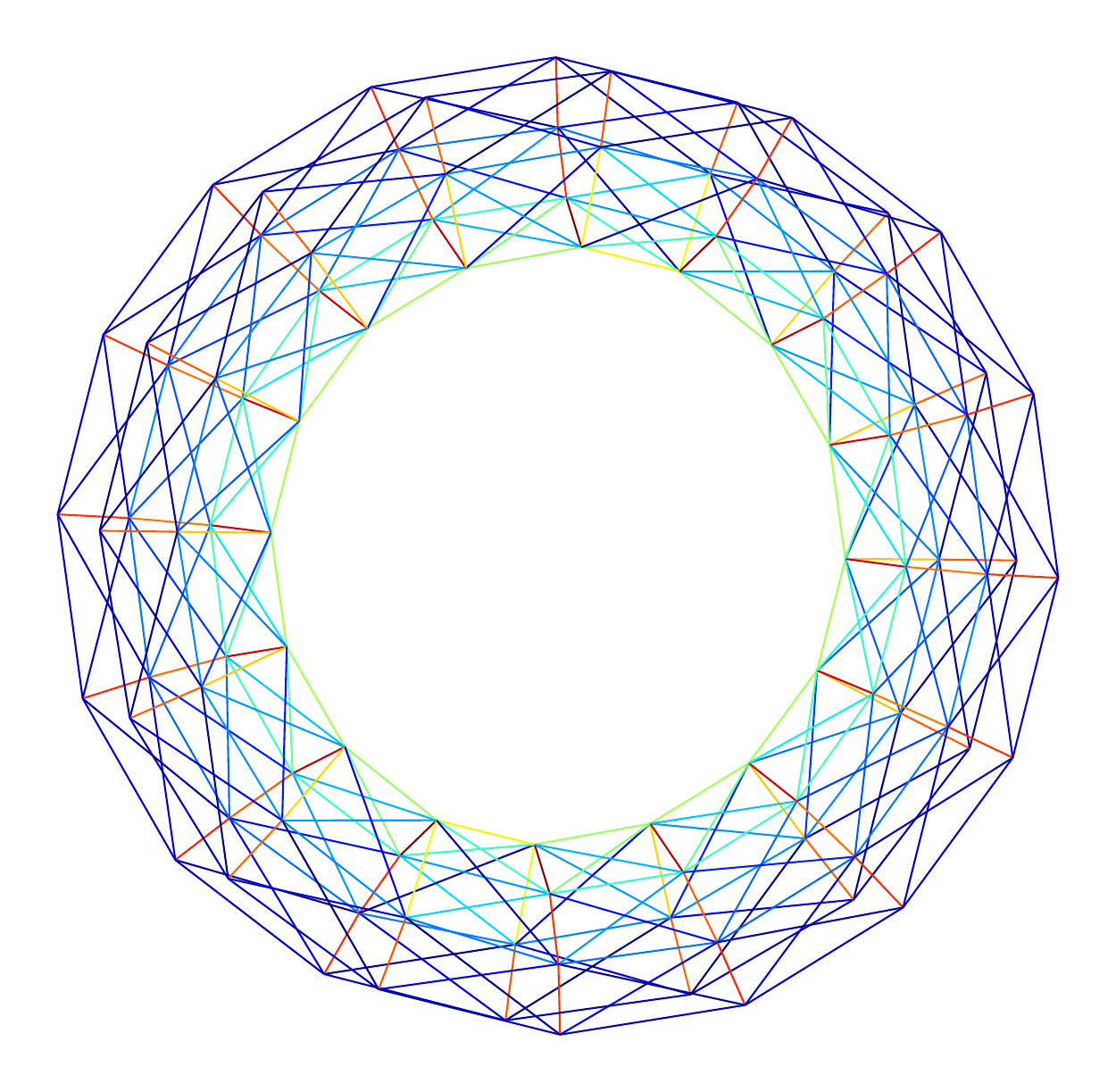}} 
      & \parbox[c]{\tabfig\textwidth}{}  \\
      
       &
      \parbox[c]{\tabfig\textwidth}{
      \taboffset\includegraphics[width=\tabfig\textwidth,height=\tabfig\textwidth]{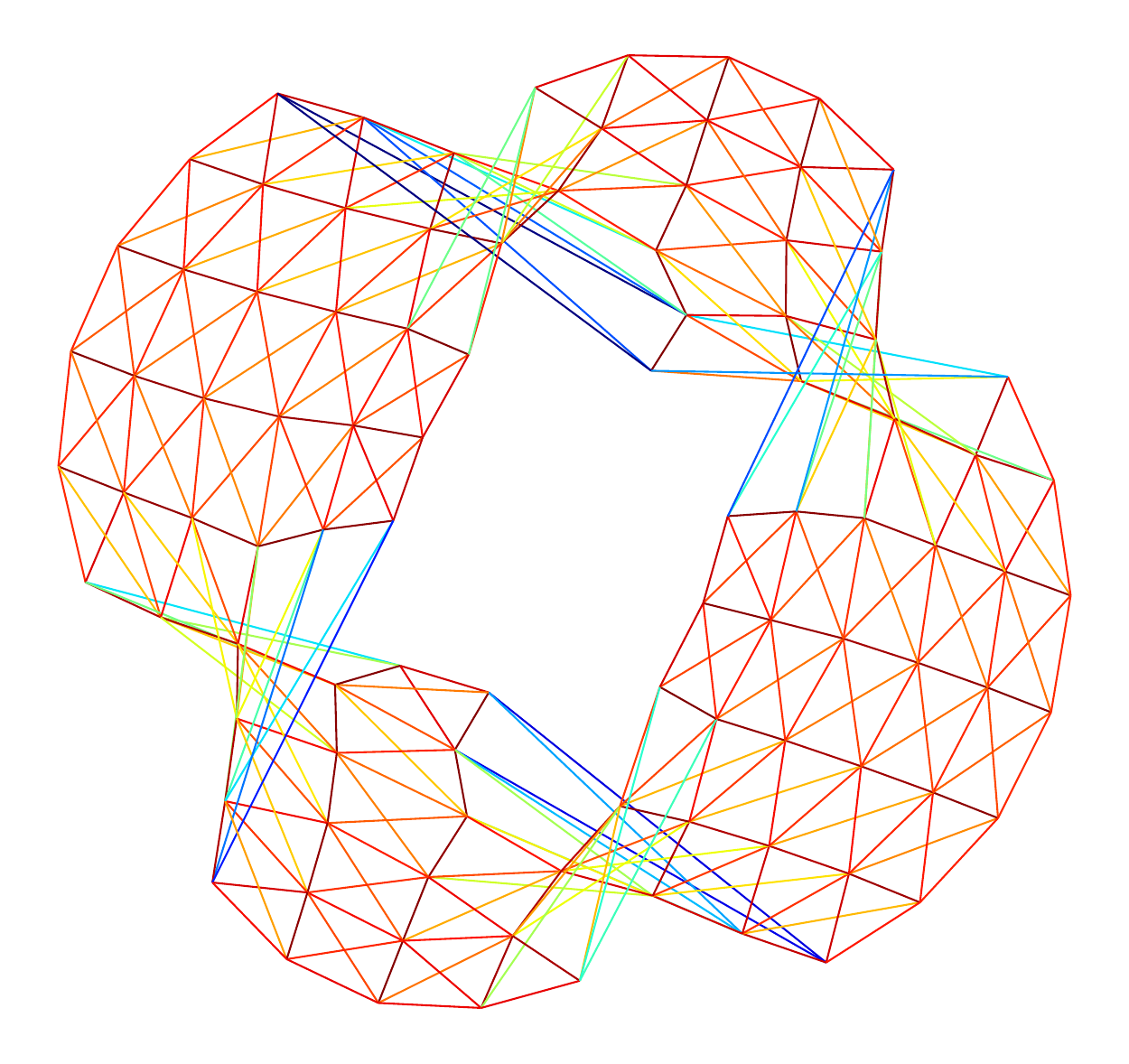}} 
      & \parbox[c]{\tabfig\textwidth}{}
      & \parbox[c]{\tabfig\textwidth}{}
      & \parbox[c]{\tabfig\textwidth}{\taboffset
      \includegraphics[width=\tabfig\textwidth,height=\tabfig\textwidth]{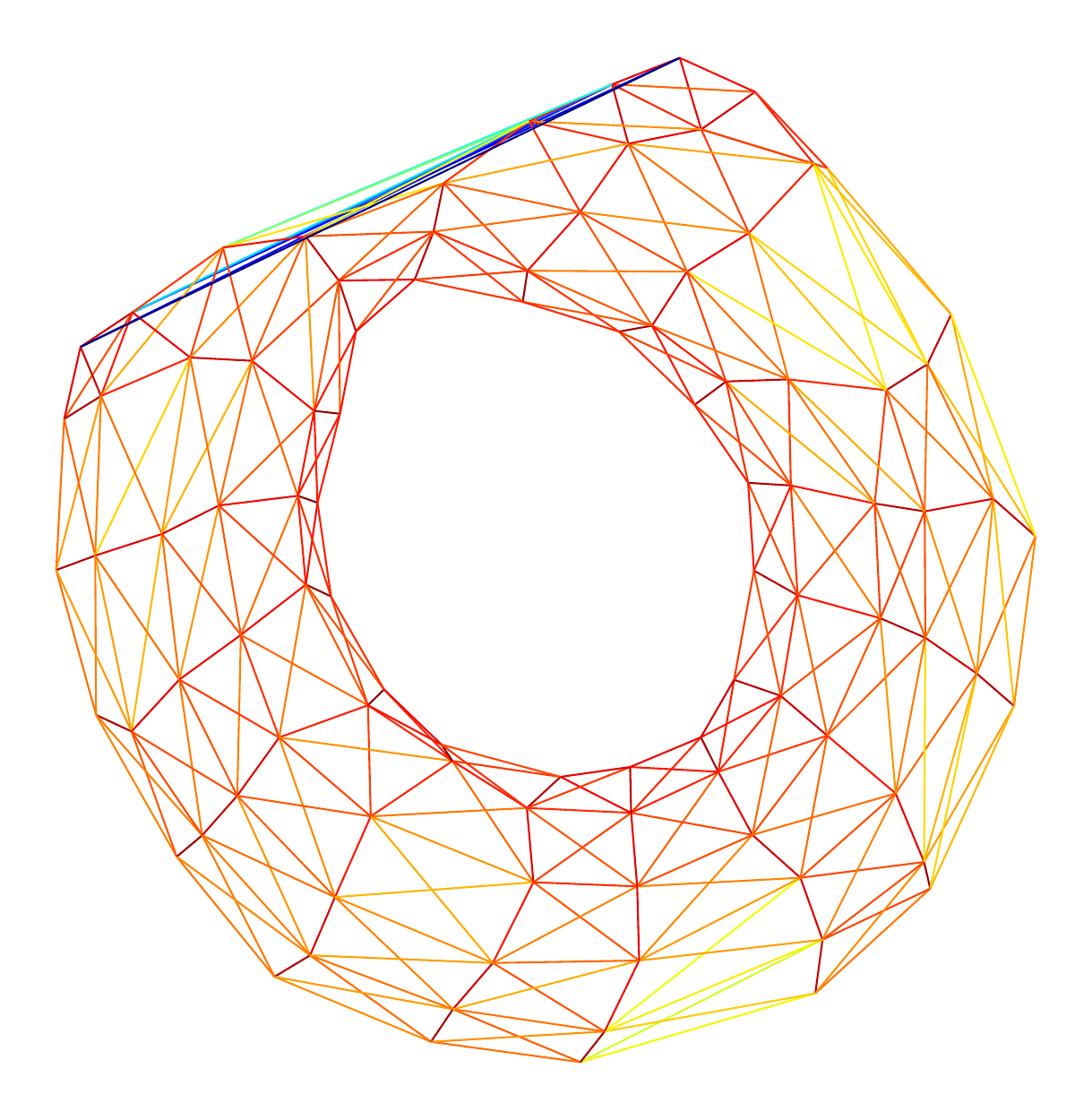}}
      & \parbox[c]{\tabfig\textwidth}{} 
      & \parbox[c]{\tabfig\textwidth}{} 
      & \parbox[c]{\tabfig\textwidth}{\taboffset
      \includegraphics[width=\tabfig\textwidth,height=\tabfig\textwidth]{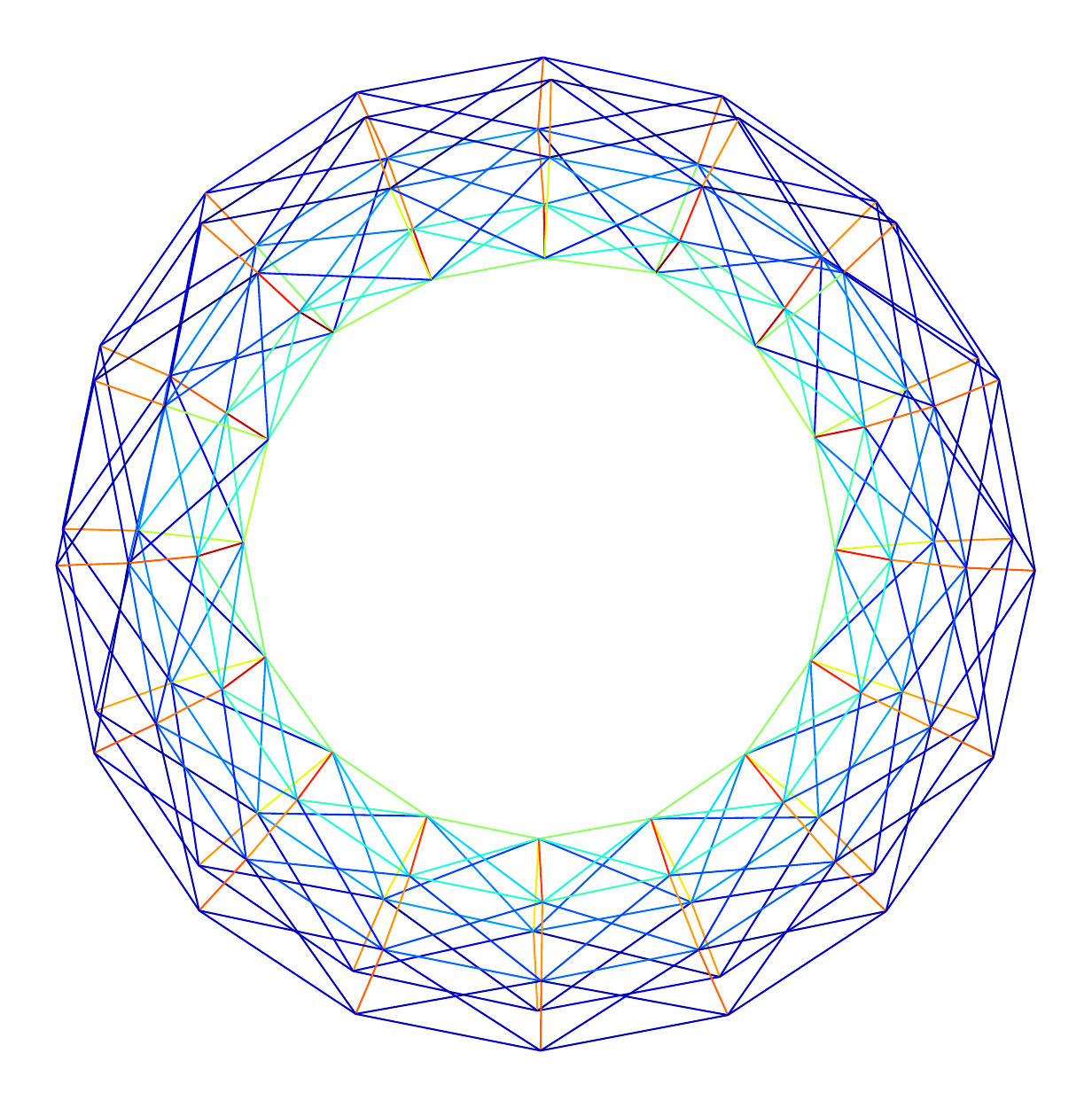}}  \\   
      \hline
      
     \multirow{2}{*}{\vspace{-1cm}\rotatebox[origin=c]{90}{rajat11}}  &
      \parbox[c]{\tabfig\textwidth}{} 
      & \parbox[c]{\tabfig\textwidth}{
      \includegraphics[width=\tabfig\textwidth,height=\tabfig\textwidth]{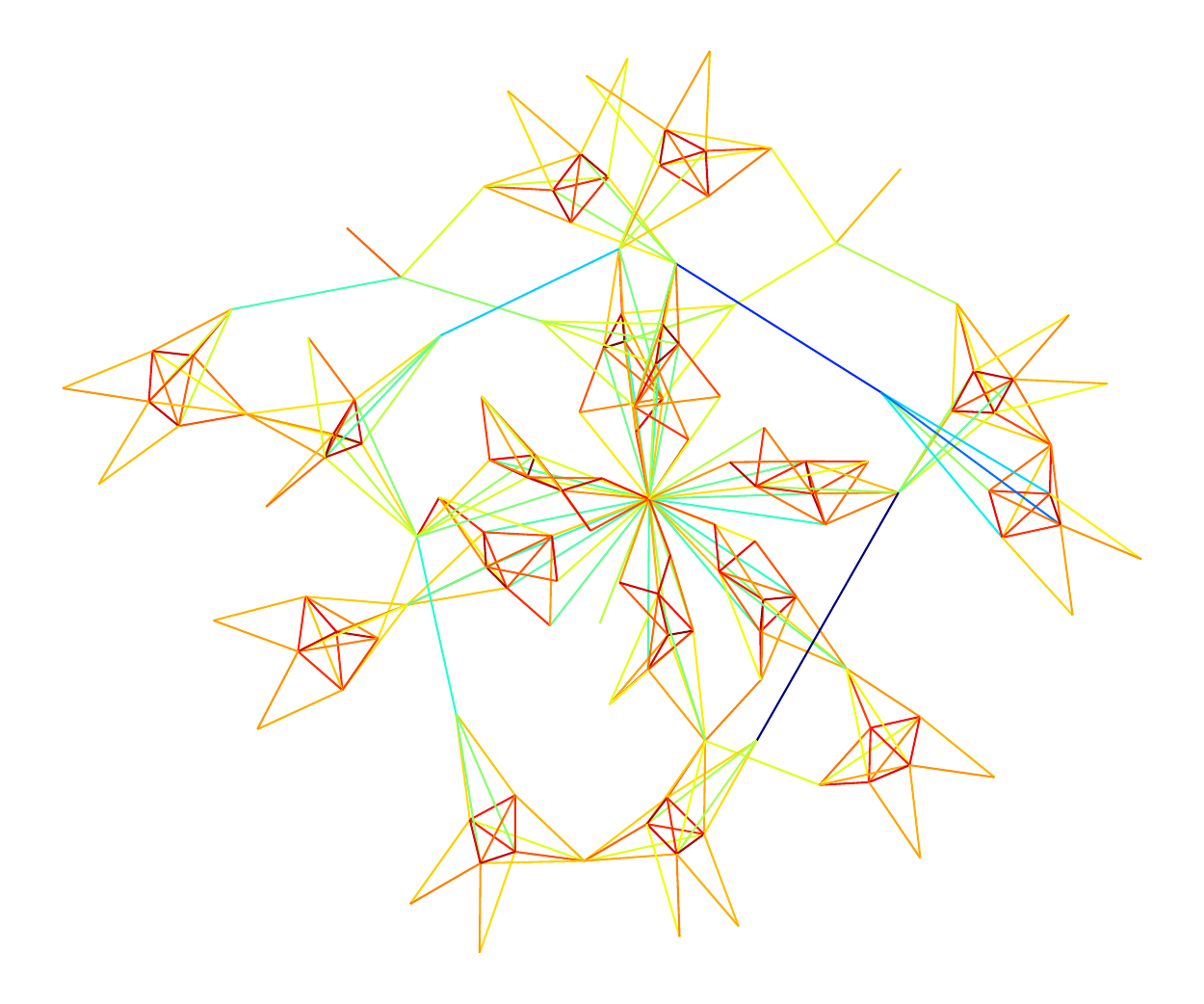}}
      & \parbox[c]{\tabfig\textwidth}{
      \includegraphics[width=\tabfig\textwidth,height=\tabfig\textwidth]{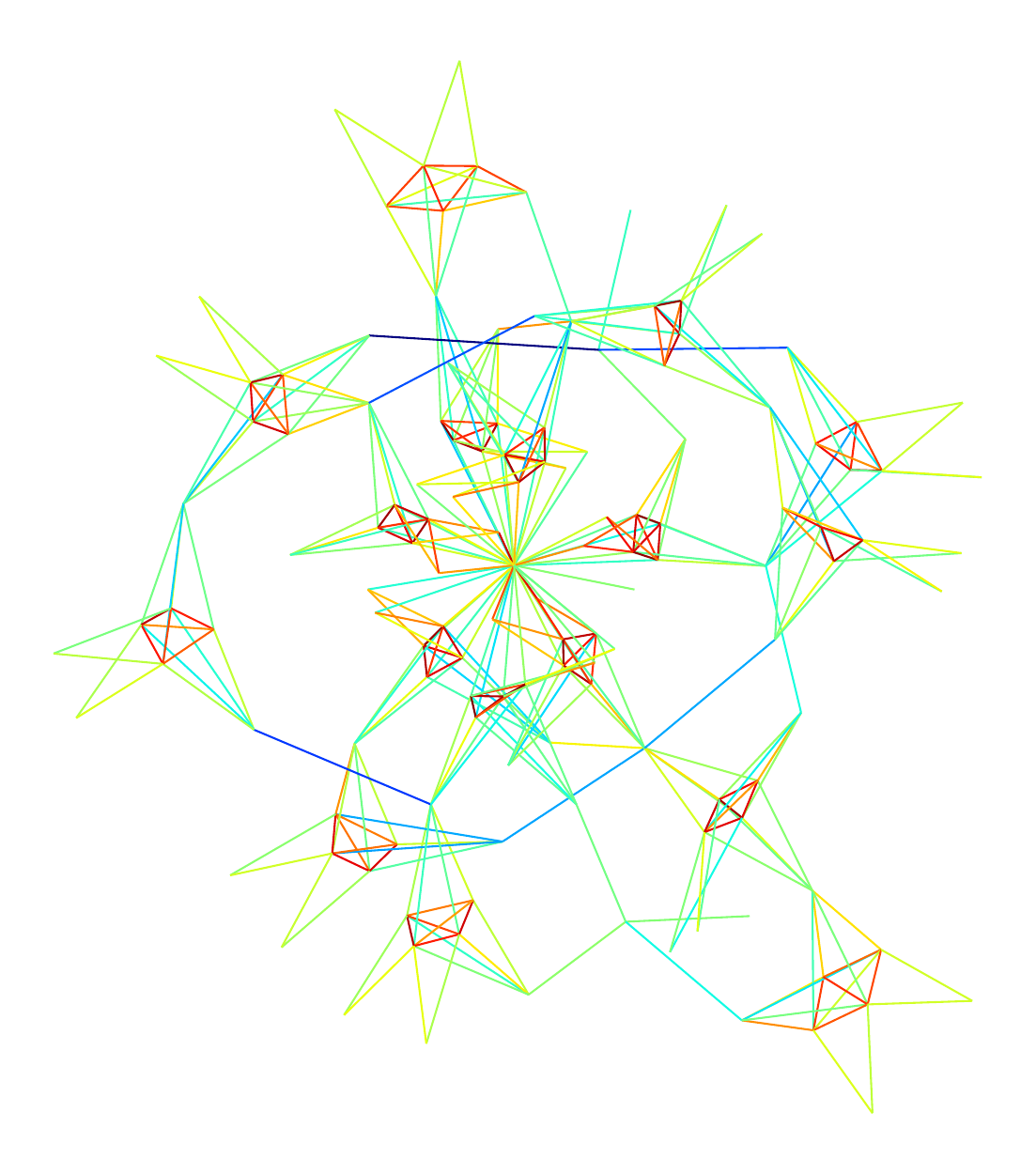}}
      & \parbox[c]{\tabfig\textwidth}{}
      & \parbox[c]{\tabfig\textwidth}{
      \includegraphics[width=\tabfig\textwidth,height=\tabfig\textwidth]{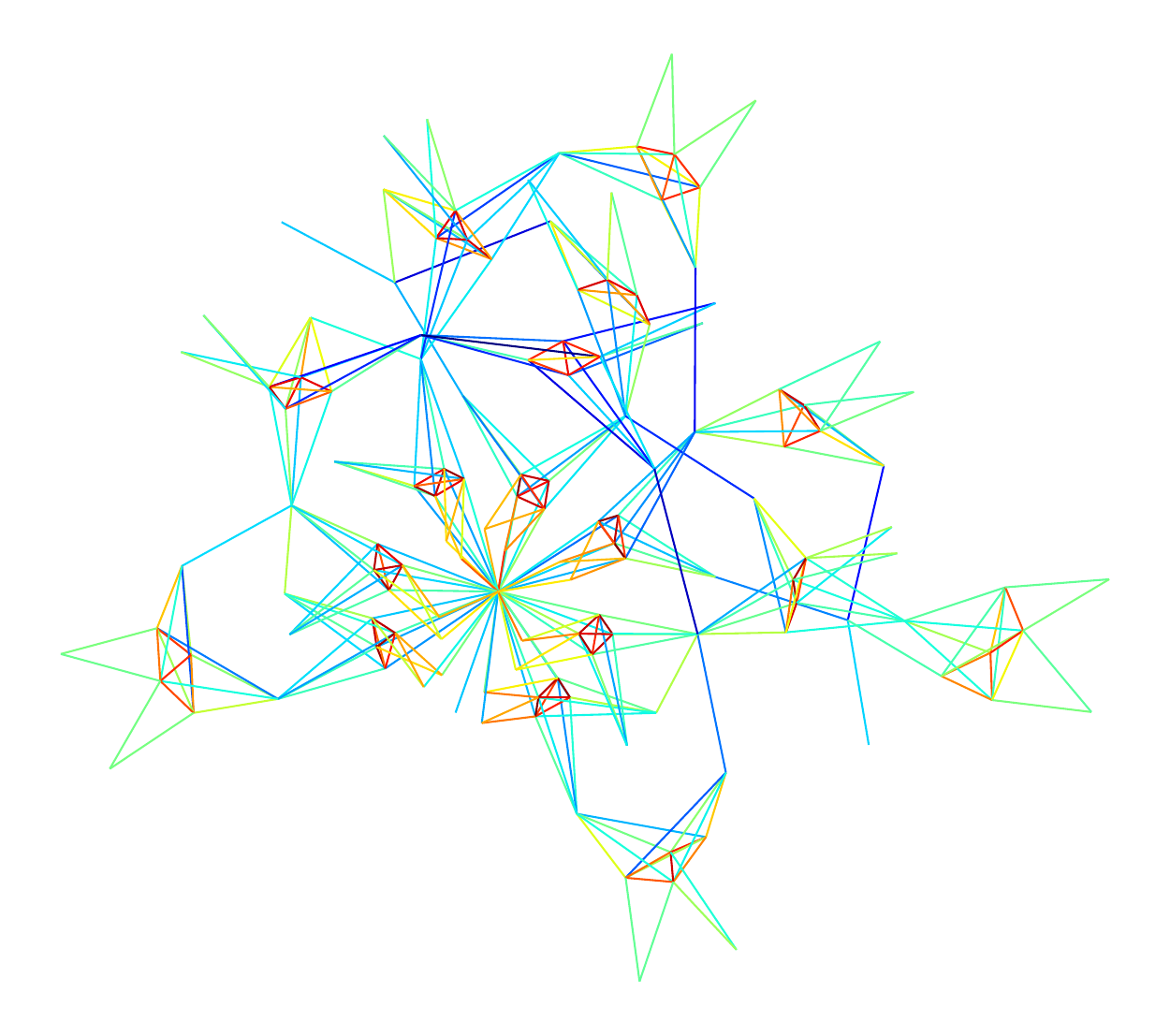}} 
      & \parbox[c]{\tabfig\textwidth}{
      \includegraphics[width=\tabfig\textwidth,height=\tabfig\textwidth]{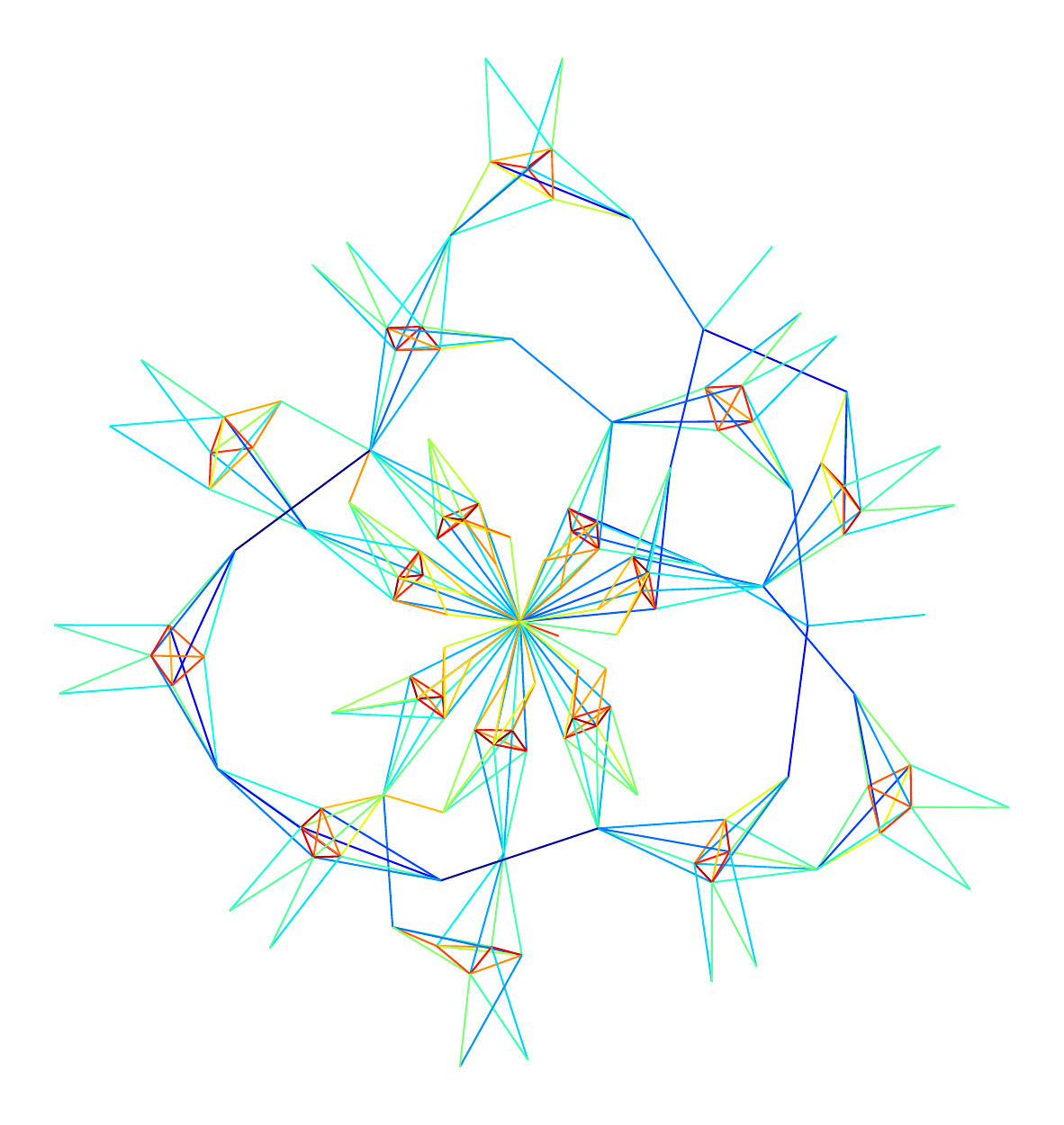}} 
      & \parbox[c]{\tabfig\textwidth}{}  \\
      
       &
      \parbox[c]{\tabfig\textwidth}{
      \taboffset\includegraphics[width=\tabfig\textwidth,height=\tabfig\textwidth]{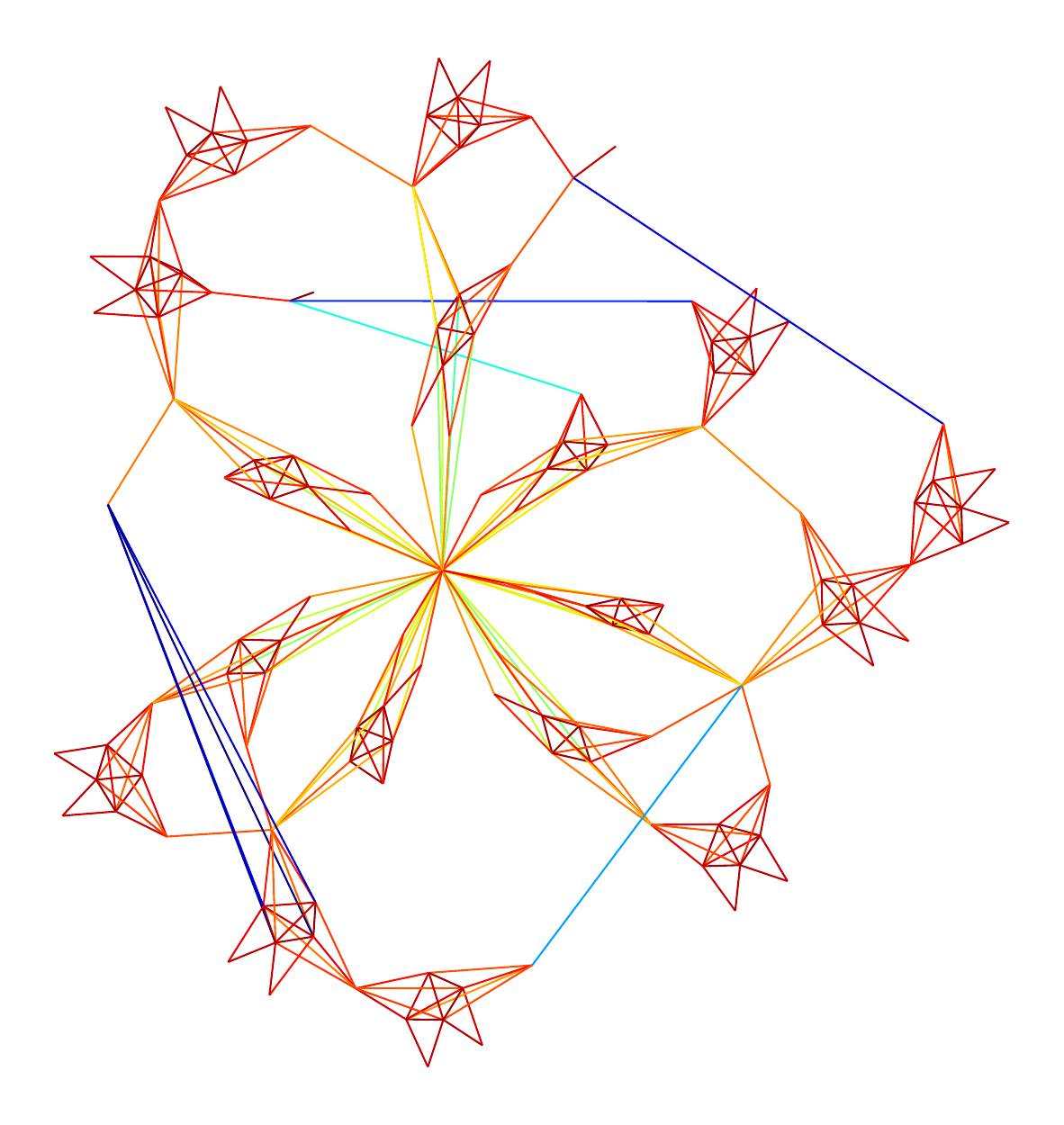}} 
      & \parbox[c]{\tabfig\textwidth}{}
      & \parbox[c]{\tabfig\textwidth}{}
      & \parbox[c]{\tabfig\textwidth}{\taboffset
      \includegraphics[width=\tabfig\textwidth,height=\tabfig\textwidth]{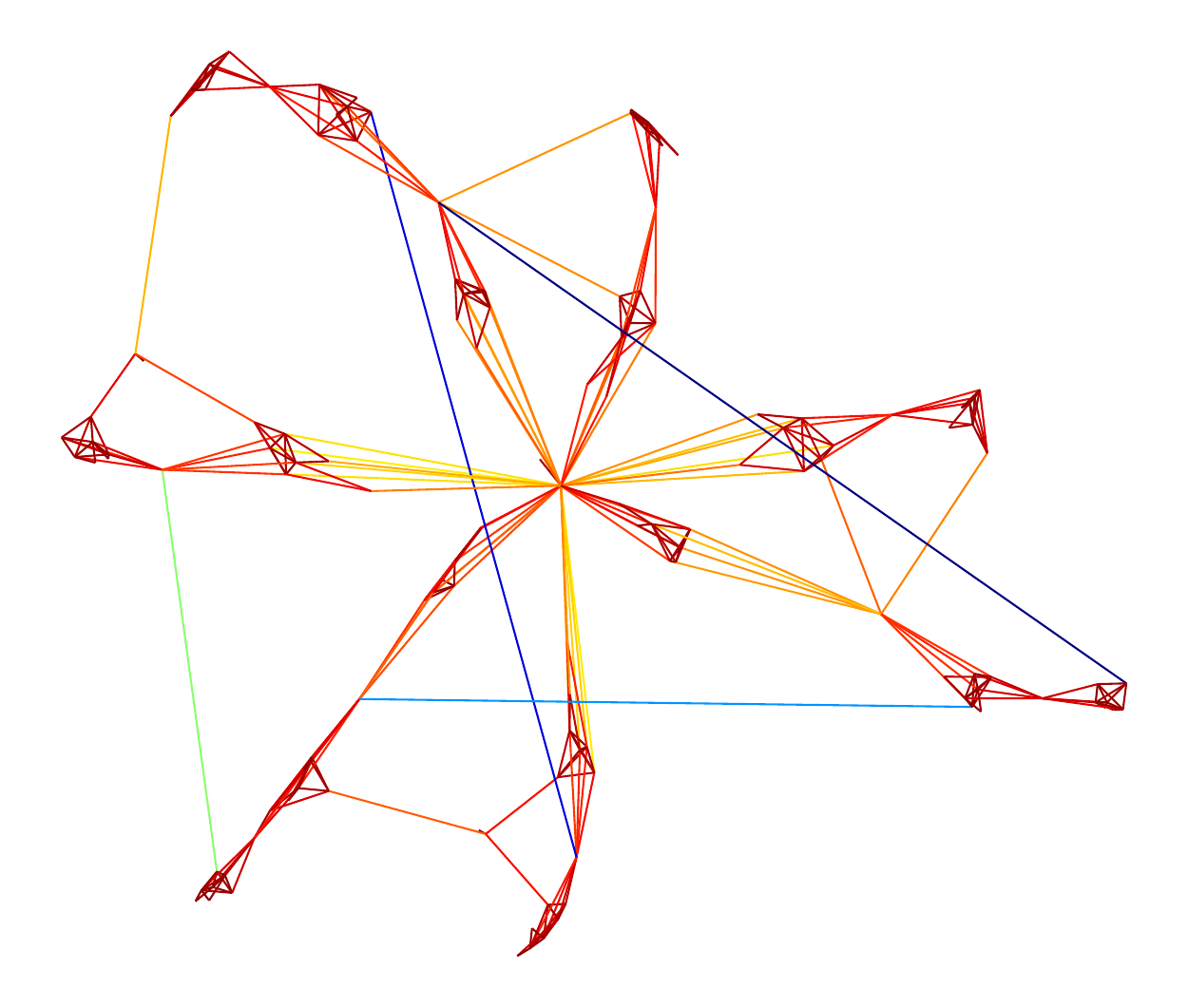}}
      & \parbox[c]{\tabfig\textwidth}{} 
      & \parbox[c]{\tabfig\textwidth}{} 
      & \parbox[c]{\tabfig\textwidth}{\taboffset
      \includegraphics[width=\tabfig\textwidth,height=\tabfig\textwidth]{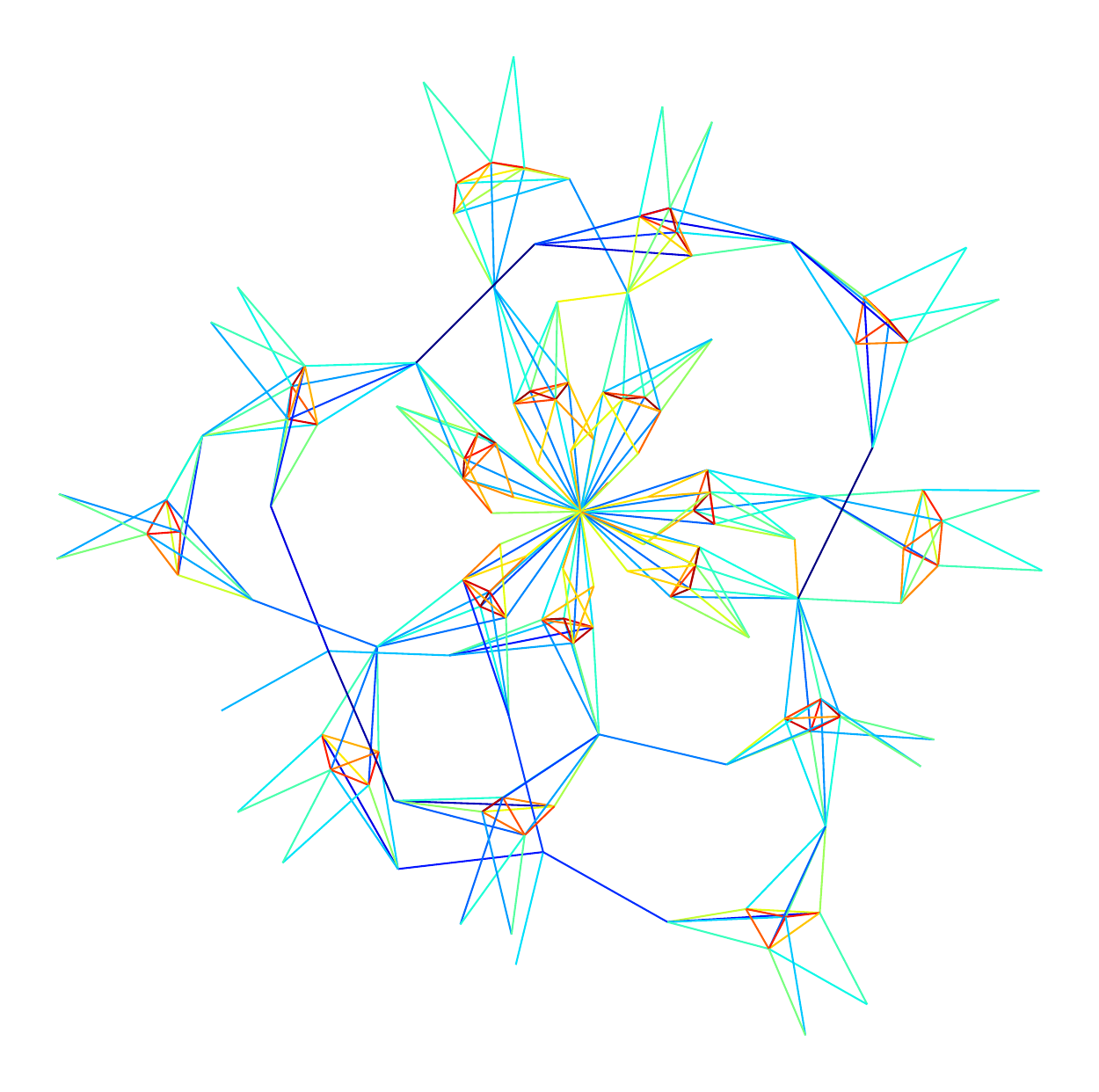}}  \\   
      \hline
      
     \multirow{2}{*}{\vspace{-1cm}\rotatebox[origin=c]{90}{mesh3e1}} &
      \parbox[c]{\tabfig\textwidth}{} 
      & \parbox[c]{\tabfig\textwidth}{
      \includegraphics[width=\tabfig\textwidth,height=\tabfig\textwidth]{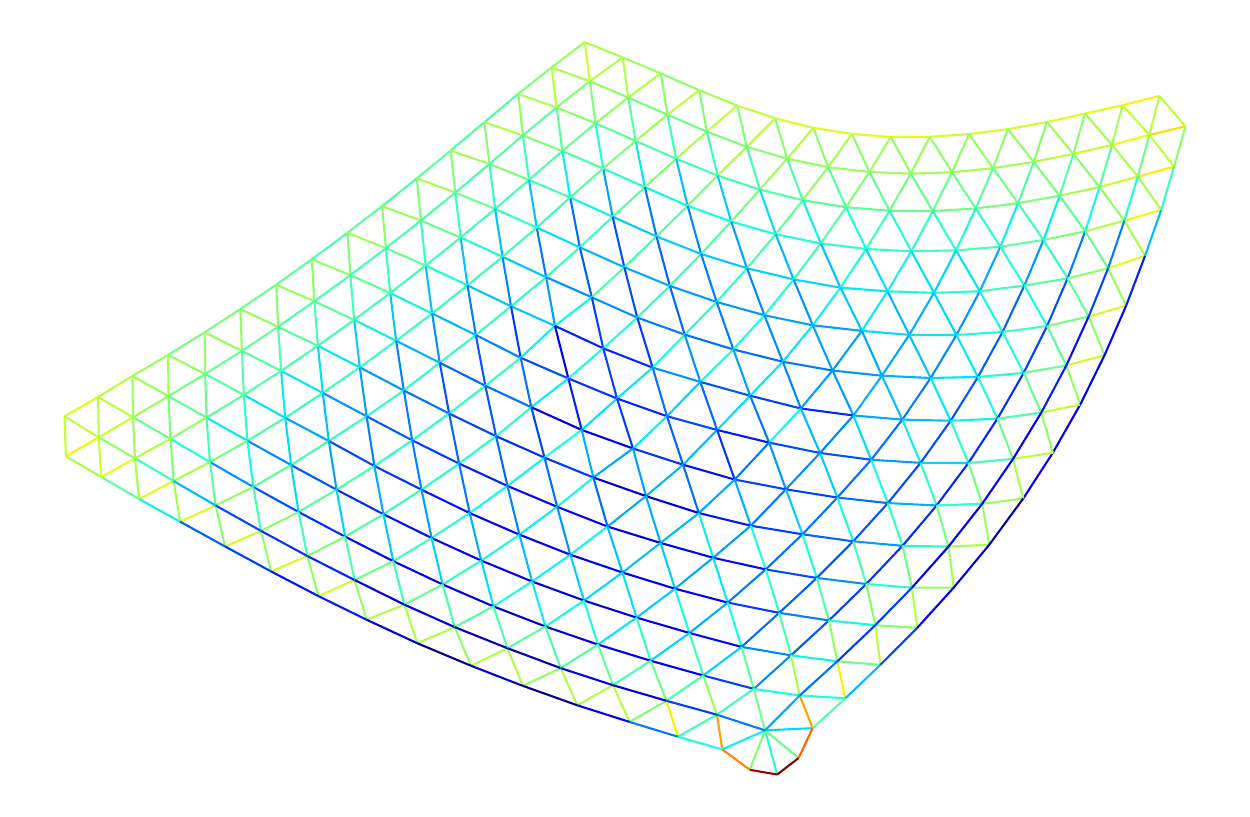}}
      & \parbox[c]{\tabfig\textwidth}{
      \includegraphics[width=\tabfig\textwidth,height=\tabfig\textwidth]{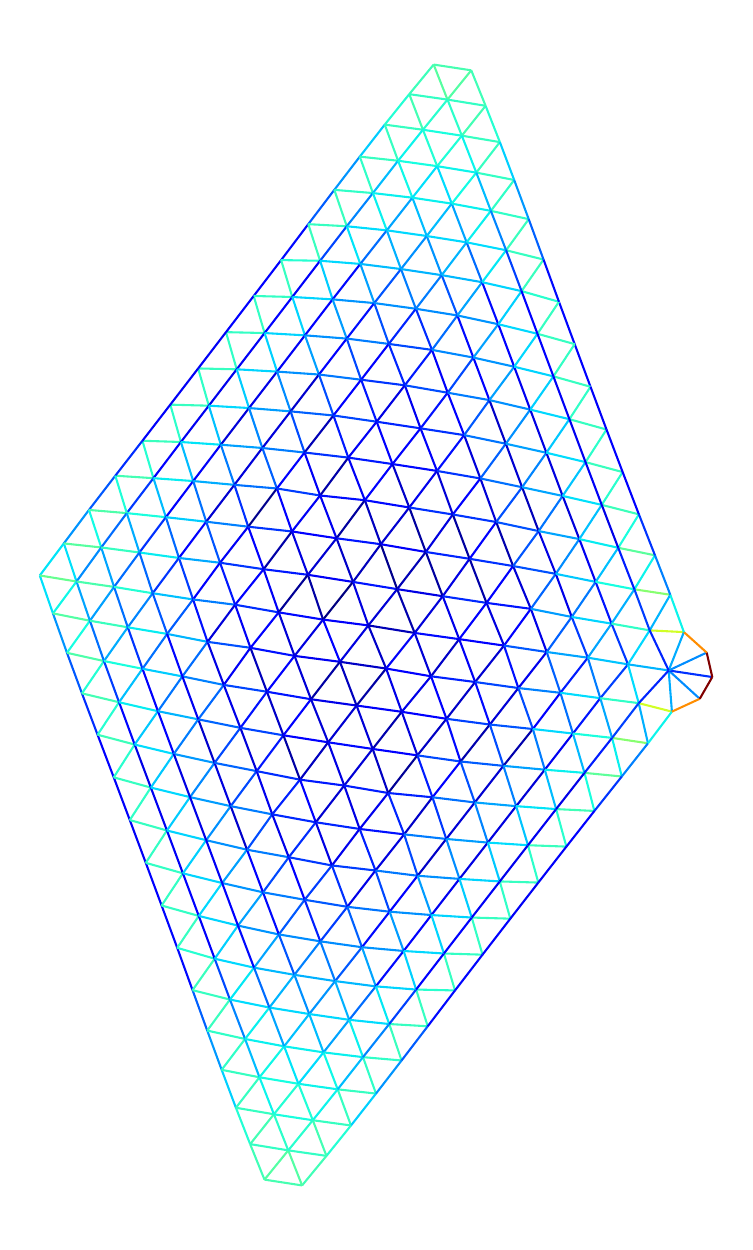}}
      & \parbox[c]{\tabfig\textwidth}{}
      & \parbox[c]{\tabfig\textwidth}{
      \includegraphics[width=\tabfig\textwidth,height=\tabfig\textwidth]{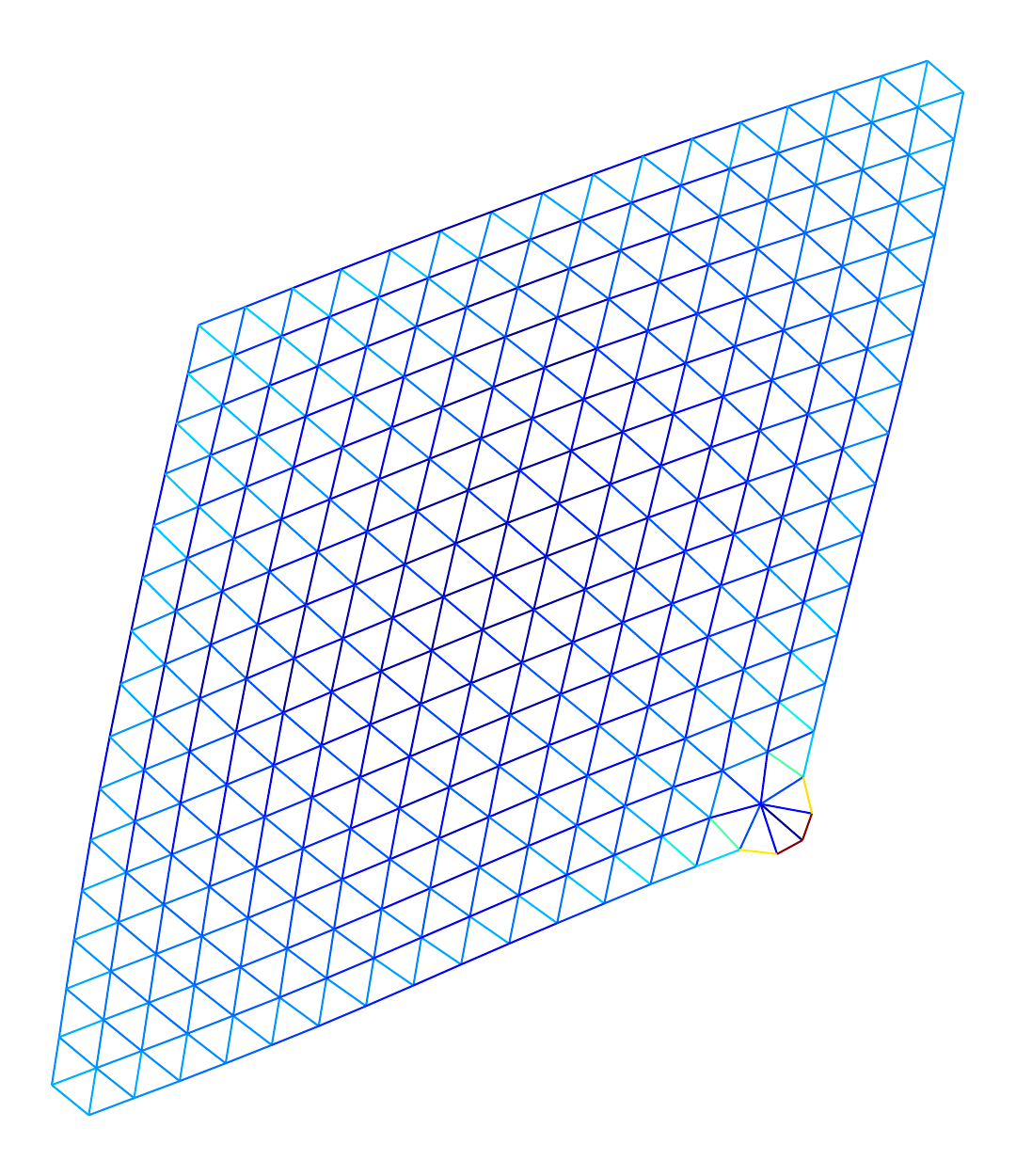}} 
      & \parbox[c]{\tabfig\textwidth}{
      \includegraphics[width=\tabfig\textwidth,height=\tabfig\textwidth]{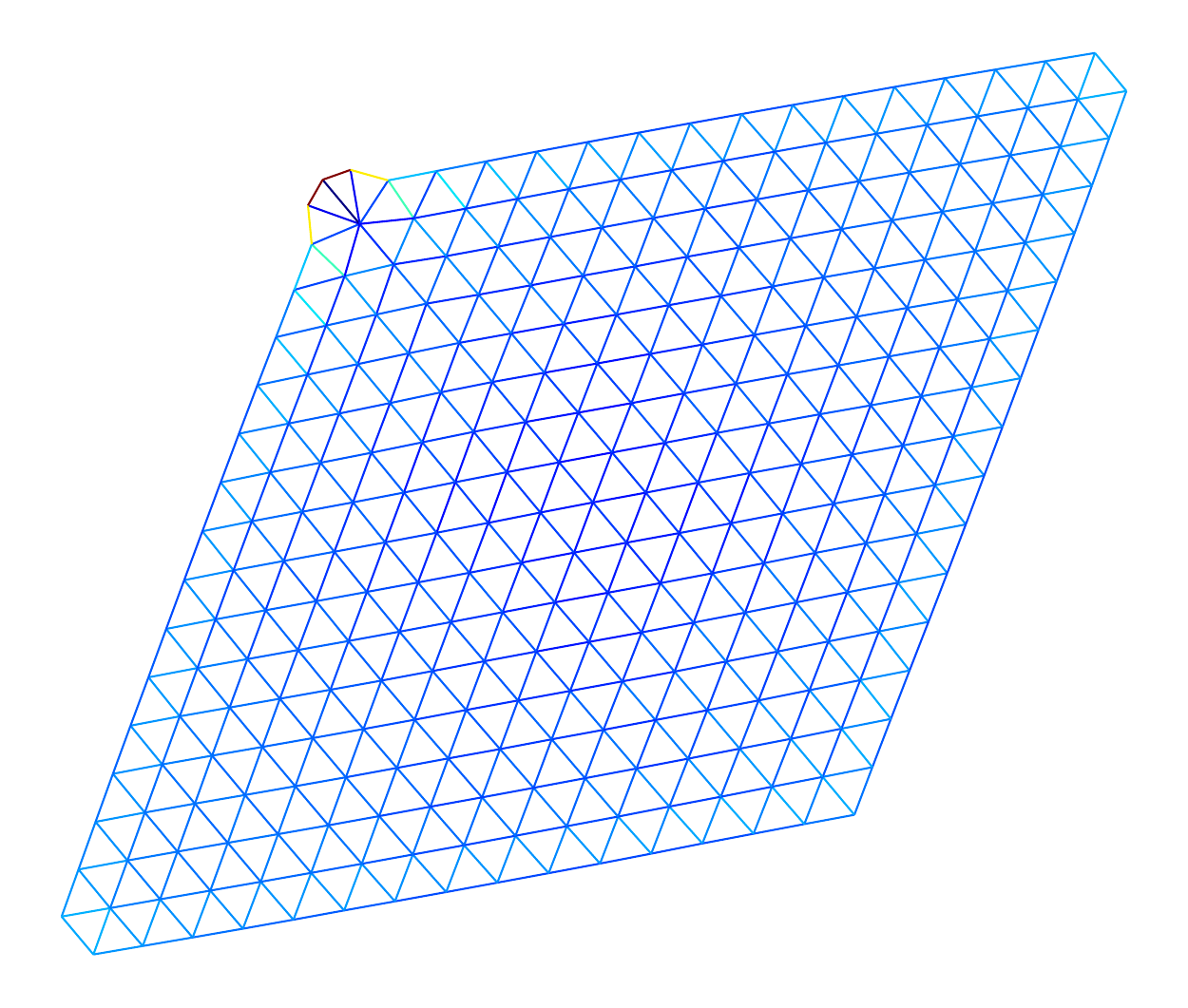}} 
      & \parbox[c]{\tabfig\textwidth}{}  \\
      
       &
      \parbox[c]{\tabfig\textwidth}{\taboffset
      \includegraphics[width=\tabfig\textwidth,height=\tabfig\textwidth]{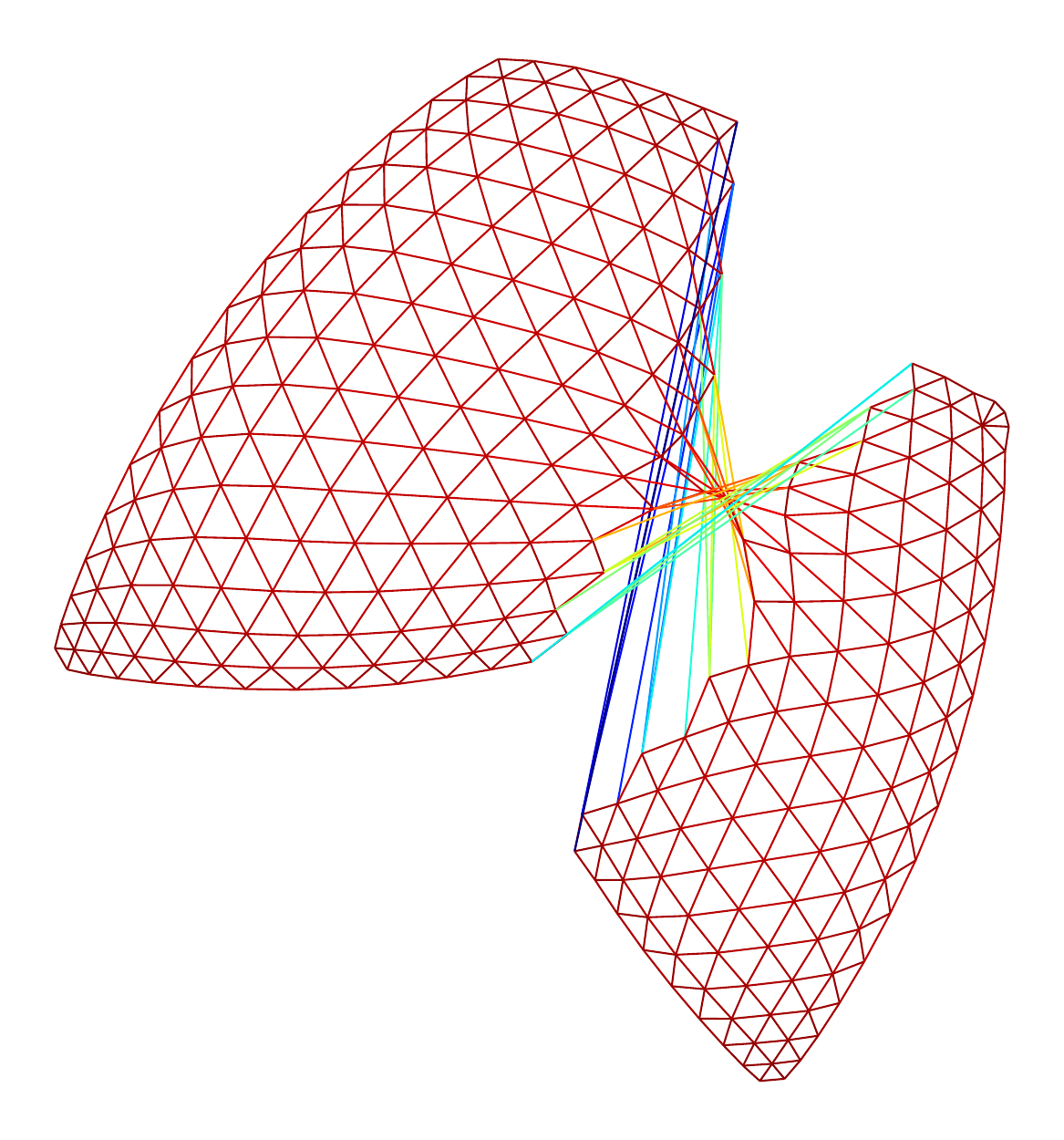}} 
      & \parbox[c]{\tabfig\textwidth}{}
      & \parbox[c]{\tabfig\textwidth}{}
      & \parbox[c]{\tabfig\textwidth}{\taboffset
      \includegraphics[width=\tabfig\textwidth,height=\tabfig\textwidth]{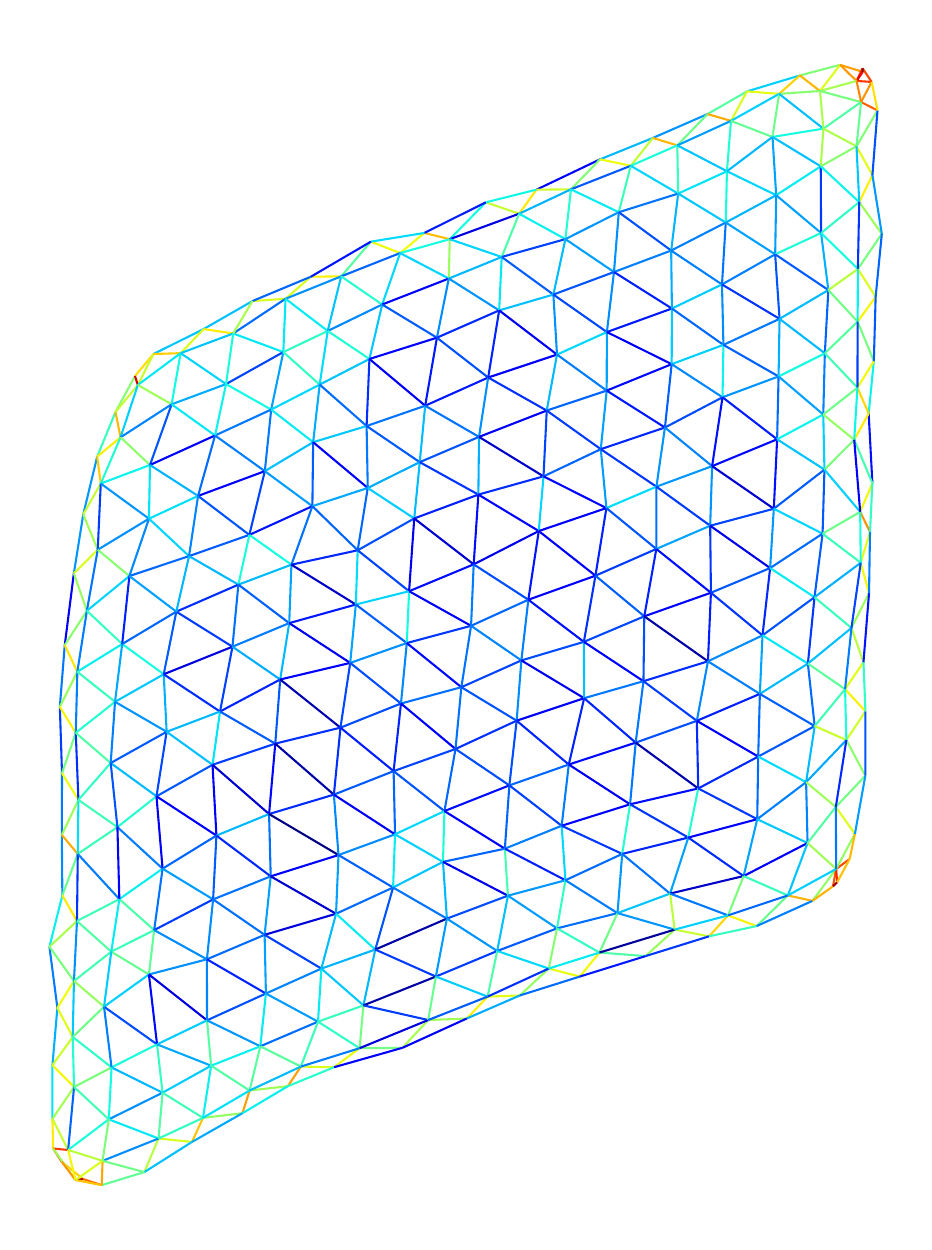}}
      & \parbox[c]{\tabfig\textwidth}{} 
      & \parbox[c]{\tabfig\textwidth}{} 
      & \parbox[c]{\tabfig\textwidth}{\taboffset
      \includegraphics[width=\tabfig\textwidth,height=\tabfig\textwidth]{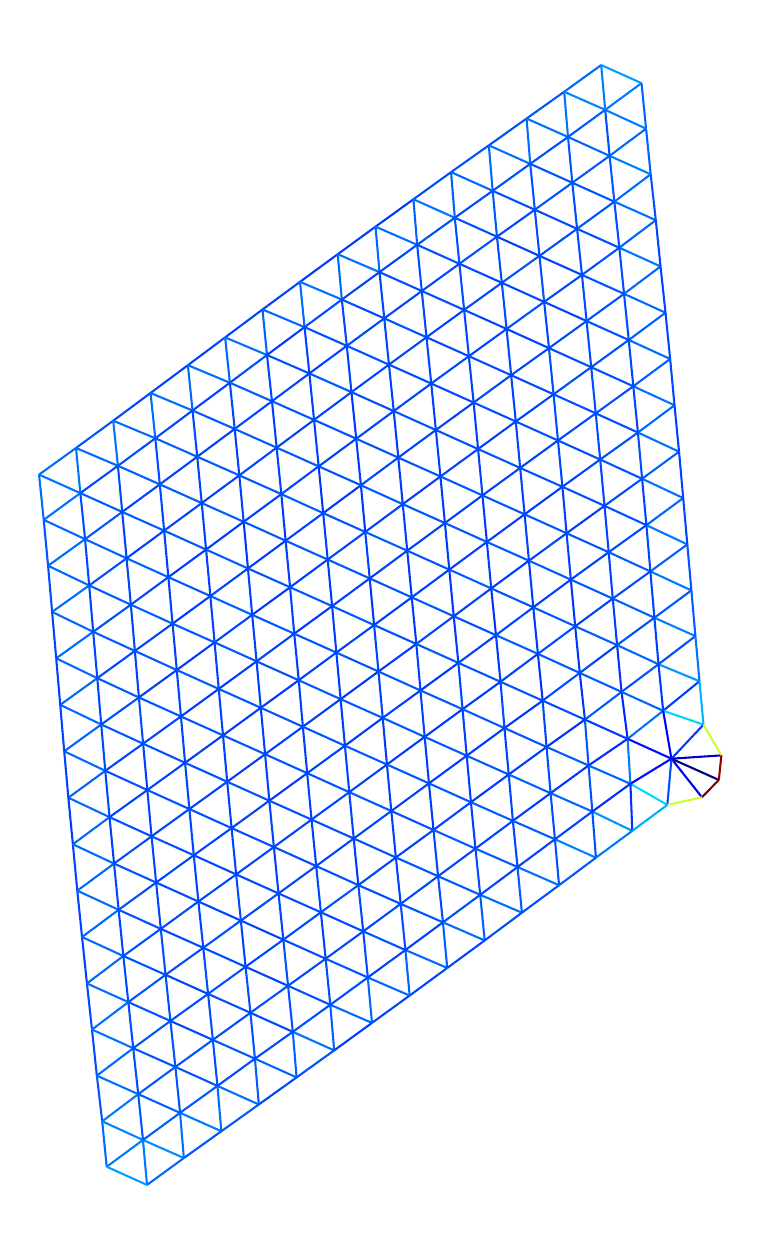}}  \\   
      \hline

      \multirow{2}{*}{\vspace{-1cm}\rotatebox[origin=c]{90}{powerlaw300}} &
      \parbox[c]{\tabfig\textwidth}{} 
      & \parbox[c]{\tabfig\textwidth}{
      \includegraphics[width=\tabfig\textwidth,height=\tabfig\textwidth]{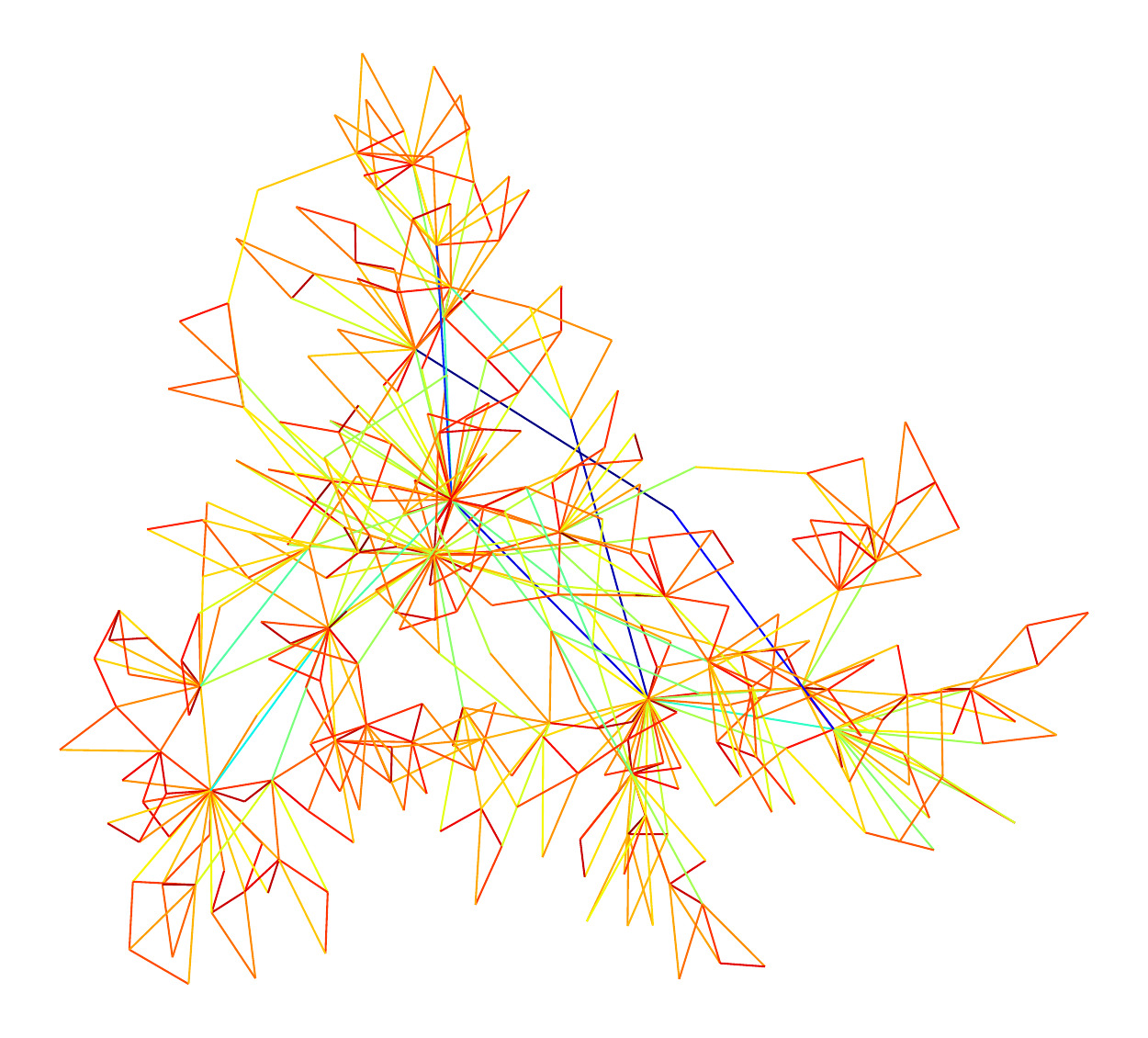}}
      & \parbox[c]{\tabfig\textwidth}{
      \includegraphics[width=\tabfig\textwidth,height=\tabfig\textwidth]{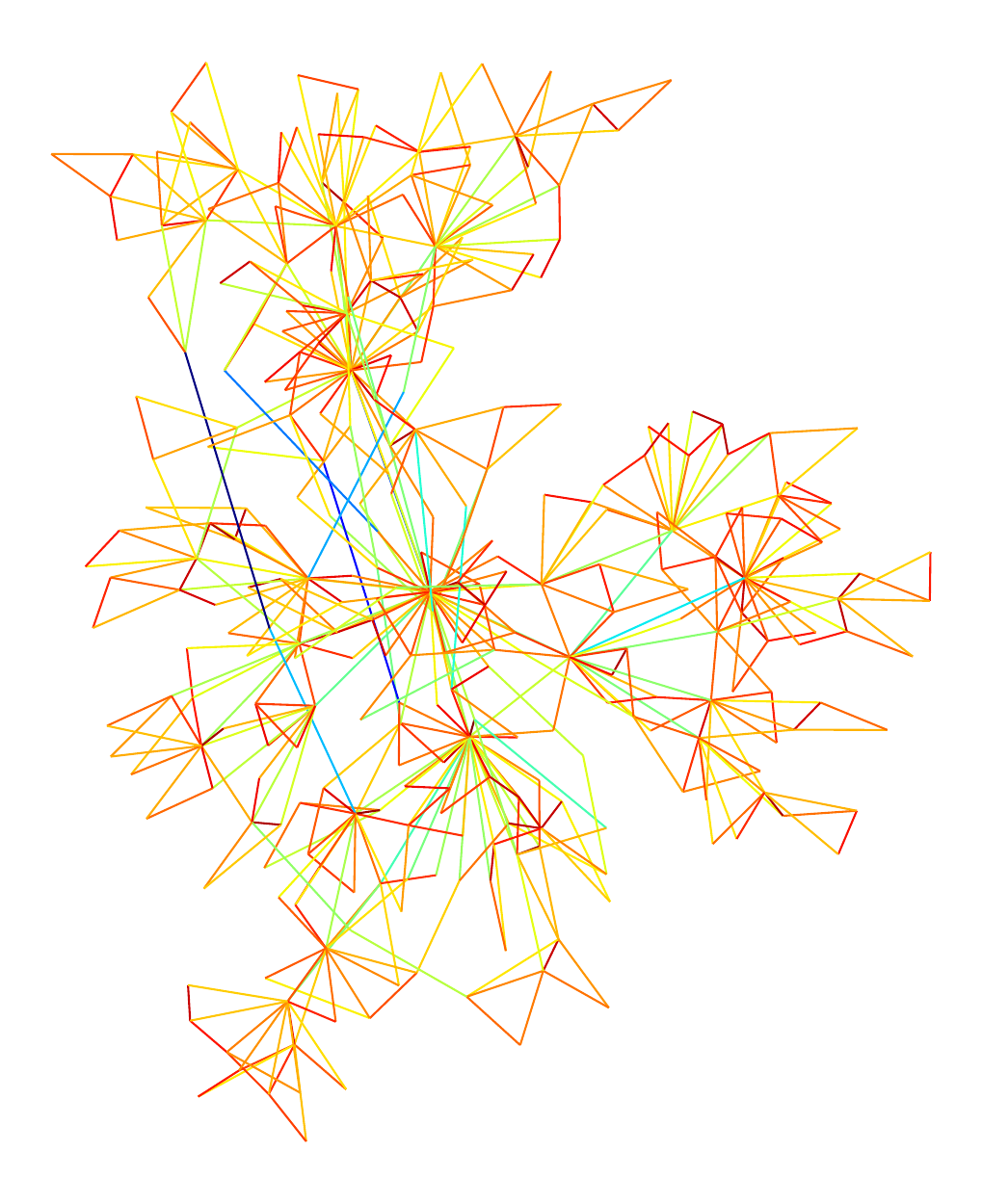}}
      & \parbox[c]{\tabfig\textwidth}{}
      & \parbox[c]{\tabfig\textwidth}{
      \includegraphics[width=\tabfig\textwidth,height=\tabfig\textwidth]{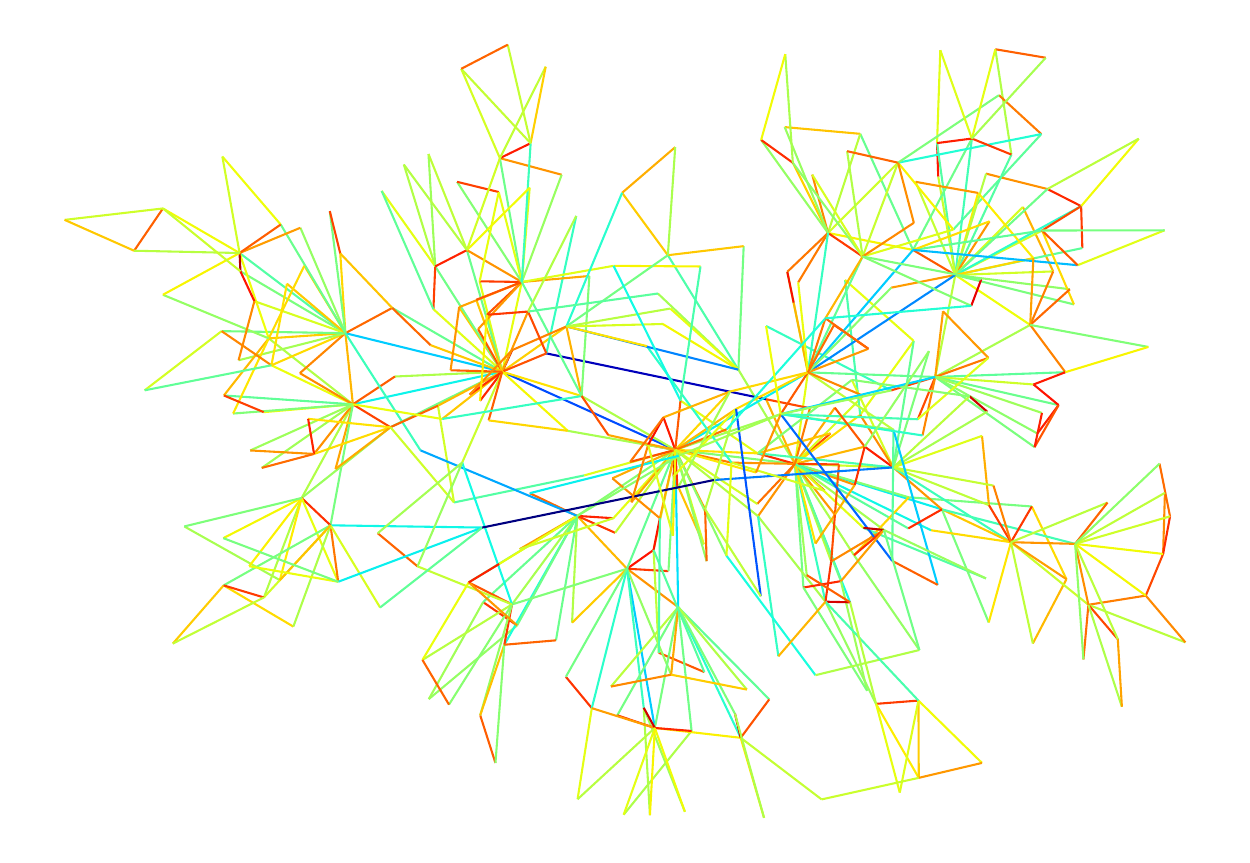}} 
      & \parbox[c]{\tabfig\textwidth}{
      \includegraphics[width=\tabfig\textwidth,height=\tabfig\textwidth]{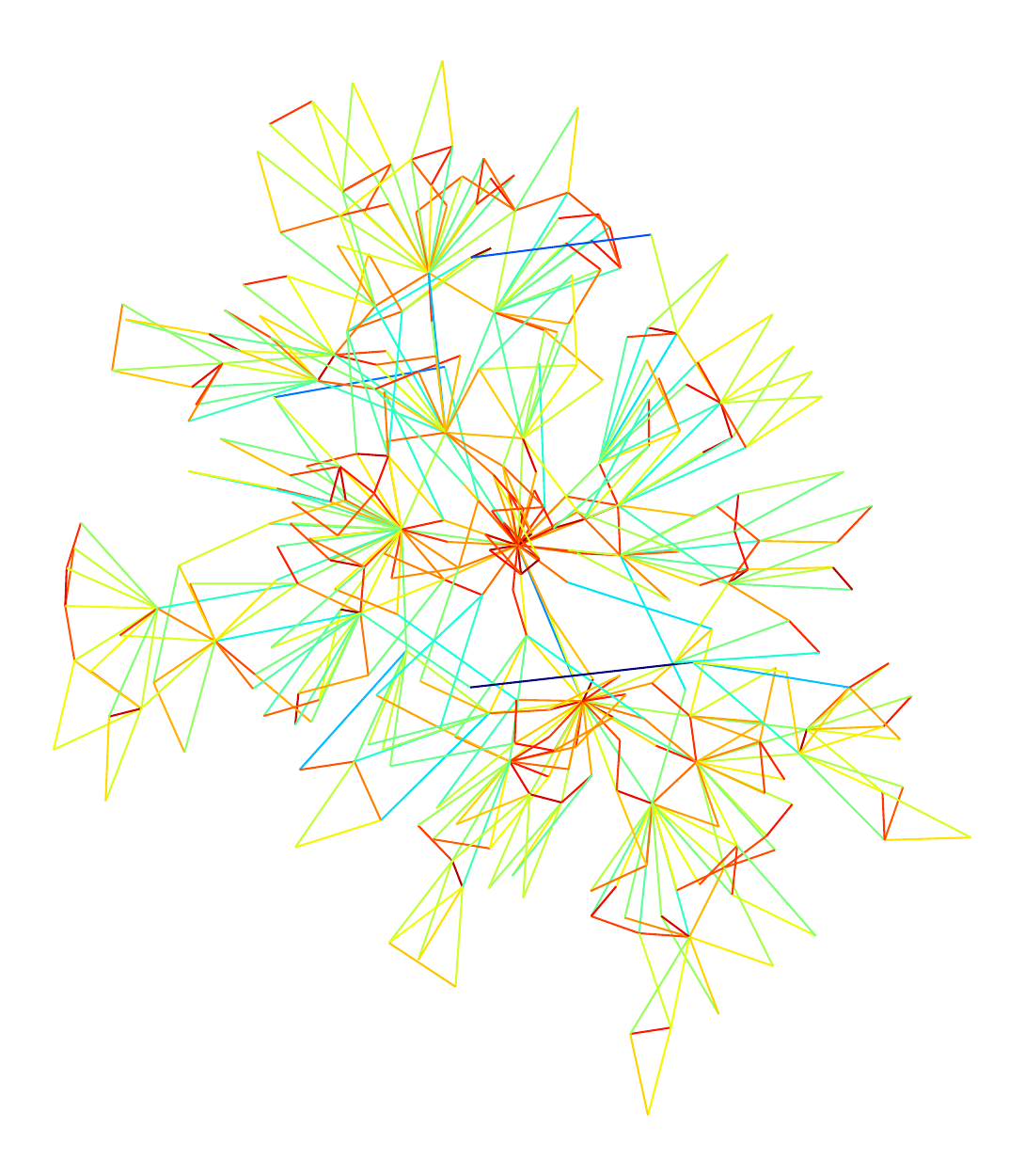}} 
      & \parbox[c]{\tabfig\textwidth}{}  \\
      
       &
      \parbox[c]{\tabfig\textwidth}{\taboffset
      \includegraphics[width=\tabfig\textwidth,height=\tabfig\textwidth]{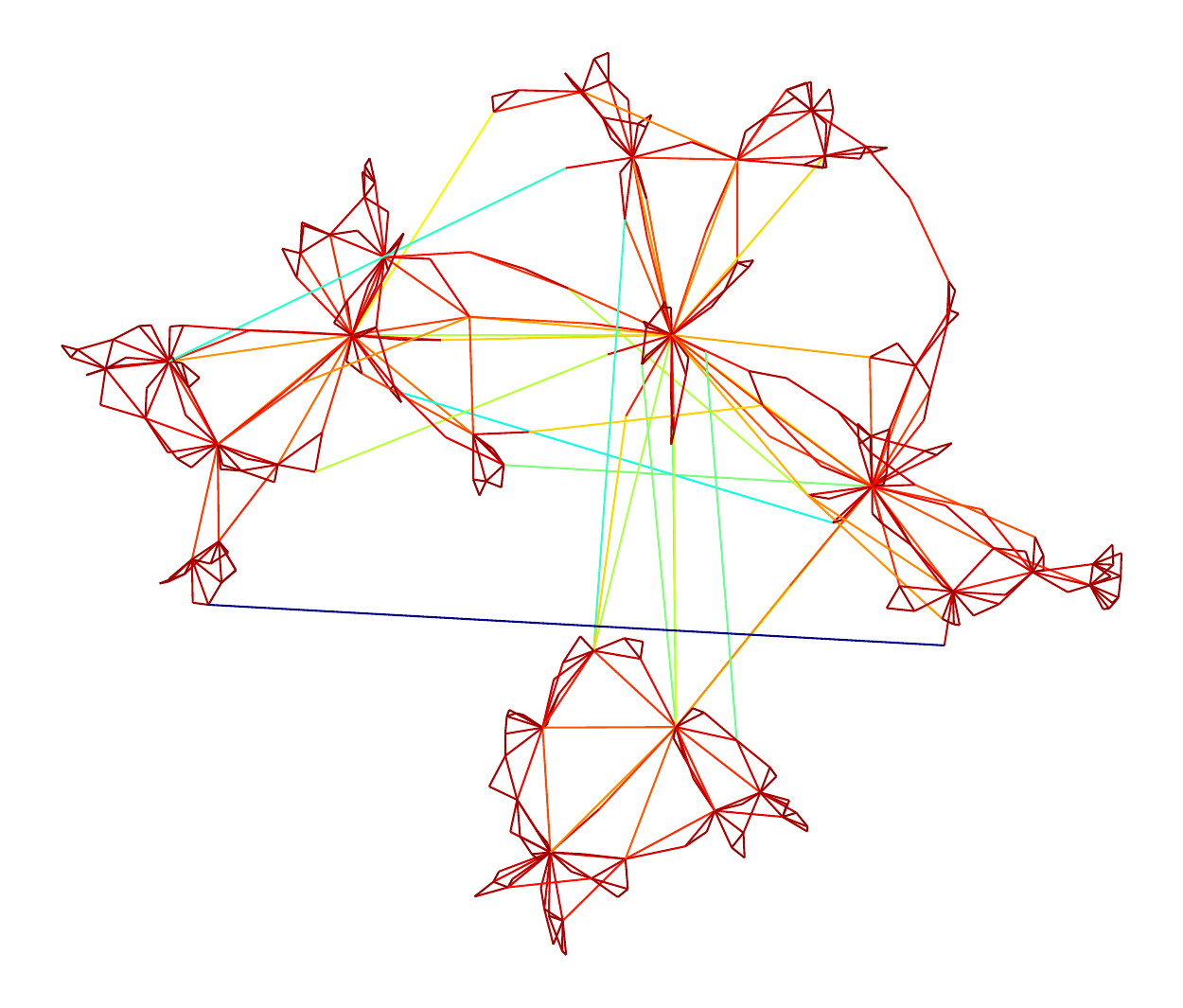}} 
      & \parbox[c]{\tabfig\textwidth}{}
      & \parbox[c]{\tabfig\textwidth}{}
      & \parbox[c]{\tabfig\textwidth}{\taboffset
      \includegraphics[width=\tabfig\textwidth,height=\tabfig\textwidth]{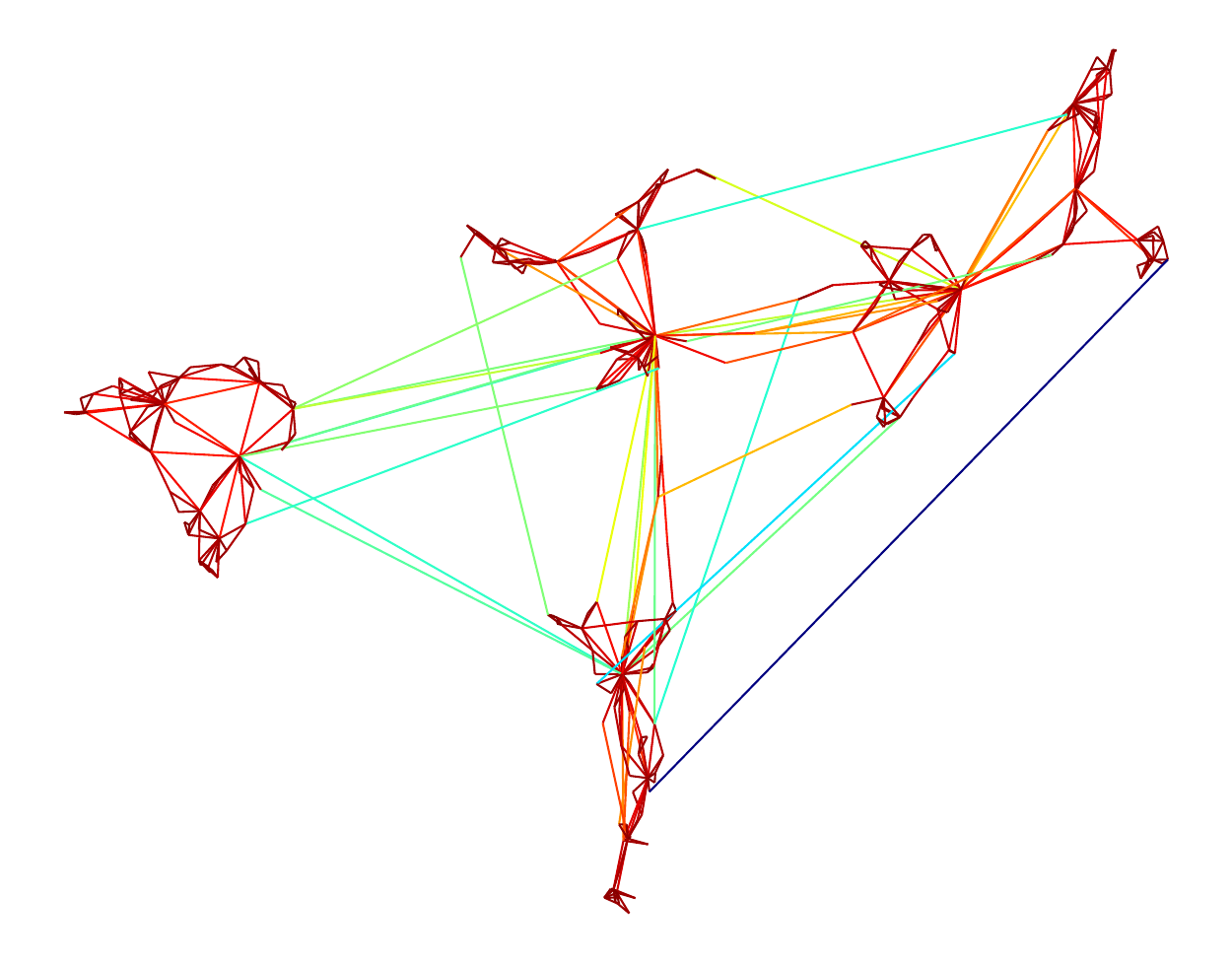}}
      & \parbox[c]{\tabfig\textwidth}{} 
      & \parbox[c]{\tabfig\textwidth}{} 
      & \parbox[c]{\tabfig\textwidth}{\taboffset
      \includegraphics[width=\tabfig\textwidth,height=\tabfig\textwidth]{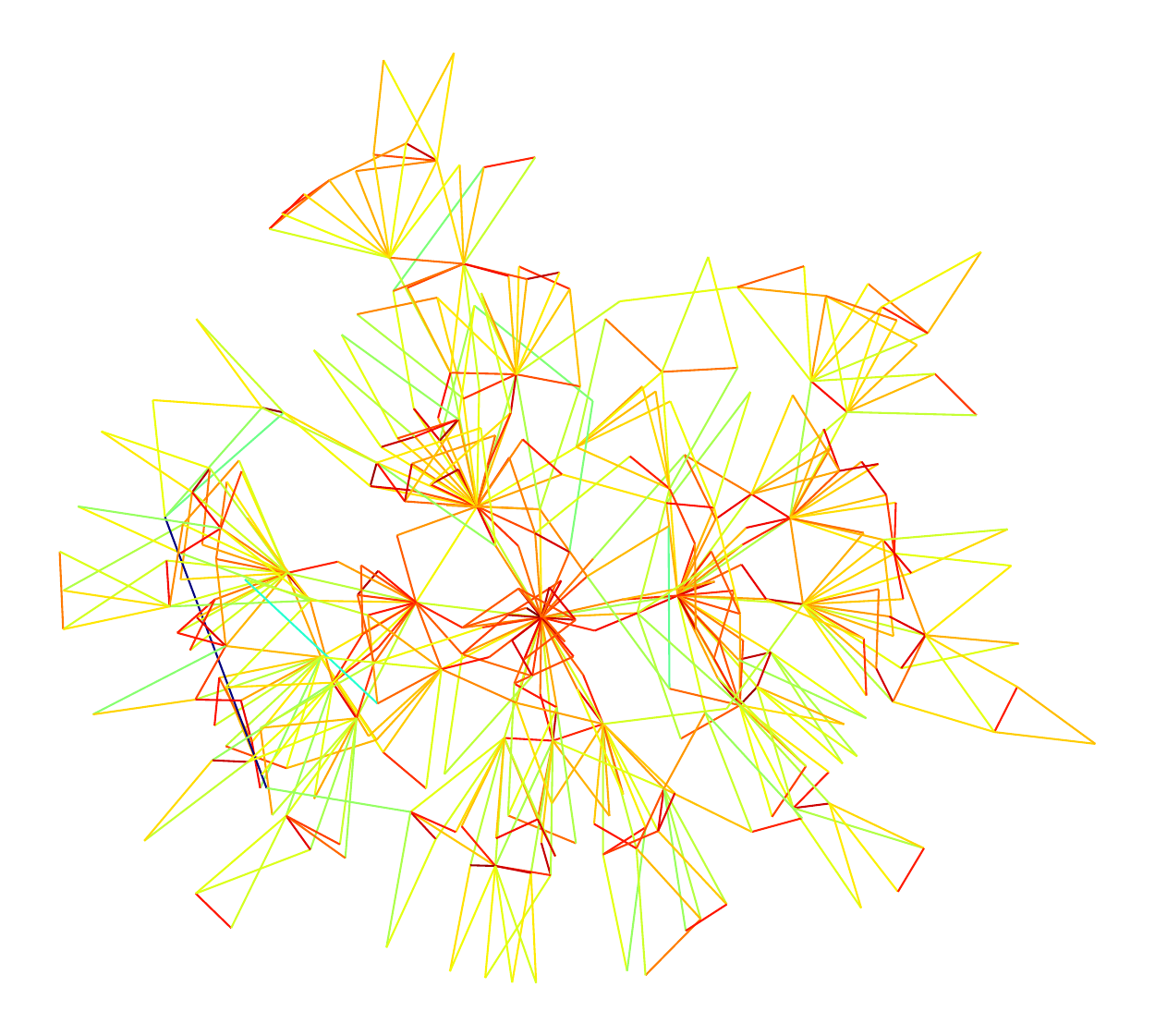}}  \\   
      \hline           

      \multirow{2}{*}{\vspace{-1cm}\rotatebox[origin=c]{90}{powerlaw500}} &
      \parbox[c]{\tabfig\textwidth}{} 
      & \parbox[c]{\tabfig\textwidth}{
      \includegraphics[width=\tabfig\textwidth,height=\tabfig\textwidth]{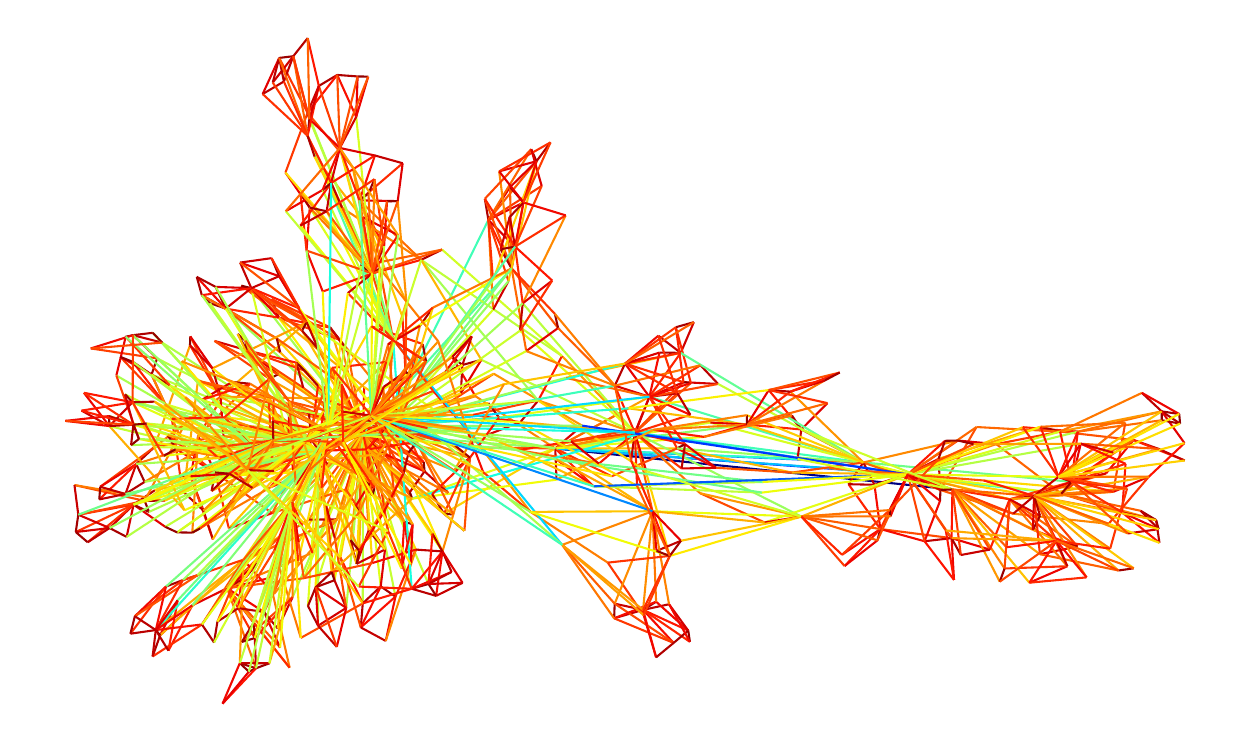}}
      & \parbox[c]{\tabfig\textwidth}{
      \includegraphics[width=\tabfig\textwidth,height=\tabfig\textwidth]{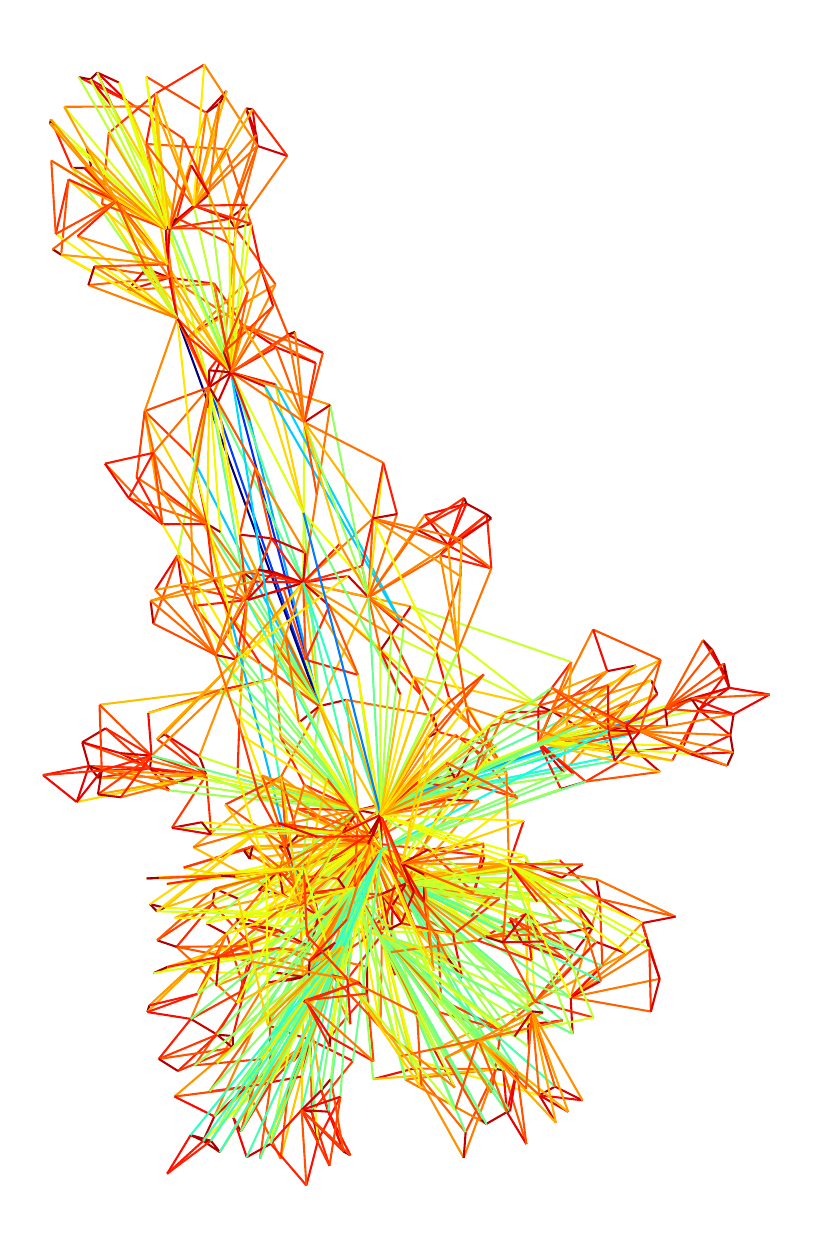}}
      & \parbox[c]{\tabfig\textwidth}{}
      & \parbox[c]{\tabfig\textwidth}{
      \includegraphics[width=\tabfig\textwidth,height=\tabfig\textwidth]{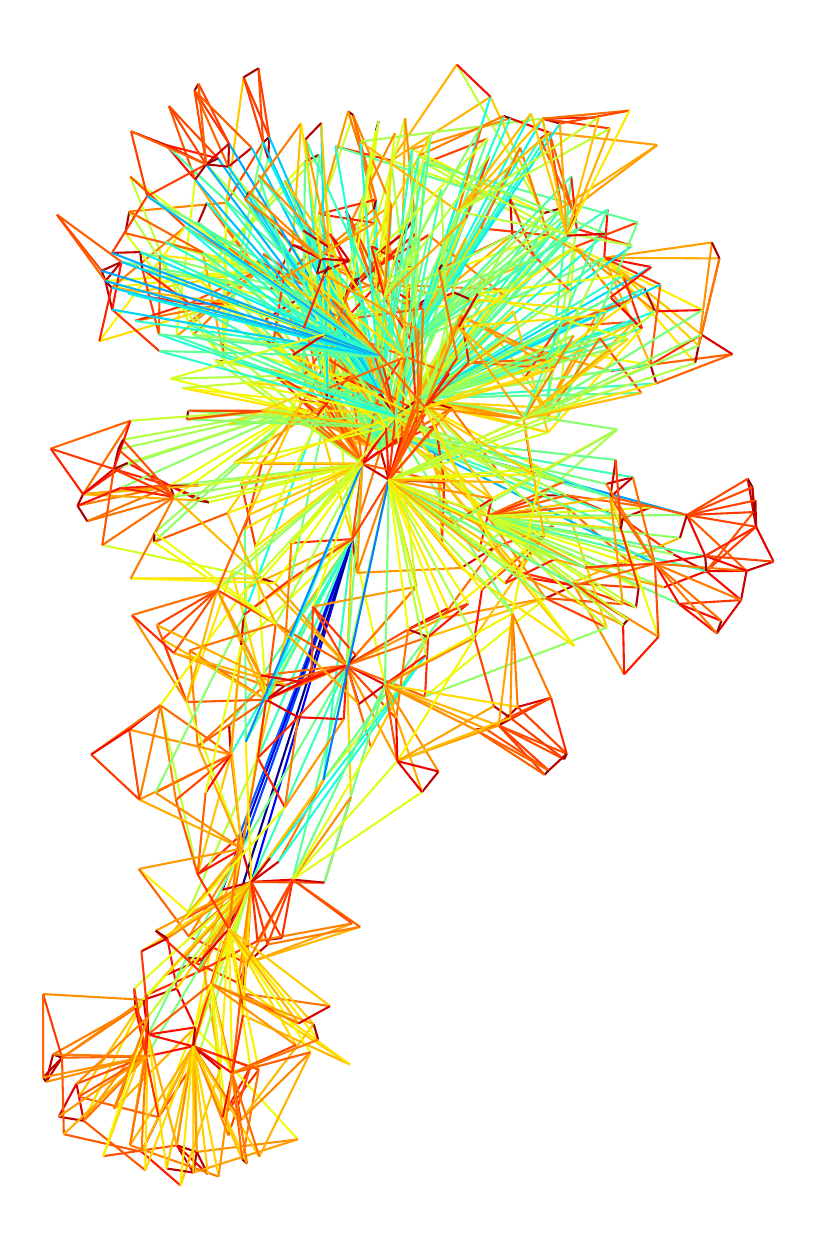}} 
      & \parbox[c]{\tabfig\textwidth}{
      \includegraphics[width=\tabfig\textwidth,height=\tabfig\textwidth]{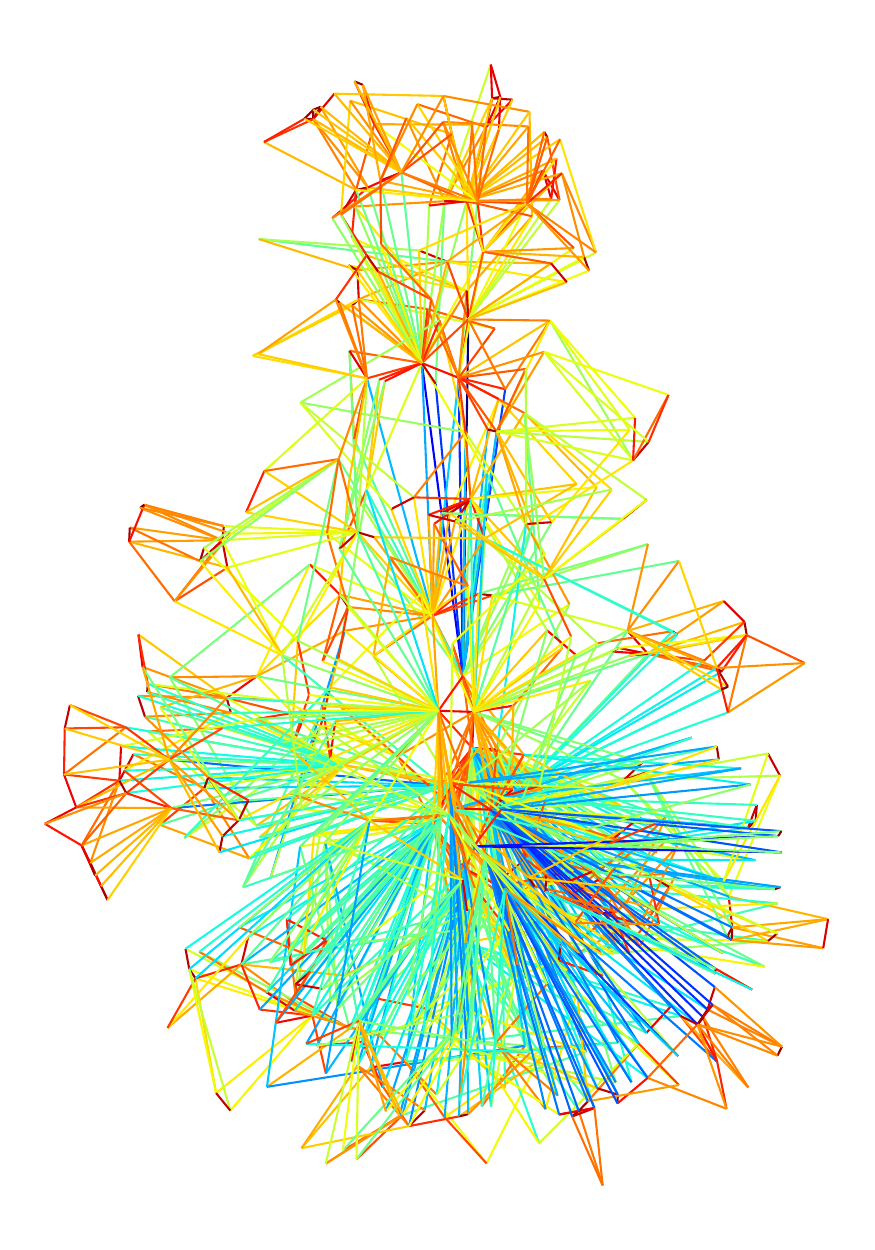}} 
      & \parbox[c]{\tabfig\textwidth}{}  \\
      
       &
      \parbox[c]{\tabfig\textwidth}{\taboffset
      \includegraphics[width=\tabfig\textwidth,height=\tabfig\textwidth]{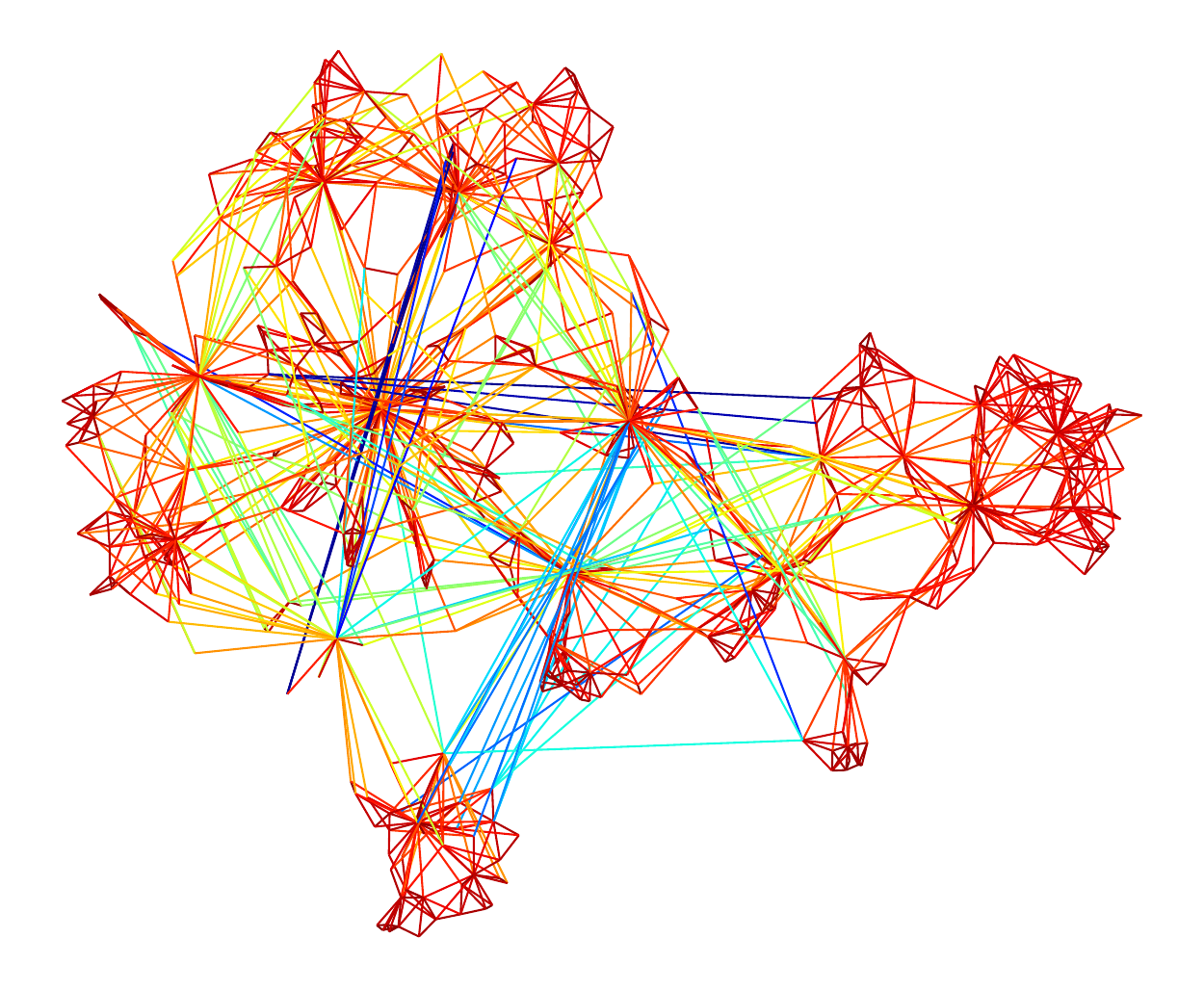}} 
      & \parbox[c]{\tabfig\textwidth}{}
      & \parbox[c]{\tabfig\textwidth}{}
      & \parbox[c]{\tabfig\textwidth}{\taboffset
      \includegraphics[width=\tabfig\textwidth,height=\tabfig\textwidth]{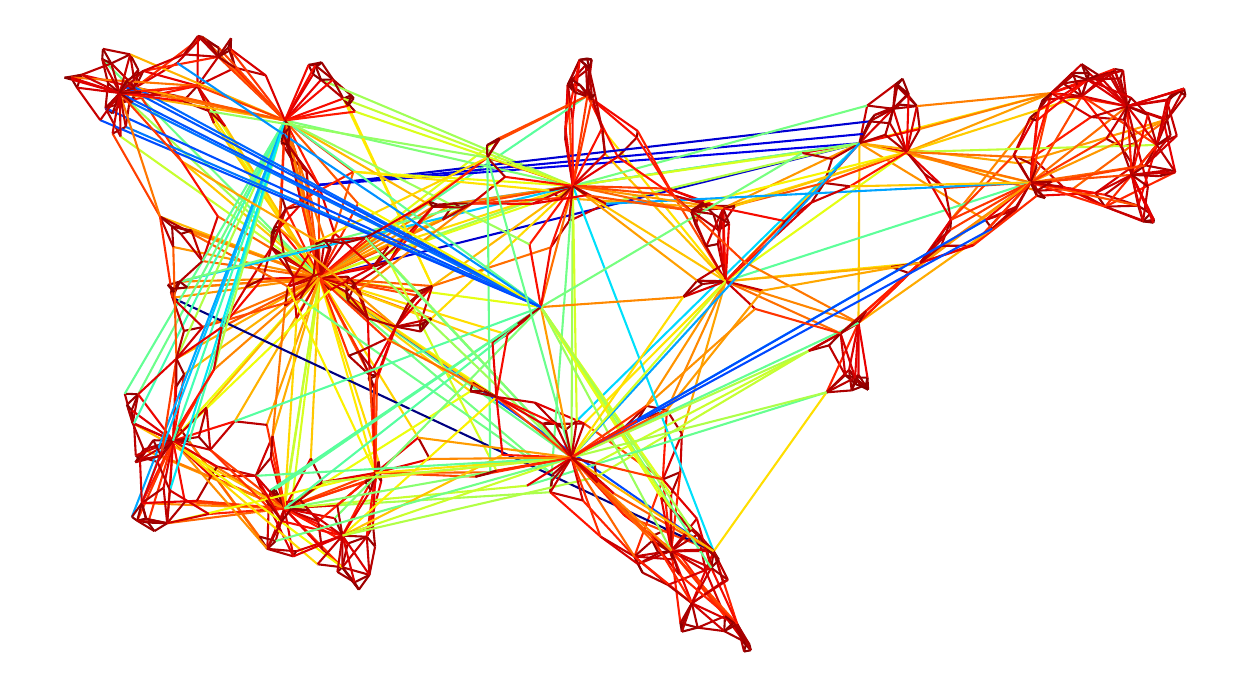}}
      & \parbox[c]{\tabfig\textwidth}{} 
      & \parbox[c]{\tabfig\textwidth}{} 
      & \parbox[c]{\tabfig\textwidth}{\taboffset
      \includegraphics[width=\tabfig\textwidth,height=\tabfig\textwidth]{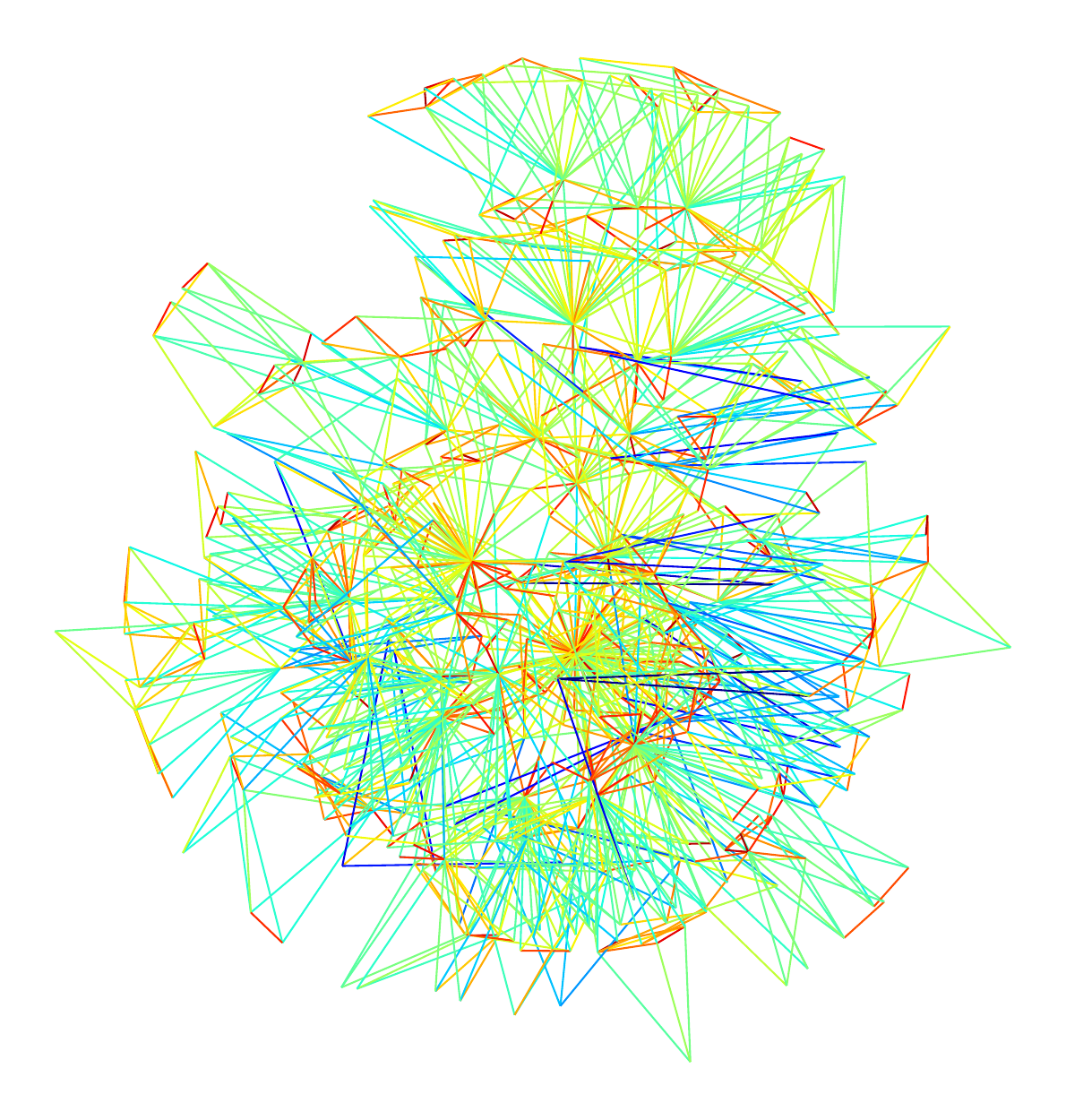}}  \\   
      \hline

     \multirow{2}{*}{\vspace{-1cm}\rotatebox[origin=c]{90}{powerlaw1000}}  &
      \parbox[c]{\tabfig\textwidth}{} 
      & \parbox[c]{\tabfig\textwidth}{
      \includegraphics[width=\tabfig\textwidth,height=\tabfig\textwidth]{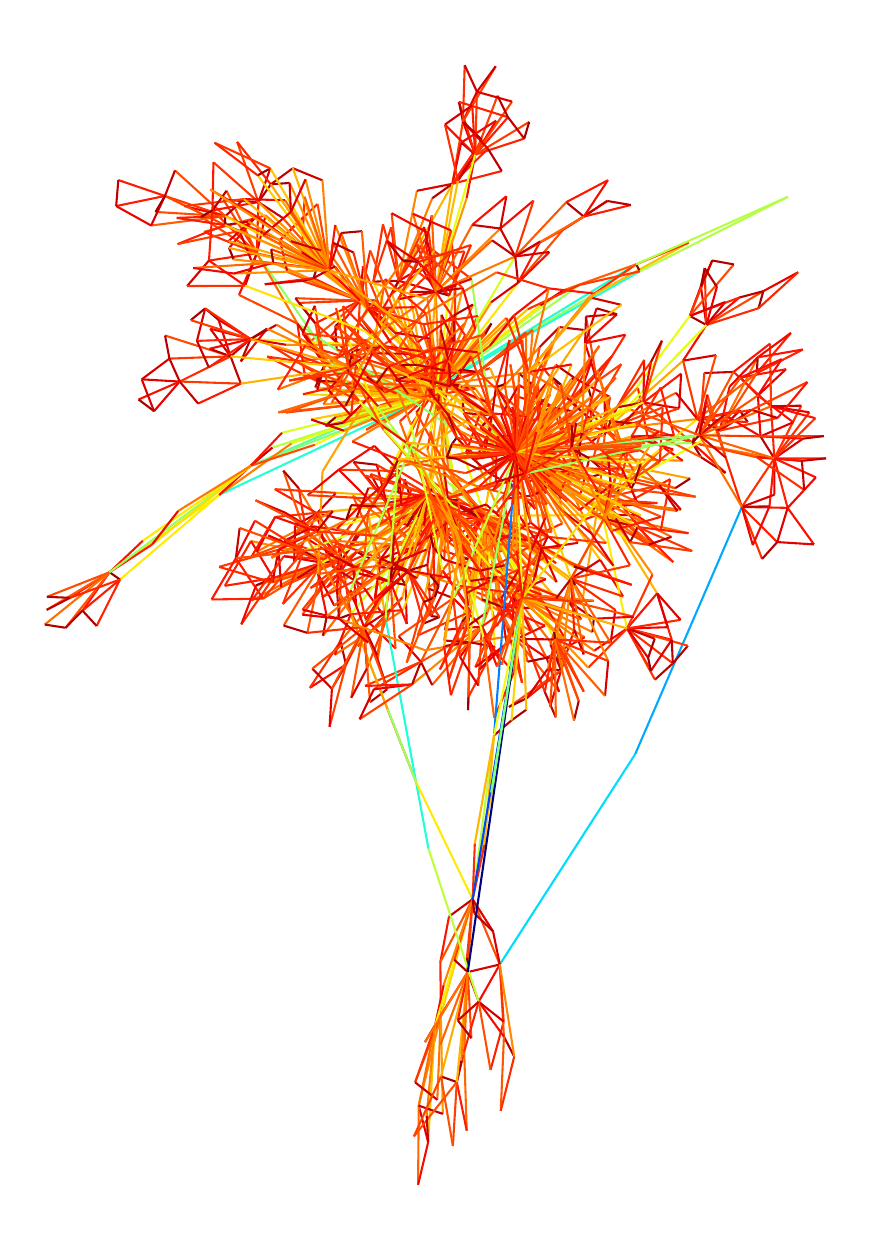}}
      & \parbox[c]{\tabfig\textwidth}{
      \includegraphics[width=\tabfig\textwidth,height=\tabfig\textwidth]{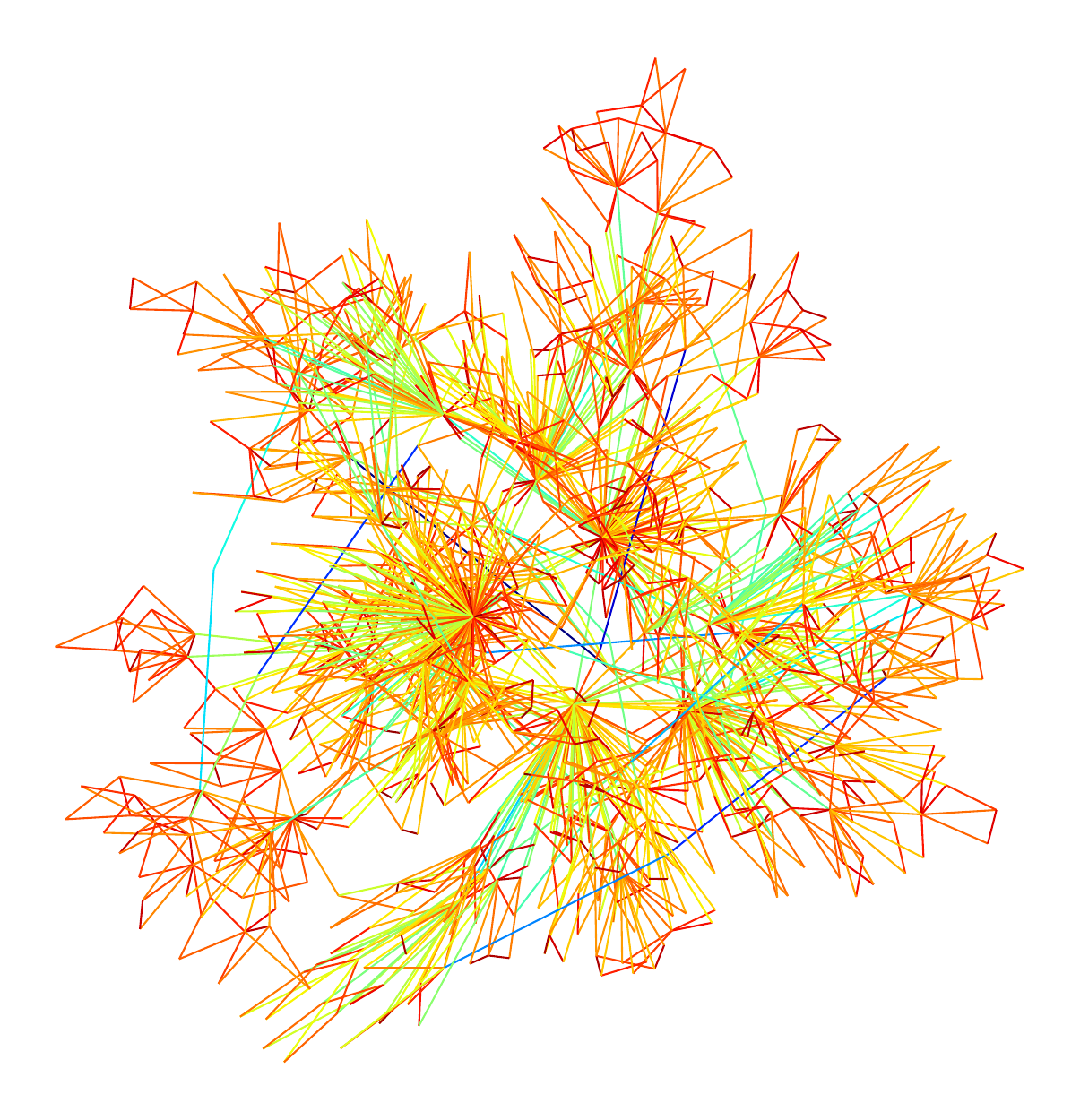}}
      & \parbox[c]{\tabfig\textwidth}{}
      & \parbox[c]{\tabfig\textwidth}{
      \includegraphics[width=\tabfig\textwidth,height=\tabfig\textwidth]{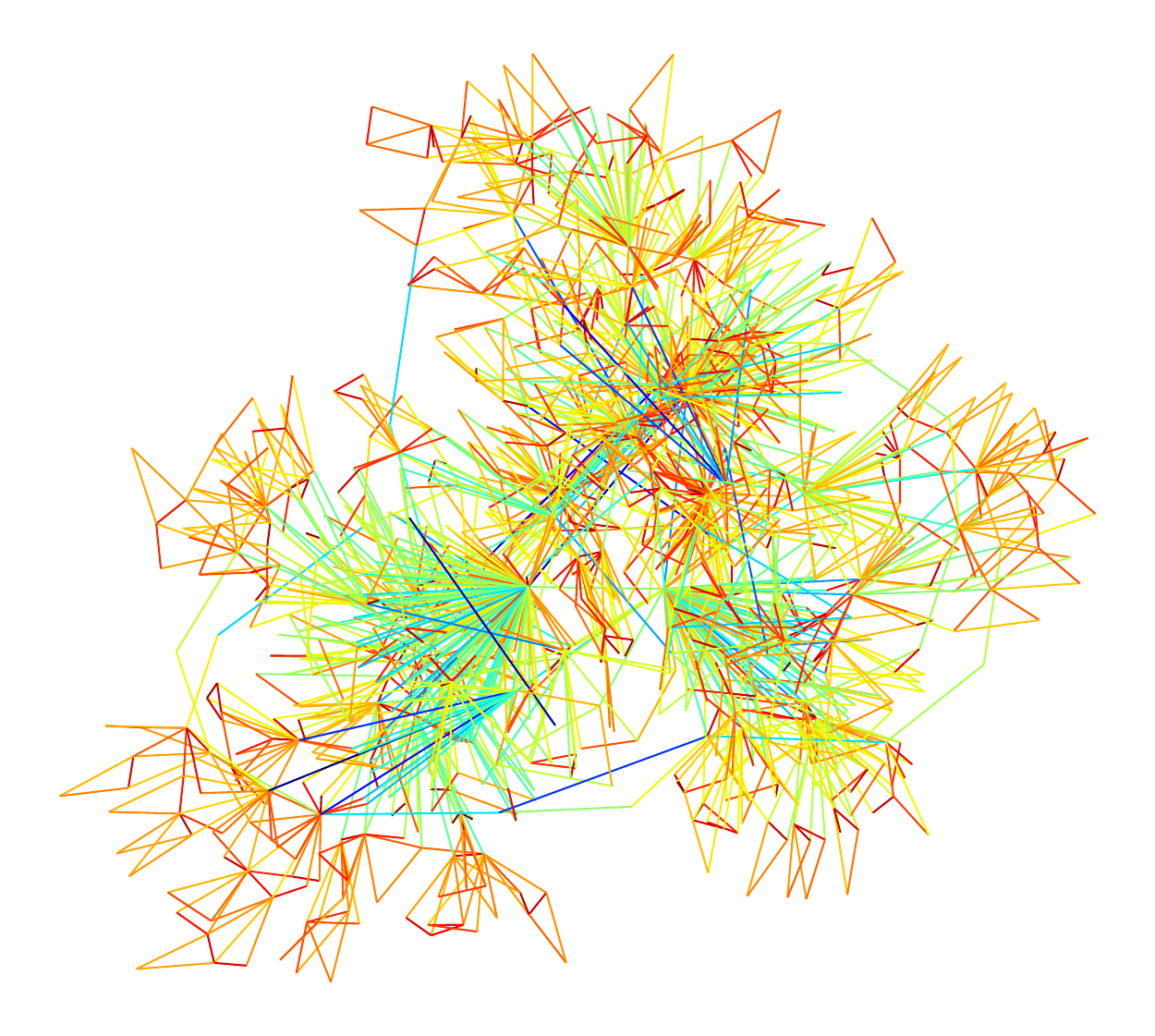}} 
      & \parbox[c]{\tabfig\textwidth}{
      \includegraphics[width=\tabfig\textwidth,height=\tabfig\textwidth]{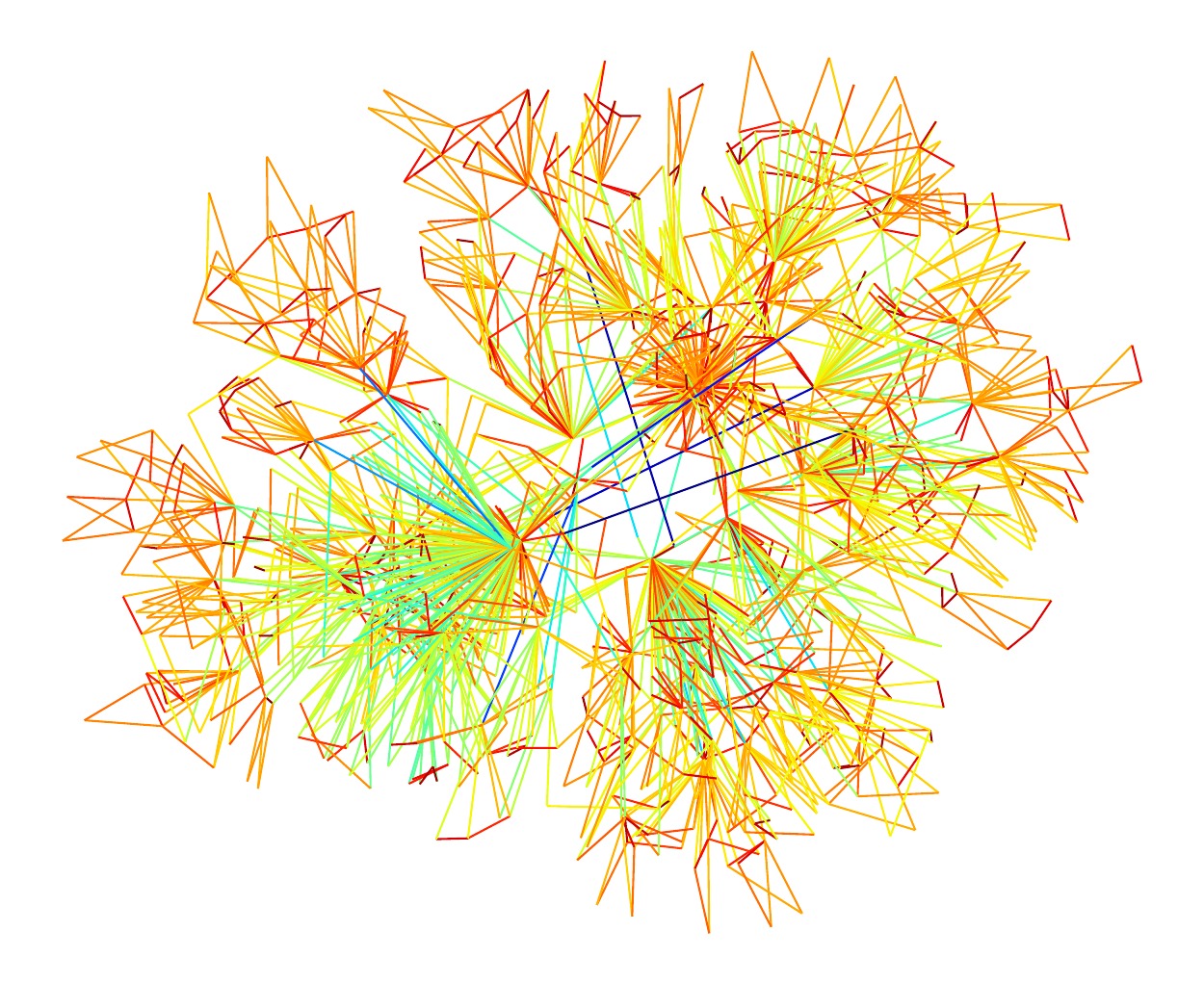}} 
      & \parbox[c]{\tabfig\textwidth}{}  \\
      
       &
      \parbox[c]{\tabfig\textwidth}{
      \taboffset\includegraphics[width=\tabfig\textwidth,height=\tabfig\textwidth]{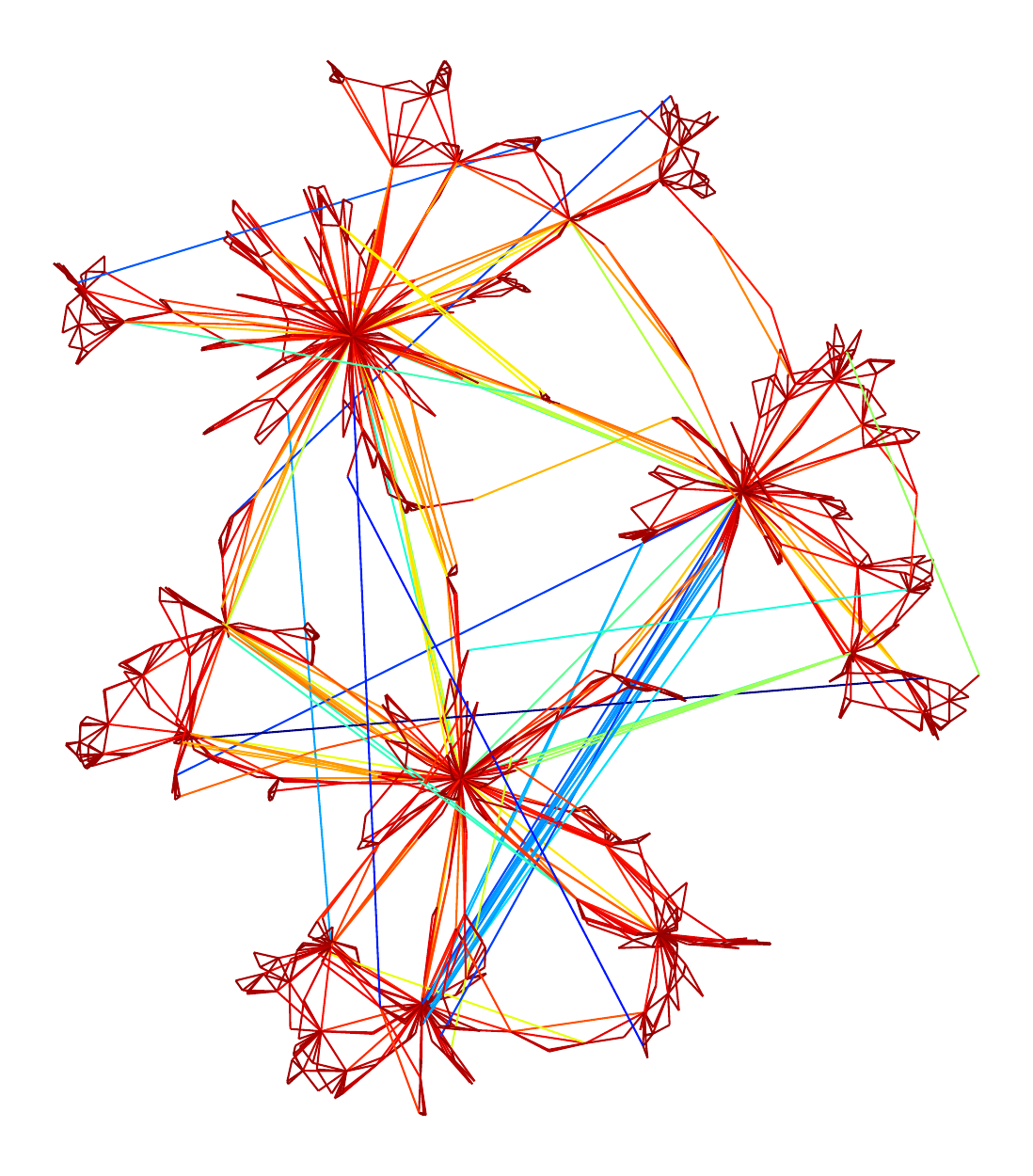}} 
      & \parbox[c]{\tabfig\textwidth}{}
      & \parbox[c]{\tabfig\textwidth}{}
      & \parbox[c]{\tabfig\textwidth}{\taboffset
      \includegraphics[width=\tabfig\textwidth,height=\tabfig\textwidth]{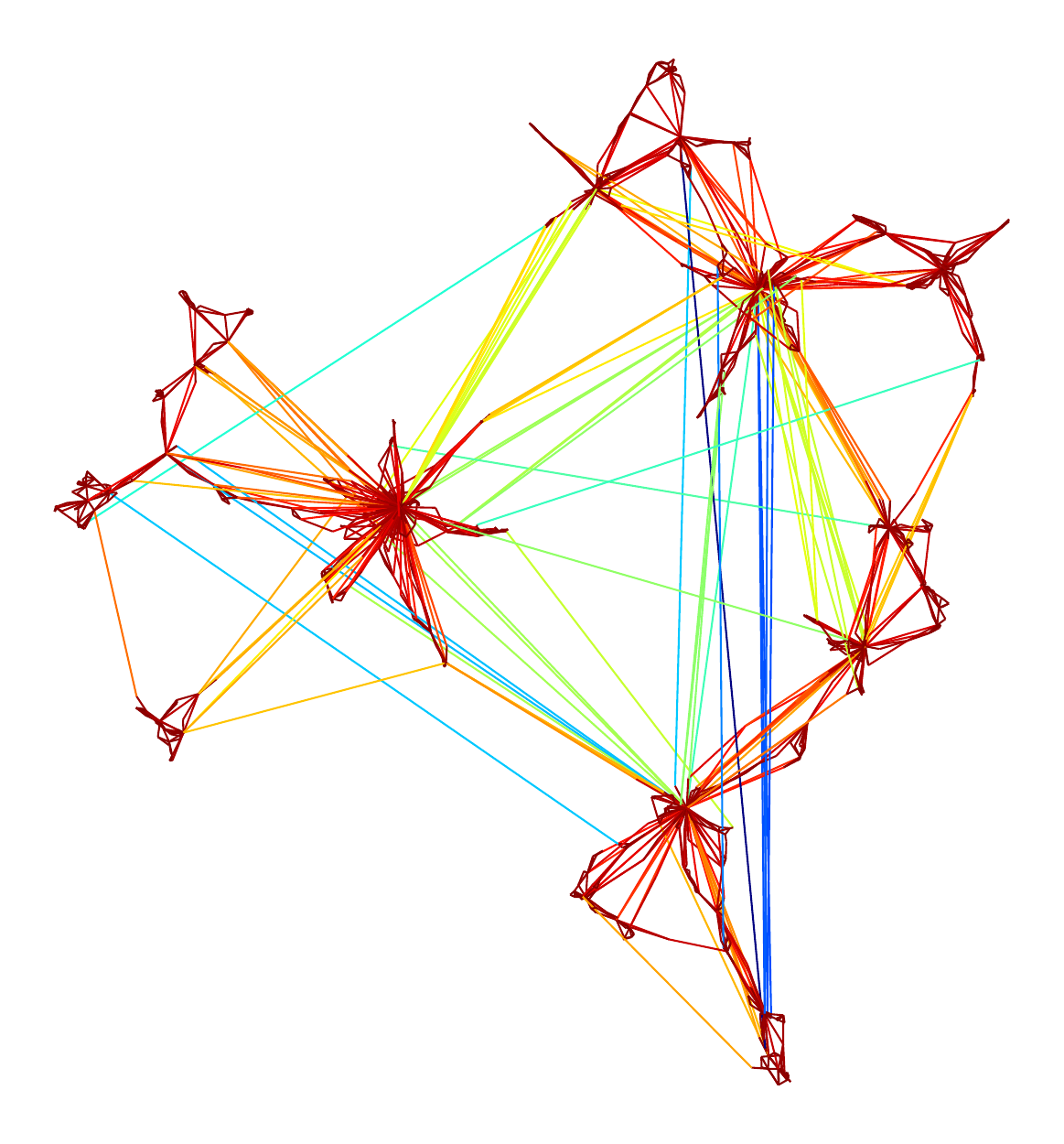}}
      & \parbox[c]{\tabfig\textwidth}{} 
      & \parbox[c]{\tabfig\textwidth}{} 
      & \parbox[c]{\tabfig\textwidth}{\taboffset
      \includegraphics[width=\tabfig\textwidth,height=\tabfig\textwidth]{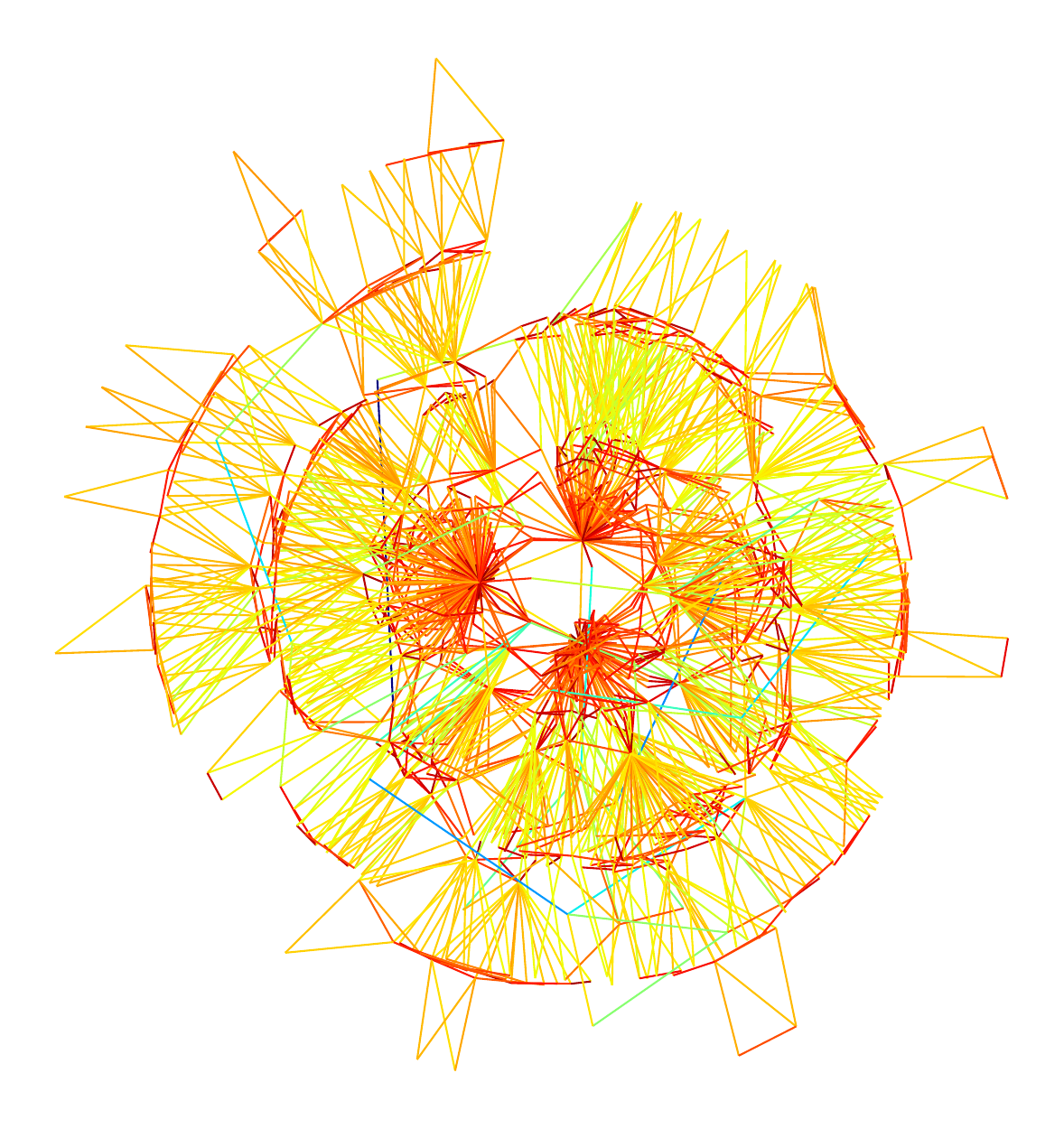}}  \\   
      \hline

     \multirow{2}{*}{\vspace{-1cm}\rotatebox[origin=c]{90}{CSphd}}  &
      \parbox[c]{\tabfig\textwidth}{} 
      & \parbox[c]{\tabfig\textwidth}{
      \includegraphics[width=\tabfig\textwidth,height=\tabfig\textwidth]{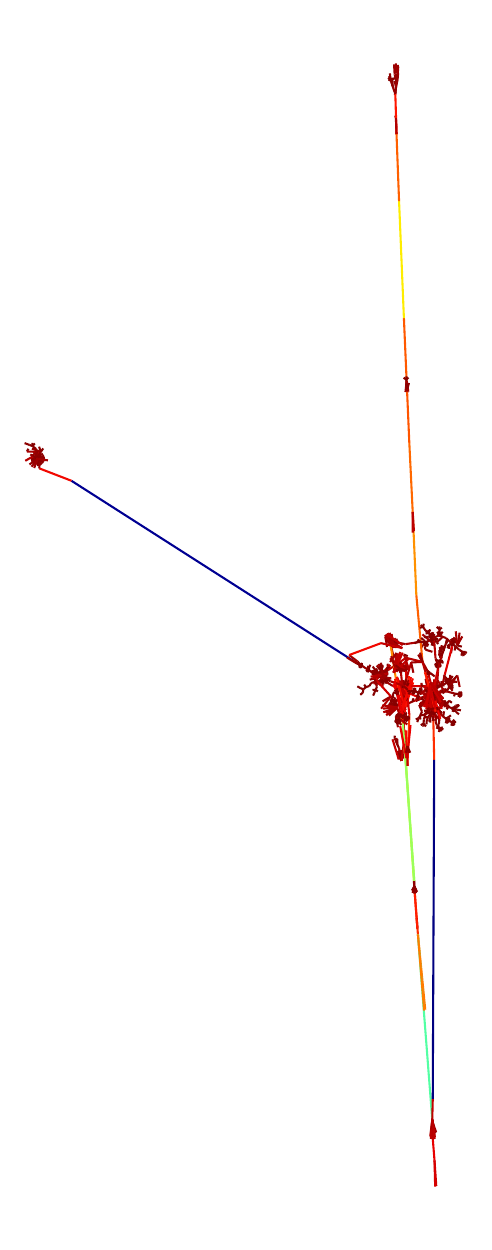}}
      & \parbox[c]{\tabfig\textwidth}{
      \includegraphics[width=\tabfig\textwidth,height=\tabfig\textwidth]{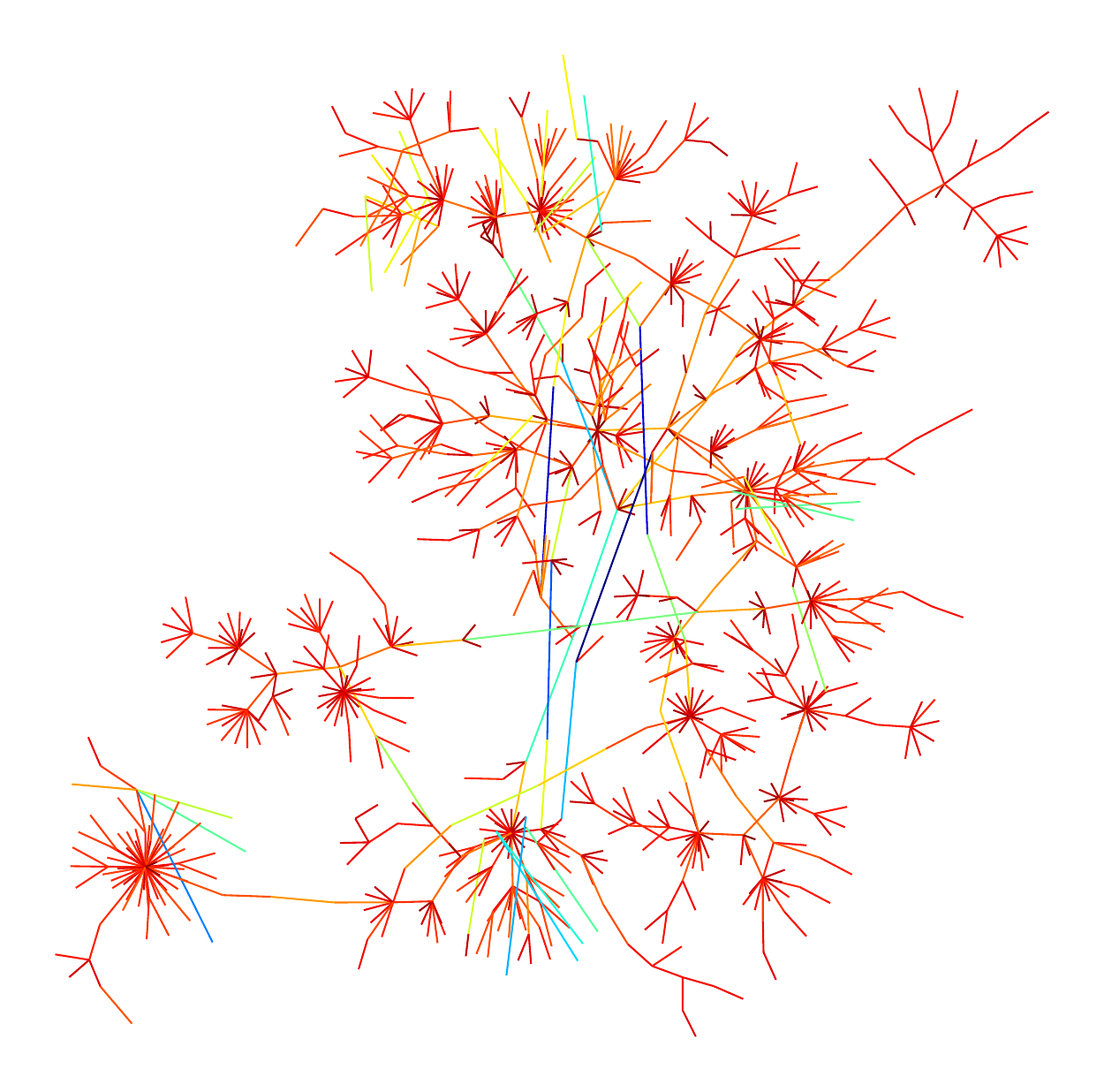}}
      & \parbox[c]{\tabfig\textwidth}{}
      & \parbox[c]{\tabfig\textwidth}{
      \includegraphics[width=\tabfig\textwidth,height=\tabfig\textwidth]{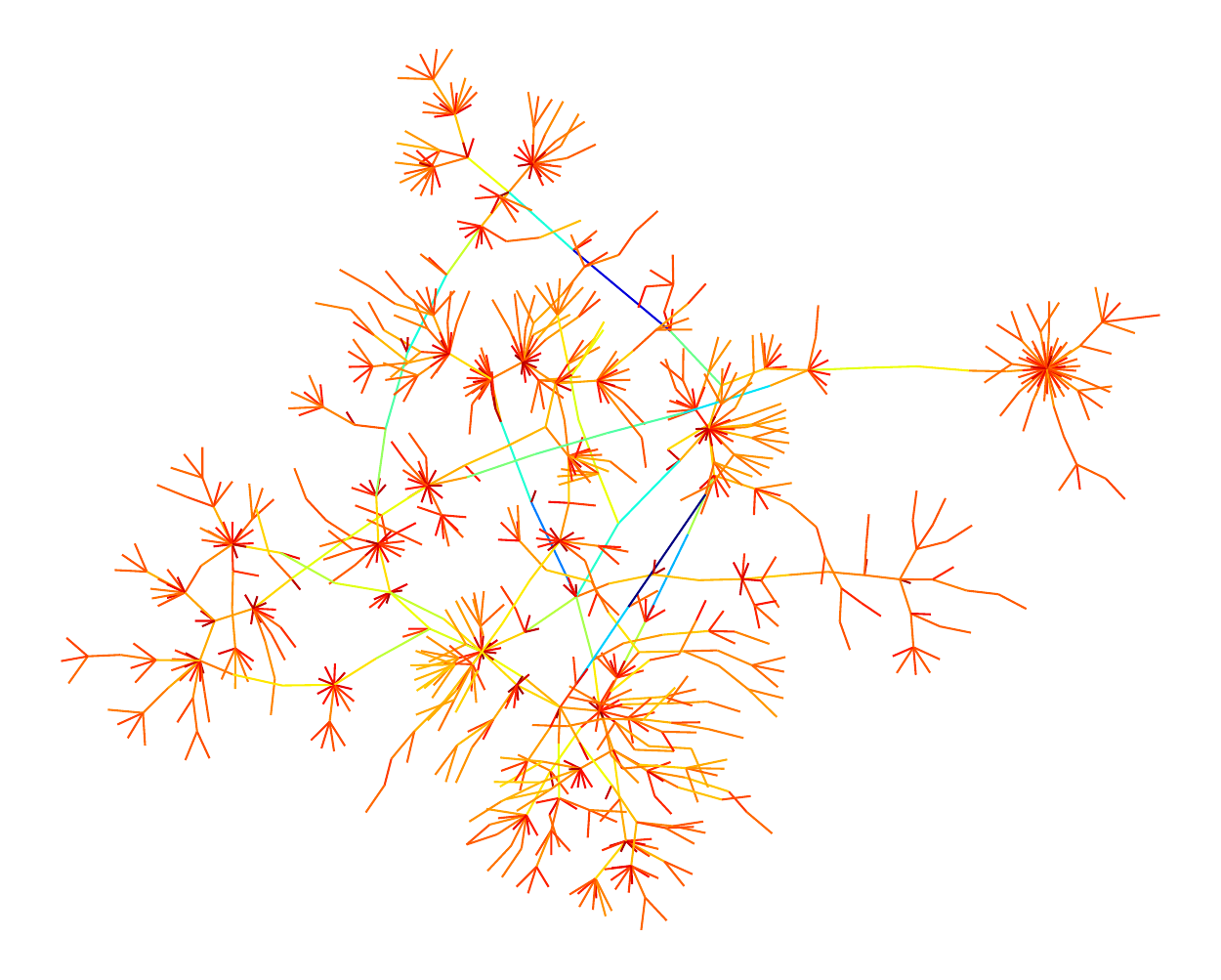}} 
      & \parbox[c]{\tabfig\textwidth}{
      \includegraphics[width=\tabfig\textwidth,height=\tabfig\textwidth]{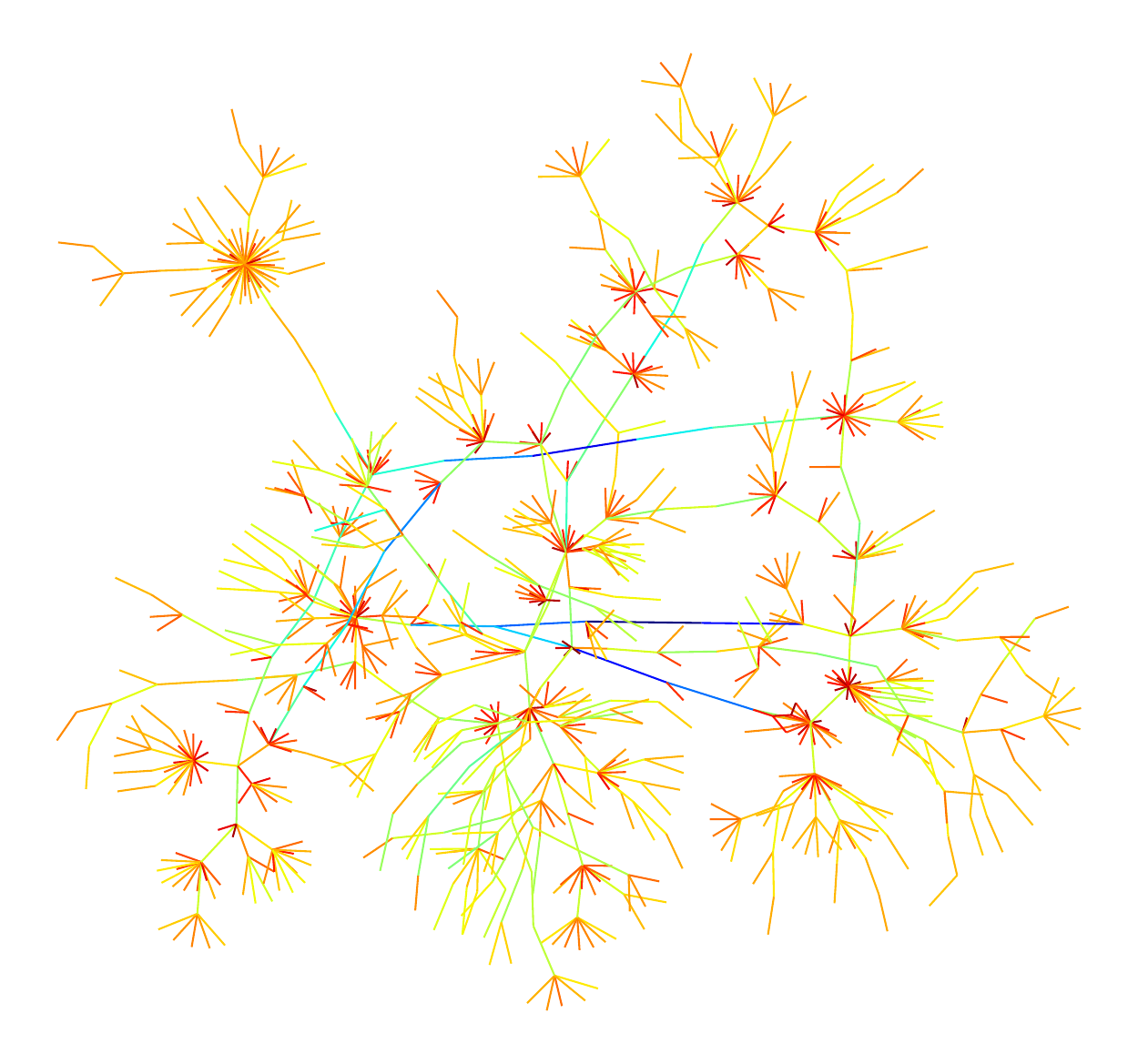}} 
      & \parbox[c]{\tabfig\textwidth}{}  \\
      
       &
      \parbox[c]{\tabfig\textwidth}{
      \taboffset\includegraphics[width=\tabfig\textwidth,height=\tabfig\textwidth]{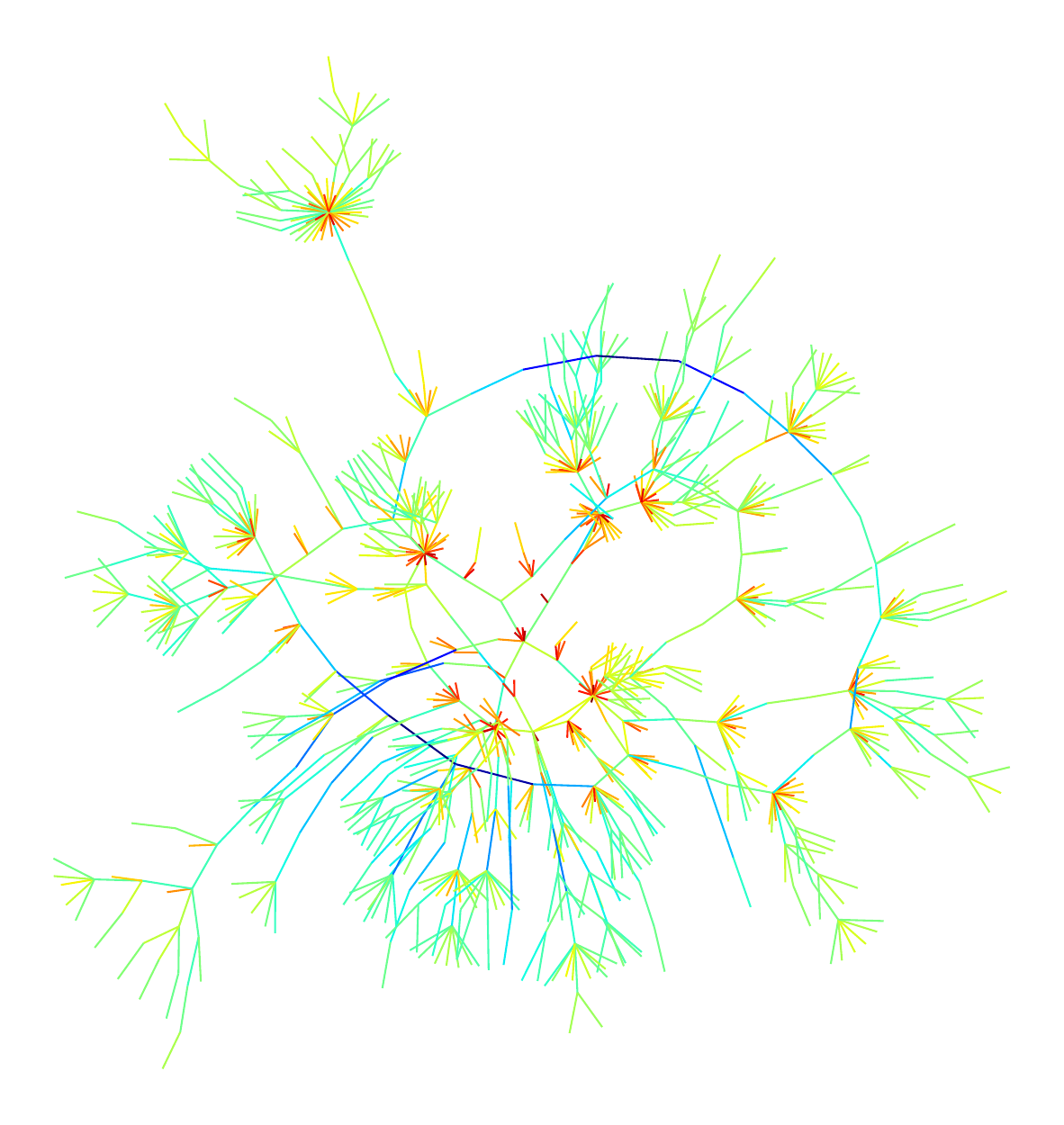}} 
      & \parbox[c]{\tabfig\textwidth}{}
      & \parbox[c]{\tabfig\textwidth}{}
      & \parbox[c]{\tabfig\textwidth}{\taboffset
      \includegraphics[width=\tabfig\textwidth,height=\tabfig\textwidth]{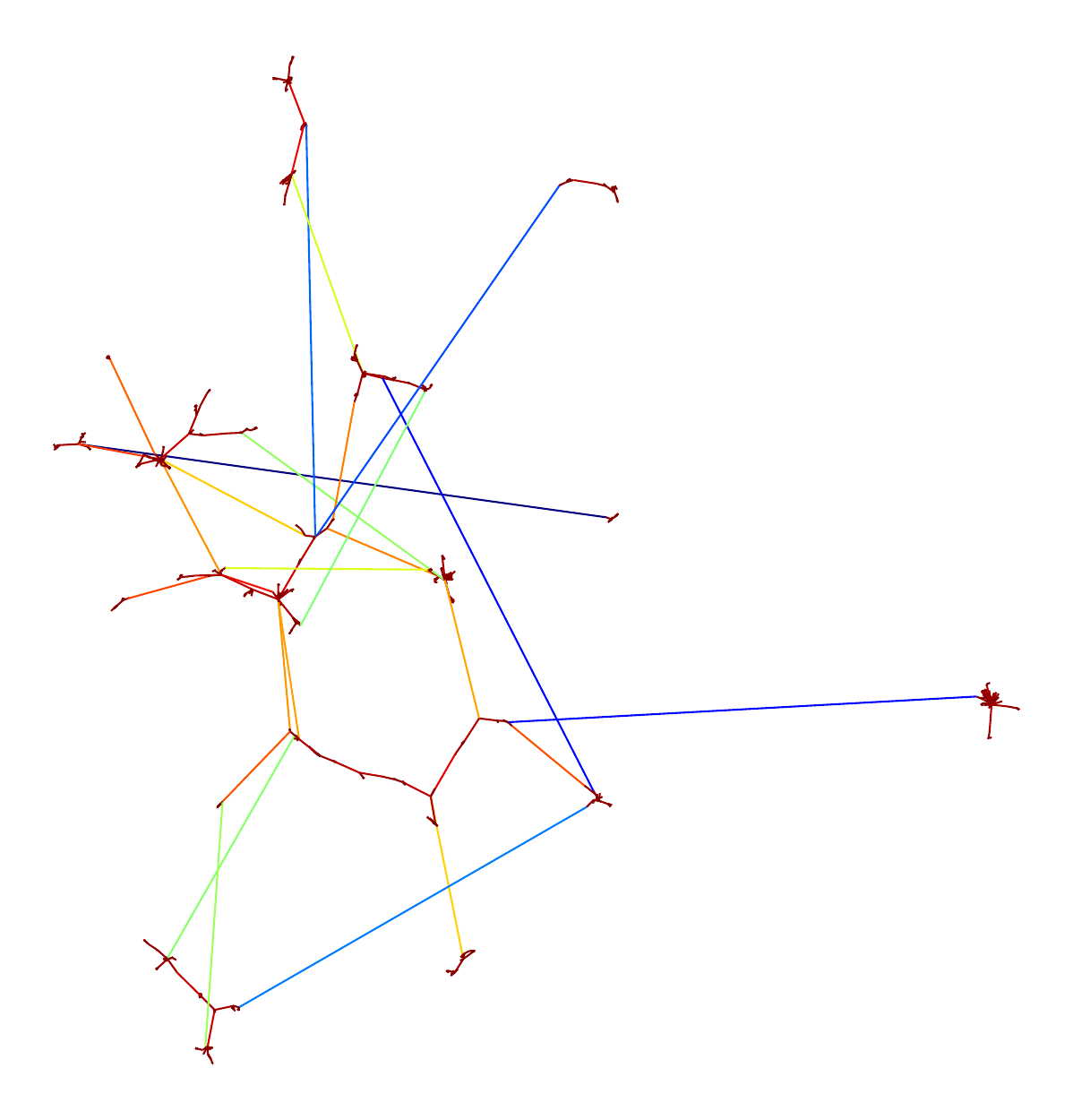}}
      & \parbox[c]{\tabfig\textwidth}{} 
      & \parbox[c]{\tabfig\textwidth}{} 
      & \parbox[c]{\tabfig\textwidth}{\taboffset
      \includegraphics[width=\tabfig\textwidth,height=\tabfig\textwidth]{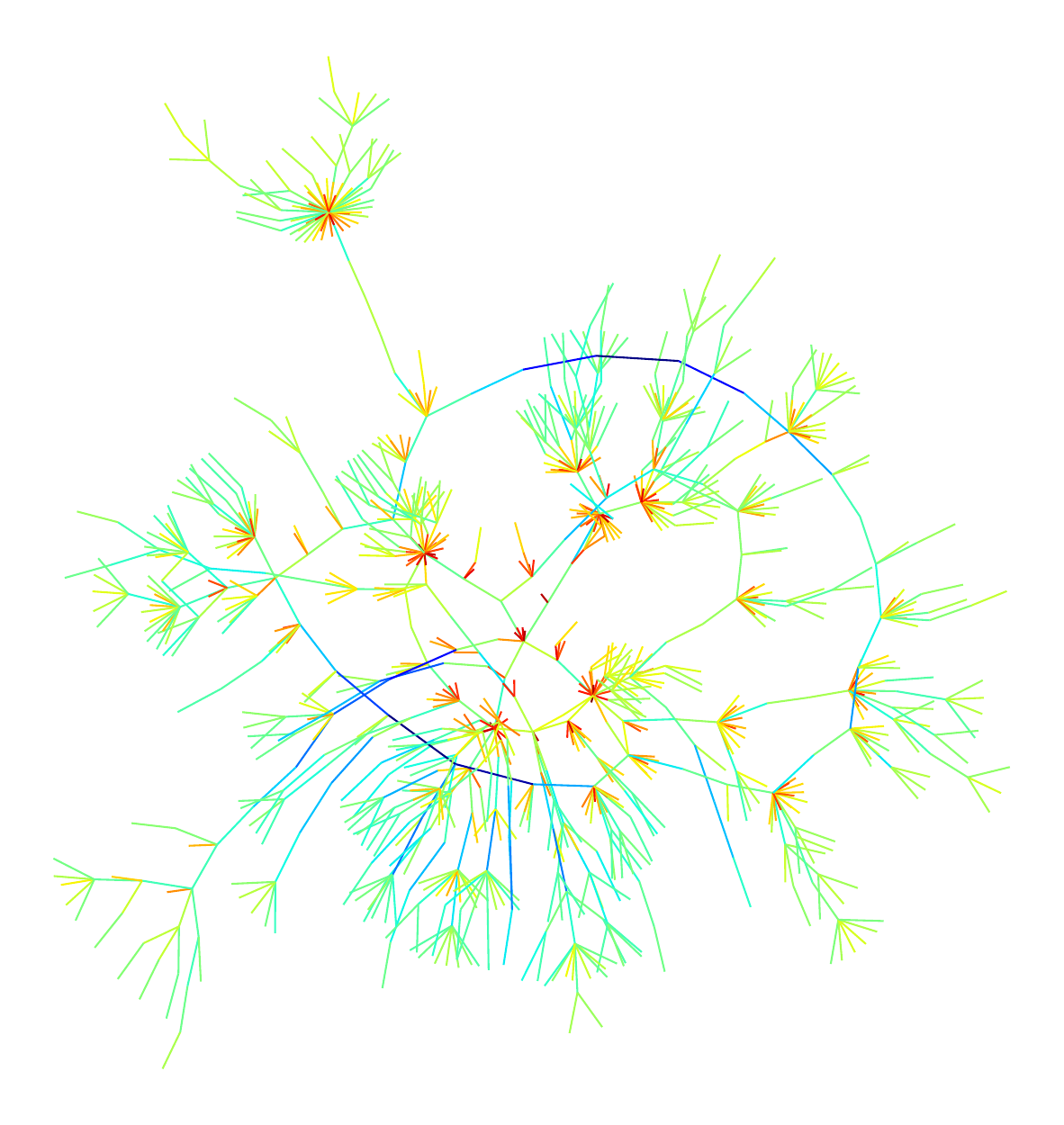}}  \\   
      \hline
      
     \multirow{2}{*}{\vspace{-1cm}\rotatebox[origin=c]{90}{price\_1000}}  &
      \parbox[c]{\tabfig\textwidth}{} 
      & \parbox[c]{\tabfig\textwidth}{
      \includegraphics[width=\tabfig\textwidth,height=\tabfig\textwidth]{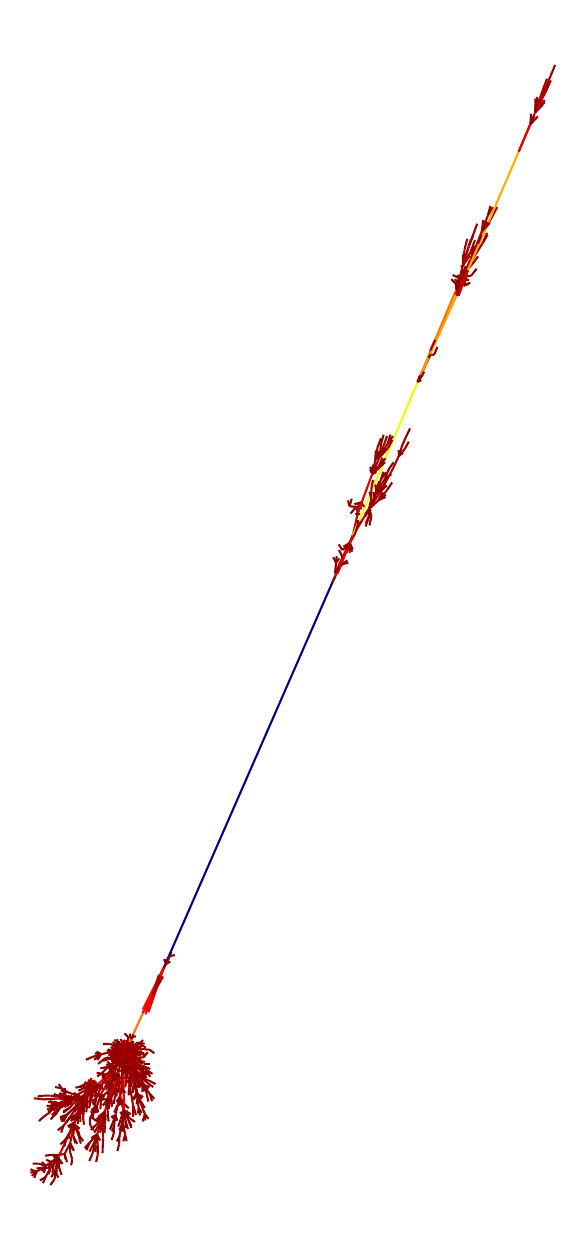}}
      & \parbox[c]{\tabfig\textwidth}{
      \includegraphics[width=\tabfig\textwidth,height=\tabfig\textwidth]{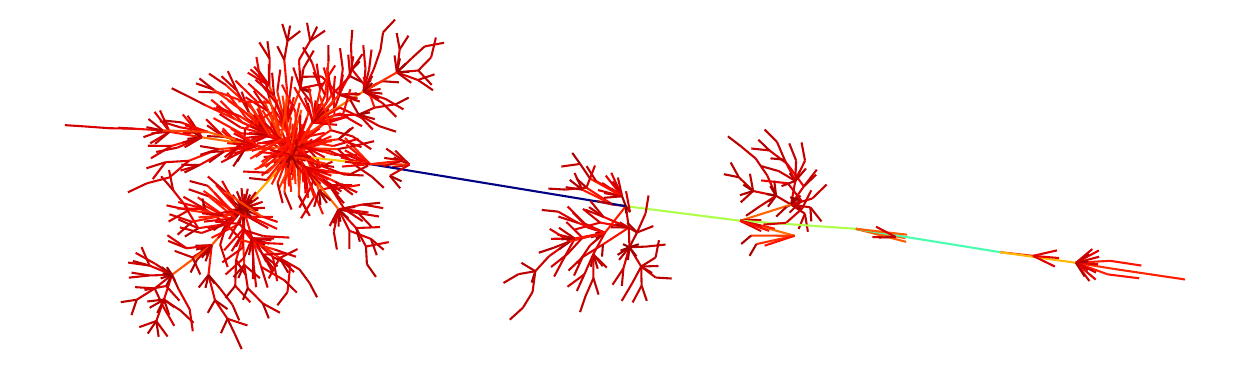}}
      & \parbox[c]{\tabfig\textwidth}{}
      & \parbox[c]{\tabfig\textwidth}{
      \includegraphics[width=\tabfig\textwidth,height=\tabfig\textwidth]{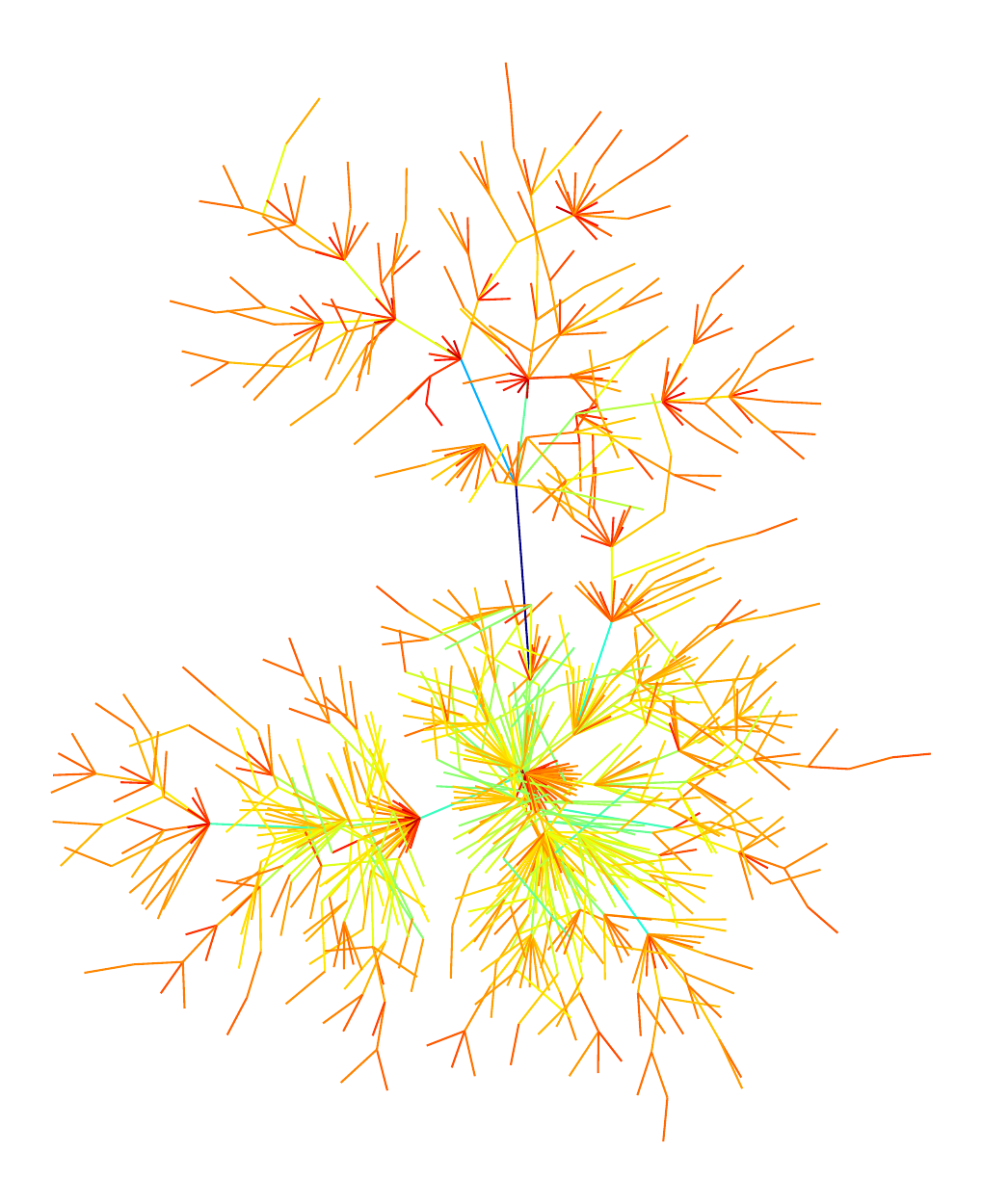}} 
      & \parbox[c]{\tabfig\textwidth}{
      \includegraphics[width=\tabfig\textwidth,height=\tabfig\textwidth]{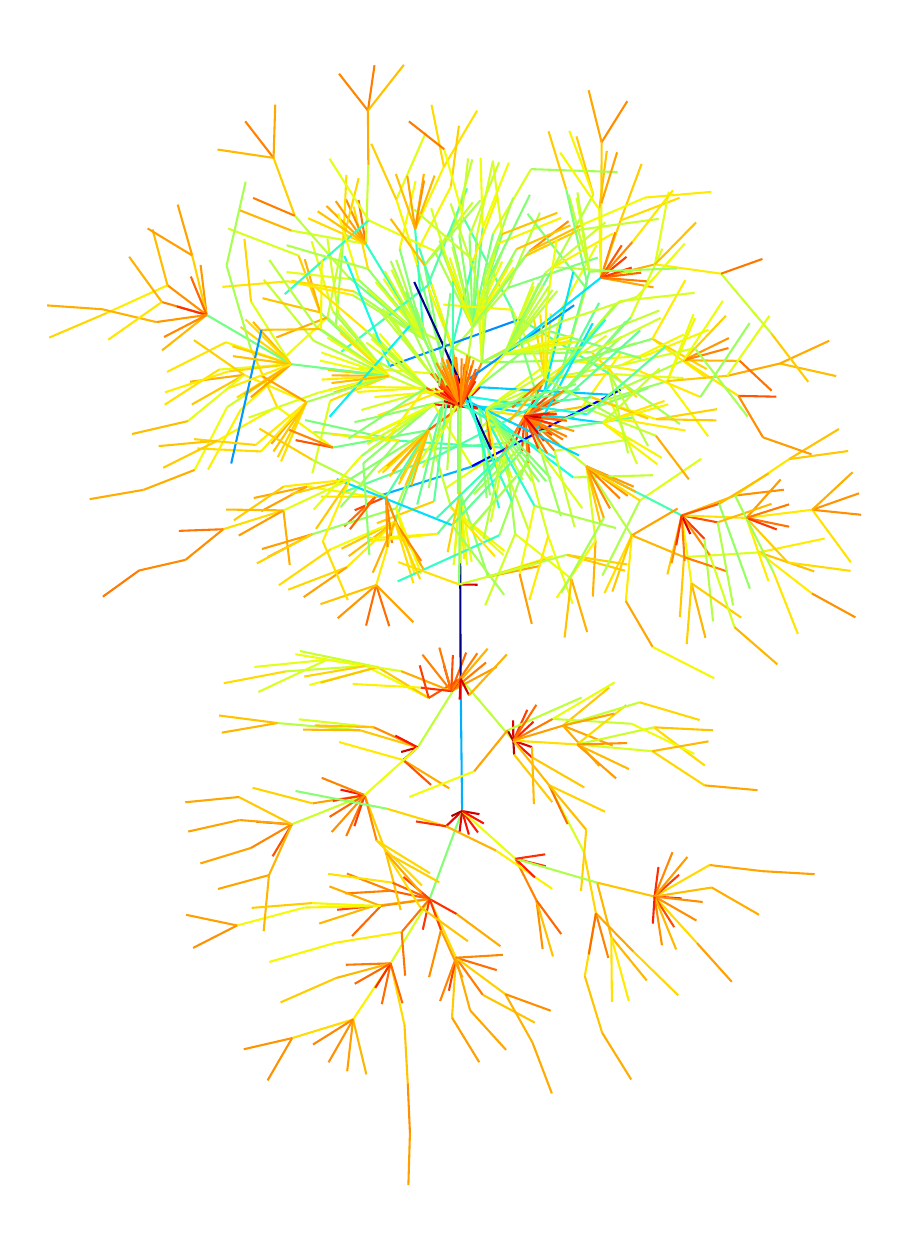}} 
      & \parbox[c]{\tabfig\textwidth}{}  \\
      
       &
      \parbox[c]{\tabfig\textwidth}{
      \taboffset\includegraphics[width=\tabfig\textwidth,height=\tabfig\textwidth]{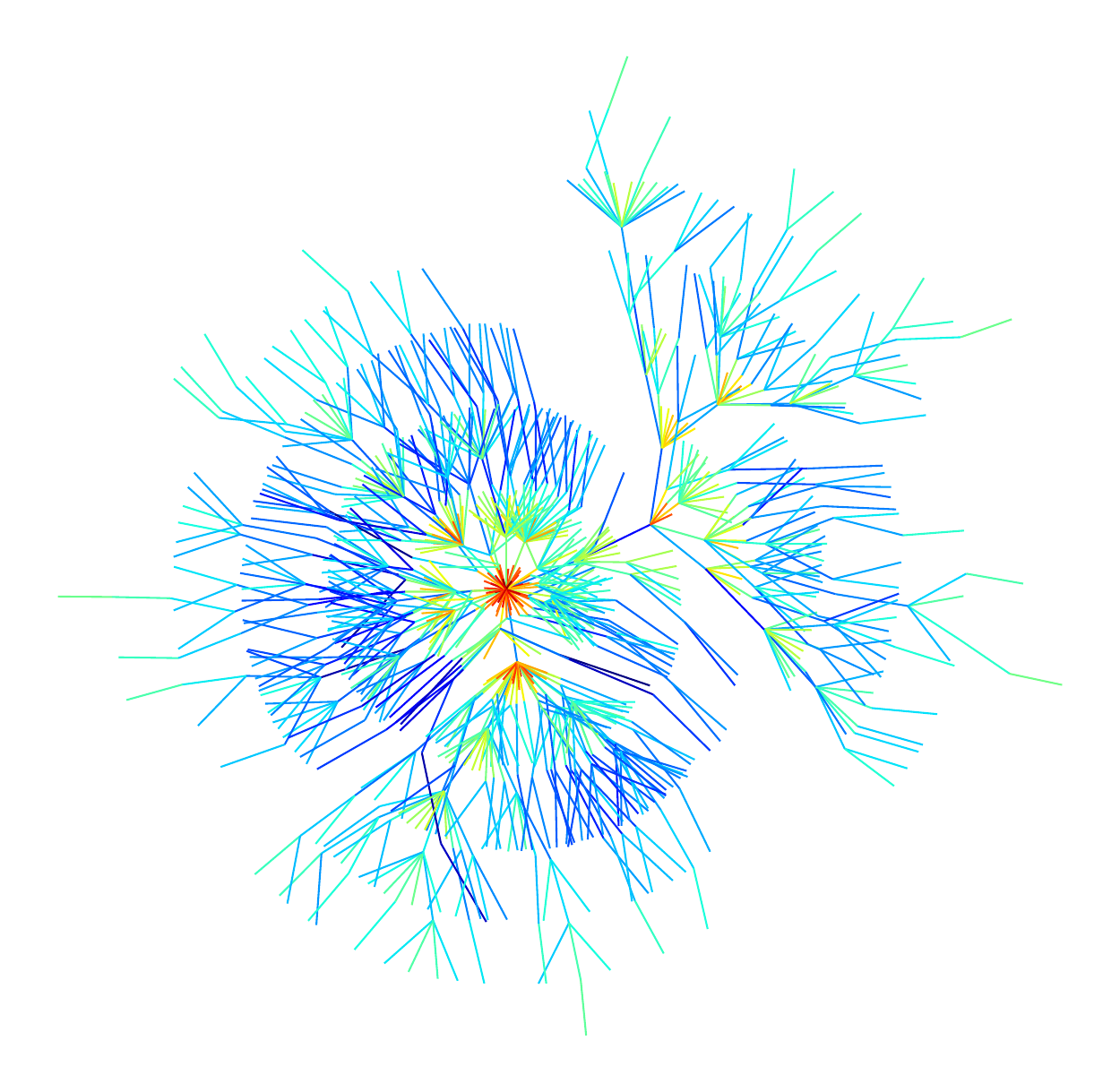}} 
      & \parbox[c]{\tabfig\textwidth}{}
      & \parbox[c]{\tabfig\textwidth}{}
      & \parbox[c]{\tabfig\textwidth}{\taboffset
      \includegraphics[width=\tabfig\textwidth,height=\tabfig\textwidth]{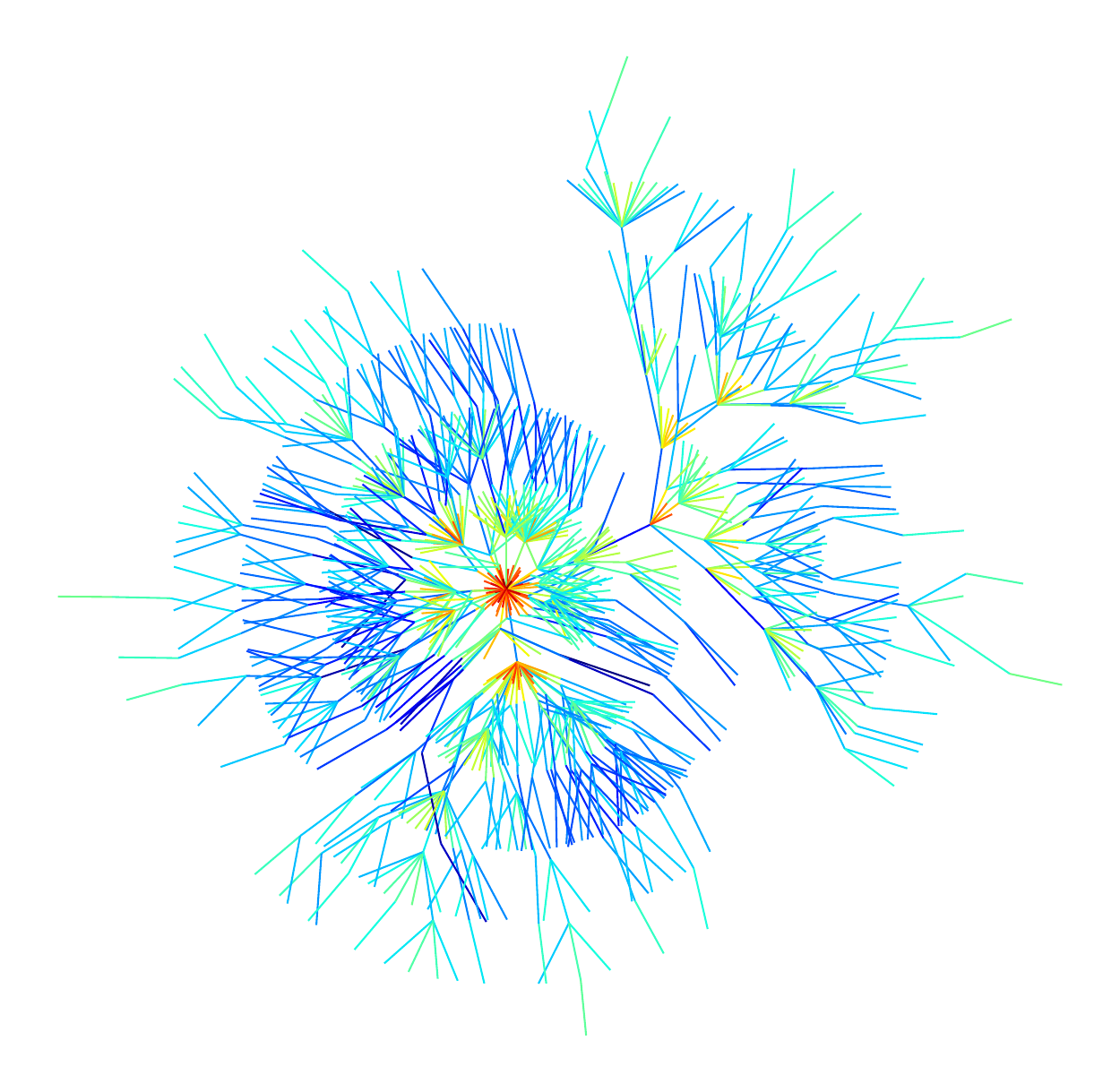}}
      & \parbox[c]{\tabfig\textwidth}{} 
      & \parbox[c]{\tabfig\textwidth}{} 
      & \parbox[c]{\tabfig\textwidth}{\taboffset
      \includegraphics[width=\tabfig\textwidth,height=\tabfig\textwidth]{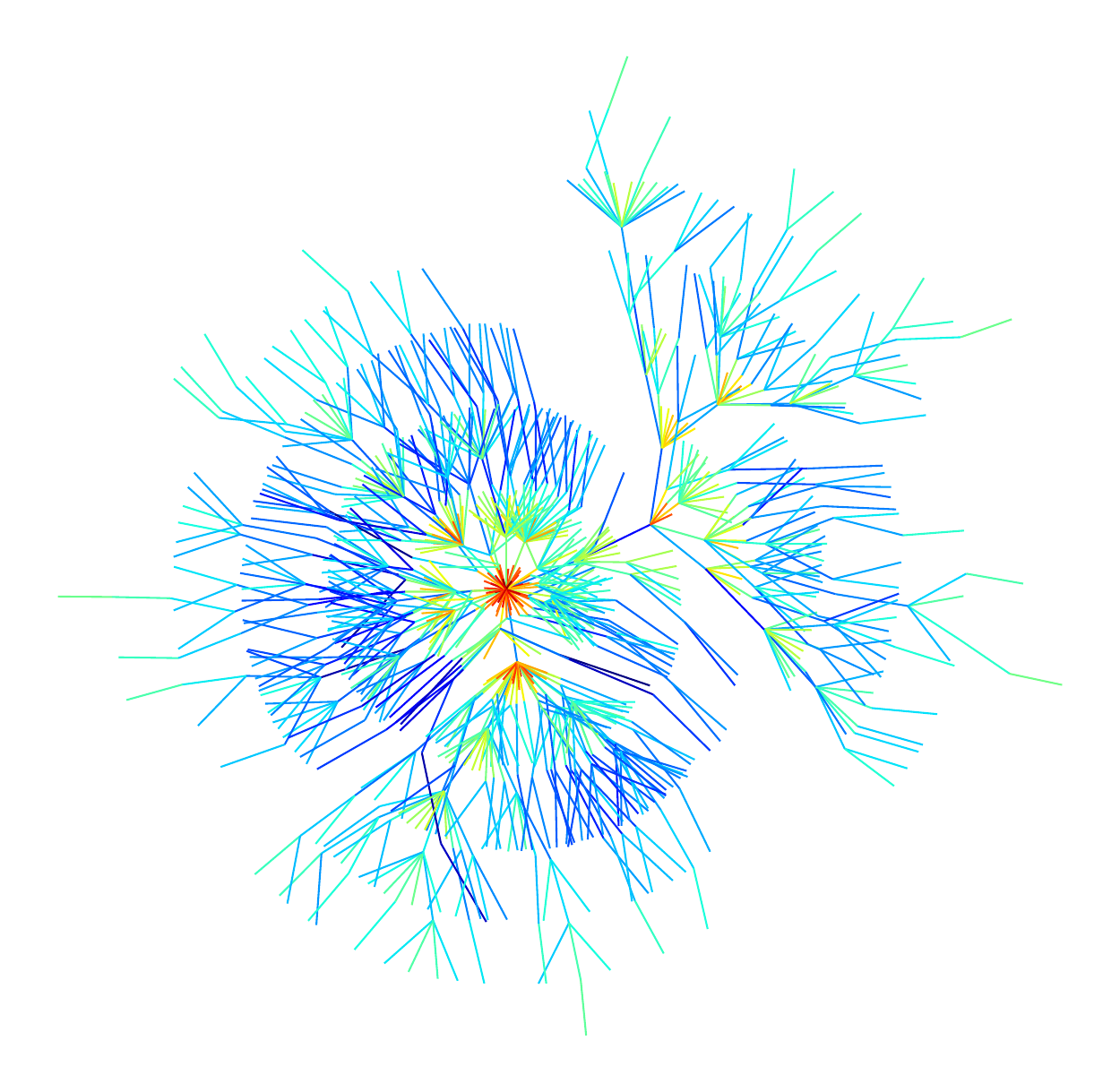}}  \\   
      \hline

  \end{tabular}
  
\end{table*}

\begin{table*}
  [ht] \caption{Additional embeddings} \label{tab:drawings3}
  \vspace{0.5cm}
  \centering
  \begin{tabular}
      {| l | c c c c c c c|} \hline & tsNET & L2G k=16 & L2G k=32 & UMAP & L2G k=64 & L2G k=100 & MDS\\
      \hline 

     \multirow{2}{*}{\vspace{-1cm}\rotatebox[origin=c]{90}{block\_300}}  &
      \parbox[c]{\tabfig\textwidth}{} 
      & \parbox[c]{\tabfig\textwidth}{
      \includegraphics[width=\tabfig\textwidth,height=\tabfig\textwidth]{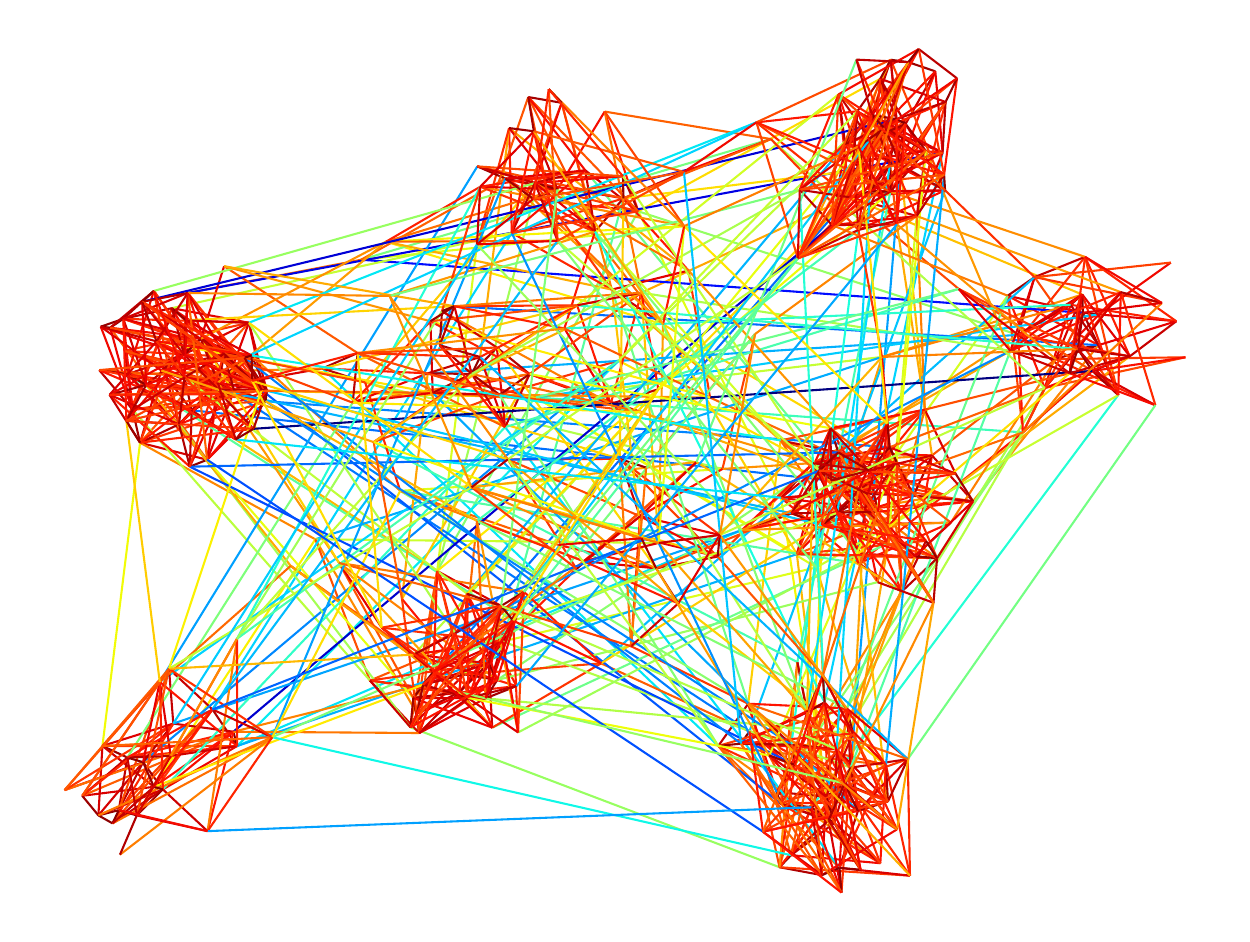}}
      & \parbox[c]{\tabfig\textwidth}{
      \includegraphics[width=\tabfig\textwidth,height=\tabfig\textwidth]{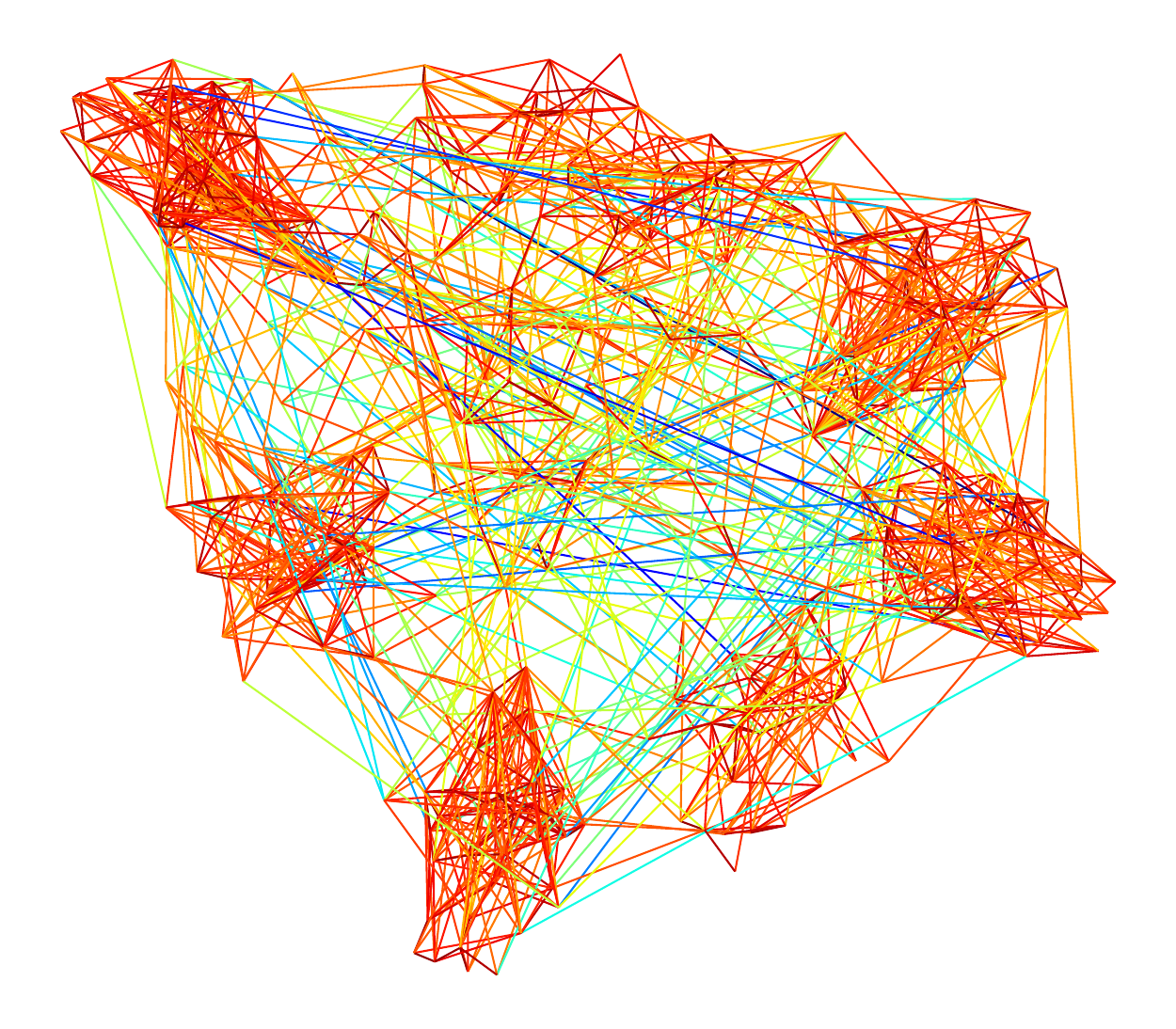}}
      & \parbox[c]{\tabfig\textwidth}{}
      & \parbox[c]{\tabfig\textwidth}{
      \includegraphics[width=\tabfig\textwidth,height=\tabfig\textwidth]{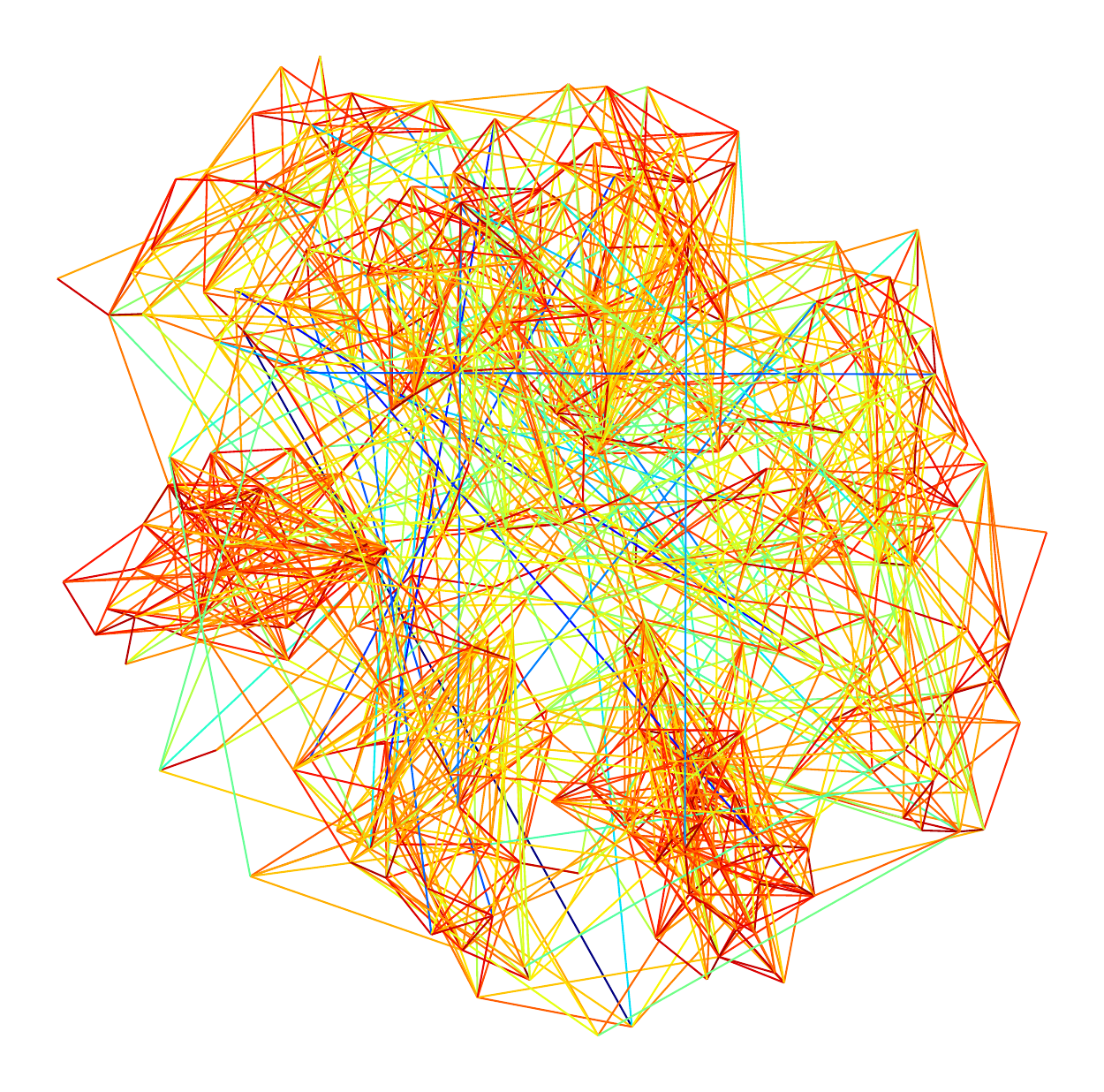}} 
      & \parbox[c]{\tabfig\textwidth}{
      \includegraphics[width=\tabfig\textwidth,height=\tabfig\textwidth]{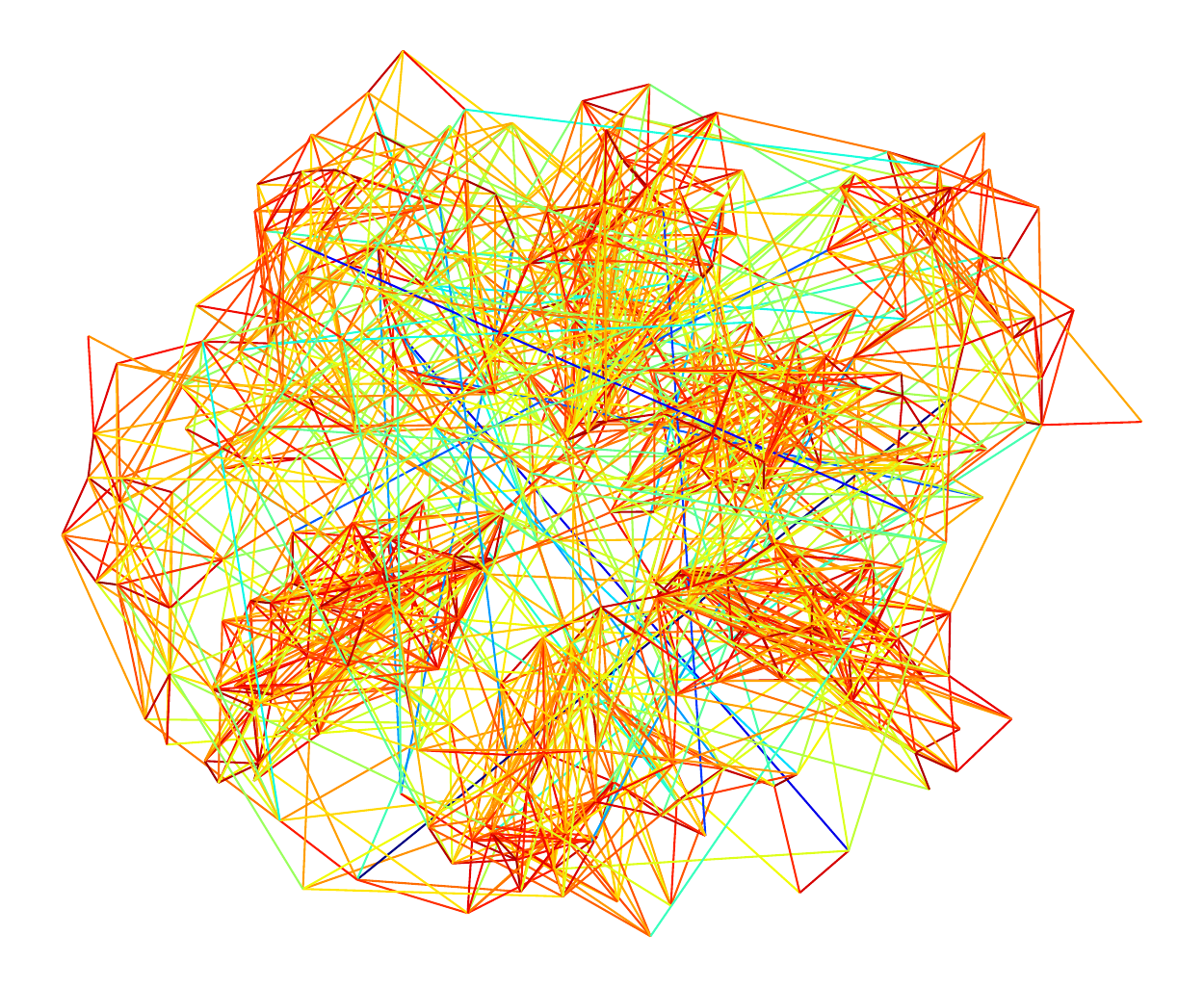}} 
      & \parbox[c]{\tabfig\textwidth}{}  \\
      
       &
      \parbox[c]{\tabfig\textwidth}{
      \taboffset\includegraphics[width=\tabfig\textwidth,height=\tabfig\textwidth]{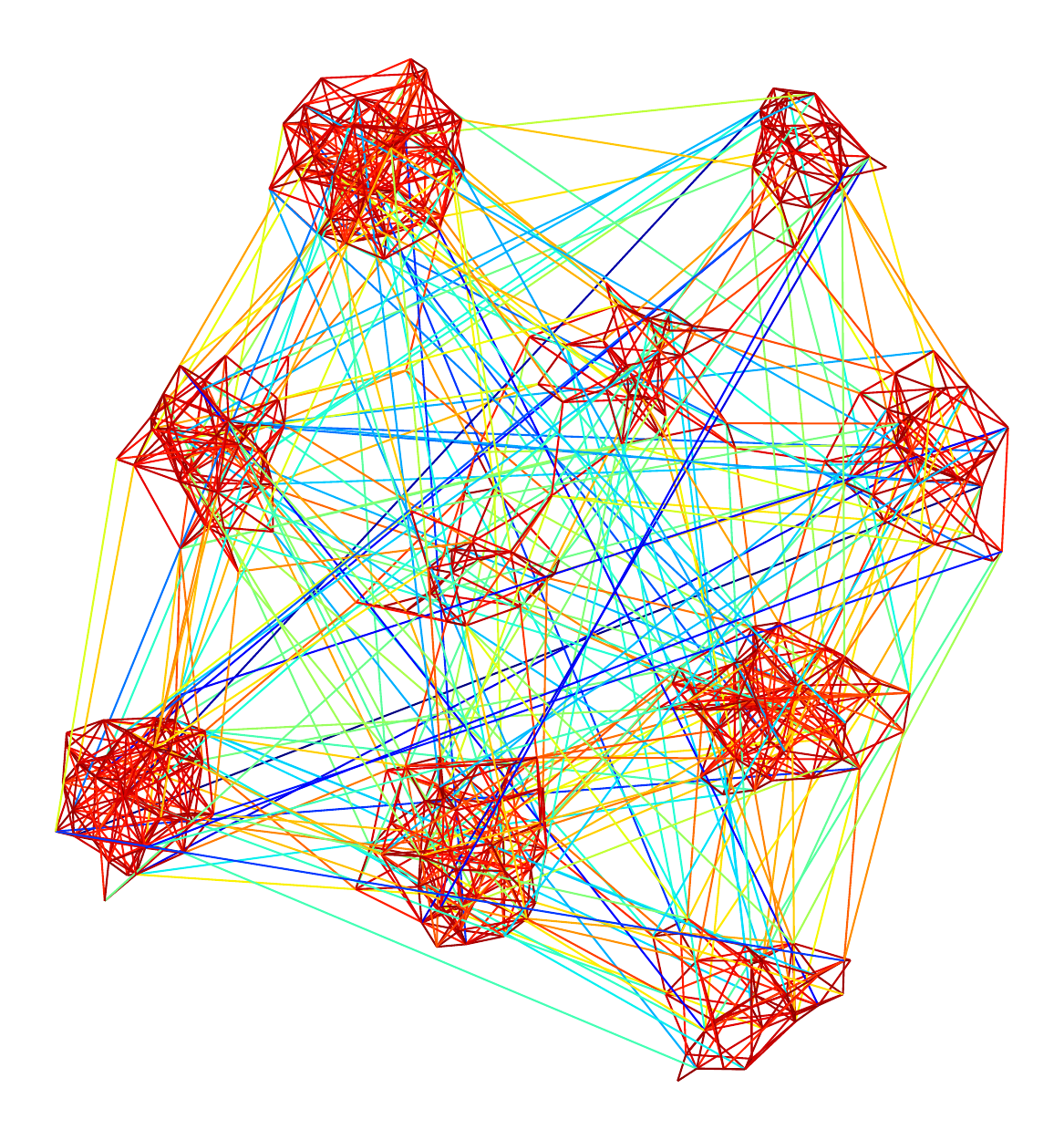}} 
      & \parbox[c]{\tabfig\textwidth}{}
      & \parbox[c]{\tabfig\textwidth}{}
      & \parbox[c]{\tabfig\textwidth}{\taboffset
      \includegraphics[width=\tabfig\textwidth,height=\tabfig\textwidth]{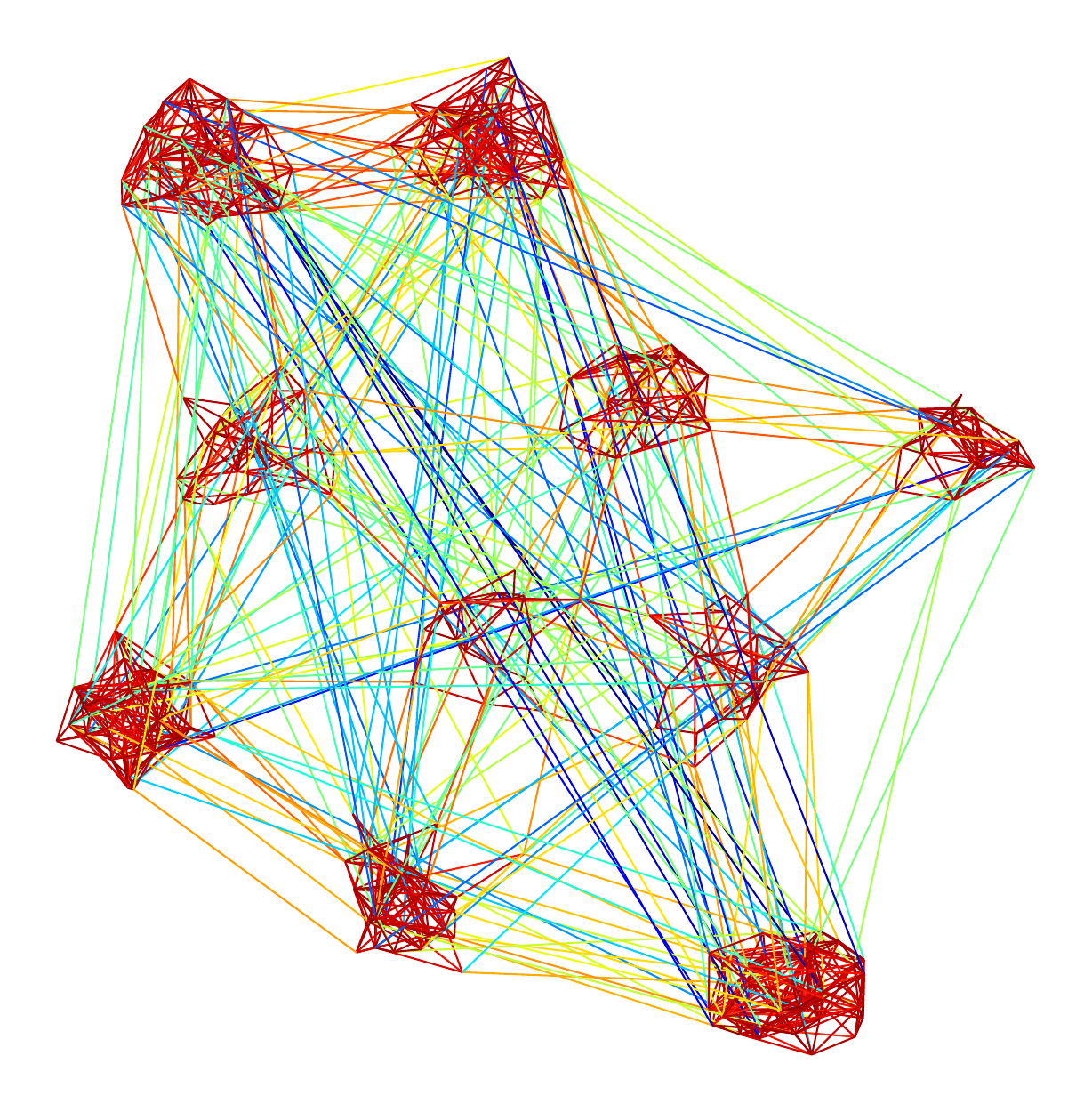}}
      & \parbox[c]{\tabfig\textwidth}{} 
      & \parbox[c]{\tabfig\textwidth}{} 
      & \parbox[c]{\tabfig\textwidth}{\taboffset
      \includegraphics[width=\tabfig\textwidth,height=\tabfig\textwidth]{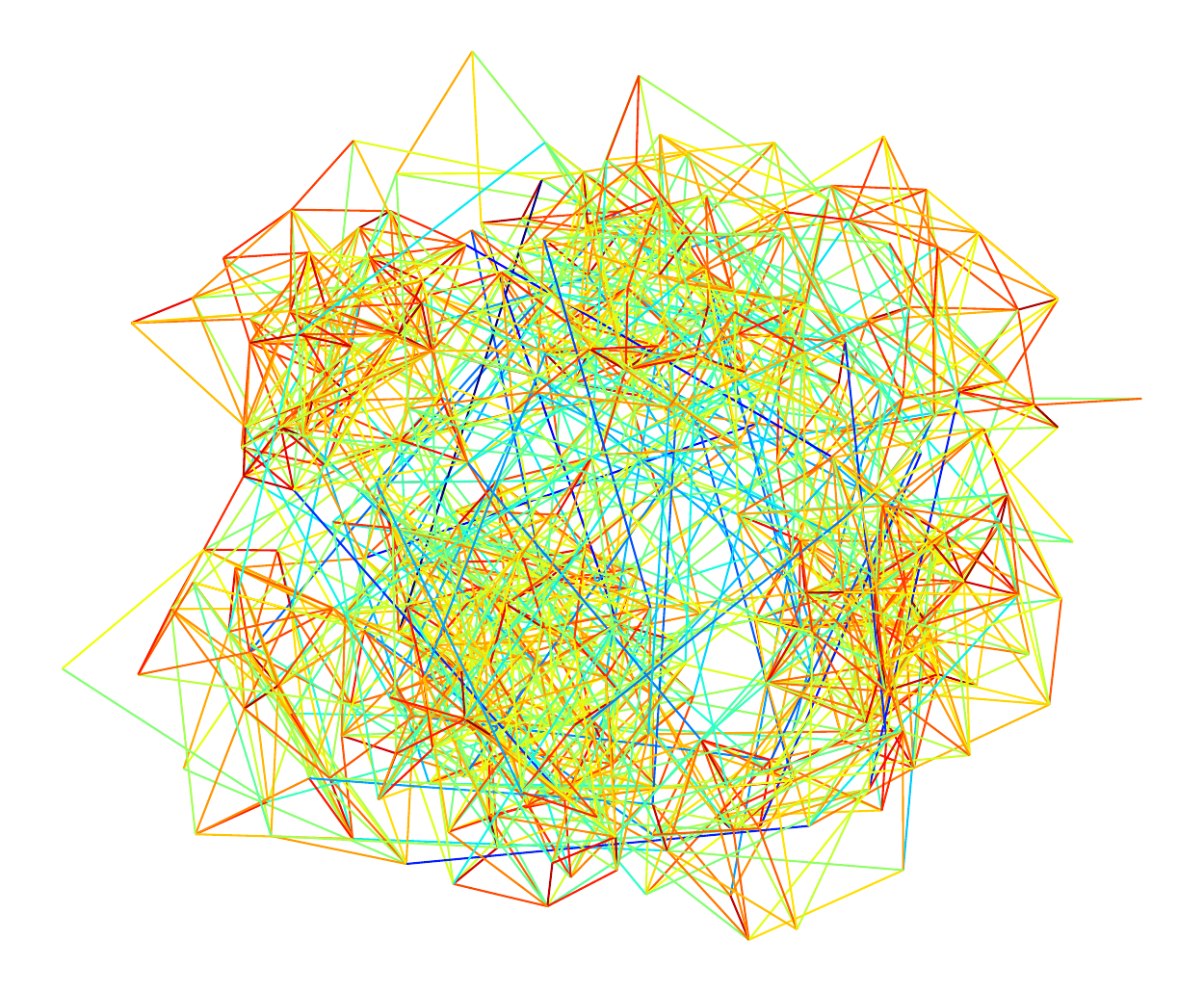}}  \\   
      \hline

     \multirow{2}{*}{\vspace{-1cm}\rotatebox[origin=c]{90}{block\_1000}}  &
      \parbox[c]{\tabfig\textwidth}{} 
      & \parbox[c]{\tabfig\textwidth}{
      \includegraphics[width=\tabfig\textwidth,height=\tabfig\textwidth]{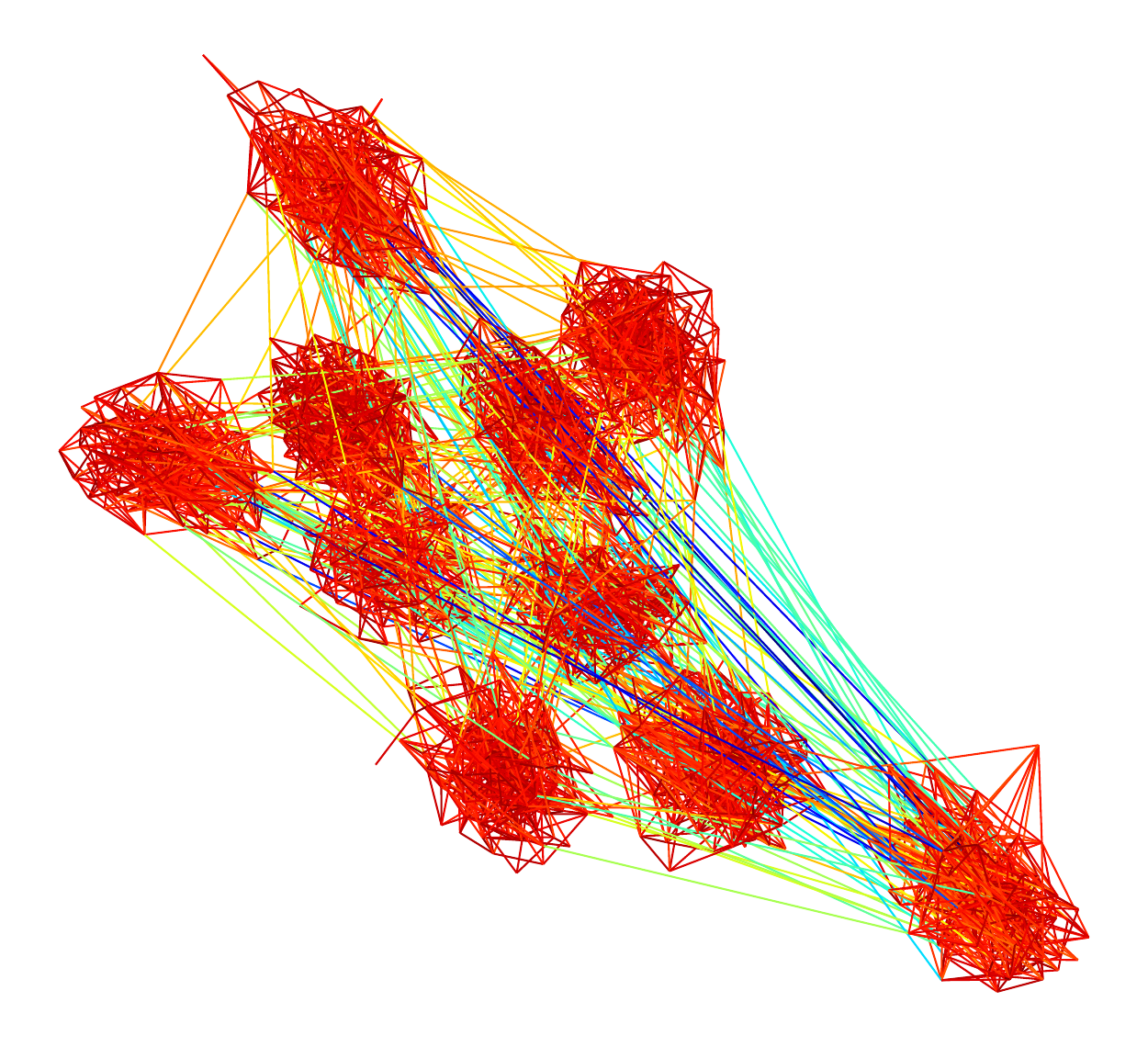}}
      & \parbox[c]{\tabfig\textwidth}{
      \includegraphics[width=\tabfig\textwidth,height=\tabfig\textwidth]{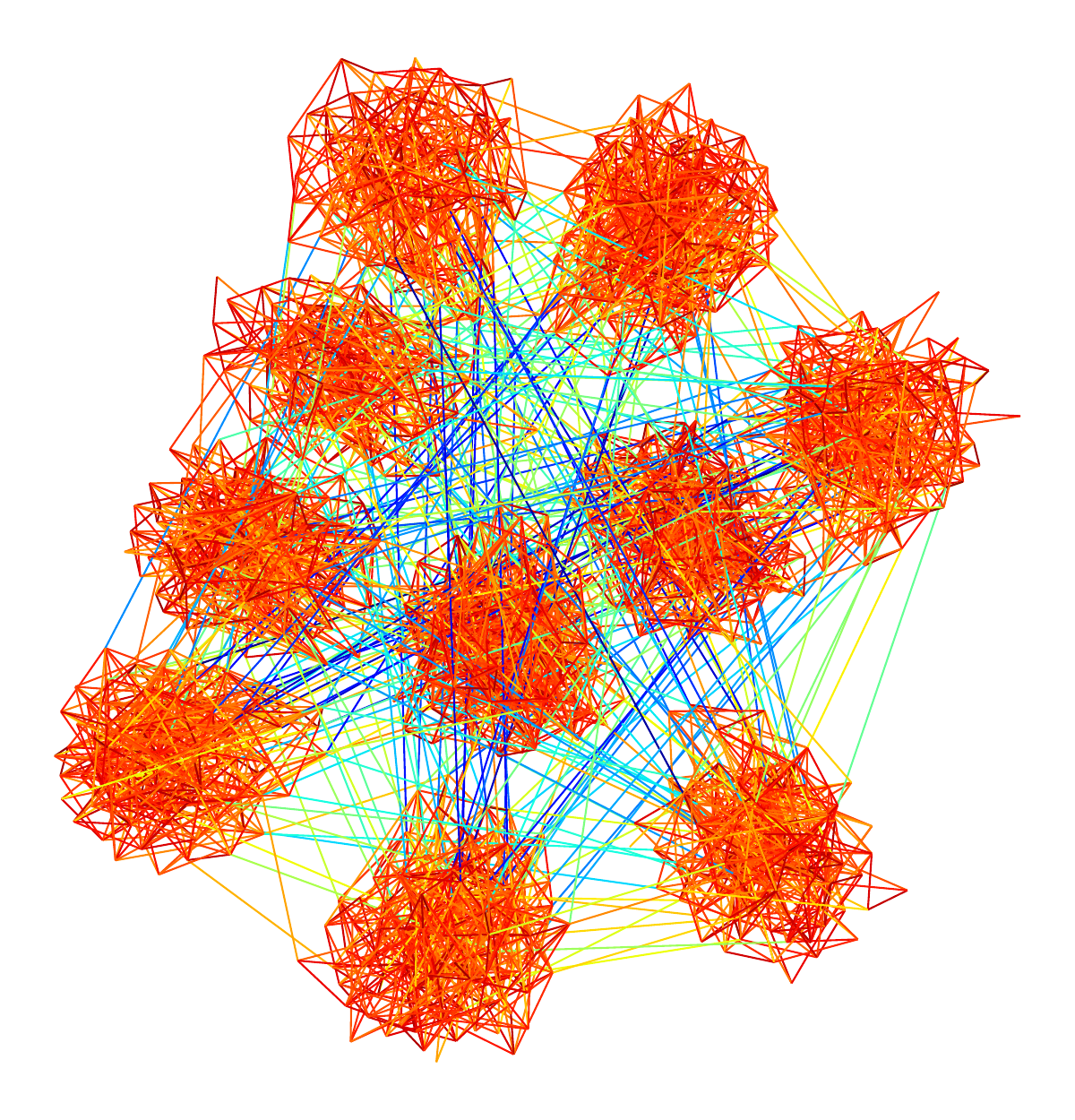}}
      & \parbox[c]{\tabfig\textwidth}{}
      & \parbox[c]{\tabfig\textwidth}{
      \includegraphics[width=\tabfig\textwidth,height=\tabfig\textwidth]{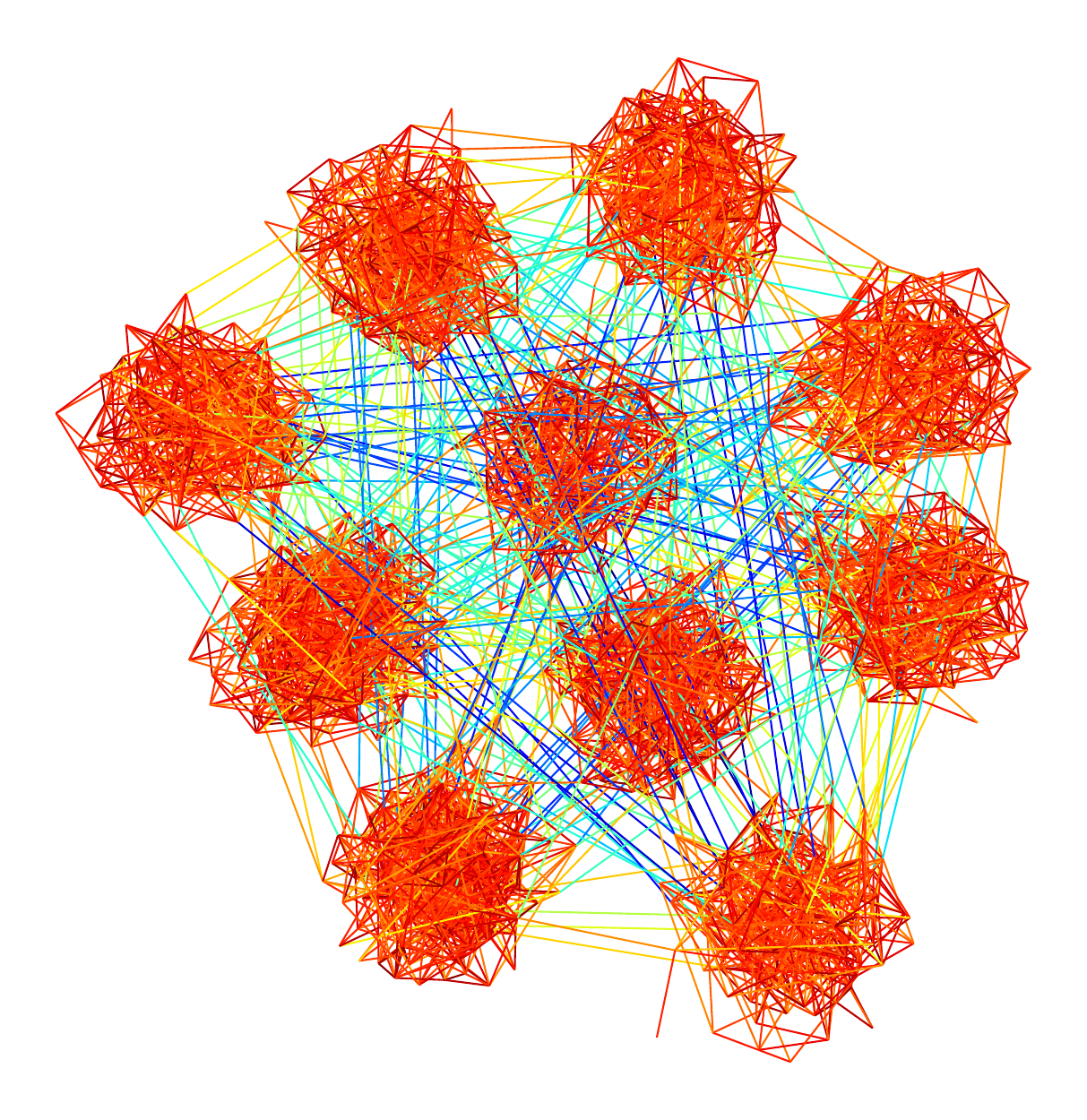}} 
      & \parbox[c]{\tabfig\textwidth}{
      \includegraphics[width=\tabfig\textwidth,height=\tabfig\textwidth]{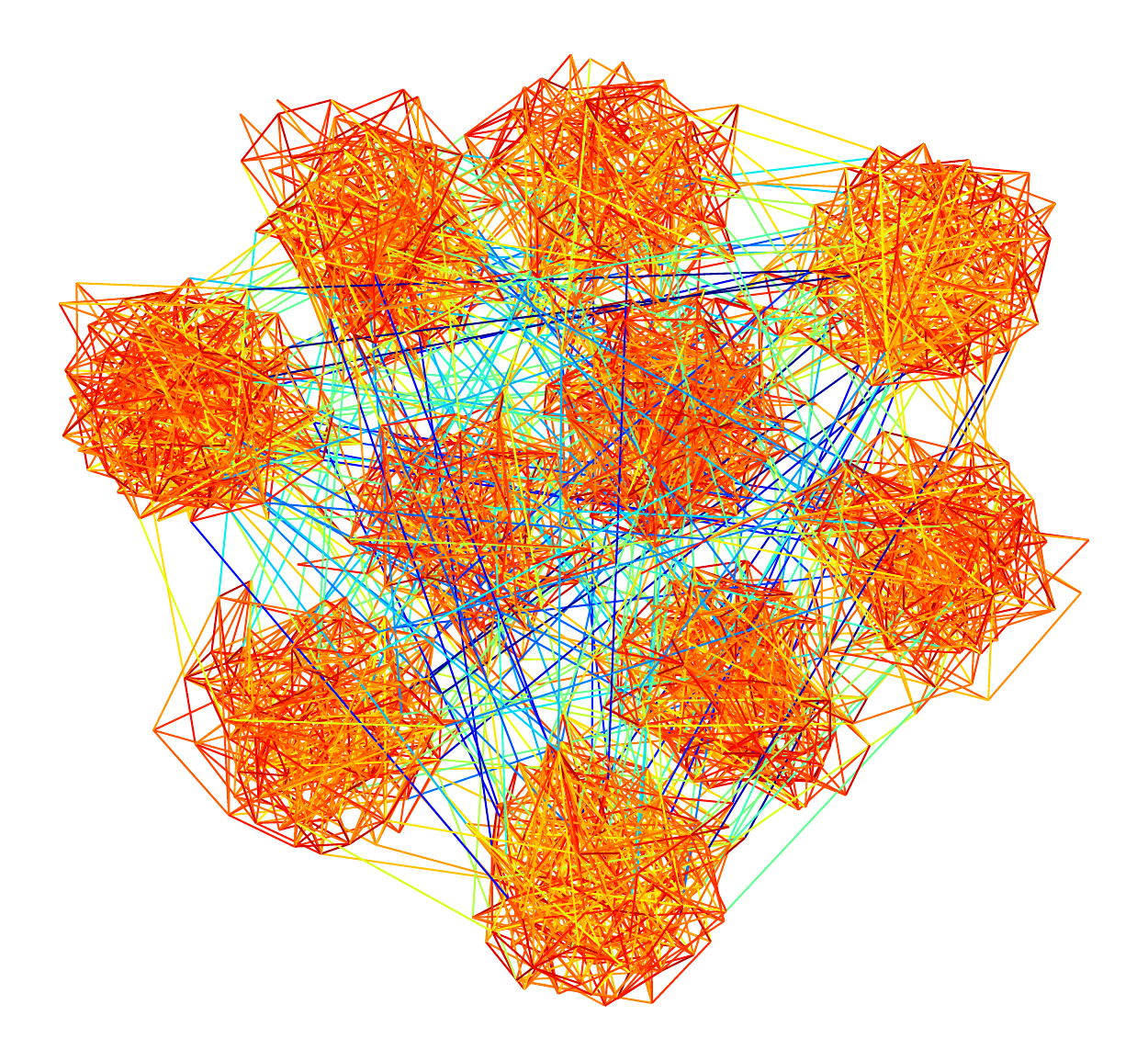}} 
      & \parbox[c]{\tabfig\textwidth}{}  \\
      
       &
      \parbox[c]{\tabfig\textwidth}{
      \taboffset\includegraphics[width=\tabfig\textwidth,height=\tabfig\textwidth]{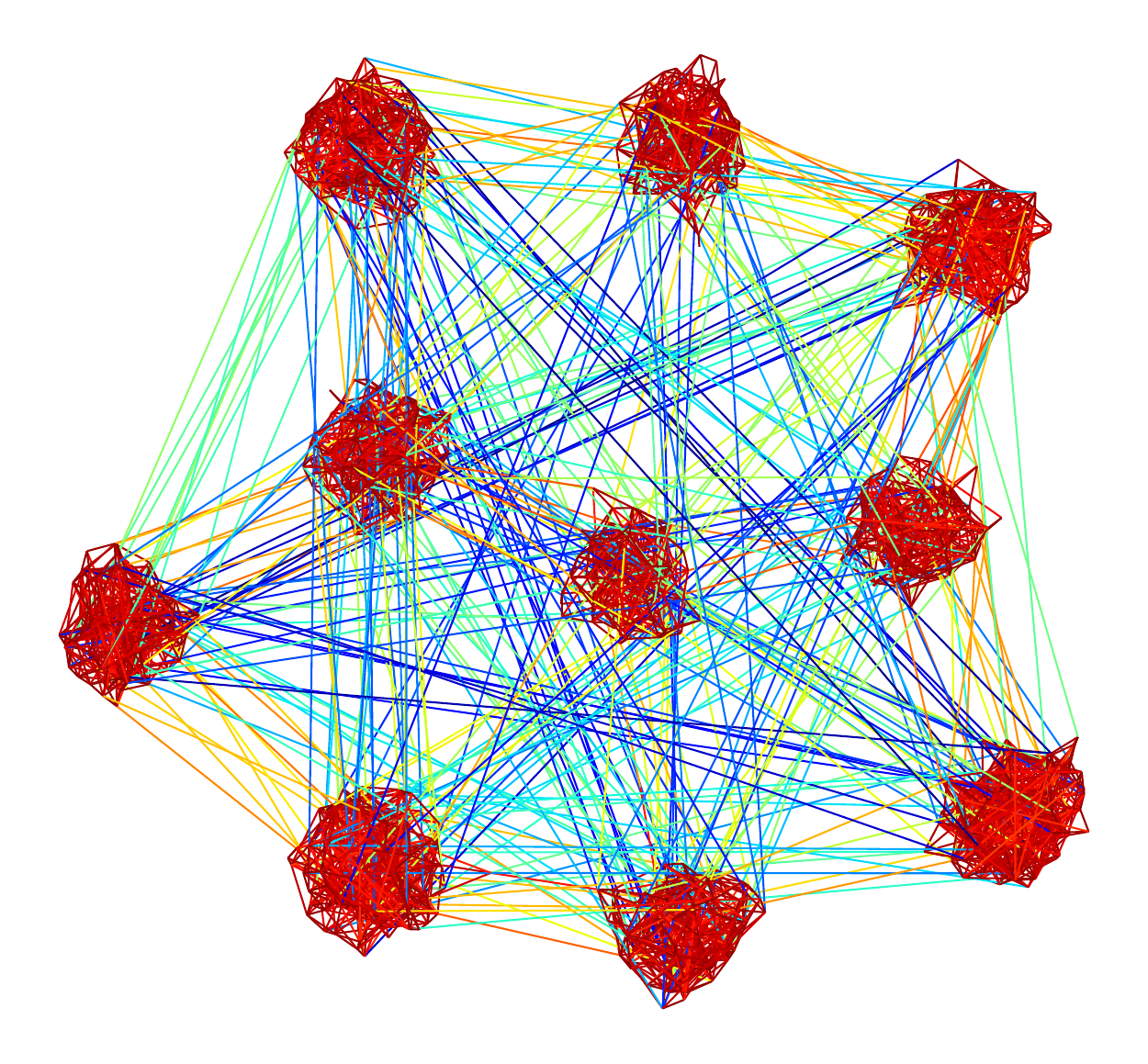}} 
      & \parbox[c]{\tabfig\textwidth}{}
      & \parbox[c]{\tabfig\textwidth}{}
      & \parbox[c]{\tabfig\textwidth}{\taboffset
      \includegraphics[width=\tabfig\textwidth,height=\tabfig\textwidth]{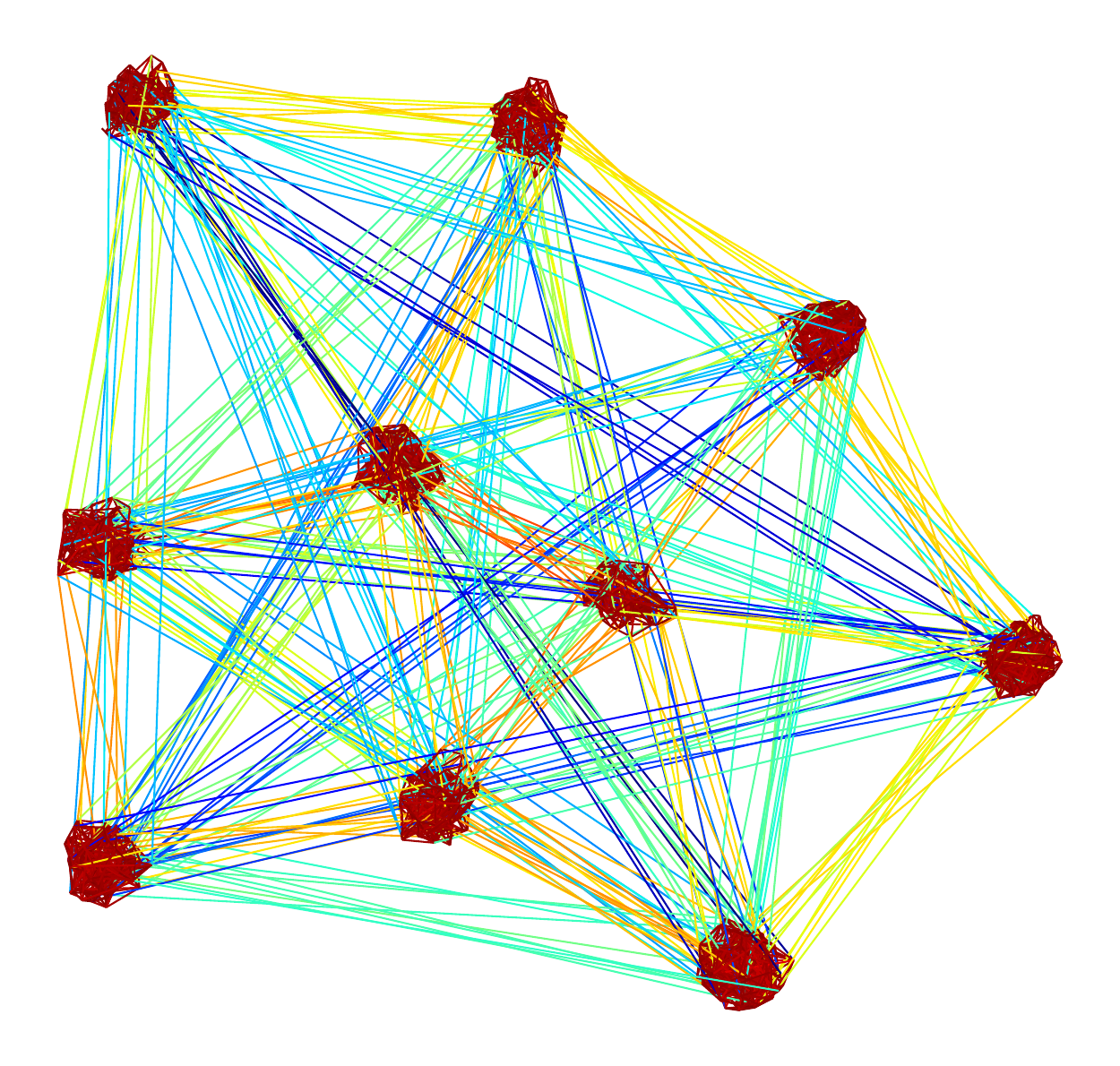}}
      & \parbox[c]{\tabfig\textwidth}{} 
      & \parbox[c]{\tabfig\textwidth}{} 
      & \parbox[c]{\tabfig\textwidth}{\taboffset
      \includegraphics[width=\tabfig\textwidth,height=\tabfig\textwidth]{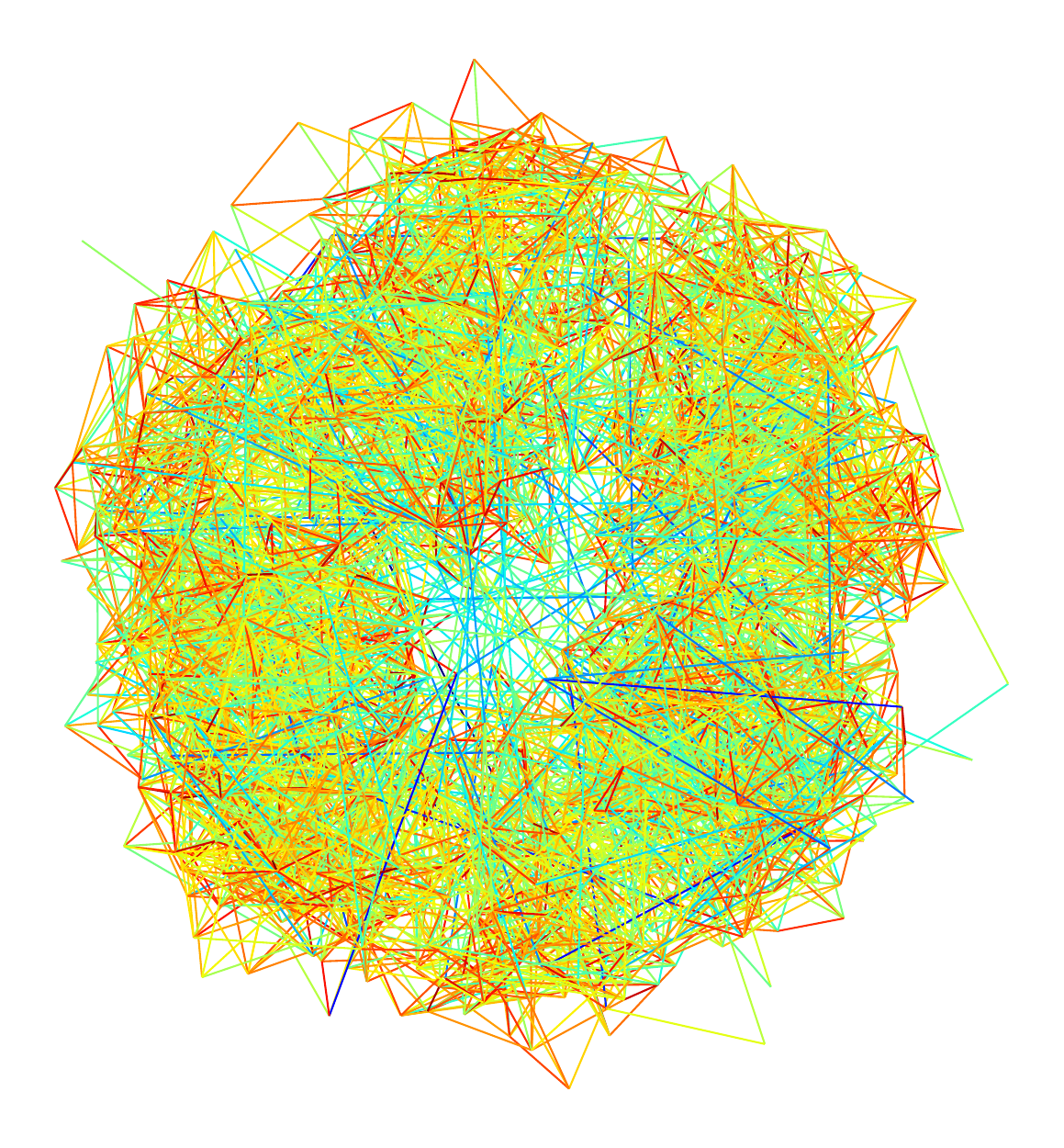}}  \\   
      \hline

     \multirow{2}{*}{\vspace{-1cm}\rotatebox[origin=c]{90}{block\_2000}}  &
      \parbox[c]{\tabfig\textwidth}{} 
      & \parbox[c]{\tabfig\textwidth}{
      \includegraphics[width=\tabfig\textwidth,height=\tabfig\textwidth]{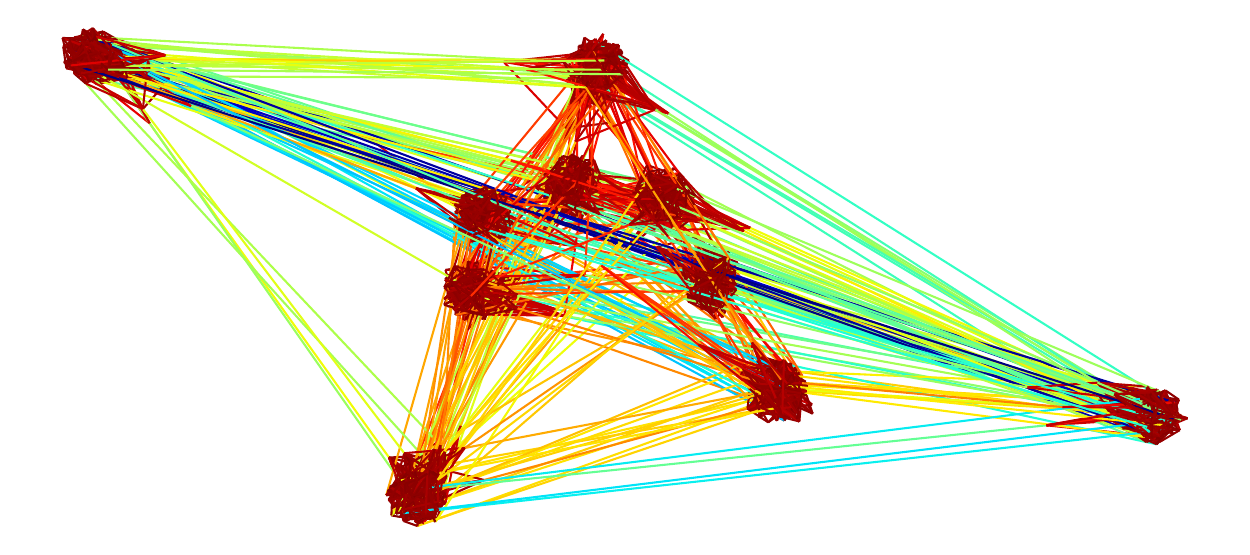}}
      & \parbox[c]{\tabfig\textwidth}{
      \includegraphics[width=\tabfig\textwidth,height=\tabfig\textwidth]{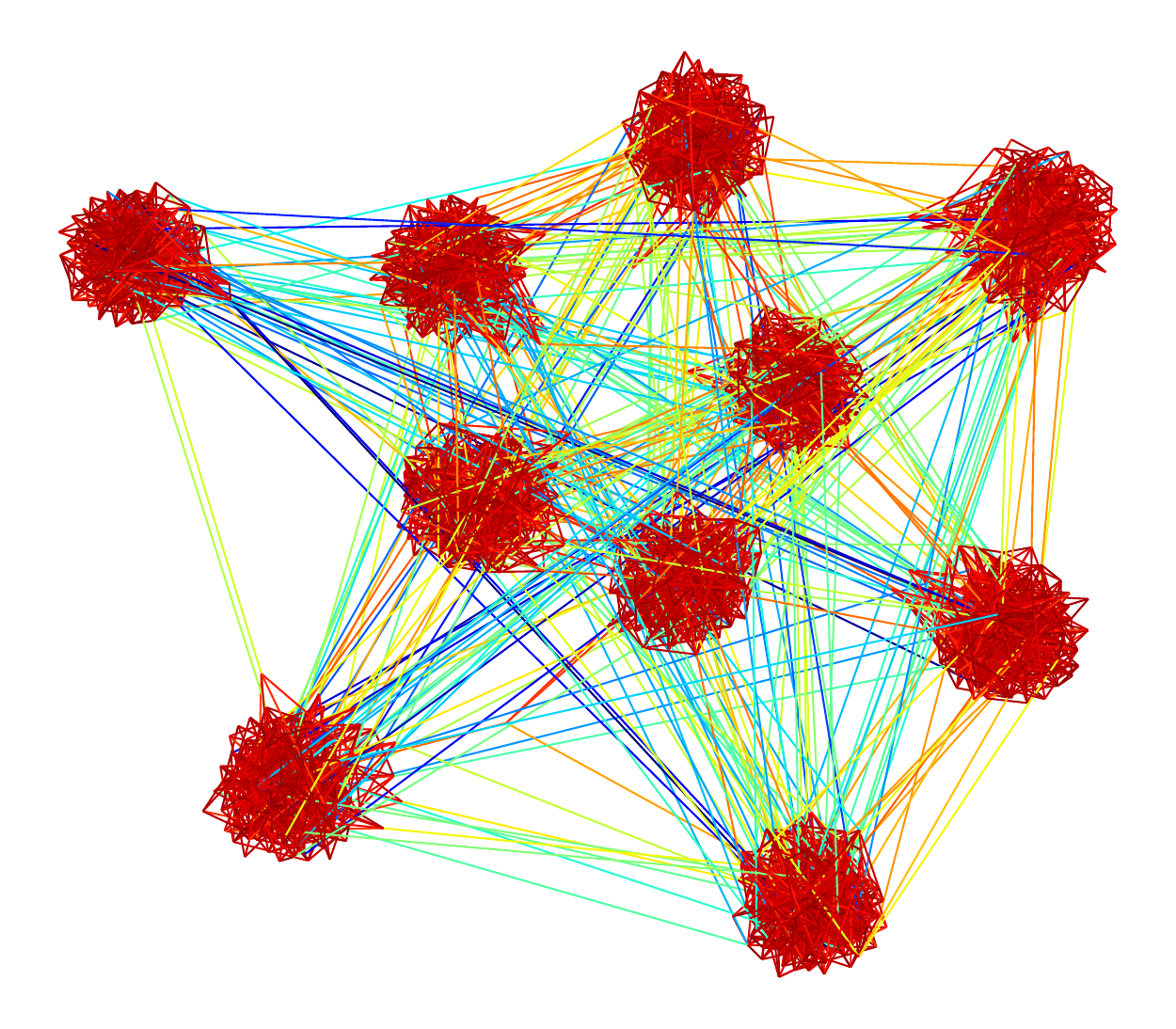}}
      & \parbox[c]{\tabfig\textwidth}{}
      & \parbox[c]{\tabfig\textwidth}{
      \includegraphics[width=\tabfig\textwidth,height=\tabfig\textwidth]{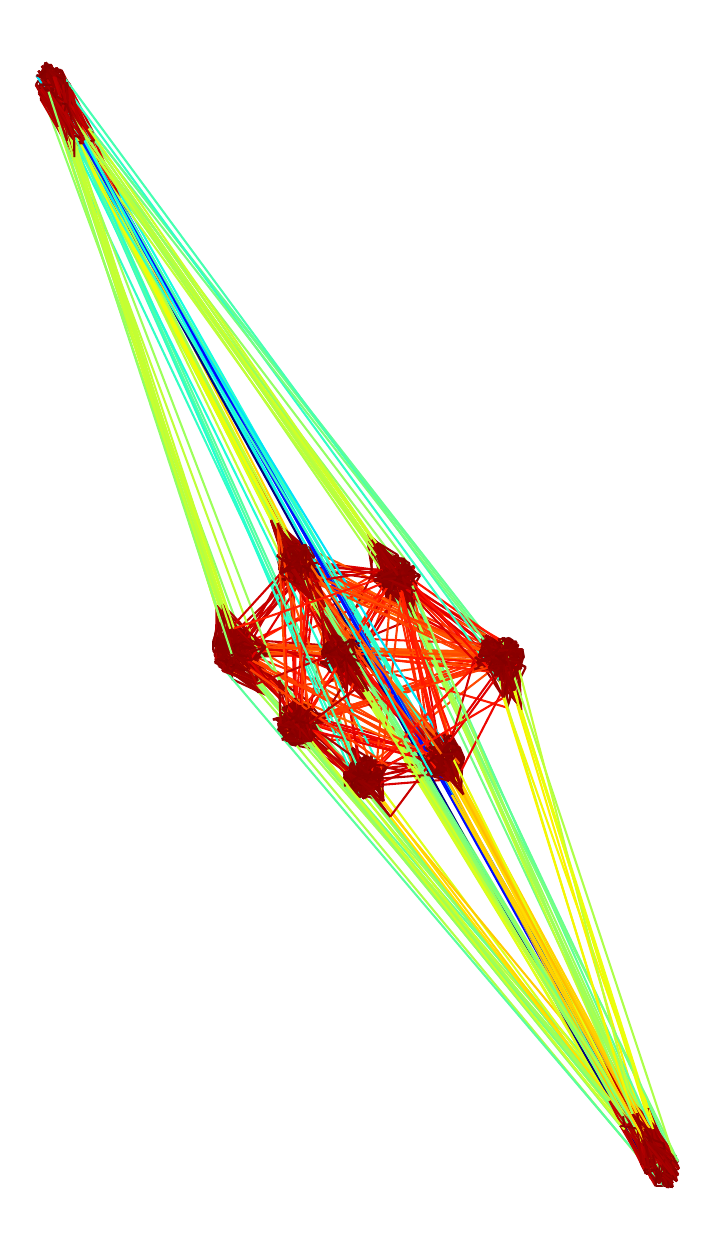}} 
      & \parbox[c]{\tabfig\textwidth}{
      \includegraphics[width=\tabfig\textwidth,height=\tabfig\textwidth]{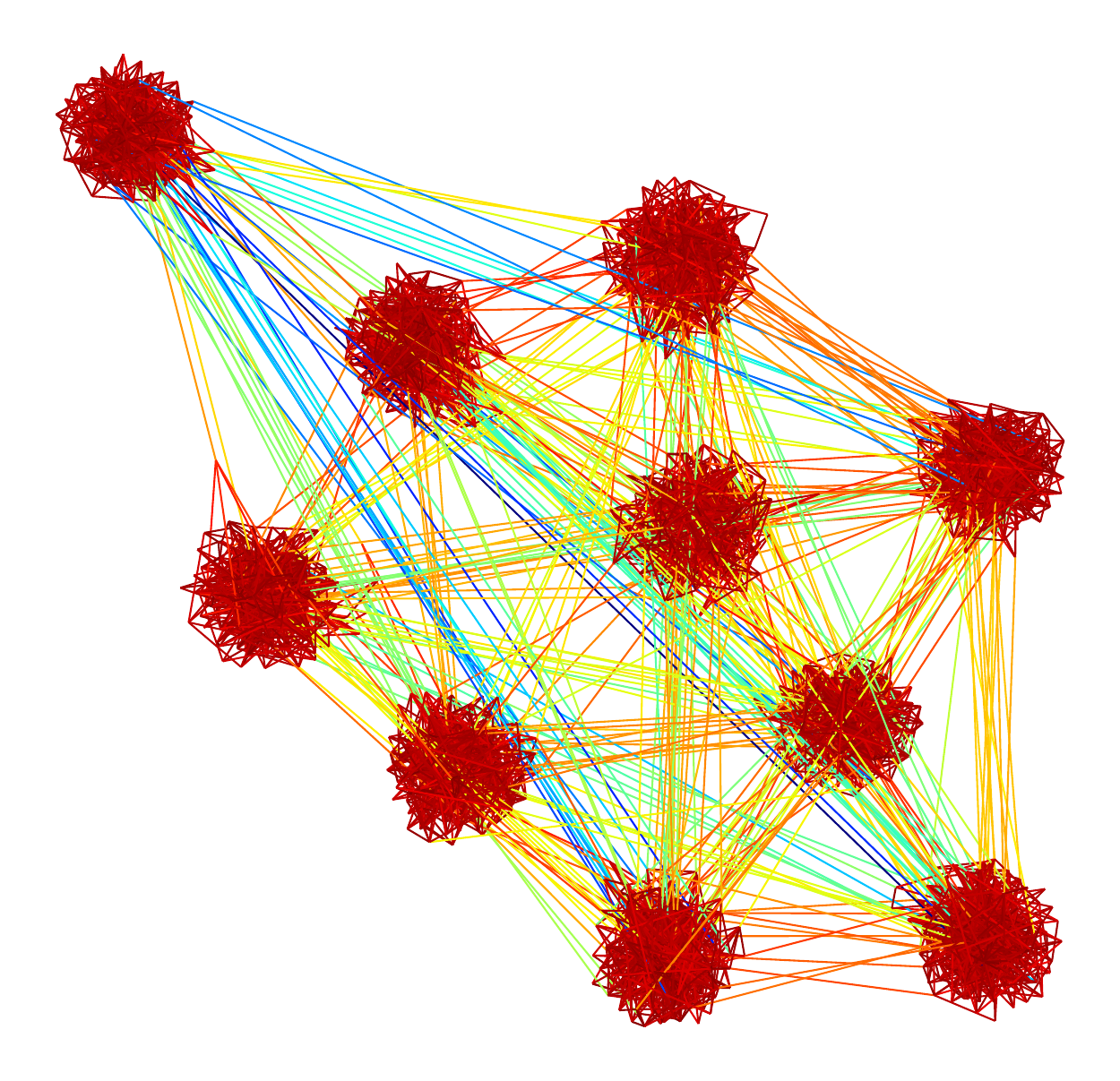}} 
      & \parbox[c]{\tabfig\textwidth}{}  \\
      
       &
      \parbox[c]{\tabfig\textwidth}{
      \taboffset\includegraphics[width=\tabfig\textwidth,height=\tabfig\textwidth]{figures/2023-pdfs/tsnet/tsnet_block_2000.pdf}} 
      & \parbox[c]{\tabfig\textwidth}{}
      & \parbox[c]{\tabfig\textwidth}{}
      & \parbox[c]{\tabfig\textwidth}{\taboffset
      \includegraphics[width=\tabfig\textwidth,height=\tabfig\textwidth]{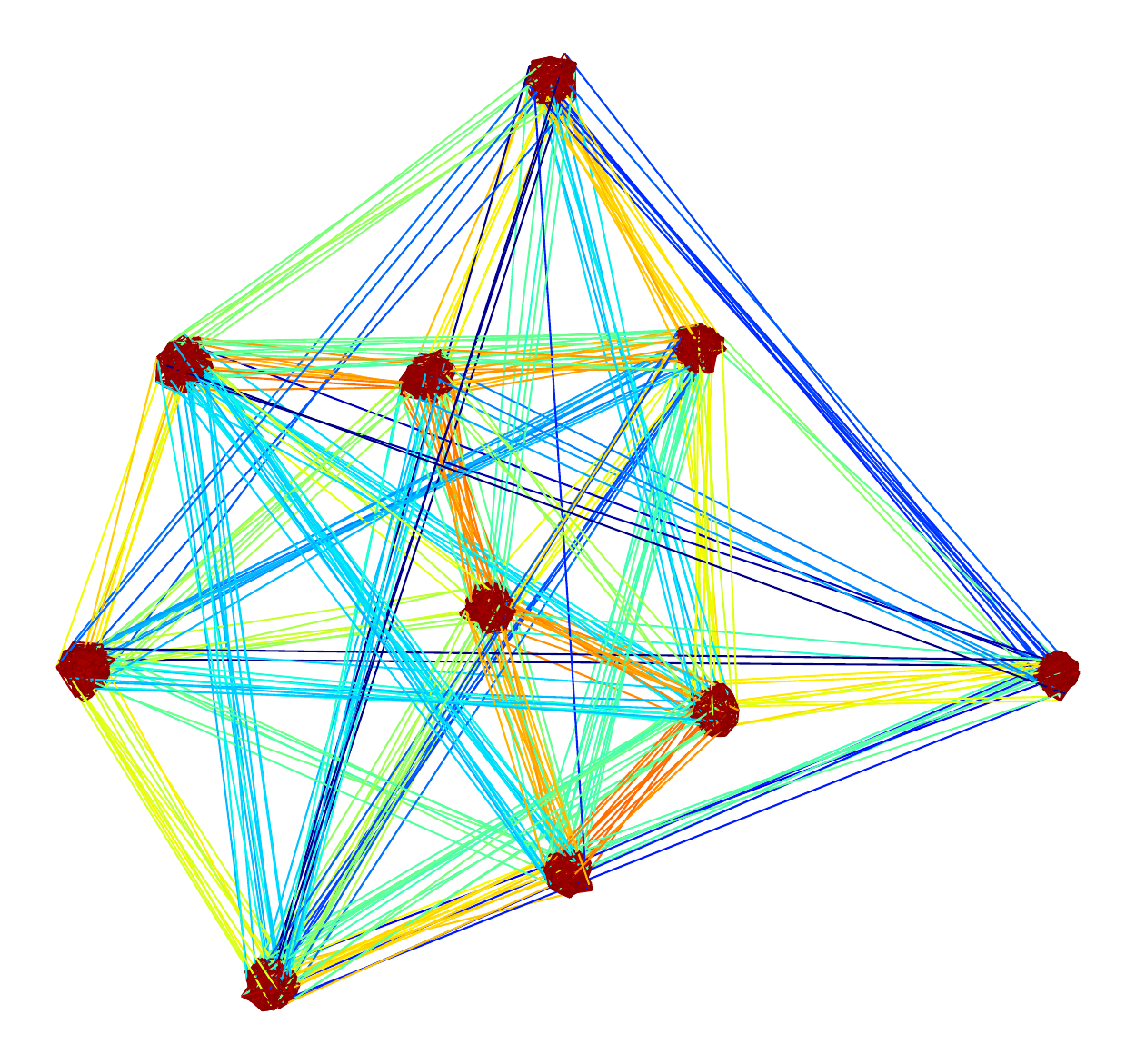}}
      & \parbox[c]{\tabfig\textwidth}{} 
      & \parbox[c]{\tabfig\textwidth}{} 
      & \parbox[c]{\tabfig\textwidth}{\taboffset
      \includegraphics[width=\tabfig\textwidth,height=\tabfig\textwidth]{figures/2023-pdfs/mds/block_2000_mds.pdf}}  \\   
      \hline

     \multirow{2}{*}{\vspace{-1cm}\rotatebox[origin=c]{90}{connected\_watts\_300}}  &
      \parbox[c]{\tabfig\textwidth}{} 
      & \parbox[c]{\tabfig\textwidth}{
      \includegraphics[width=\tabfig\textwidth,height=\tabfig\textwidth]{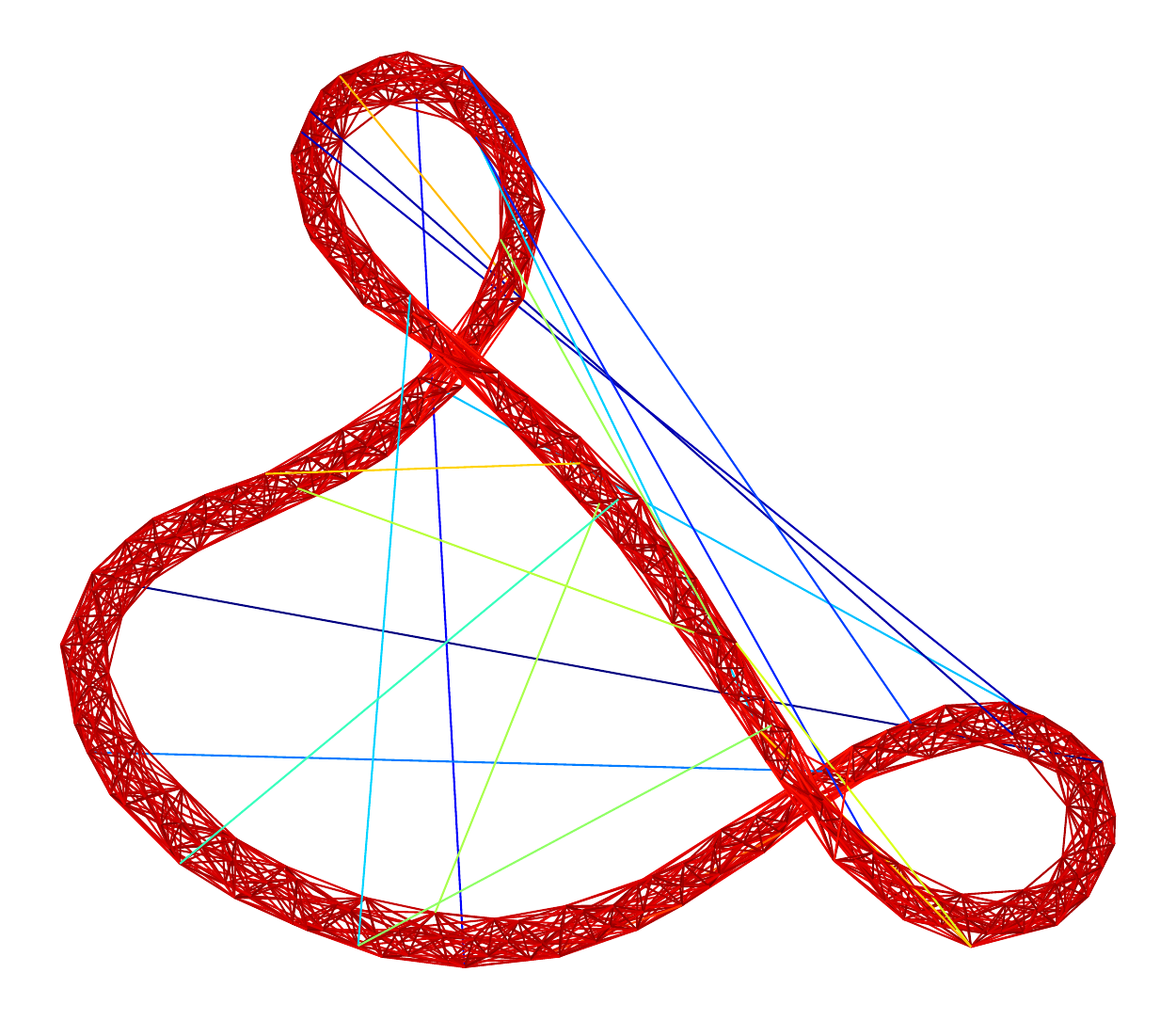}}
      & \parbox[c]{\tabfig\textwidth}{
      \includegraphics[width=\tabfig\textwidth,height=\tabfig\textwidth]{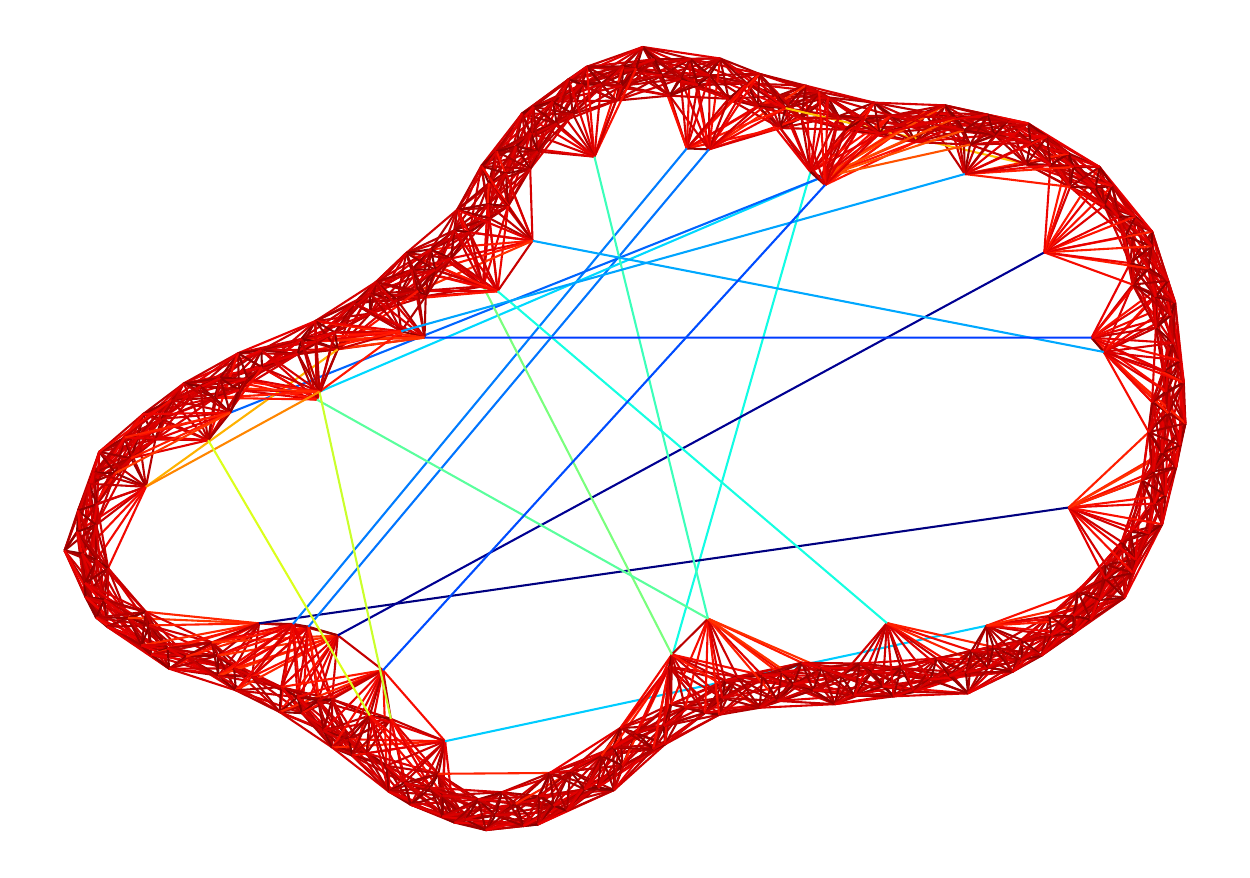}}
      & \parbox[c]{\tabfig\textwidth}{}
      & \parbox[c]{\tabfig\textwidth}{
      \includegraphics[width=\tabfig\textwidth,height=\tabfig\textwidth]{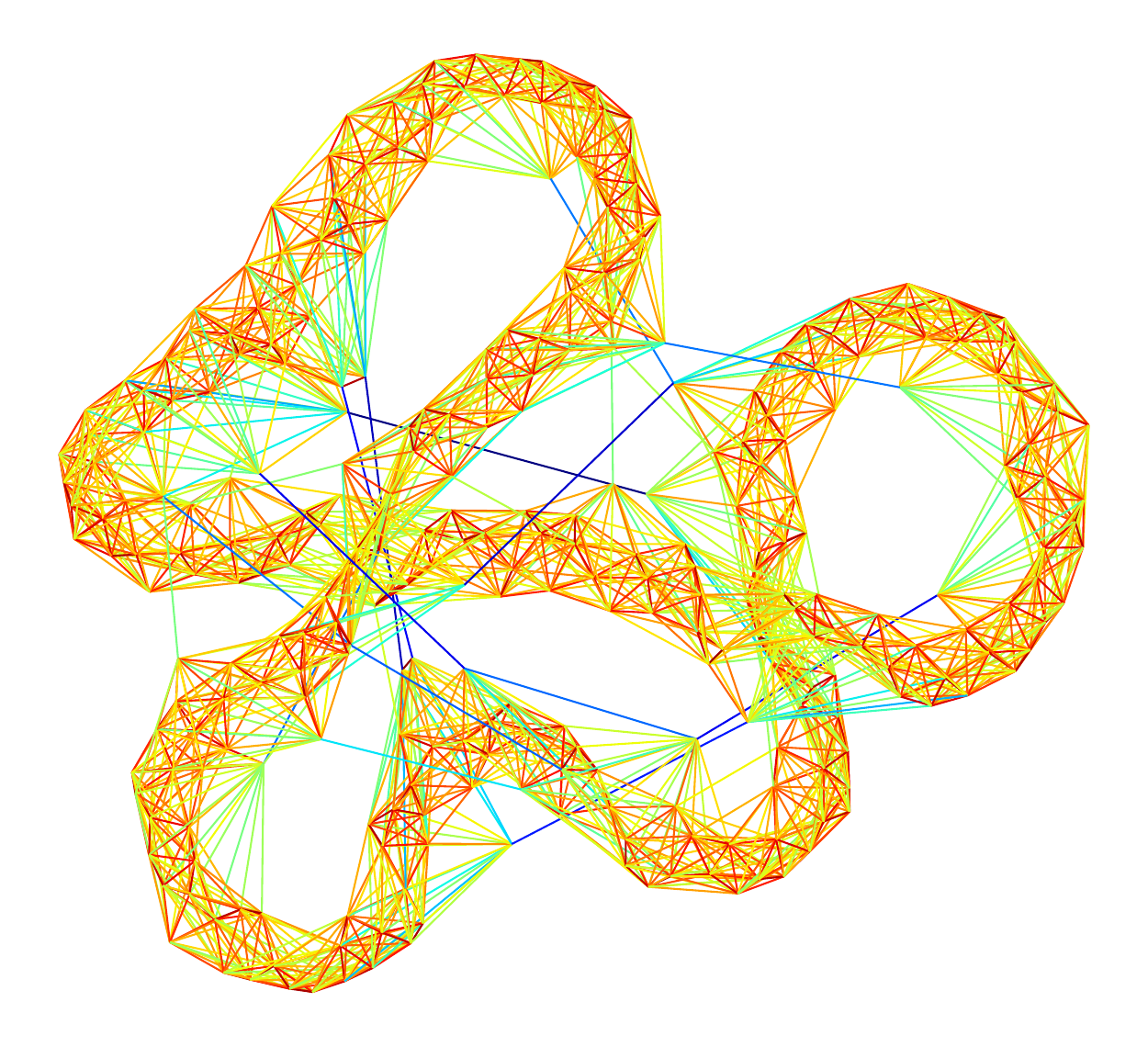}} 
      & \parbox[c]{\tabfig\textwidth}{
      \includegraphics[width=\tabfig\textwidth,height=\tabfig\textwidth]{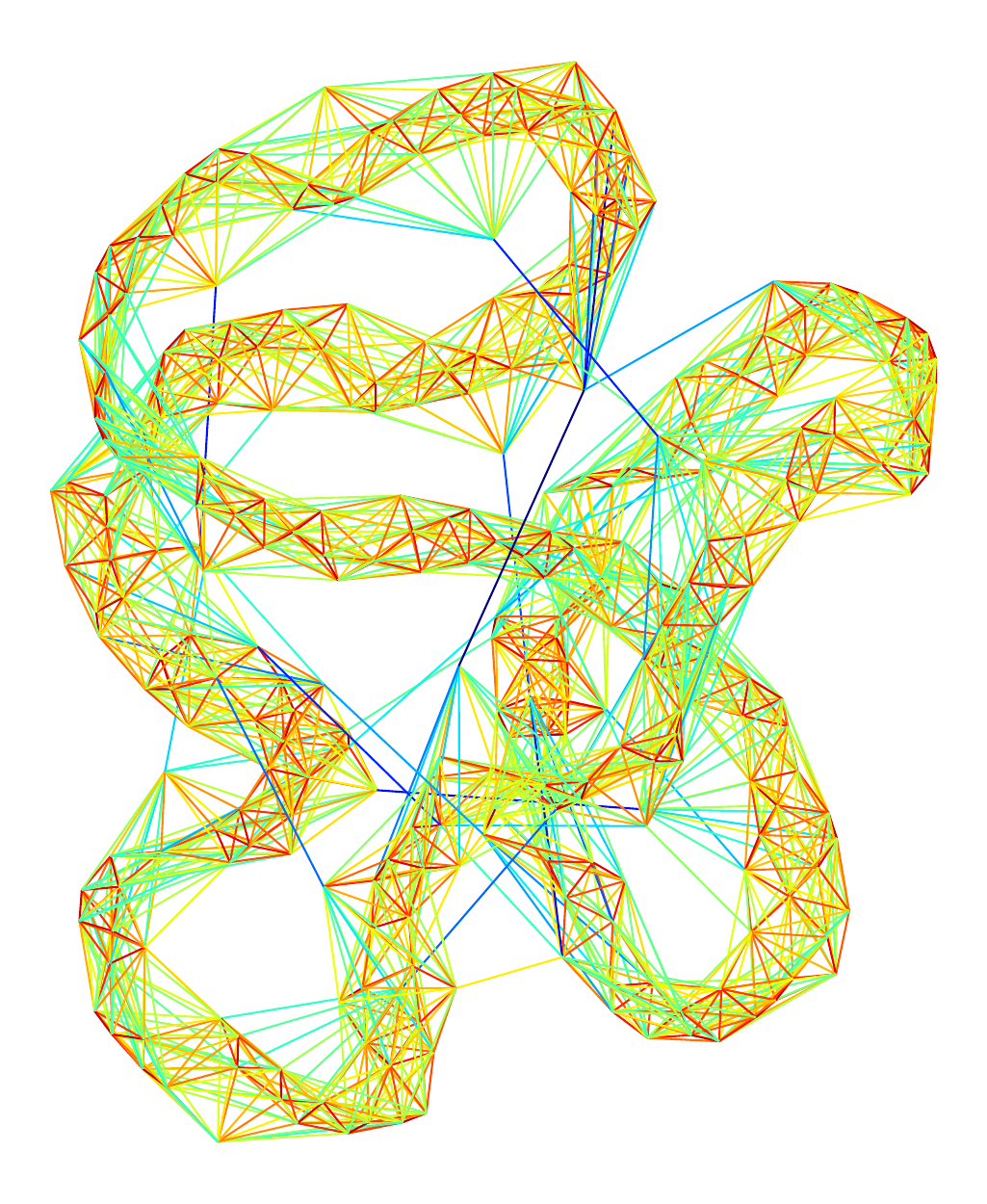}} 
      & \parbox[c]{\tabfig\textwidth}{}  \\
      
       &
      \parbox[c]{\tabfig\textwidth}{
      \taboffset\includegraphics[width=\tabfig\textwidth,height=\tabfig\textwidth]{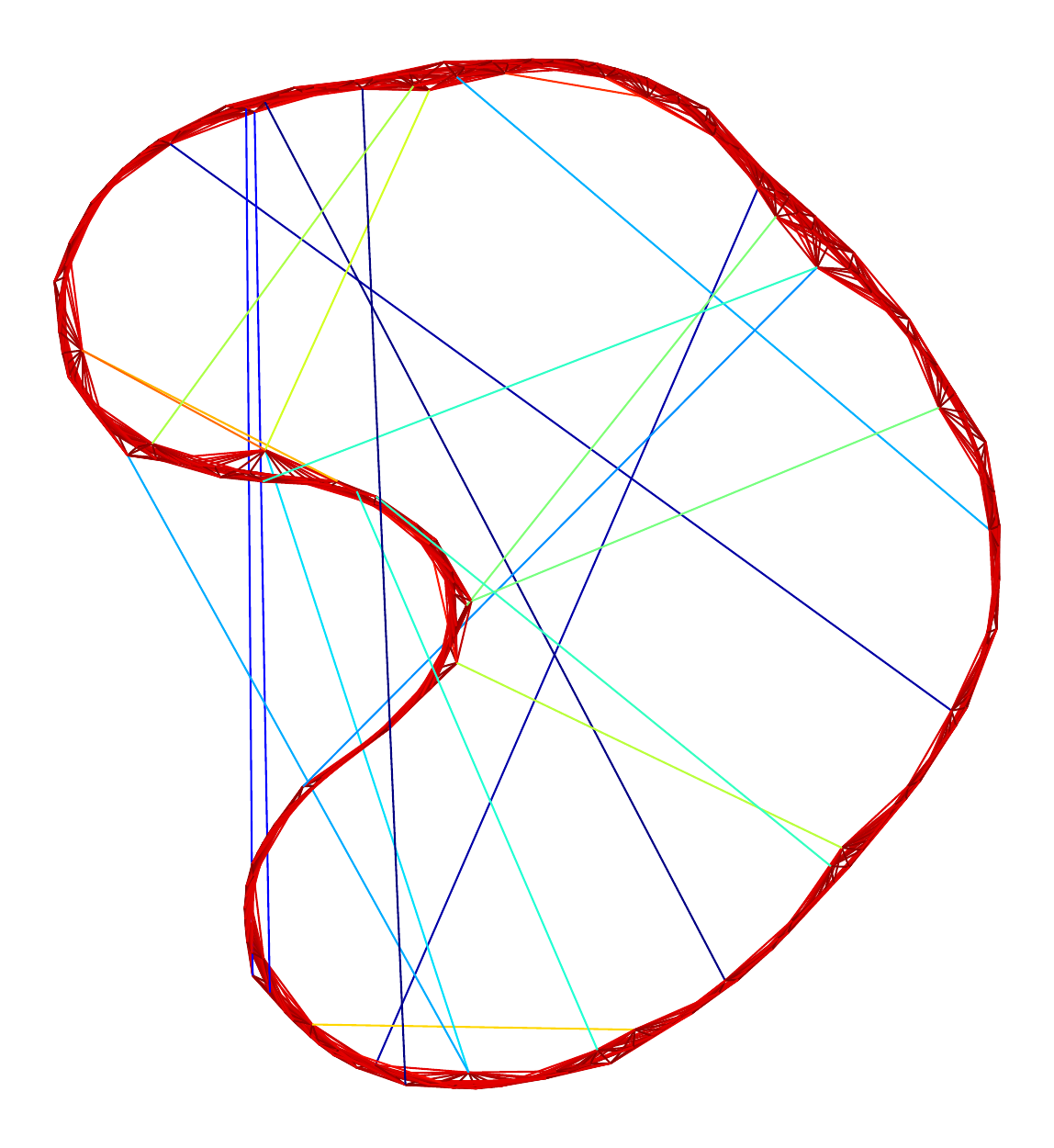}} 
      & \parbox[c]{\tabfig\textwidth}{}
      & \parbox[c]{\tabfig\textwidth}{}
      & \parbox[c]{\tabfig\textwidth}{\taboffset
      \includegraphics[width=\tabfig\textwidth,height=\tabfig\textwidth]{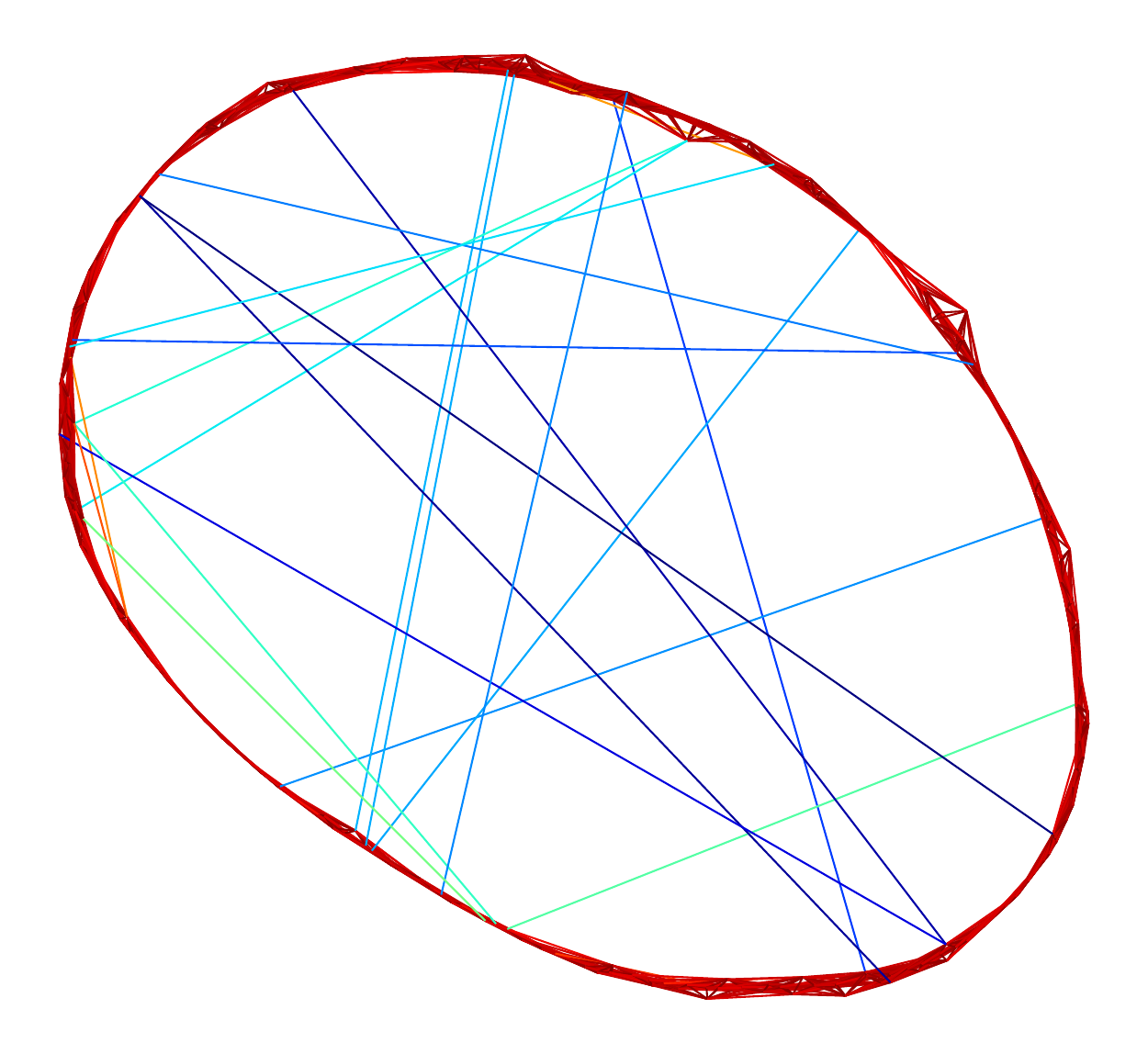}}
      & \parbox[c]{\tabfig\textwidth}{} 
      & \parbox[c]{\tabfig\textwidth}{} 
      & \parbox[c]{\tabfig\textwidth}{\taboffset
      \includegraphics[width=\tabfig\textwidth,height=\tabfig\textwidth]{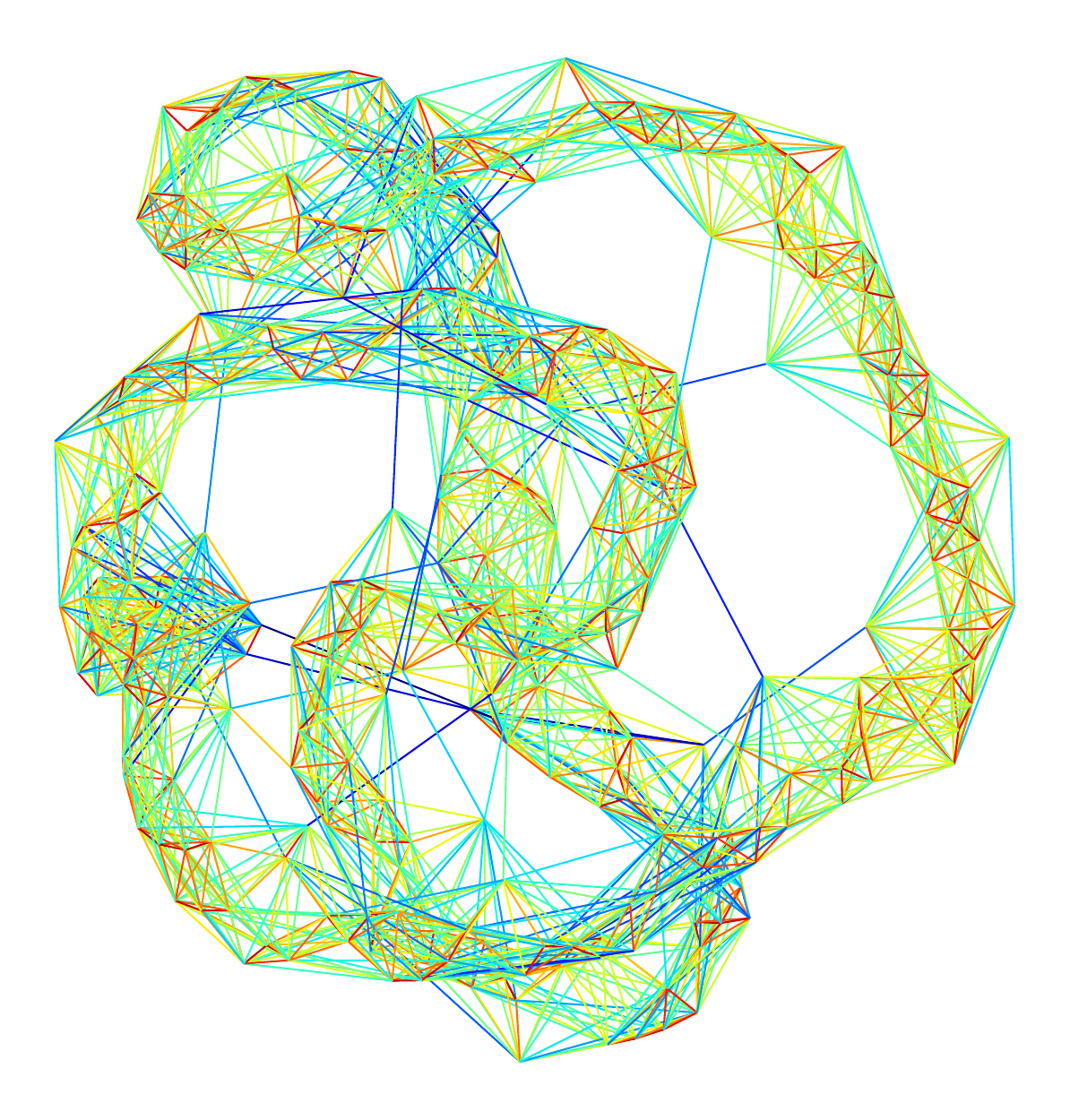}}  \\   
      \hline

     \multirow{2}{*}{\vspace{-1cm}\rotatebox[origin=c]{90}{connected\_watts\_300}}  &
      \parbox[c]{\tabfig\textwidth}{} 
      & \parbox[c]{\tabfig\textwidth}{
      \includegraphics[width=\tabfig\textwidth,height=\tabfig\textwidth]{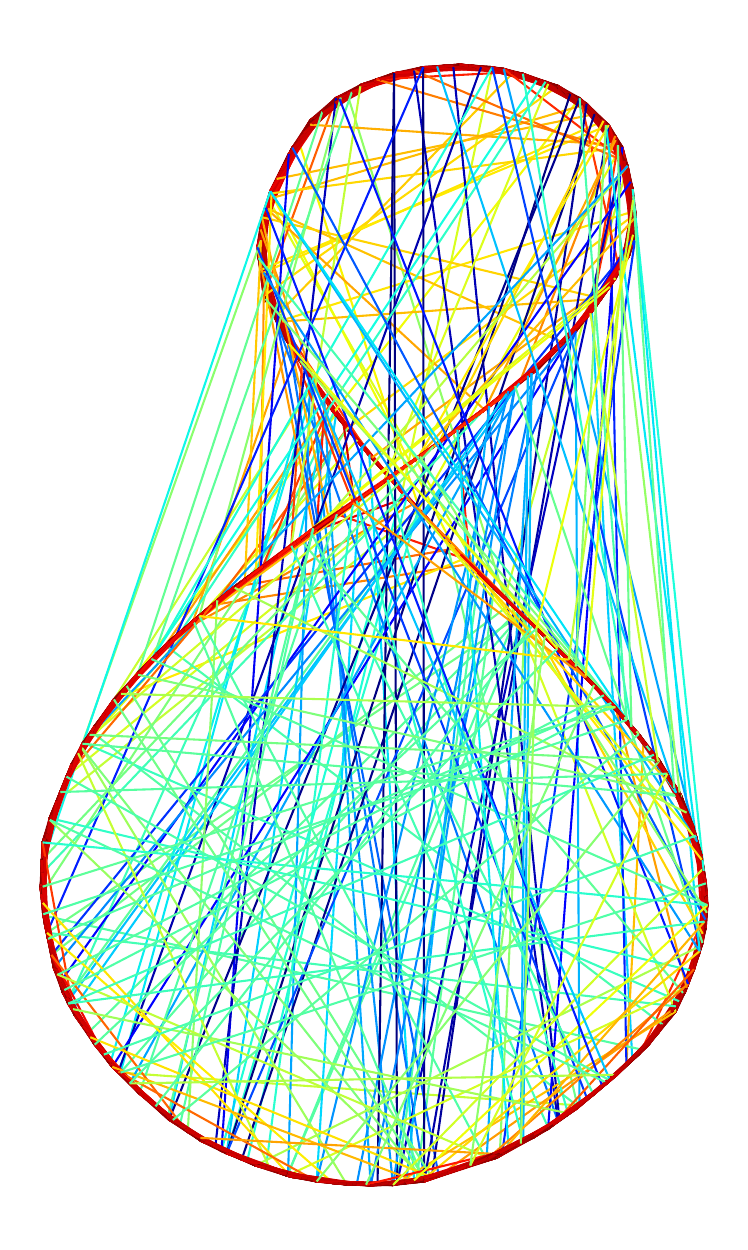}}
      & \parbox[c]{\tabfig\textwidth}{
      \includegraphics[width=\tabfig\textwidth,height=\tabfig\textwidth]{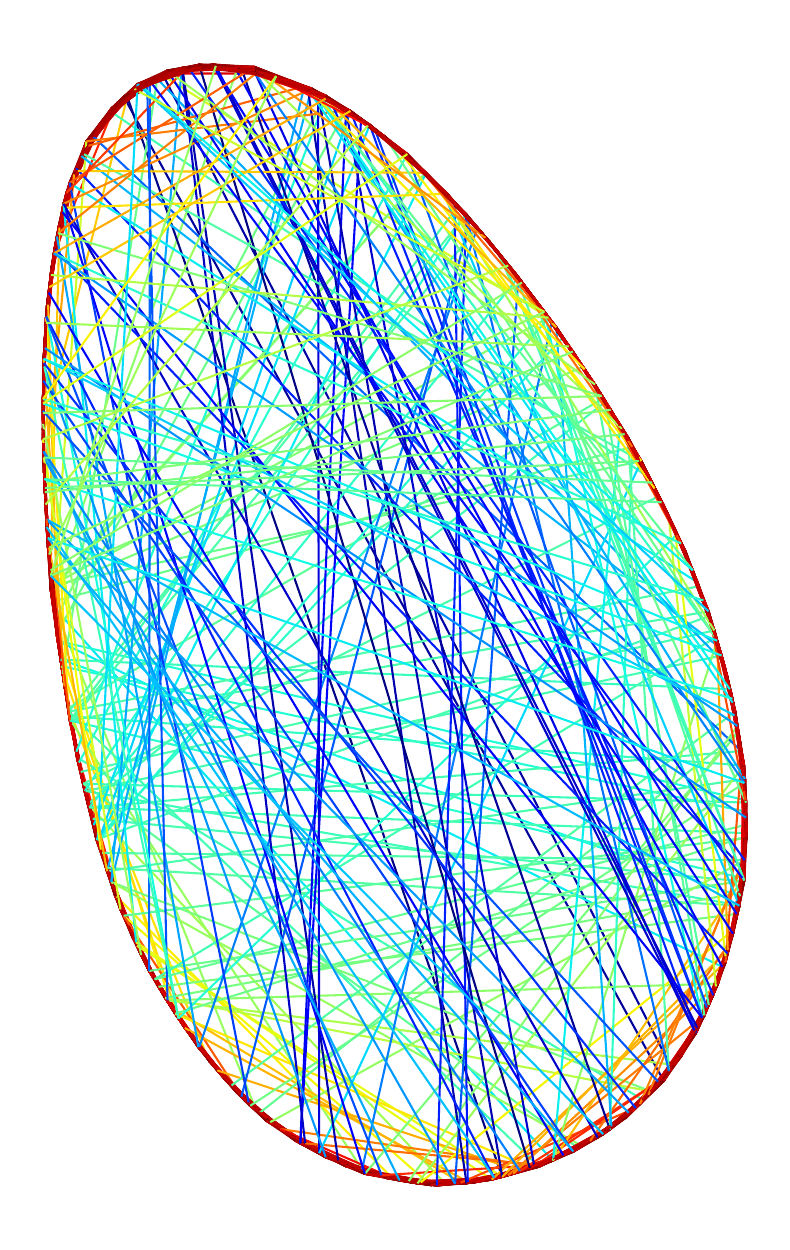}}
      & \parbox[c]{\tabfig\textwidth}{}
      & \parbox[c]{\tabfig\textwidth}{
      \includegraphics[width=\tabfig\textwidth,height=\tabfig\textwidth]{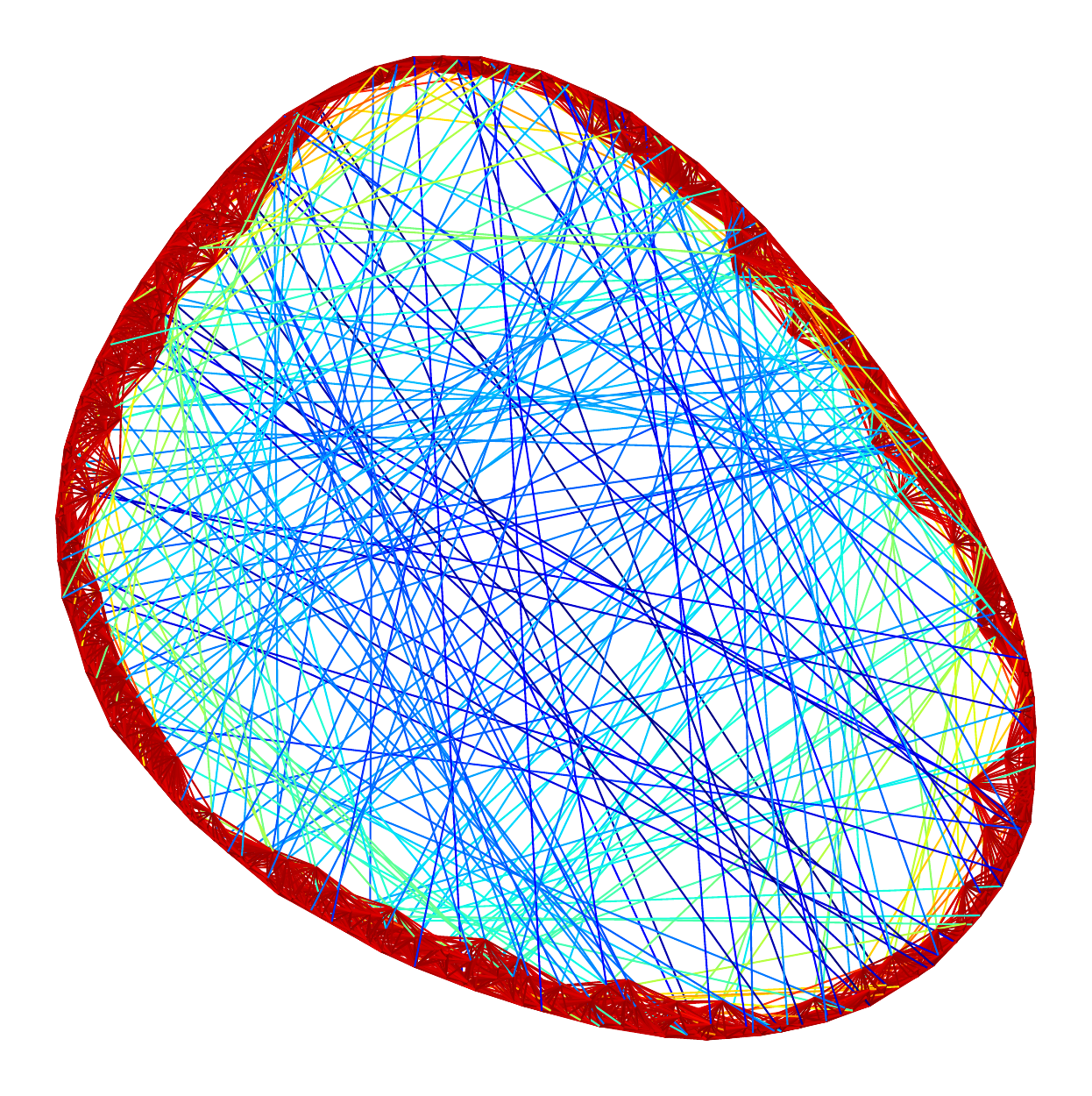}} 
      & \parbox[c]{\tabfig\textwidth}{
      \includegraphics[width=\tabfig\textwidth,height=\tabfig\textwidth]{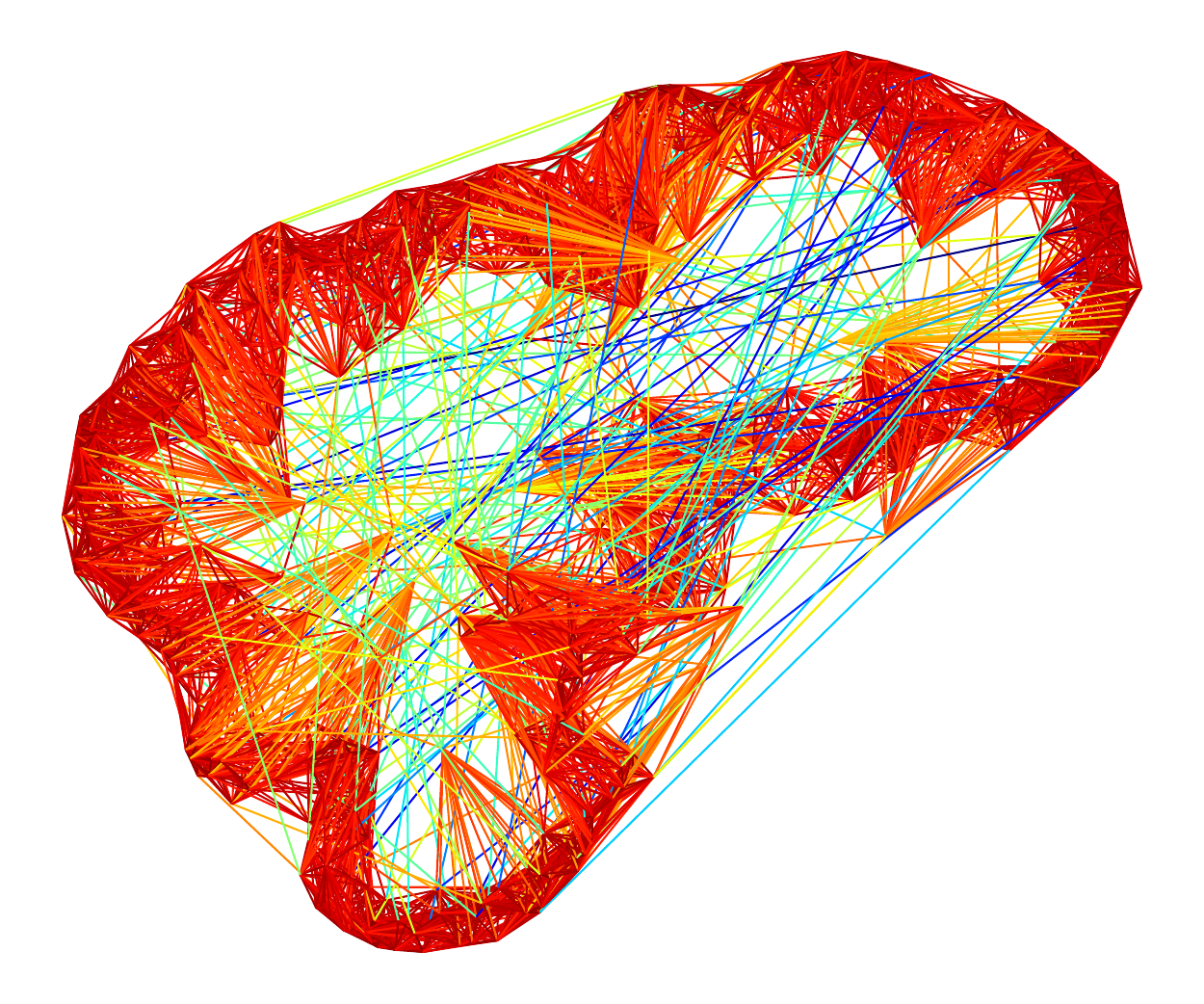}} 
      & \parbox[c]{\tabfig\textwidth}{}  \\
      
       &
      \parbox[c]{\tabfig\textwidth}{
      \taboffset\includegraphics[width=\tabfig\textwidth,height=\tabfig\textwidth]{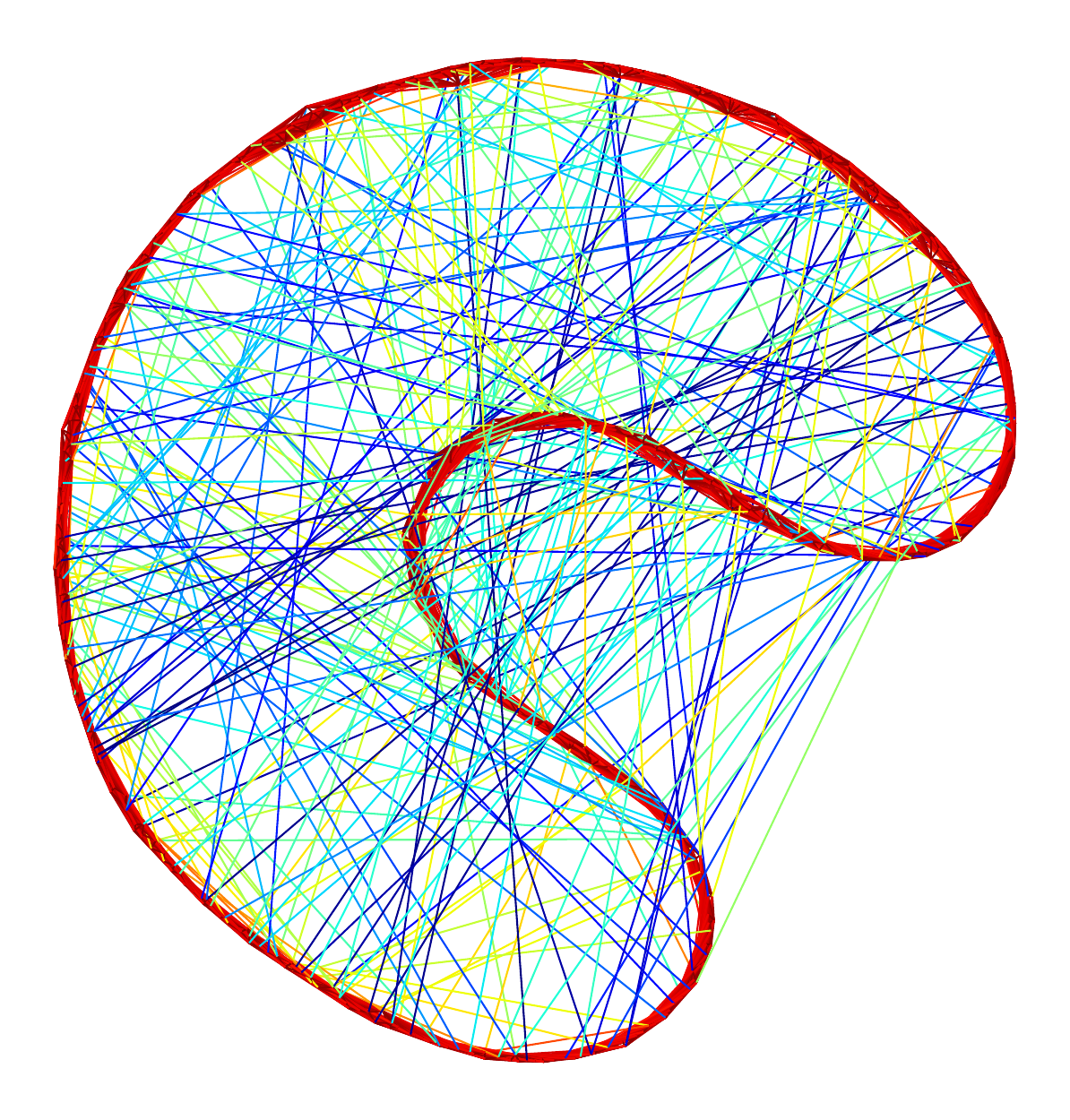}} 
      & \parbox[c]{\tabfig\textwidth}{}
      & \parbox[c]{\tabfig\textwidth}{}
      & \parbox[c]{\tabfig\textwidth}{\taboffset
      \includegraphics[width=\tabfig\textwidth,height=\tabfig\textwidth]{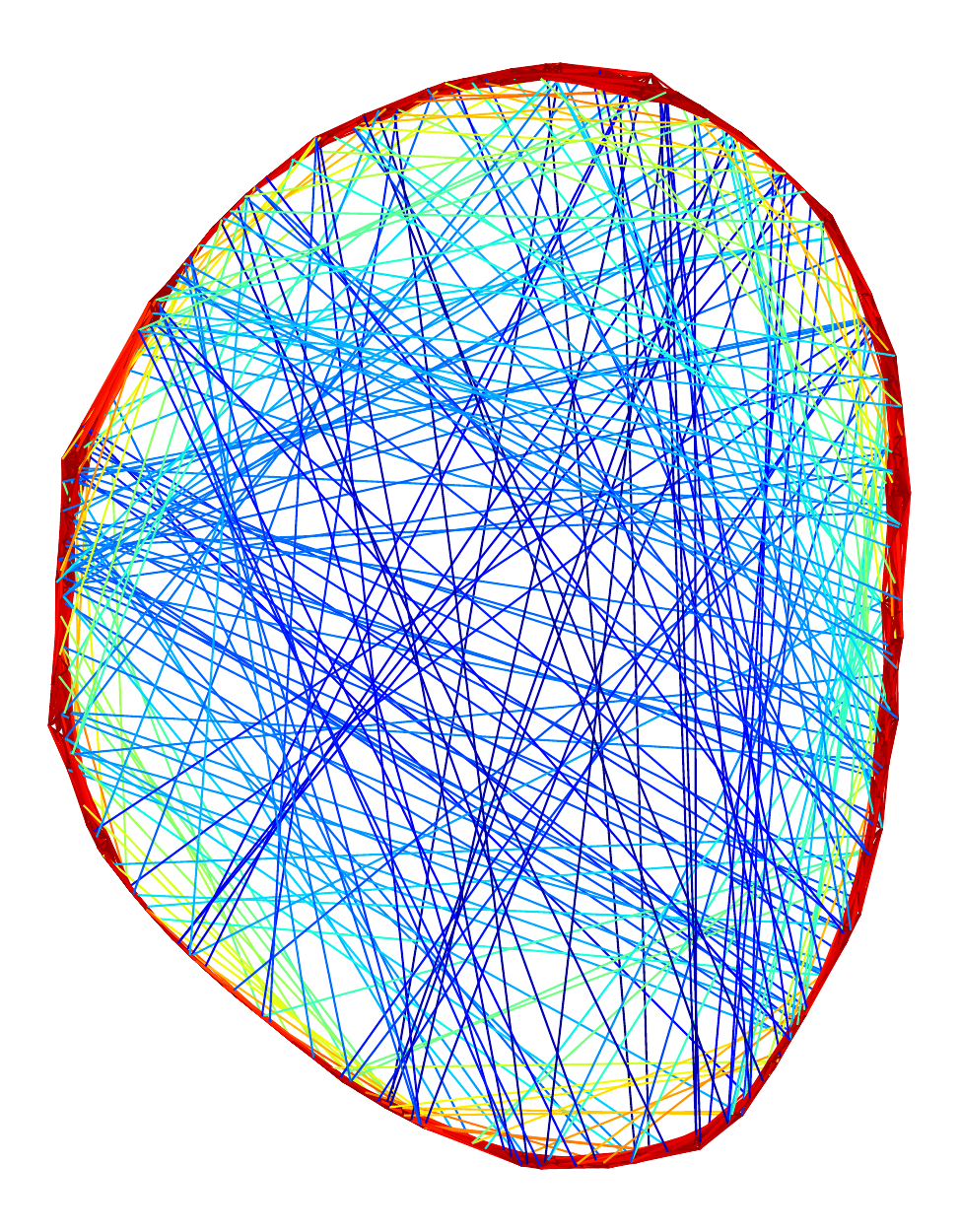}}
      & \parbox[c]{\tabfig\textwidth}{} 
      & \parbox[c]{\tabfig\textwidth}{} 
      & \parbox[c]{\tabfig\textwidth}{\taboffset
      \includegraphics[width=\tabfig\textwidth,height=\tabfig\textwidth]{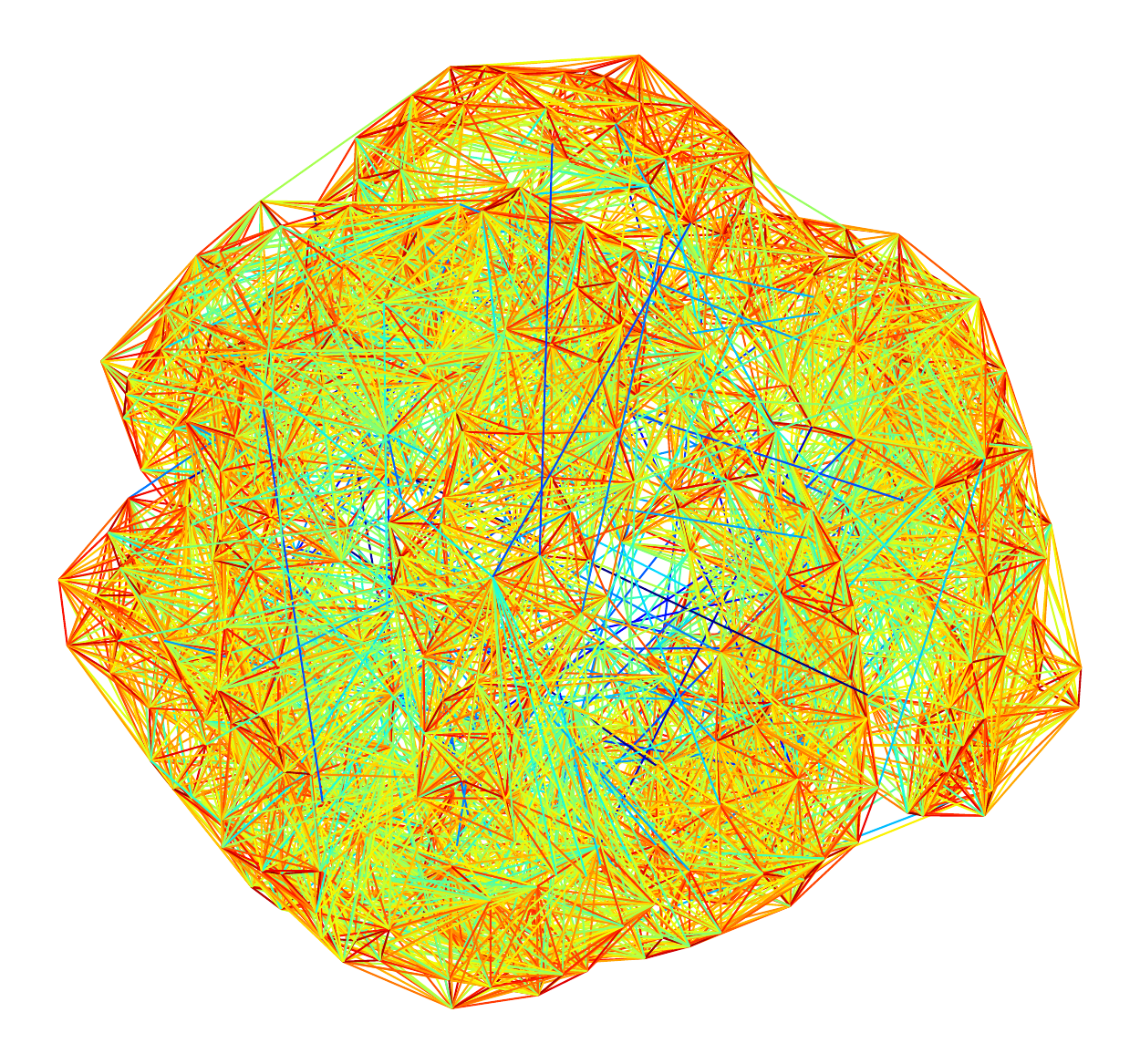}}  \\   
      \hline

     \multirow{2}{*}{\vspace{-1cm}\rotatebox[origin=c]{90}{EVA}}  &
      \parbox[c]{\tabfig\textwidth}{} 
      & \parbox[c]{\tabfig\textwidth}{
      \includegraphics[width=\tabfig\textwidth,height=\tabfig\textwidth]{figures/2023-pdfs/L2G/EVA_16.pdf}}
      & \parbox[c]{\tabfig\textwidth}{
      \includegraphics[width=\tabfig\textwidth,height=\tabfig\textwidth]{figures/2023-pdfs/L2G/EVA_32.pdf}}
      & \parbox[c]{\tabfig\textwidth}{}
      & \parbox[c]{\tabfig\textwidth}{
      \includegraphics[width=\tabfig\textwidth,height=\tabfig\textwidth]{figures/2023-pdfs/L2G/EVA_64.pdf}} 
      & \parbox[c]{\tabfig\textwidth}{
      \includegraphics[width=\tabfig\textwidth,height=\tabfig\textwidth]{figures/2023-pdfs/L2G/EVA_100.pdf}} 
      & \parbox[c]{\tabfig\textwidth}{}  \\
      
       &
      \parbox[c]{\tabfig\textwidth}{
      \taboffset\includegraphics[width=\tabfig\textwidth,height=\tabfig\textwidth]{figures/2023-pdfs/tsnet/EVA.pdf}} 
      & \parbox[c]{\tabfig\textwidth}{}
      & \parbox[c]{\tabfig\textwidth}{}
      & \parbox[c]{\tabfig\textwidth}{\taboffset
      \includegraphics[width=\tabfig\textwidth,height=\tabfig\textwidth]{figures/2023-pdfs/umap/EVA.pdf}}
      & \parbox[c]{\tabfig\textwidth}{} 
      & \parbox[c]{\tabfig\textwidth}{} 
      & \parbox[c]{\tabfig\textwidth}{\taboffset
      \includegraphics[width=\tabfig\textwidth,height=\tabfig\textwidth]{figures/2023-pdfs/mds/EVA.pdf}}  \\   
      \hline
      
  \end{tabular}
  
\end{table*}

\end{document}